%% file: report.tex
\documentclass{elsart}

\usepackage{amsmath,amssymb}
\usepackage{graphicx}
\usepackage{rotating}
\usepackage{array}
\usepackage{cite}

\newcommand{\Tr}{\mathop{\mathrm{Tr}}}
\newcommand{\sgn}{\mathop{\mathrm{sgn}}}
\newcommand{\diag}{\mathop{\mathrm{diag}}}
\newcommand{\eVq}{\ensuremath{\text{eV}^2}}
\newcommand{\Nuc}[2]{\ensuremath{^{#1}\text{#2}}}
\newcommand{\Turn}[1]{\begin{sideways}\makebox[0pt]{#1}\end{sideways}}

\allowdisplaybreaks

\begin{document}


\begin{frontmatter}

\title{Phenomenology with Massive Neutrinos}

\author{M.~C.~Gonzalez-Garcia}
\address{%
  C.N.~Yang Institute for Theoretical Physics, SUNY at Stony Brook,
  Stony Brook, NY 11794-3840, USA}
\address{%
  Instituci\'o Catalana de Recerca i Estudis Avan\c{c}ats (ICREA) \&
  Departament d'Estructura i Constituents de la Mat\`eria, Universitat
  de Barcelona, Diagonal 647, E-08028 Barcelona, Spain} 

\author{Michele Maltoni}
\address{%
  Departamento de F\'{\i}sica Te\'orica \& Instituto de F\'{\i}sica
  Te\'orica UAM/CSIC, Facultad de Ciencias C-XI, Universidad
  Aut\'onoma de Madrid, Cantoblanco, E-28049 Madrid, Spain}

\begin{abstract}
    The current status and some perspectives of the phenomenology of 
    massive neutrinos is reviewed.  We start with the phenomenology of
    neutrino oscillations in vacuum and in matter. We summarize the
    results of neutrino experiments using solar, atmospheric, reactor
    and accelerator neutrino beams. We update the leptonic parameters
    derived from the three-neutrino oscillation interpretation of this
    data.  We describe the method and present results on our
    understanding of the solar and atmospheric neutrino fluxes by
    direct extraction from the corresponding neutrino event rates. We
    present some tests of different forms of new physics which induce
    new sources of leptonic flavor transitions in vacuum and in
    matter which can be performed with the present neutrino data.  The
    aim and potential of future neutrino experiments and facilities to
    further advance in these fronts is also briefly summarized. Last,
    the implications of the LSND observations are discussed, and the
    status of extended models which could accommodate all
    flavor-mixing signals is presented in the light of the recent
    results from MiniBooNE.  
\end{abstract}


\end{frontmatter}

\tableofcontents
\newpage

\input{sec.intro}
\input{sec.theory}
\input{sec.expe}
\input{sec.3nu}
\input{sec.fluxes}
\input{sec.npatm}
\input{sec.npsolar}

\input{sec.future}
\input{sec.numass}
\input{sec.lsnd}
\input{sec.conclu}

\section*{Acknowledgments}

We are grateful to T.~Schwetz for sharing with us the results of his
analysis of the latest KamLAND data in a three-neutrino context. 
This review is based on work with a number of collaborators, with whom
we have had many stimulating discussions. The list includes
E.K.~Akhmedov, J.N.~Bahcall, J.E.~Campagne, M.~Cirelli, A.~Donini,
N.~Fornengo, A.~Friedland, C.~Giunti, W.~Grimus, M.M.~Guzzo,
F.~Halzen, P.C.~de Holanda, P.~Huber, P.I.~Krastev, C.~Lunardini,
E.~Masso, D.~Meloni, M.~Mezzetto, Y.~Nir, H.~Nunokawa,
C.~Pe\~na-Garay, O.L.G.~Peres, V.~Pleitez, T.~Rashba, J.~Rojo,
T.~Schwetz, V.B.~Semikoz, A.Yu.~Smirnov, R.~Tom\`as, M.A.~T\'ortola,
J.W.F.~Valle and R.~Zukanovich Funchal.
This work is supported by National Science Foundation grant
PHY-0354776 and by Spanish Grants FPA-2004-00996 and FPA2006-01105. MM
is supported by MCYT through the Ram\'on y Cajal program and by the
Comunidad Aut\'onoma de Madrid through the project P-ESP-00346.

For updated results, see \verb"http://dark.ft.uam.es/~maltoni/neutrino/".

\appendix
\input{sec.appatm}
\input{sec.biblio}

\end{document}

%% file: sec.intro.tex
It is already five decades since the first neutrino was observed by
Cowan and Reines~\cite{reines} in 1956 in a reactor experiment, and
more than seventy five years since its existence was postulated by
Wolfgang Pauli~\cite{pauli}, in 1930, in order to reconcile the
observed continuous spectrum of nuclear beta decay with energy
conservation. It has been a long and winding road that has lead us
from these pioneering times to the present overwhelming proof that
neutrinos are massive and leptonic flavors are not symmetries of
Nature. A road in which both theoretical boldness and experimental
ingenuity have walked hand by hand to provide us with the first
evidence of physics beyond the Standard Model~\cite{review}. From the
desperate solution of Pauli to the cathedral-size detectors built to
capture and study in detail the elusive particle.

Neutrinos are copiously produced in natural sources: in the burning of
the stars, in the interaction of cosmic rays\ldots even as relics of
the Big Bang. Starting from the 1960's, neutrinos produced in the sun
and in the atmosphere were observed. In 1987, neutrinos from a
supernova in the Large Magellanic Cloud were also detected.  Indeed an
important leading role in this story was played by the neutrinos
produced in the sun and in the atmosphere.  The experiments that
measured the flux of atmospheric neutrinos found results that
suggested the disappearance of muon-neutrinos when propagating over
distances of order hundreds (or more) kilometers.  Experiments that
measured the flux of solar neutrinos found results that suggested the
disappearance of electron-neutrinos while propagating within the Sun
or between the Sun and the Earth. 

These results called back to 1968 when Gribov and
Pontecorvo~\cite{pontecorvo1,pontecorvo2} realized that flavor
oscillations arise if neutrinos are massive and mixed. The
disappearance of both atmospheric $\nu_\mu$'s and solar $\nu_e$'s was
most easily explained in terms of neutrino oscillations.  The emerging
picture was that at least two neutrinos were massive and mixed, unlike
what it is predicted in the Standard Model.  

In the last decade this picture became fully established with the
upcome of a set of precise experiments. In particular, during the last
five years the results obtained with solar and atmospheric neutrinos
have been confirmed in experiments using terrestrial beams in which
neutrinos produced in nuclear reactors and accelerators facilities
have been detected at distances of the order of hundred kilometers.

Neutrinos were introduced in the Standard Model as truly massless 
fermions, for which no gauge invariant renormalizable mass term can be
constructed. Consequently, in the Standard Model there is neither
mixing nor CP violation in the leptonic sector. Therefore, the
experimental evidence for neutrino masses and mixing provided an
unambiguous signal of new physics.

At present the phenomenology of massive neutrinos is in a very
interesting moment.  On the one hand many extensions of the Standard
Model anticipated ways in which neutrinos may have small, but
definitely non-vanishing masses. The better determination of the
flavor structure of the leptons at low energies is of vital
importance as, at present, it is our only source of positive
information to pin-down the high energy dynamics implied by the
neutrino masses. Needless to say that its potential will be further
expanded and complemented if a positive signal on the absolute value
of the mass scale is observed in kinematic searches or or in
neutrinoless double beta decay as well as if the observations from a
positive evidence in the precision cosmological data. 

However as we stand now, even the minimal picture of three massive
neutrino, although satisfactory, is still not complete. We do not have
direct evidence of one of the three mixing angles and we are far from
a precise determination of the other two. Also, although oscillations
have allowed us to establish that neutrinos have mass, they do not
probe their absolute mass scale. Finally, we ignore if there is CP
violation in the leptonic sector, and we do not know if neutrinos are
their own antiparticle.  Different experiments have been proposed and
several techniques are being explored to answer all these fundamental
questions.

On the other hand, the attained precision in the observed signals is
already good enough to allow us the use of the existing data to probe
physics beyond neutrino masses and mixings. In particular it is
possible to test more exotic neutrino properties and/or interactions
which can induce new sources of leptonic flavor mixing and affect the
established oscillation pattern.  Also the independent determination
of neutrino masses and mixing in experiments performed with
terrestrial beams opens up the possibility of testing the theoretical
predictions of the natural neutrino fluxes, produced either in the Sun
or in the atmosphere, directly from the corresponding neutrino data.

The purpose of this review is to quantitatively summarize the present
status of the phenomenology of massive neutrinos on some of these
fronts.  In Sec.~\ref{sec:theory} we present the low energy formalism
for adding neutrino masses to the SM and the induced leptonic mixing,
and then we describe the phenomenology associated with neutrino
oscillations in vacuum and in matter.  In Sec.~\ref{sec:expe} we
briefly summarize the present experimental results from solar,
atmospheric, reactor and accelerator neutrinos, as independently as
possible of any particle physics interpretation. Section~\ref{sec:3nu}
contains an update of the three-neutrino oscillation interpretation of
the existing bulk of neutrino data (with the exception of the LSND
result).  Section~\ref{sec:fluxes} describes the method and present
results of test of our understanding of the solar and atmospheric
neutrino fluxes by direct extraction from the corresponding neutrino
data. Sections~\ref{sec:npatm} and~\ref{sec:npsolar} are devoted to
tests of different forms of new physics which induce new sources of
leptonic flavor transitions in vacuum and in matter and which can be
performed with the present neutrino data.  The aim and potential of
future neutrino experiments and facilities to further advance in these
fronts is briefly summarized in Sec.~\ref{sec:fut}. In
Sec.~\ref{sec:numass} we describe the status of the existing probes to
the absolute neutrino mass scale.
For years the most troublesome piece of experimental evidence in
neutrino physics was that of the LSND experiment which observed a
small appearance of electron anti-neutrinos in a muon anti-neutrino
beam at a value of $L/E$ different from that of either solar and
atmospheric neutrinos.  Very recently, the MiniBooNE experiment has
presented their negative results on the search for $\nu_\mu\rightarrow
\nu_e$ oscillations in the same $L/E$ region.  In Sec.~\ref{sec:lsnd}
we describe the extensions proposed to accommodate the LSND result and
discuss their phenomenological status in the light of the recent
MiniBooNE result. Our conclusions are summarized in
Sec.~\ref{sec:conclu}. This review is complemented by an Appendix in
which we update the details of the atmospheric neutrino analysis
presented in this report.

The field of neutrino phenomenology and its forward-looking
perspectives is rapidly evolving. The overview presented in this
review is only partial and lacks of many aspects. For other excellent
reviews see Refs.~\cite{giunti, dolgov, review, reviewbarger,
npreview, reviewmichele, reviewlisi, reviewsmirnov, pastor,
reviewstrumia, reviewjose} and the books~\cite{bahcall, vogel, kayser,
gaisser, kim, mohapatra, fukugita}.  An exhaustive list of related
references can be found in Ref.~\cite{reviewstrumia}.

%% file: sec.theory.tex
\section{Neutrino Masses and Mixing}
\label{sec:theory}

The greatest success of modern particle physics has been the
establishment of the connection between forces mediated by spin-1
particles and local (gauge) symmetries. Within the Standard Model, the
strong, weak and electromagnetic interactions are connected to,
respectively, $SU(3)$, $SU(2)$ and $U(1)$ gauge groups. The
characteristics of the different interactions are explained by the
symmetry to which they are related.  For example, the way in which the
fermions exert and experience each of the forces is determined by
their representation under the corresponding symmetry group (or simply
their charges in the case of Abelian gauge symmetries).  

Once the gauge invariance is elevated to the level of fundamental 
physics principle, it must be verified by all terms in the Lagrangian,
including the mass terms. This, as we will see, has important
implications for the neutrino. 


\subsection{Standard Model of Massless Neutrinos}
\label{sec:standard}

The Standard Model (SM) is based on the gauge group
\begin{equation}
    \label{eq:smgroup}
    G_\text{SM} =
    SU(3)_\text{C} \times SU(2)_\text{L}\times U(1)_\text{Y},
\end{equation}
with three matter fermion generations. Each generation consists of
five different representations of the gauge group:
\begin{equation}
    \left( 1, 2, -\frac{1}{2} \right), \quad
    \left( 3, 2, \frac{1}{6} \right), \quad
    \left( 1, 1, -1 \right), \quad
    \left( 3, 1, \frac{2}{3} \right), \quad
    \left( 3, 1, -\frac{1}{3} \right)
\end{equation}
where the numbers in parenthesis represent the corresponding charges
under the group~\eqref{eq:smgroup}. In this notation the electric
charge is given by
\begin{equation}
    \label{eq:emcharge}
    Q_\text{EM} = T_{L3} + Y \,.
\end{equation}
The matter content is shown in Table~\ref{tab:smrep}, and together
with the corresponding gauge fields it constitutes the full list of
fields required to describe the observed elementary particle
interactions. In fact, these charge assignments have been tested to
better than the percent level for the light fermions~\cite{pdg}. 
The model also contains a single Higgs boson doublet, $\phi$ with
charges $(1, 2, 1/2)$, whose vacuum expectation value breaks the gauge
symmetry,
\begin{equation}
    \label{eq:spsybr}
    \langle\phi\rangle = 
    \begin{pmatrix} 0\\ \frac{v}{\sqrt2} \end{pmatrix}
    \quad\Longrightarrow\quad
    G_\text{SM}\rightarrow SU(3)_\text{C}\times U(1)_\text{EM}.
\end{equation}
This is the only piece of the SM model which still misses experimental
confirmation. Indeed, the search for the Higgs boson, remains one of
the premier tasks of present and future high energy collider
experiments.

\begin{table}\centering
    \caption{\label{tab:smrep}%
      Matter contents of the SM.}
    \vspace{1mm}
    \begin{tabular}{|cc|ccc|}
	\hline
	$L_L( 1, 2, -\frac{1}{2})$ & 
	$Q_L( 3, 2, \frac{1}{6})$ & 
	$E_R( 1, 1, -1)$ & 
	$U_R( 3, 1, \frac{2}{3})$ & 
	$D_R( 3, 1, -\frac{1}{3})$
	\\[+1mm]
	\hline
	$\begin{pmatrix} \nu_e\\[-2mm] e\end{pmatrix}_L$
	& $\begin{pmatrix} u\\[-2mm] d\end{pmatrix}_L$
	& $e_R$ & $u_R$ & $d_R$
	\\[+5mm]
	$\begin{pmatrix} \nu_\mu\\[-2mm] \mu\end{pmatrix}_L$
	& $\begin{pmatrix} c\\[-2mm] s\end{pmatrix}_L$
	& $\mu_R$ & $ c_R$ & $s_R$
	\\[+5mm]
	$\begin{pmatrix} \nu_\tau \\[-2mm] \tau \end{pmatrix}_L$
	& $\begin{pmatrix} t\\[-2mm] b\end{pmatrix}_L$
	& $\tau_R$ & $t_R$ & $b_R$
	\\[+5mm]
	\hline
    \end{tabular}
\end{table}

As can be seen in Table~\ref{tab:smrep} neutrinos are fermions that
have neither strong nor electromagnetic interactions (see
Eq.~\eqref{eq:emcharge}), \textit{i.e.} they are singlets of
$SU(3)_\text{C}\times U(1)_\text{EM}$.  We will refer as \emph{active}
neutrinos to neutrinos that, such as those in Table~\ref{tab:smrep},
reside in the lepton doublets, that is, that have weak interactions.
Conversely \emph{sterile} neutrinos are defined as having no SM gauge
interactions (their charges are $(1,1,0)$), that is, they are singlets
of the full SM gauge group. 

The SM has three active neutrinos accompanying the charged lepton mass
eigenstates, $e$, $\mu$ and $\tau$, thus there are weak charged
current (CC) interactions between the neutrinos and their
corresponding charged leptons given by 
\begin{equation}
    \label{eq:CCleptons}  
    -{\mathcal{L}}_\text{CC} = 
    \frac{g}{\sqrt{2}} \sum_\ell \bar{\nu}_{L\ell}
    \gamma^\mu
    \ell^-_L W_\mu^+ + \text{h.c.}.
\end{equation}
In addition, the SM neutrinos have also neutral current (NC)
interactions,
\begin{equation}
    \label{eq:NCneut}  
    - {\mathcal{L}}_\text{NC}
    = \frac{g}{2\cos\theta_W} \sum_\ell \bar{\nu}_{L\ell}
    \gamma^\mu\nu_{L\ell} Z_\mu^0.
\end{equation}
The SM as defined in Table~\ref{tab:smrep}, contains no sterile 
neutrinos.

Thus, within the SM, Eqs.~\eqref{eq:CCleptons} and~\eqref{eq:NCneut} 
describe all the neutrino interactions.  From Eq.~\eqref{eq:NCneut}
one can determine the decay width of the $Z^0$ boson into neutrinos
which is proportional to the number of light (that is, $m_\nu\leq
m_Z/2$) left-handed neutrinos. At present the measurement of the
invisible Z width yields $N_\nu=2.984\pm0.008$~\cite{pdg} which
implies that whatever the extension of the SM we want to consider, it
must contain three, and only three, light active neutrinos. 

An important feature of the SM, which is relevant to the question of
the neutrino mass, is the fact that the SM with the gauge symmetry of
Eq.~\eqref{eq:smgroup} and the particle content of
Table~\ref{tab:smrep} presents an accidental global symmetry: 
\begin{equation}
    \label{eq:smglob}
    G_\text{SM}^\text{global} = U(1)_B \times U(1)_{L_e}
    \times U(1)_{L_\mu} \times U(1)_{L_\tau}.
\end{equation}
$U(1)_B$ is the baryon number symmetry, and $U(1)_{L_e, L_\mu,
L_\tau}$ are the three lepton flavor symmetries, with total lepton
number given by $L=L_e + L_\mu + L_\tau$. It is an accidental symmetry
because we do not impose it. It is a consequence of the gauge symmetry
and the representations of the physical states.

In the SM, fermions masses arise from the Yukawa interactions which 
couple a right-handed fermion with its left-handed doublet and the
Higgs field,
\begin{equation}
    \label{eq:yukawa}  
    -\mathcal{L}_\text{Yukawa} = Y^d_{ij}\bar{Q}_{Li} \phi D_{Rj}
    + Y^u_{ij}\bar{Q}_{Li} \tilde\phi U_{Rj}
    + Y^\ell_{ij}\bar{L}_{Li} \phi E_{Rj} + \text{h.c.},
\end{equation}
(where $\tilde\phi = i\tau_2\phi^\star$) which after spontaneous
symmetry breaking lead to charged fermion masses 
\begin{equation}
    \label{eq:diracmass}
    m^f_{ij} = Y^f_{ij} \frac{v}{\sqrt{2}} \,.
\end{equation}
However, since no right-handed neutrinos exist in the model, the 
Yukawa interactions of Eq.~\eqref{eq:yukawa} leave the neutrinos
massless.

In principle neutrino masses could arise from loop corrections.  In
the SM, however, this cannot happen because the only possible neutrino
mass term that can be constructed with the SM fields is the bilinear
$\bar L_{L} L^C_L$ which violates the total lepton symmetry by two
units.  As mentioned above total lepton number is a global symmetry of
the model and therefore $L$-violating terms cannot be induced by loop
corrections.  Furthermore, the $U(1)_{B-L}$ subgroup of
$G_\text{SM}^\text{global}$ is non-anomalous. and therefore
$B-L$-violating terms cannot be induced even by nonperturbative
corrections.

It follows that the SM predicts that neutrinos are precisely massless.
In order to add a mass to the neutrino the SM has to be extended. 


\subsection{Introducing Massive Neutrinos}

As discussed above, with the fermionic content and gauge symmetry of
the SM one cannot construct a renormalizable mass term for the
neutrinos.  So in order to introduce a neutrino mass one must either
extend the particle contents of the model or abandon gauge invariance
and/or renormalizability.  

In what follows we illustrate the different types of neutrino mass
terms by assuming that we keep the gauge symmetry and we explore the
possibilities that we have to introduce a neutrino mass term if one
adds to the SM an arbitrary number $m$ of sterile neutrinos
$\nu_{si}(1, 1, 0)$.

With the particle contents of the SM and the addition of an arbitrary
$m$ number of sterile neutrinos one can construct two types mass terms
that arise from gauge invariant renormalizable operators:
\begin{equation}
    \label{eq:massnu1}
    -\mathcal{L}_{M_\nu} 
    = {M_D}_{ij} \bar{\nu}_{si} \nu_{Lj} +
    \frac{1}{2} {M_N}_{ij} \bar{\nu}_{si} \nu^c_{sj} + \text{h.c.}.
\end{equation}
Here $\nu^c$ indicates a charge conjugated field, $\nu^c = C
\bar{\nu}^T$ and $C$ is the charge conjugation matrix.  $M_D$ is a
complex $m\times 3$ matrix and $M_N$ is a symmetric matrix of
dimension $m\times m$.

The first term is a Dirac mass term. It is generated after spontaneous
electroweak symmetry breaking from Yukawa interactions 
\begin{equation}
    Y^\nu_{ij}\bar{\nu}_{si}
    \tilde\phi^\dagger L_{Lj} \Rightarrow {M_D}_{ij} =
    Y^\nu_{ij}\frac{v}{\sqrt{2}}
\end{equation}
similarly to the charged fermion masses. It conserves total lepton
number but it breaks the lepton flavor number symmetries.

The second term in Eq.~\eqref{eq:massnu1} is a Majorana mass term.  It
is different from the Dirac mass terms in many important aspects. It
is a singlet of the SM gauge group. Therefore, it can appear as a bare
mass term. Furthermore, since it involves two neutrino fields, it
breaks lepton number by two units. More generally, such a term is 
allowed only if the neutrinos carry no additive conserved charge. 

In general Eq.~\eqref{eq:massnu1} can be rewritten as:
\begin{equation}
    \label{eq:mnu}
    -\mathcal{L}_{M_\nu}
    = \frac{1}{2} \overline{\vec\nu^c} M_\nu \vec\nu + \text{h.c.} \,,
\end{equation}
where 
\begin{equation}
    M_\nu = 
    \begin{pmatrix}
	0 & M^T_D \\
	M_D  & M_N
    \end{pmatrix}, 
\end{equation}
and $\vec\nu = (\vec \nu_{L}, \, \vec{\nu^c_{s}} )^T$ is a 
$(3+m)$-dimensional vector.  The matrix $M_\nu$ is complex and
symmetric. It can be diagonalized by a unitary matrix of dimension
$(3+m)$, $V^\nu$, so that
\begin{equation}
    ({V^\nu})^T M_\nu {V^\nu} = \diag(m_1,m_2,\dots,m_{3+m}) \,.
\end{equation}
In terms of the resulting $3+m$ mass eigenstates
\begin{equation} 
    \vec\nu_\text{mass} = ({V^\nu})^\dagger \vec \nu \,,
\end{equation}
Eq.~\eqref{eq:mnu} can be rewritten as:
\begin{equation}
    -\mathcal{L}_{M_\nu}=
    \frac{1}{2}\sum_{k=1}^{3+m}
    m_k \left( \bar{\nu}^c_{\text{mass},k} \nu_{\text{mass},k}
    + \bar{\nu}_{\text{mass},k} \nu^c_{\text{mass},k} \right)
    = \frac{1}{2}\sum_{k=1}^{3+m}
    m_k \bar{\nu}_{Mk} \nu_{Mk} \,,
\end{equation}
where 
\begin{equation}
    \label{eq:diagm1}
    \nu_{Mk}
    = \nu_{\text{mass},k} + \nu^c_{\text{mass},k} =
    ({V^\nu}^\dagger \vec\nu)_k + ({V^\nu}^\dagger \vec\nu)^c_k
\end{equation}
which obey the Majorana condition
\begin{equation}
    \label{eq:majcon}
    \nu_M = \nu_M^c
\end{equation}
and are refereed to as Majorana neutrinos.
Notice that this condition implies that there is only one field which
describes both neutrino and antineutrino states. Thus a Majorana
neutrino can be described by a two-component spinor unlike the charged
fermions, which are Dirac particles, and are represented by
four-component spinors.

From Eq.~\eqref{eq:diagm1} we find that the weak-doublet components of
the neutrino fields are:
\begin{equation}
    \label{eq:diagm}
    \nu_{Li} = L \, \sum_{j=1}^{3+m} V^\nu_{ij} \nu_{Mj}
    \quad {i=1,3} \,,
\end{equation}
where $L$ is the left-handed projector. 

In the rest of this section we will discuss three interesting cases.

\subsubsection{$M_N=0$: Dirac Neutrinos} 

Forcing $M_N=0$ is equivalent to imposing lepton number symmetry on
the model.  In this case, only the first term in
Eq.~\eqref{eq:massnu1}, the Dirac mass term, is allowed. For $m=3$ we
can identify the three sterile neutrinos with the right-handed
component of a four-spinor neutrino field. In this case the Dirac mass
term can be diagonalized with two $3\times 3$ unitary matrices,
$V^\nu$ and $V^\nu_R$ as:
\begin{equation}
    \label{eq:diracmassdiag}
    {V^\nu_R}^\dagger M_D V^\nu = \diag(m_1,m_2,m_3) \,.
\end{equation}
The neutrino mass term can be written as:
\begin{equation}
    -\mathcal{L}_{M_\nu} = \sum_{k=1}^{3} m_k \bar{\nu}_{Dk} \nu_{Dk}
\end{equation} 
where 
\begin{equation}
    {\nu_D}_k = ({V^\nu}^\dagger \vec\nu_L)_k 
    + ({V^\nu_R}^\dagger \vec\nu_s)_k \,,
\end{equation}
so the weak-doublet components of the neutrino fields are
\begin{equation}
    \label{eq:diagd}
    \nu_{Li} = L \sum_{j=1}^{3} V^\nu_{ij} \nu_{Dj} \,,
    \qquad i=1,3 \,.
\end{equation}
Let's point out that in this case the SM is not even a good 
low-energy effective theory since both the matter content and the 
assumed symmetries are different. Furthermore there is no explanation 
to the fact that neutrino masses happen to be much lighter than the
corresponding charged fermion masses as in this case all acquire their
mass via the same mechanism.

\subsubsection{$M_N\gg M_D$: The see-saw mechanism} 

In this case the scale of the mass eigenvalues of $M_N$ is much higher
than the scale of electroweak symmetry breaking $\langle\phi\rangle$.
The diagonalization of $M_\nu$ leads to three light, $\nu_l$, and $m$
heavy, $N$, neutrinos: 
\begin{equation}
    -\mathcal{L}_{M_\nu}
    = \frac{1}{2}\bar{\nu}_{l}  M^{l} \nu_{l} +
    \frac{1}{2}\bar{N} M^h {N} 
\end{equation}
with 
\begin{equation}
    \label{eq:mlseesaw}
    M^l\simeq -V_l^T M_D^T M_N^{-1} M_D V_l,
    \qquad M^h \simeq V_h^T M_N V_h
\end{equation}
and
\begin{equation}
    \label{eq:Useesaw}
    V^\nu \simeq
    \begin{bmatrix}
	\left(1 - \frac{1}{2}M_D^\dagger {M^*_N}^-1 M_N^{-1} M_D
	\right) V_l & M_D^\dagger {M^*_N}^{-1} V_h
	\\
	-M_N^{-1} M_D V_l
	& \left(1 - \frac{1}{2}{M_N}^{-1} M_D M_D^\dagger
	{M^*_N}^{-1} \right) V_h
    \end{bmatrix}
\end{equation}
where $V_l$ and $V_h$ are $3\times 3$ and $m\times m$ unitary matrices
respectively. 
So the heavier are the heavy states, the lighter are the light ones.
This is the \emph{see-saw
mechanism}~\cite{seesaw,seesaw1,seesaw2,seesaw3,seesaw4}.  Also as
seen from Eq.~\eqref{eq:Useesaw} the heavy states are mostly
right-handed while the light ones are mostly left-handed. Both the
light and the heavy neutrinos are Majorana particles.  Two well-known
examples of extensions of the SM that lead to a see-saw mechanism for
neutrino masses are SO(10) GUTs~\cite{seesaw1,seesaw2,seesaw3} and
left-right symmetry~\cite{seesaw4}.

In this case the SM is a good effective low energy theory. Indeed the
see-saw mechanism is a particular realization of the general case of a
full theory which leads to the SM with three light Majorana neutrinos
as its low energy effective realization as we discuss next.

\subsubsection{Neutrino Masses from Non-renormalizable Operators}

In general, if the SM is an effective low energy theory valid up to
the scale $\Lambda_\text{NP}$, the gauge group, the fermionic
spectrum, and the pattern of spontaneous symmetry breaking of the SM 
are still valid ingredients to describe Nature at energies
$E\ll\Lambda_\text{NP}$. But because it is an effective theory, one
must also consider non-renormalizable higher dimensional terms in the
Lagrangian whose effect will be suppressed by powers
$1/\Lambda_\text{NP}^\text{dim-4}$.  In this approach the largest
effects at low energy are expected to come from dim$=5$ operators. 

There is no reason for generic NP to respect the accidental symmetries
of the SM~\eqref{eq:smglob}. Indeed, there is a single set of 
dimension-five terms that is made of SM fields and is consistent with
the gauge symmetry, and this set violates~\eqref{eq:smglob}. It is
given by
\begin{equation}
    \label{eq:dimfiv}  
    \mathcal{O}_5 = \frac{Z^\nu_{ij}}{\Lambda_\text{NP}}
    {\left(\bar L_{Li} \tilde \phi\right)}
    {\left( {\tilde \phi}^T L^C_{Lj}\right)} + \text{h.c.},
\end{equation}
which violate total lepton number by two units and leads, upon 
spontaneous symmetry breaking, to:
\begin{equation}
    -\mathcal{L}_{M_\nu} =
    \frac{Z^\nu_{ij}}{ 2}\frac{v^2}{\Lambda_\text{NP}}
    \bar{\nu}_{Li} \nu^c_{Lj} +\text{h.c.} \,.
\end{equation}
Comparing with Eq.~\eqref{eq:mnu} we see that this is a Majorana mass
term built with the left-handed neutrino fields and with:
\begin{equation}
    \label{eq:nrmass}
    (M_\nu)_{ij} = {Z^\nu_{ij}}
    \frac{v^2}{\Lambda_\text{NP}}.
\end{equation}
Since Eq.~\eqref{eq:nrmass} would arise in a generic extension of the
SM, we learn that neutrino masses are very likely to appear if there
is NP. As mentioned above, a theory with SM plus $m$ heavy sterile
neutrinos leads to three light mass eigenstates and an effective low
energy interaction of the form~\eqref{eq:dimfiv}.  In particular, the
scale $\Lambda_\text{NP}$ is identified with the mass scale of the
heavy sterile neutrinos, that is the typical scale of the eigenvalues
of $M_N$. 

Furthermore, comparing Eq.~\eqref{eq:nrmass} and
Eq.~\eqref{eq:diracmass}, we find that the scale of neutrino masses is
suppressed by $v/\Lambda_\text{NP}$ when compared to the scale of
charged fermion masses providing an explanation not only for the
existence of neutrino masses but also for their smallness.  Finally,
Eq.~\eqref{eq:nrmass} breaks not only total lepton number but also the
lepton flavor symmetry $U(1)_e\times U(1)_\mu\times U(1)_\tau$.
Therefore, as we shall see in Sec.~\ref{sec:lepmix}, we should expect
lepton mixing and CP violation unless additional symmetries are
imposed on the coefficients $Z_{ij}$. 

\subsubsection{Light sterile neutrinos}

This appears if the scale of some eigenvalues of $M_N$ is not higher
than the electroweak scale. As in the case with $M_N=0$, the SM is not
even a good low energy effective theory: there are more than three
light neutrinos, and they are admixtures of doublet and singlet
fields. Again both light and heavy neutrinos are Majorana particles.

As we will see the analysis of neutrino oscillations is the same 
whether the light neutrinos are of the Majorana- or Dirac-type. From
the phenomenological point of view, only in the discussion of
neutrinoless double beta decay the question of Majorana versus Dirac
neutrinos is crucial. However, as we have tried to illustrate above,
from the theoretical model building point of view, the two cases are 
very different.


\subsection{Lepton Mixing}
\label{sec:lepmix}

The possibility of arbitrary mixing between two massive neutrino 
states was first introduced in Ref.~\cite{MNS}.  In the general case,
we denote the neutrino mass eigenstates by $(\nu_1, \nu_2, \nu_3,
\dots, \nu_n)$ and the charged lepton mass eigenstates by
$(e,\mu,\tau)$. The corresponding interaction eigenstates are denoted
by $(e^I,\mu^I,\tau^I)$ and $\vec\nu = (\nu_{Le}, \nu_{L\mu},
\nu_{L\tau}, \nu_{s1}, \dots, \nu_{sm})$. In the mass basis, leptonic
charged current interactions are given by
\begin{equation}
    \label{eq:CClepmas}  
    -\mathcal{L}_\text{CC} = \frac{g}{\sqrt{2}}
    (\bar{e}_L,\, \bar{\mu}_L,\, \bar{\tau}_L) \gamma^\mu U
    \begin{pmatrix}
	\nu_1 \\
	\nu_2 \\
	\nu_3 \\
	\vdots \\
	\nu_n
    \end{pmatrix} W_\mu^+ - \text{h.c.}.
\end{equation}
Here $U$ is a $3\times n$
matrix~\cite{Schechter:Vmatrix1,Schechter:Vmatrix2,Schechter:cpphase}
which verifies
\begin{equation}
    U U^\dagger = I_{3\times 3}  
\end{equation}
but in general $U^\dagger U\neq I_{n\times n}$. 

The charged lepton and neutrino mass terms and the neutrino mass in
the interaction basis are:
\begin{equation}
    -{\mathcal{L}}_{M} = 
    [ (\bar{e}_L^I,\, \bar{\mu}_L^I,\, \bar{\tau}_L^I) M_\ell
    \begin{pmatrix}
	e_R^I \\
	\mu_R^I \\
	\tau_R^I
    \end{pmatrix} + \text{h.c.}]
    - \mathcal{L}_{M_\nu}
\end{equation}
with $\mathcal{L}_{M_\nu}$ given in Eq.~\eqref{eq:mnu}.  One can find
two $3\times 3$ unitary diagonalizing matrices for the charge leptons,
$V^\ell$ and $V^\ell_R$, such that 
\begin{equation}
    {V^\ell}^\dagger M_\ell V^\ell_R =
    \diag(m_e,m_\mu,m_\tau) \,.
\end{equation}
The charged lepton mass term can be written as:
\begin{equation}
    -\mathcal{L}_{M_\ell} = \sum_{k=1}^{3} m_{\ell_k} 
    \bar{\ell}_{k} \ell_{k}
\end{equation} 
where
\begin{equation}
    \ell_k = ({V^\ell}^\dagger\ell^I_L)_k + ({V^\ell_R}^\dagger\ell^I_R)_k
\end{equation}
so the weak-doublet components of the charge lepton fields are
\begin{equation}
    \label{eq:diagl}
    \ell^I_{Li} = L \, \sum_{j=1}^{3} V^\ell_{ij} \ell_j \,,
    \qquad i=1,3
\end{equation}
From Eqs.~\eqref{eq:diagm}, \eqref{eq:diagd} and~\eqref{eq:diagl} we
find that $U$ is:
\begin{equation}
    \label{eq:diamat}
    U_{ij} = P_{\ell,ii}\, {V^\ell_{ik}}^\dagger \,
    V^\nu_{kj}\, (P_{\nu,jj}).
\end{equation}
$P_\ell$ is a diagonal $3\times 3$ phase matrix, that is
conventionally used to reduce by three the number of phases in $U$.
$P_\nu$ is a diagonal $n\times n$ phase matrix with additional
arbitrary phases which can chosen to reduce the number of phases in
$U$ by $n-1$ only for Dirac states. For Majorana neutrinos, this
matrix is simply a unit matrix.  The reason for that is that if one
rotates a Majorana neutrino by a phase, this phase will appear in its
mass term which will no longer be real. Thus, the number of phases
that can be absorbed by redefining the mass eigenstates depends on
whether the neutrinos are Dirac or Majorana particles. Altogether for
Majorana [Dirac] neutrinos the $U$ matrix contains a total of $6(n-2)$
[$5n-11$] real parameters, of which $3(n-2)$ are angles and $3(n-2)$
[$2n-5$] can be interpreted as physical phases.  

In particular, if there are only three Majorana neutrinos, $U$ is a
$3\times 3$ matrix analogous to the CKM matrix for the
quarks~\cite{ckm} but due to the Majorana nature of the neutrinos it
depends on six independent parameters: three mixing angles and three
phases. In this case the mixing matrix can be conveniently
parametrized as:
\begin{equation}
    \label{eq:U3m}
    U =
    \begin{pmatrix}
	1 & 0 & 0 \\
	0 & c_{23}  & {s_{23}} \\
	0 & -s_{23} & {c_{23}}
    \end{pmatrix}
    \cdot
    \begin{pmatrix}
	c_{13} & 0 & s_{13} e^{-i\delta_\text{CP}} \\
	0 & 1 & 0 \\
	-s_{13} e^{i\delta_\text{CP}} & 0 & c_{13}
    \end{pmatrix}
    \cdot
    \begin{pmatrix}
	c_{21} & s_{12} & 0 \\
	-s_{12} & c_{12} & 0 \\
	0 & 0 & 1
    \end{pmatrix}
    \cdot
    \begin{pmatrix}
	e^{i \eta_1} & 0 & 0 \\
	0 & e^{i \eta_2} & 0 \\
	0 & 0 & 1
    \end{pmatrix},
\end{equation}
where $c_{ij} \equiv \cos\theta_{ij}$ and $s_{ij} \equiv
\sin\theta_{ij}$.  The angles $\theta_{ij}$ can be taken without loss
of generality to lie in the first quadrant, $\theta_{ij} \in
[0,\pi/2]$ and the phases $\delta_\text{CP},\; \eta_i\in [0,2\pi]$.
This is to be compared to the case of three Dirac neutrinos, where the
Majorana phases, $\eta_1$ and $\eta_2$, can be absorbed in the
neutrino states and therefore the number of physical phases is one
(similarly to the CKM matrix).  In this case the mixing matrix $U$
takes the form~\cite{pdg}:
\begin{equation}
    \label{eq:U3d}
    U =
    \begin{pmatrix}
	c_{12} \, c_{13}
	& s_{12} \, c_{13}
	& s_{13} \, e^{-i\delta_\text{CP}}
	\\
	- s_{12} \, c_{23} - c_{12} \, s_{13} \, s_{23} \, e^{i\delta_\text{CP}}
	& \hphantom{+} c_{12} \, c_{23} - s_{12} \, s_{13} \, s_{23} \, e^{i\delta_\text{CP}}
	& c_{13} \, s_{23} \hspace*{5.5mm}
	\\
	\hphantom{+} s_{12} \, s_{23} - c_{12} \, s_{13} \, c_{23} \, e^{i\delta_\text{CP}}
	& - c_{12} \, s_{23} - s_{12} \, s_{13} \, c_{23} \, e^{i\delta_\text{CP}}
	& c_{13} \, c_{23} \hspace*{5.5mm}
    \end{pmatrix}.
\end{equation}
Note, however, that the two extra Majorana phases are very hard to
measure since they are only physical if neutrino mass is non-zero and
therefore the amplitude of any process involving them is suppressed a
factor $m_\nu/E$ to some power where $E$ is the energy involved in the
process which is typically much larger than the neutrino mass.  The
most sensitive experimental probe of Majorana phases is the rate of
neutrinoless $\beta\beta$ decay. 

If no new interactions for the charged leptons are present we can
identify their interaction eigenstates with the corresponding mass
eigenstates after phase redefinitions. In this case the charged
current lepton mixing matrix $U$ is simply given by a $3\times n$
sub-matrix of the unitary matrix $V^\nu$.  

It worth noticing that while for the case of 3 light Dirac neutrinos
the procedure leads to a fully unitary $U$ matrix for the light
states, generically for three light Majorana neutrinos this is not the
case when the full spectrum contains heavy neutrino states which have
been integrated out as can be seen, from Eq.~\eqref{eq:Useesaw}. Thus,
strictly speaking, the parametrization in Eq.~\eqref{eq:U3m} does not
hold to describe the flavor mixing of the three light Majorana
neutrinos in the see-saw mechanism.  However, as seen in
Eq.~\eqref{eq:Useesaw}, the unitarity violation is of the order
${\mathcal O}(M_D/M_N)$ and it is expected to be very small (at it is
also severely constrained experimentally). Consequently in what
follows we will ignore this effect.


\subsection{Neutrino Oscillations in Vacuum}
\label{sec:oscvac}

If neutrinos have masses, the weak eigenstates, $\nu_\alpha$, produced
in a weak interaction are, in general, linear combinations of the mass
eigenstates $\nu_i$
\begin{equation}
    |\nu_\alpha\rangle = \sum_{i=1}^{n} U^*_{\alpha i} |\nu_i\rangle
\end{equation}
where $n$ is the number of light neutrino species and $U$ is the the
mixing matrix.  (Implicit in our definition of the state $|\nu\rangle$
is its energy-momentum and space-time dependence). After traveling a
distance $L$ (or, equivalently for relativistic neutrinos, time $t$),
a neutrino originally produced with a flavor $\alpha$ evolves as: 
\begin{equation}
    |\nu_\alpha (t)\rangle
    = \sum_{i=1}^{n} U^*_{\alpha i} |\nu_i(t)\rangle \,,
\end{equation}
and it can be detected in the charged-current (CC) interaction
$\nu_\alpha(t) N^\prime\to\ell_\beta N$ with a probability
\begin{equation}
    \label{eq:palbe}
    P_{\alpha\beta} = |\langle\nu_\beta|\nu_\alpha(t)\rangle|^2 =
    |\sum_{i=1}^n \sum_{j=1}^n U^*_{\alpha i} U_{\beta j}
    \langle\nu_j|\nu_i(t)\rangle|^2 \,,
\end{equation}
where $E_i$ and $m_i$ are, respectively, the energy and the mass of
the neutrino mass eigenstate $\nu_i$. 

Using the standard approximation that $|\nu\rangle$ is a plane wave 
$|\nu_i(t)\rangle = e^{-i \,E_i t} |\nu_i(0)\rangle$, that neutrinos
are relativistic with $p_i\simeq p_j\equiv p\simeq E$ 
\begin{equation}
    E_i = \sqrt{p_i^2 + m_i^2} \simeq
    p + \frac{{m_i^2}}{2E}
\end{equation}
and the orthogonality relation $\langle\nu_j|\nu_i\rangle =
\delta_{ij}$, we get the following transition probability 
\begin{multline}
    \label{eq:pab}
    P_{\alpha\beta} =
    \delta_{\alpha\beta} - 4\sum_{i< j}^n
    \mbox{Re}[U_{\alpha i}U^*_{\beta i} U^*_{\alpha j} U_{\beta j}]
    \sin^2 X_{ij}
    \\
    + 2 \sum_{i<j}^n
    \mbox{Im}[U_{\alpha i}U^*_{\beta i} U^*_{\alpha j} U_{\beta j}]
    \sin 2 X_{ij} \,,
\end{multline}
where 
\begin{equation}
    \label{eq:deltaij}
    X_{ij} = \frac{(m_i^2-m_j^2) L}{4 E} = 1.27 \,
    \frac{\Delta m^2_{ij}}{\eVq} \, \frac{L/E}{\text{m/MeV}} \,.
\end{equation}
Here $L=t$ is the distance between the production point of
$\nu_\alpha$ and the detection point of $\nu_\beta$. The first line in
Eq.~\eqref{eq:pab} is CP conserving while the second one is CP
violating and has opposite sign for neutrinos and antineutrinos.

The transition probability, Eq.~\eqref{eq:pab}, has an oscillatory
behavior, with oscillation lengths 
\begin{equation}
    \label{eq:L0}
    L_{0,ij}^\text{osc}=\frac{4 \pi E}{\Delta m_{ij}^2}
\end{equation}
and amplitudes that are proportional to elements in the mixing matrix.
Thus, in order to undergo flavor oscillations, neutrinos must have
different masses ($\Delta m^2_{ij}\neq 0$) and they must mix
($U_{\alpha_i} U_{\beta i} \neq 0$).  Also, as can be seen from
Eq.~\eqref{eq:pab}, the Majorana phases cancel out in the oscillation
probability as expected because flavor oscillation is a total lepton
number conserving process. 

A neutrino oscillation experiment is characterized by the typical 
neutrino energy $E$ and by the source-detector distance $L$. But in
general, neutrino beams are not monoenergetic and, moreover, detectors
have finite energy resolution. Thus, rather than measuring $P_{\alpha
\beta}$, the experiments are sensitive to the average probability 
\begin{equation} \begin{split}
    \langle P_{\alpha \beta}\rangle
    &= \dfrac{\int dE \frac{d\Phi}{d E}
      \sigma_{CC}(E) P_{\alpha\beta}(E) \epsilon(E)}
    {\int dE \frac{d\Phi}{d E} \sigma_{CC}(E) \epsilon(E)}
    \\
    &= \delta_{\alpha\beta} - 4\sum_{i<j}^n
    \mbox{Re}[ U_{\alpha i}U^*_{\beta i} U^*_{\alpha j} U_{\beta j}] 
    \langle\sin^2 X_{ij}\rangle
    \\
    &\hspace{11.5mm} + 2\sum_{i<j}^n
    \mbox{Im}[U_{\alpha i}U^*_{\beta i} U^*_{\alpha j} U_{\beta j}]
    \langle\sin 2 X_{ij}\rangle \,,
\end{split} \end{equation}
where $\Phi$ is the neutrino energy spectrum, $\sigma_{CC}$ is the
cross section for the process in which the neutrino is detected (in
general, a CC interaction), and $\epsilon(E)$ is the detection
efficiency. The range of the energy integral depends on the energy
resolution of the experiment.

In order to be sensitive to a given value of $\Delta m^2_{ij}$, the
experiment has to be set up with $E/L\approx \Delta m^2_{ij}$ ($L\sim
L_{0,ij}^\text{osc}$). The typical values of $L/E$ for different types
of neutrino sources and experiments and the corresponding ranges of 
$\Delta m^2$ to which they can be most sensitive are summarized in 
Table~\ref{tab:lovere}.

Generically if $(E/L) \gg \Delta m^2_{ij}$ ($L \ll
L_{0,ij}^\text{osc}$), the oscillation phase does not have time to
give an appreciable effect because $\sin^2 X_{ij}\ll1$. Conversely if
$L\gg L_{0,ij}^\text{osc}$, the oscillating phase goes through many
cycles before the detection and is averaged to $\langle \sin^2 X_{ij}
\rangle=1/2$. Maximum sensitivity to the oscillation phase --~and
correspondingly to $\Delta m^2$~-- is obtained when the set up is such
that:
\begin{itemize}
  \item $E/L\approx \Delta m^2_{ij}$,
  \item the energy resolution of the experiment is good enough,
    $\Delta E\ll L \Delta m^2_{ij}$,
  \item the experiment is sensitive to different values of $L$ with
    $\Delta L \ll E/\Delta m^2$.
\end{itemize}

\begin{table}\centering
    \caption{\label{tab:lovere}%
      Characteristic values of $L$ and $E$ for various neutrino 
      sources and experiments and the corresponding ranges of $\Delta
      m^2$ to which they can be most sensitive.}
    \vspace{1mm}
    \begin{tabular} {|l|lc|c|c|}
	\hline
	Experiment  & & L (m) & E (MeV) &  $\Delta m^2$ (\eVq) \\ \hline 
	Solar  & &$10^{10}$ & 1 & $10^{-10}$\\ \hline 
	Atmospheric & & $10^4-10^7$ & $10^2$--$10^5$ & $10^{-1}-10^{-4}$ \\ \hline 
	Reactor   & SBL& $10^2-10^3$ & 1 & $10^{-2}-10^{-3}$   \\ 
	& LBL &$10^4-10^5$ &   & $10^{-4}-10^{-5}$   \\ \hline 
	Accelerator &SBL &  $10^2$   & $10^3$--$10^4$ & $> 0.1$ \\   
	&LBL & $10^5-10^{6}$  & $10^4$ &  $10^{-2}-10^{-3}$\\ 
	\hline
    \end{tabular} 
\end{table} 

For a two-neutrino case, the mixing matrix depends on a single
parameter,
\begin{equation} 
    \label{eq:mixU2}
    U = 
    \begin{pmatrix}
	\cos\theta & \sin\theta \\
	-\sin\theta & \cos\theta
    \end{pmatrix} \,,
\end{equation}
and there is a single mass-squared difference $\Delta m^2$. Then
$P_{\alpha\beta}$ of Eq.~\eqref{eq:pab} takes the well known form 
\begin{equation}
    \label{eq:ptwo}
    P_{\alpha\beta} = 
    \delta_{\alpha\beta}- (2\delta_{\alpha\beta}-1) \sin^22\theta 
    \sin^2 X \,.
\end{equation}
The physical parameter space is covered with $\Delta m^2\geq 0$ and
$0\leq\theta\leq\frac{\pi}{2}$ (or, alternatively,
$0\leq\theta\leq\frac{\pi}{4}$ and either sign for $\Delta m^2$).

Changing the sign of the mass difference, $\Delta m^2\to-\Delta m^2$,
and changing the octant of the mixing angle,
$\theta\to\frac{\pi}{2}-\theta$, amounts to redefining the mass
eigenstates, $\nu_1\leftrightarrow\nu_2$: $P_{\alpha\beta}$ must be
invariant under such transformation.  Eq.~\eqref{eq:ptwo} reveals,
however, that $P_{\alpha\beta}$ is actually invariant under each of
these transformations separately. This situation implies that there is
a two-fold discrete ambiguity in the interpretation of
$P_{\alpha\beta}$ in terms of two-neutrino mixing: the two different
sets of physical parameters, ($\Delta m^2, \theta$) and ($\Delta m^2,
\frac{\pi}{2} -\theta$), give the same transition probability in
vacuum. One cannot tell from a measurement of, say, $P_{e\mu}$ in
vacuum whether the larger component of $\nu_e$ resides in the heavier
or in the lighter neutrino mass eigenstate.  This symmetry is lost
when neutrinos travel through regions of dense matter and/or for when
there are more than two neutrinos mixed in the neutrino evolution. 


\subsection{Propagation of Massive Neutrinos in Matter}
\label{sec:matterosc}

When neutrinos propagate in dense matter, the interactions with the
medium affect their properties. These effects can be either coherent
or incoherent.  For purely incoherent inelastic ${\nu}$-p scattering,
the characteristic cross section is very small: 
\begin{equation}
    \label{eq:sigmanp}
    \sigma\sim \frac{G_F^2 s}{\pi}\sim 10^{-43}~\text{cm}^2
    {\left( \frac{E}{\text{MeV}} \right)^2} \,.
\end{equation}
On the contrary, in coherent interactions, the medium remains
unchanged and it is possible to have interference of scattered and
unscattered neutrino waves which enhances the effect. Coherence
further allows one to decouple the evolution equation of the neutrinos
from the equations of the medium. In this approximation, the effect of
the medium is described by an effective potential which depends on the
density and composition of the matter~\cite{wolf}.

Taking this into account, the evolution equation for $n$ 
ultrarelativistic neutrinos propagating in matter written in the mass
basis can be casted in the following form (there are several
derivations in the literature of the evolution equation of a neutrino
system in matter, see for instance 
Ref.~\cite{halprin86,mannhein88,baltz88}):
\begin{equation}
    \label{eq:evol.1}
    i \frac{d\vec\nu}{dx} = H \, \vec\nu, \qquad
    H = H_m + {U^\nu}^\dagger \, V \, U^\nu \,,
\end{equation}
where $\vec\nu \equiv (\nu_1,\, \nu_2,\, \dots,\, \nu_n)^T$, $H_m$ is
the Hamiltonian for the kinetic energy, 
\begin{equation}
    \label{eq:evol.3}
    H_m = \frac{1}{2E}
    \diag(m_1^2,\, m_2^2,\, \dots,\, m_n^2 ),
\end{equation}
and $V$ is the effective potential that describes the coherent forward
interactions of the neutrinos with matter in the interaction basis. 
$U^\nu$ is the $n\times n$ submatrix of the unitary $V^\nu$ matrix
corresponding to the $n$ ultrarelativistic neutrino states.

Let's consider the evolution of $\nu_e$ in a medium with electrons,
protons and neutrons with corresponding $n_e$, $n_p$ and $n_n$ number
densities.  The effective low-energy Hamiltonian describing the
relevant neutrino interactions is given by
\begin{equation}
    H_W = \frac{G_F}{\sqrt{2}} \left[ 
    J^{(+)\alpha}(x) J^{(-)}_\alpha (x)
    + \frac{1}{4} J^{(N)\alpha}(x) J^{(N)}_\alpha (x) \right] ,
\end{equation}
where the $J_\alpha$'s are the standard fermionic currents
\begin{align}
    J^{(+)}_\alpha(x)
    &= \bar{\nu}_e(x)\gamma_\alpha(1-\gamma_5)e(x) \,,
    \\
    J^{(-)}_\alpha(x)
    &= \bar{e}(x)\gamma_\alpha(1-\gamma_5)\nu_e(x) \,,
    \\
    \begin{split}
	J^{(N)}_\alpha(x)
	&= \bar{\nu}_e(x)\gamma_\alpha(1-\gamma_5)\nu_e(x)
	\\
	& \hspace{5mm} - \bar{e}(x)
	[\gamma_\alpha(1-\gamma_5) - 4\sin^2\theta_W\gamma_\alpha] e(x)
	\\
	& \hspace{5mm} + \bar{p}(x) [\gamma_\alpha (1 - g_A^{(p)} \gamma_5)
	- 4\sin^2\theta_W \gamma_\alpha] p(x)
	\\
	& \hspace{5mm} - \bar{n}(x)
	\gamma_\alpha (1 - g_A^{(n)} \gamma_5) n(x) \,,
    \end{split}
\end{align}
and $g^{(n,p)}_A$ are the axial couplings for neutrons and protons,
respectively.  

Consider first the effect of the charged current interactions. The
effective CC Hamiltonian due to electrons in the medium is
\begin{equation} \begin{split}
    \label{eq:HCC}
    H_C^{(e)}
    &= \frac{G_F}{\sqrt{2}}\int d^3p_e { f(E_e,T)}
    \\
    & \hspace{5mm} \times \pmb{\Big\langle}
    \langle e(s,p_e) \,|\, \bar{e}(x)\gamma^\alpha 
    (1 - \gamma_5) \nu_e(x) \bar{\nu}_e(x) \gamma_\alpha
    (1 - \gamma_5) e(x) \,|\, e(s,p_e) \rangle \pmb{\Big\rangle}
    \\
    &= \frac{G_F}{\sqrt{2}} \bar{\nu}_e(x) \gamma_\alpha
    (1 - \gamma_5) \nu_e(x)
    \\
    & \hspace{5mm} \int d^3p_e { f(E_e,T)} \pmb{\Big\langle}
    \langle e(s,p_e) \,|\, \bar{e}(x) \gamma_\alpha
    (1 - \gamma_5) e(x) \,|\, e(s,p_e) \rangle \pmb{\Big\rangle} \,,
\end{split} \end{equation}
where $s$ is the electron spin and $p_e$ its momentum. The energy
distribution function of the electrons in the medium, $f(E_e,T)$, is
assumed to be homogeneous and isotropic and is normalized as
\begin{equation}
    \int d^3 p_e f(E_e,T) = 1 \,.
\end{equation}
By $\pmb{\Big\langle} \ldots \pmb{\Big\rangle}$ we denote the
averaging over electron spinors and summing over all electrons in the
medium. Notice that coherence implies that $s,\, p_e$ are the same for
initial and final electrons. To calculate the averaging we notice that
the axial current reduces to the spin in the non-relativistic limit
and therefore averages to zero for a background of non-relativistic
electrons. The spatial components of the vector current cancel because
of isotropy and therefore the only non trivial average is
\begin{equation}
    \int d^3 p_e f(E_e,T)
    \pmb{\Big\langle} \langle e(s,p_e) \,|\, \bar{e}(x) \gamma_0
    e(x) \,|\, e(s,p_e) \rangle \pmb{\Big\rangle} = n_e(x)
\end{equation}
which gives a contribution to the effective Hamiltonian
\begin{equation}
    H_C^{(e)} = \sqrt{2} G_F n_e \bar{\nu}_{eL}(x)
    \gamma_0 \nu_{eL}(x) \,.
\end{equation}
This can be interpreted as a contribution to the $\nu_{eL}$ potential
energy
\begin{equation}
    \label{eq:effV}
    V_C = \sqrt{2} G_F n_e \,.
\end{equation}
A more detailed derivation of the matter potentials can be found, for
example, in Ref.~\cite{kim}.

For $\nu_\mu$ and $\nu_\tau$, the potential due to its CC interactions
is zero for most media since neither $\mu$'s nor $\tau's$ are present.

In the same fashion one can derive the effective potential for any 
active neutrino due to the neutral current interactions to be
\begin{equation}
    V_{NC} = \frac{\sqrt{2}}{2} G_F \left[
    - n_e (1 - 4\sin^2\theta_w)
    + n_p (1 - 4\sin^2\theta_w) - n_n \right] \,. 
\end{equation}
For neutral matter $n_e=n_p$ so the contribution from electrons and
protons cancel each other and we are left only with the neutron
contribution
\begin{equation}
    \label{eq:ncpot}
    V_{NC} = -1/\sqrt{2} G_F n_{n}
\end{equation}

Altogether we can write the evolution equation for the three SM active
neutrinos with purely SM interactions in a neutral medium with
electrons, protons and neutrons as Eq.~\eqref{eq:evol.1} with $U^\nu
\equiv U$, and the effective potential:
\begin{equation}
    \label{eq:evol.4}
    V= \diag \left( \pm \sqrt{2} G_F n_e(x),\, 0,\, 0 \right)
    \equiv \diag \left( V_e,\, 0,\, 0 \right).
\end{equation}
In Eq.~\eqref{eq:evol.4}, the sign $+$ ($-$) refers to neutrinos 
(antineutrinos), and $n_e(x)$ is the electron number density in the
medium, which in general changes along the neutrino trajectory and so
does the potential.  For example, at the Earth core $V_e\sim
10^{-13}$~eV while at the solar core $V_e\sim 10^{-12}$~eV. Notice
that the neutral current potential Eq.~\eqref{eq:ncpot} is flavor
diagonal and therefore it can be eliminated from the evolution
equation as it only contributes to an overall phase which is
unobservable.

The instantaneous mass eigenstates in matter, $\nu^m_i$, are the
eigenstates of $H$ for a fixed value of $x$, which are related to the
interaction basis by
\begin{equation}
    \vec\nu = \tilde U(x) \vec{\nu^m} \,, 
\end{equation}
while $\mu_i(x)^2/(2E)$ are the corresponding instantaneous
eigenvalues with $\mu_i(x)$ being the instantaneous effective neutrino
masses.

For the simplest case of the evolution of a neutrino state which is 
an admixture of only two neutrino species ${|\nu_\alpha\rangle}$ and 
${|\nu_\beta\rangle}$ 
\begin{multline}
    \label{eq:effmass}
    \mu_{1,2}^2(x) = \frac { m_1^2+m_2^2 }{2} 
    + E [V_\alpha + V_\beta]
    \\
    \mp\frac{1}{2} \sqrt{\left[{\Delta m^2\cos 2\theta} - A\right]^2
      + \left[{\Delta m^2\sin 2\theta}\right]^2 } \,,
\end{multline}
and $\tilde U(x)$ can be written as Eq.~\eqref{eq:mixU2} with the
instantaneous mixing angle in matter given by
\begin{equation}
    \label{eq:effmix}
    \tan 2\theta_{m} =
    \frac{\Delta{m}^2\sin2\theta}{\Delta{m}^2\cos 2\theta-A}.
\end{equation}
The quantity $A$ is defined by
\begin{equation}
    A\equiv 2 E ({ V_\alpha-V_\beta}).
\end{equation}
Notice that for a given sign of $A$ (which depends on the composition
of the medium and on the flavor composition of the neutrino state)
the mixing angle in matter is larger or smaller than in vacuum 
depending on whether this last one lies on the first or the second
octant. Thus the symmetry present in vacuum oscillations is broken by
matter potentials.  

Generically matter effects are important when for some of the states
the corresponding potential difference factor, $A$, is comparable to
their mass difference term $\Delta{m}^2\cos 2\theta$. Most relevant,
as seen in Eq.~\eqref{eq:effmix}, the mixing angle $\tan\theta_m$
changes sign if in some point along its path the neutrino passes by
some matter density region verifying the \emph{resonance condition}
\begin{equation}
    \label{eq:AR}
    A_R = {\Delta m^2\cos 2\theta} \,.
\end{equation}
Thus if the neutrino is created in a region where the relevant
potential verifies $A_0> A_R$, then the effective mixing angle in
matter at the production point verifies that $\sgn(\cos 2\theta_{m,0})
= -\sgn(\cos 2\theta)$, this is, the flavor component of the mass
eigenstates is inverted as compared to their composition in vacuum.
For example for $A_0 = 2 A_R$ $\theta_{m,0} = \frac{\Pi}{2} - \theta$.
Asymptotically, for $A_0\gg A_R$, $\theta_{m,0} \rightarrow
\frac{\pi}{2}$. 

In other words, if in vacuum the lightest mass eigenstate has a larger
projection on the flavor $\alpha$ while the heaviest has it on the 
flavor $\beta$, once inside a matter potential with $A> A_R$ the 
opposite holds. Thus for a neutrino system which is traveling across 
a monotonically varying matter potential the dominant flavor
component of a given mass eigenstate changes when crossing the region
with $A=A_R$.  This phenomenon is known as \emph{level crossing}. 
  
In the instantaneous mass basis the evolution equation reads:
\begin{equation}
    \label{eq:evol.5} 
    i \frac{d\vec\nu^m}{dx} = \left[
    \frac{1}{2E} \, 
    \diag \left(\mu^2_1(x),\, \mu^2_2(x),\, \dots,\, \mu^2_n(x) \right) 
    - i \, \tilde{U}^\dagger(x) \,
    \frac{d \tilde U(x)}{dx} \right] \vec\nu^m \,.
\end{equation}
Because of the last term, Eq.~\eqref{eq:evol.5} constitute a system of
coupled equations which implies that the instantaneous mass
eigenstates, ${\nu_i^m}$, mix in the evolution and are not energy
eigenstates.  For constant or slowly enough varying matter potential
this last term can be neglected. In this case the instantaneous mass
eigenstates, $\nu^m_i$, behave approximately as energy eigenstates and
they do not mix in the evolution.  This is the \emph{adiabatic}
transition approximation. On the contrary, when the last term in
Eq.~\eqref{eq:evol.5} cannot be neglected, the instantaneous mass
eigenstates mix along the neutrino path so there can be 
\emph{level-jumping}~\cite{Landau,Zener} and the evolution is
\emph{non-adiabatic}. 

The oscillation probability takes a particularly simple form for
adiabatic evolution in matter and it can be cast very similarly to the
vacuum oscillation expression, Eq.~\eqref{eq:pab}.  For example,
neglecting CP violation:
\begin{equation}
    \label{eq:adigen}
    P_{\alpha\beta} = \left| \sum_{i}
    \tilde U_{\alpha i} (0) \tilde U_{\beta i}(L)
    \exp\left( -\frac{i}{2E} \int^L_{0} \mu^2_i(x')
    dx' \right) \right|^2 \,.
\end{equation}

In general $P_{\alpha\beta}$ has to be evaluated numerically although
there exist in the literature several analytical approximations for
specific profiles of the matter potential~\cite{kuo}. 

\subsubsection{The MSW Effect for Solar Neutrinos}

As an illustration of the matter effects discussed in the previous
section we describe now the propagation of a $\nu_e-\nu_X$ neutrino
system in the matter density of the Sun where $X$ is some
superposition of $\mu$ and $\tau$. 

The solar density distribution decreases monotonically with the
distance $R$ to the center of the Sun. For $R<0.9 R_\odot$ it can be
approximated by an exponential
\begin{equation}
    n_e(R) = n_e(0)\exp\left(-R/r_0\right)
\end{equation}
with $r_0 = R_\odot / 10.54 = 6.6\times 10^7$ m $=3.3\times
10^{14}~\text{eV}^{-1}$.  After traversing this density the dominant
component of the exiting neutrino state depends on the value of the
mixing angle in vacuum, and on the relative size of ${\Delta m^2 \cos
2\theta}$ versus ${A_0}=2\, E \, G_F \, n_{e,0}$ (at the neutrino
production point) as we describe next:
\begin{itemize}
  \item If ${\Delta m^2\cos2\theta}\gg A_0$ matter effects are
    negligible and the propagation occurs as in vacuum with the
    oscillating phase averaged out due to the large value of $L$.  In
    this case the survival probability at the sunny surface of the
    Earth is
    \begin{equation}
	P_{ee}(\Delta m^2\cos2\theta\gg A_0) =
	1-\frac{1}{2}{\sin^22\theta} > \frac{1}{2} \,.
    \end{equation}
    
  \item If ${\Delta m^2 \cos 2\theta}\gtrsim {A_0}$ the neutrino does not
    pass any resonance region but its mixing is affected by the solar
    matter. This effect is well described by an adiabatic propagation,
    Eq.~\eqref{eq:adigen}.  Using 
    \begin{equation}
	\tilde U(0) =
	\begin{pmatrix}
	    \cos\theta_{m,0} &\sin\theta_{m,0}
	    \\
	    -\sin\theta_{m,0} & \cos\theta_{m,0}
	\end{pmatrix} \,,
	\qquad
	\tilde U(L) =
	\begin{pmatrix}
	    \cos\theta &\sin\theta 
	    \\
	    -\sin\theta &\cos\theta
	\end{pmatrix} \,,
    \end{equation}
    (where $\theta_{m,0}$ is the mixing angle in matter at the
    production point) we get
    \begin{multline} 
	\label{eq:peeadiab1}
	P_{ee} = \cos^2\theta_{m,0} \cos^2\theta
	+ \sin^2\theta_{m,0} \sin^2\theta
	\\
	+ \frac{1}{2}\sin^2 2\theta_{m,0} \sin^2 2\theta
	\cos\left(\frac{\int^L_{0} \mu^2_2(x') - \mu_1^2(x')}{2E}dx' \right) \,.
    \end{multline}
    For all practical purposes, the oscillation term in
    Eq.~\eqref{eq:peeadiab1} is averaged out in the regime $\Delta m^2
    \cos 2\theta \gtrsim {A_0}$ and then the resulting probability
    reads
    \begin{equation} \begin{split}
	\label{eq:peeadiab}
	P_{ee}(\Delta m^2\cos2\theta\geq A_0) &=
	\cos^2\theta_{m,0} \cos^2\theta+\sin^2\theta_{m,0} \sin^2\theta
	\\ 
	&= \frac{1}{2}\left[1 + \cos 2\theta_{m,0} \cos 2\theta \right] \,.
    \end{split} \end{equation}
    The physical interpretation of this expression is straightforward.
    An electron neutrino produced at $A_0$ consists of an admixture of
    ${\nu_1}$ with fraction $P_{e1,0} = \cos^2\theta_{m,0}$ and
    ${\nu_2}$ with fraction $P_{e2,0} = \sin^2\theta_{m,0}$.  At the
    exit ${\nu_1}$ consists of ${\nu_e}$ with fraction $P_{1e} =
    \cos^2\theta$ and $\nu_2$ consists of $\nu_e$ with fraction
    $P_{2e} = \sin^2\theta$ so~\cite{parke86,haxton86,petcov87}
    \begin{equation}
	P_{ee} = P_{e1,0} P_{1e}+ P_{e2,0} P_{2e}
    \end{equation}
    which reproduces Eq.~\eqref{eq:peeadiab}. Notice that as long as
    $A_0 < A_R$ the resonance is not crossed and consequently $\cos
    2\theta_{m,0}$ has the same sign as $\cos 2\theta$ and the
    corresponding survival probability is also larger than 1/2. 
    
  \item If $\Delta m^2 \cos 2\theta < A_0$  the neutrino can cross
    the resonance on its way out if, in the convention of positive
    $\Delta m^2$, $\cos 2\theta > 0$ ($\theta < \pi/4$). In this case,
    at the production point $\nu_e$ is a combination of $\nu_1^m$ and
    $\nu_2^m$ with larger $\nu_2^m$ component while outside of the Sun
    the opposite holds. More quantitatively for $\Delta m^2 \cos
    2\theta \ll A_0$ (density at the production point much higher than
    the resonant density), 
    \begin{equation} 
	\label{eq:tm}
	\theta_{m,0} = \frac{\pi}{2}
	\quad\Rightarrow\quad
	\cos 2\theta_{m,0} = -1 \,.
    \end{equation}
    Depending on the particular values of $\Delta m^2$ and the mixing
    angle, the evolution can be adiabatic or non-adiabatic. As we will
    see in Sec.~\ref{sec:3nu} presently we know that the oscillation
    parameters are such that the transition is indeed adiabatic for
    all ranges of solar neutrino energies. Thus the survival
    probability at the sunny surface of the Earth is
    \begin{equation}
	P_{ee}(\Delta m^2\cos2\theta< A_0)=\frac{1}{2} \left[
	1 + \cos 2\theta_{m,0}\cos 2\theta \right] = \sin^2\theta
    \end{equation}
    where we have used Eq.~\eqref{eq:tm}. Thus in this regime $P_{ee}$
    can be much smaller than $1/2$ because $\cos 2\theta_{m,0}$ and
    $\cos2\theta$ have opposite signs. This is the MSW
    effect~\cite{wolf,ms} which plays a crucial role in the
    interpretation of the solar neutrino data.
\end{itemize}

%% file: sec.expe.tex
\section{Present Experimental Tests of Neutrino Oscillations}
\label{sec:expe}


\subsection{Solar Neutrinos}
\label{sec:expsolar}

Solar neutrinos are electron neutrinos produced in the thermonuclear
reactions which generate the solar energy. These reactions occur via
two main chains, the $pp$ chain and the CNO cycle.  There are five
reactions which produce $\nu_e$ in the $pp$ chain and three in the CNO
cycle.  Both chains result in the overall fusion of protons into
\Nuc{4}{He}:
\begin{equation}
    4p \to \Nuc{4}{He} + 2 e^+ + 2\nu_e + \gamma,
\end{equation}
where the energy released in the reaction, $Q = 4m_p - m_{\Nuc{4}{He}}
- 2 m_e \simeq 26$ MeV, is mostly radiated through the photons and
only a small fraction is carried by the neutrinos, $\langle
E_{2\nu_e}\rangle = 0.59$ MeV.

\begin{figure}\centering
    \includegraphics[angle=-90,width=3.73in]{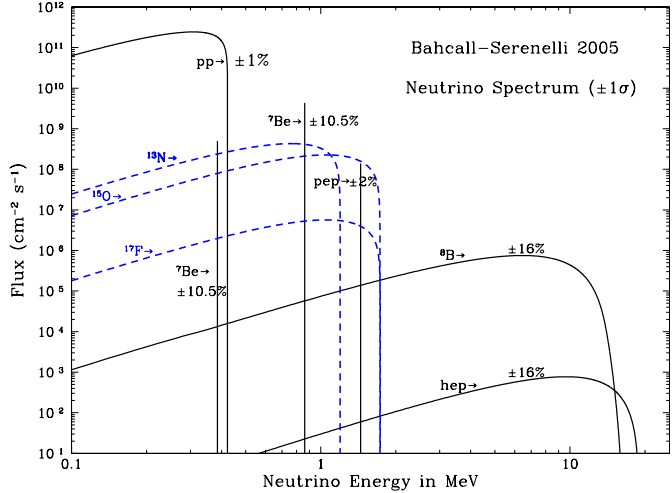}
    \caption{\label{fig:bs05}%
      Neutrino fluxes predicted by the SSM~\cite{bs05} as a function
      of the neutrino energy.}
\end{figure}

Along this review we use as Standard Solar Model (SSM) the most
updated version of the model developed by Bahcall and
Serenelli~\cite{bs05}. In Fig.~\ref{fig:bs05} we show the energy
spectrum of the neutrino fluxes from the eight reactions as predicted
by the SSM. In what follows we refer to the neutrino fluxes by the
corresponding source reaction, so, for instance, the neutrinos
produced from \Nuc{8}{B} decay are called \Nuc{8}{B} neutrinos. Most
reactions produce a neutrino spectrum characteristic of $\beta$ decay.
For \Nuc{8}{B} neutrinos the energy distribution presents deviations
with respect to the maximum allowed energy because the final state,
\Nuc{8}{Be}, is a wide resonance. On the other hand, the \Nuc{7}{Be}
neutrinos are almost monochromatic, with an energy width of about 2
keV which is characteristic of the temperature in the core of the Sun.
 
In order to precisely determine the rates of the different reactions
in the two chains which would give us the final neutrino fluxes and
their energy spectrum, a detailed knowledge of the Sun and its
evolution is needed. Solar Models describe the properties of the Sun
and its evolution after entering the main sequence. The models are
based on a set of observational parameters: the surface luminosity
($3.844 \times 10^{26}$ W), the age ($4.5 \times 10^9$ years), the
radius ($6.961 \times 10^8$ m) and the mass ($1.989 \times 10^{30}$
kg), and on several basic assumptions: spherical symmetry, hydrostatic
and thermal equilibrium, equation of state of an ideal gas, and
present surface abundances of elements similar to the primordial
composition. Over the past four decades, the solar models have been 
steadily refined as the result of increased observational and
experimental information about the input parameters (such as nuclear
reaction rates and the surface abundances of different elements), more
accurate calculations of constituent quantities (such as radiative
opacity and equation of state), the inclusion of new physical effects
(such as element diffusion), and the development of faster computers
and more precise stellar evolution codes. Other important elements of
the model which are relevant to the evolution of neutrinos in the
solar matter are the density and composition of solar matter and the
production point distribution for the different neutrino fluxes. 

\subsubsection{Experiments}

$\bullet$ {\bf Chlorine experiment: Homestake.} The first result on
the detection of solar neutrinos was announced by Ray Davis Jr and his
collaborators from Brookhaven in 1968~\cite{homestake}. In the gold
mine of Homestake in Lead, South Dakota, they installed a detector
consisting of $\sim$ 615 Tons of C$_2$Cl$_4$. Solar $\nu_e$'s are
captured via $\Nuc{37}{Cl}~(\nu,e^-)~\Nuc{37}{Ar}$. The energy
threshold for this reaction is 0.814 MeV, so the relevant fluxes are
the \Nuc{7}{Be} and \Nuc{8}{B} neutrinos.  For the SSM fluxes, 78\% of
the expected number of events are due to \Nuc{8}{B} neutrinos while
13\% arise from \Nuc{7}{Be} neutrinos.  

The average event rate measured during the more than 20 years of 
operation is~\cite{chlast} 
\begin{equation}
    \label{eq:rate_cl}
    R_\text{Cl} = 2.56 \pm 0.16 \pm 0.16~\text{SNU}
    \quad\Rightarrow\quad
    \frac{R_\text{Cl}}{\text{SSM}} = 0.30 \pm 0.03
\end{equation}
(1 SNU = $10^{-36}$ captures/atom/sec).

\medskip

$\bullet$ {\bf Gallium experiments: SAGE and GALLEX/GNO.} In January
1990 and May 1991, two new radiochemical experiments using a 
\Nuc{71}{Ga} target started taking data, SAGE~\cite{sage} and 
GALLEX~\cite{gallex}. The SAGE detector is located in Baksan,
Kaberdino-Balkaria, Russia, with 30 Tons (increased to 57 Tons from
July 1991) of liquid metallic Ga. GALLEX is located in Gran Sasso,
Italy, and consists of 30 Tons of GaCl$_3$-HCl. In these experiments
the solar neutrinos are captured via $\Nuc{71}{Ga}(\nu,e^-)
\Nuc{71}{Ge}$. The special properties of this target include a low
threshold (0.233 MeV) and a strong transition to the ground level of
\Nuc{71}{Ge}, which gives a large cross section for the lower energy
$pp$ neutrinos.  According to the SSM, approximately 54\% of the
events are due to $pp$ neutrinos, while 26\% and 11\% arise from
\Nuc{7}{Be} and \Nuc{8}{B} neutrinos, respectively.  The GALLEX
program was completed in fall 1997 and its successor GNO started
taking data in spring 1998 and it ended in April 2003.
 
The averaged event rates measured by SAGE and GALLEX+GNO
are~\cite{sagegnolast}
\begin{equation}
    R_\text{GALLEX+GNO+SAGE} = 68.1 \pm 3.75~\text{SNU} 
    \quad\Rightarrow\quad
    \frac{R_\text{Ga}}{\text{SSM}}= 0.52 \pm 0.03 \,.
\end{equation}
Since the $pp$ flux is directly constrained by the solar luminosity,
in all stationary solar models there is a theoretical minimum of the
expected number of events of 79 SNU. 

\medskip

$\bullet$ {\bf Water Cherenkov: Kamiokande and Super-Kamiokande.}
Kamiokande~\cite{kamsun} and its successor
Super-Kamiokande~\cite{sksun} (SK) in Japan are water Cherenkov
detectors that are able to detect in real time the electrons which are
emitted from the water by the elastic scattering (ES) of the solar
neutrinos, $\nu_a + e^- \to \nu_a + e^-$.

The scattered electrons produce Cherenkov light which is detected by
photomultipliers. Notice that, while the detection process in
radiochemical experiments is purely a CC ($W$-exchange) interaction,
the detection ES process goes through both CC NC ($Z$-exchange)
interactions. Consequently, the ES detection process is sensitive to
all active neutrino flavors, although $\nu_e$'s (which are the only
ones to scatter via $W$-exchange) give a contribution that is about 6
times larger than that of $\nu_\mu$'s or $\nu_\tau$'s.

Kamiokande, with 2140 tons of water, started taking data in January
1987 and was terminated in February 1995. SK, with 45000 tons of water
(of which 22500 are usable in solar neutrino measurements) started in
May 1996 and it has analyzed so far the full SK-I low energy data
corresponding to 1496 live days.  The detection threshold in
Kamiokande was 7.5 MeV while SK late runs were at 5 MeV.  This means
that these experiments are able to measure only the \Nuc{8}{B}
neutrinos (and the very small hep neutrino flux). Their results are
presented in terms of measured \Nuc{8}{B} flux:
\begin{equation} \begin{aligned}
    \hspace{-3mm}
    \Phi_\text{Kam} &= (2.80 \pm 0.19 \pm 0.33) \times 10^6 ~
    \text{cm}^{-2}\text{s}^{-1} \,,
    \\
    \Phi_\text{SK} &= (2.35 \pm 0.02 \pm 0.08) \times 10^6 ~
    \text{cm}^{-2}\text{s}^{-1}
    ~\Rightarrow~
    \frac{\Phi_\text{SK}}{\Phi_\text{SSM}} = 0.413 \pm 0.014 \,.
\end{aligned} \end{equation}

\medskip

$\bullet$ {\bf SNO.} The Sudbury Neutrino Observatory (SNO) was first
proposed in 1987 and it started taking data in November
1999~\cite{SNOIa,SNOIb,SNOIc}. The detector, a great sphere surrounded
by photomultipliers, contains approximately 1000 Tons of heavy water,
D$_2$O, and is located at the Creighton mine, near Sudbury in Canada.
SNO was designed to give a model independent test of the possible
explanations of the observed deficit in the solar neutrino flux by
having sensitivity to all flavors of active neutrinos and not just to 
$\nu_e$. This sensitivity is achieved because energetic neutrinos can
interact in the D$_2$O of SNO via three different reactions. Electron
neutrinos may interact via the CC reaction $\nu_e + d \to p + p +
e^-$, and can be detected above an energy threshold of a few MeV
(presently $T_e>$ 5 MeV). All active neutrinos ($\nu_a = \nu_e$,
$\nu_\mu$, $\nu_\tau$) interact via the NC reaction $\nu_a + d \to n +
p + \nu_a $ with an energy threshold of 2.225 MeV. The non-sterile
neutrinos can also interact via ES, $\nu_a + e^- \to \nu_a + e^- $,
but with smaller cross section.

SNO has also performed measurements of the energy spectrum and time
variation of the event rates. But the uniqueness of SNO lied in its
ability to directly test if the deficit of solar $\nu_e$ is due to
changes in the flavor composition of the solar neutrino beam, since
the ratio CC/NC compares the number of $\nu_e$ interactions with those
from all active flavors. This comparison is independent of the overall
flux normalization.

The experimental plan of SNO consisted of three phases.  In its first
year of operation, SNO concentrated on the measurement of the CC
reaction rate~\cite{SNOIa,SNOIb,SNOIc} while in a following phase,
after the addition of $\text{MgCl}_2$ salt to enhance the NC signal,
it also performed a precise measurement of the NC
rate~\cite{SNOIIa,SNOIIb}.  In the present third phase, starting
taking data in November 2004, the salt was eliminated and a network of
proportional counters filled with \Nuc{3}{He} was added with the
purpose of directly measuring the NC rate
$\Nuc{3}{He}~(n,p)~\Nuc{3}{H}$.

At present their most precise determination of the solar fluxes
yields:
\begin{equation} \begin{aligned}
    \label{eq:snodat}    
    \Phi^\text{CC}_\text{SNO}
    &= (1.68 \,^{+0.06}_{-0.06}\, ^{+0.08}_{-0.09}) \times 10^6
    \text{ cm}^{-2}\text{s}^{-1}
    & \Rightarrow~
    \frac{\Phi^\text{CC}_\text{SNO}}{\Phi_\text{SSM}}
    &= 0.29 \pm 0.02 \,,
    \\
    \Phi^\text{ES}_\text{SNO}
    &= (2.35\pm 0.22\pm 0.15) \times 10^6
    \text{ cm}^{-2}\text{s}^{-1} 
    & \Rightarrow~
    \frac{\Phi^\text{ES}_\text{SNO}}{\Phi_\text{SSM}}
    &= 0.41 \pm 0.05 \,,
    \\
    \Phi^\text{NC}_\text{SNO}
    &= (4.94 \pm 0.21 \,^{+0.38}_{-0.34})
    \times 10^6 \text{ cm}^{-2} \text{s}^{-1}
    & \Rightarrow~
    \frac{\Phi^\text{NC}_\text{SNO}}{\Phi_\text{SSM}}
    &= 0.87\pm 0.08 \,.
\end{aligned} \end{equation}

There are three features unique to the Cherenkov detectors,
Kamiokande, Super-Kamiokande and SNO. First, they are real time
experiments. Each event is individually recorded. Second, for each ES
event the scattered electron keeps the neutrino direction within an
angular interval which depends on the neutrino energy as $\sqrt{2
m_e/E_{\nu}}$. Thus, it is possible, for example, to correlate the
neutrino detection with the position of the Sun. Third, the amount of 
Cherenkov light produced allows a measurement of the energy.  In
summary, the experiments can provide information on the time, 
direction and energy for each event. Signatures of neutrino
oscillations might include distortion of the recoil electron energy 
spectrum, difference between the night-time solar neutrino flux and
the day-time flux, or a seasonal variation in the neutrino flux.
Observation of these effects were searched as strong evidence in
support of solar neutrino oscillations independent of absolute flux
calculations.

Over the years the SK and SNO collaborations have provided us with 
information on the energy and time dependence of their event rates in
different forms. At present their most precise data is presented in
form of a zenith-energy spectrum with 44 data points for SK, the CC
day-night spectrum measured in the pure D$_2$O phase of SNO with 34
data points, and the NC and ES event rates during the day and during
the night (4 data points) plus the CC day-night spectral data (34 data
points) corresponding to the SNO Salt Phase.  These results show no
significant energy or time dependence of the event rates beyond the
expected ones in the SSM. 

\medskip

$\bullet$ {\bf Borexino.} The Borexino experiment~\cite{borexino} is
currently taking data in the Laboratori Nazionali del Gran Sasso in
Italy. Its main goal is to measure the flux from the 0.86 MeV
monoenergetic line of \Nuc{7}{Be} solar neutrinos in real-time.

Borexino employs a liquid scintillator that produces sufficient light
to observe low energy neutrino events via elastic scattering by
electrons.  The reaction is sensitive to all neutrino flavors by the
neutral current interaction, but the cross section for $\nu_e$ is
larger due to the combination of charged and neutral currents.  

Monochromatic 862~keV neutrinos from \Nuc{7}{Be} offer two signatures
in Borexino.  The first is a recoil electron profile with clear
Compton edge at 665~keV.  The second possible signature is the
$\pm3.5\%$ annual variation of the flux due to the Earth orbit
eccentricity.

In August 2007 Borexino release their first data~\cite{borex07},
collected for 47.4~live days between May and July~2007.  The fiducial
exposure accumulated during this live time was 4136~day~$\cdot$~ton.
They found that the best fit for their observed rate is
\begin{equation}
    R_{\Nuc{7}{Be}}
    = (47 \pm 7 \pm 12)~\text{counts/day} \times \text{100 ton}
    \quad\Rightarrow\quad
    \frac{R_{\Nuc{7}{Be}}}{\text{SSM}} = 0.63\pm 0.18 \,.
\end{equation}

\subsubsection{Evidence of Flavor Conversion of Solar Neutrinos}

From the experimental results described above one can conclude that: 
\begin{itemize}
  \item Before the NC measurement at SNO, all experiments observed a flux 
    that was smaller than the SSM predictions, $\Phi^\text{obs} /
    \Phi^\text{SSM} \sim 0.3-0.6$.
    
  \item The deficit is not the same for the various experiments, 
    which may indicate that the effect is energy dependent.
\end{itemize}
These two statements constituted the solar neutrino
problem~\cite{SNPa,SNPb}. 

The results of SNO provided further model independent evidence of the 
problem. Both SNO and SK are sensitive mainly to the \Nuc{8}{B} flux. 
Consequently, in the absence of new physics, the measured fluxes in
any reaction at these two experiments should be equal. Conversely in
presence of flavor conversion
\begin{equation} \begin{aligned}
    \label{eq:snoev}
    \Phi^\text{CC} &= \Phi_e \,,
    \\
    \Phi^\text{ES} &= \Phi_e + r \, \Phi_{\mu\tau} \,,
    \\
    \Phi^\text{NC} &= \Phi_e + \Phi_{\mu\tau} \,,
\end{aligned} \end{equation}
where $r\equiv \sigma_{\mu}/\sigma_{e}\simeq 0.15$ is the ratio of the
the $\nu_e - e$ and $\nu_{\mu} - e$ elastic scattering cross-sections.
The flux $\Phi_{\mu\tau}$ of active non-electron neutrinos is zero in
the SSM. Thus, the three observed rates should be equal, an hypothesis
which is now ruled out at more than 7 $\sigma$ CL by the latest SNO
data establishing the evidence for neutrino flavor transition 
independently of the solar model.

\begin{figure}\centering
    \includegraphics[width=3.73in]{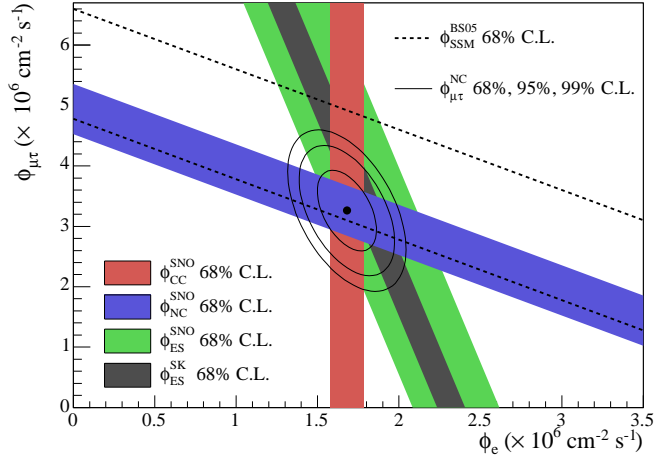}
    \caption{\label{fig:snoev}%
      Flux of $\mu+\tau$ neutrinos versus flux of electron neutrinos.
      CC, NC and ES flux measurements are indicated by the filled
      bands.  The total \Nuc{8}{B} solar neutrino flux predicted by
      the Standard Solar Model~\cite{bs05} is shown as dashed lines,
      and that measured with the NC channel is shown as the solid band
      parallel to the model prediction.} 
\end{figure}

This evidence is graphically displayed in Fig.~\ref{fig:snoev} (from
Ref.~\cite{SNOIIa,SNOIIb}) which shows the flux of non-electron flavor
active neutrinos ($\phi_{\mu\tau}$) versus the flux of electron
neutrinos ($\phi_e$) obtained by comparing Eq.~\eqref{eq:snoev} with
the data in Eq.~\eqref{eq:snodat}. The error ellipses shown are the
68\%, 95\% and 99\% joint probability contours for $\phi_{\mu\tau}$
and $\phi_e$ from the combined analysis. 

The simplest mechanism for the solar neutrino flavor transition is
that of oscillations of $\nu_e$ into $\nu_\mu$ and/or $\nu_\tau$.
Because of the importance played by the solar matter in the neutrino
evolution, the interpretation of the data in terms of oscillation
parameters lead during many years to a rather degenerate set of
possible solutions with mass differences and mixing angles ranges
varying over more than 7 and 5 orders of magnitude respectively.
Fortunately with the upcome of the most precise SK and SNO data the
situation became much more clear as we will describe in
Sec.~\ref{sec:2nusolar}.


\subsection{Atmospheric Neutrinos}
\label{sec:expatmos}

Cosmic rays interacting with the nitrogen and oxygen in the Earth's 
atmosphere at an average height of 15 kilometers produce mostly pions
and some kaons that decay into electron and muon neutrinos and
anti-neutrinos. 

Atmospheric neutrinos are observed in underground experiments using
different techniques and leading to different type of events depending
on their energy. They can be detected by the direct observation of
their CC interaction inside the detector. These are the
\emph{contained} events.  Contained events can be further classified
into \emph{fully contained} events, when the charged lepton (either
electron or muon) that is produced in the neutrino interaction does
not escape the detector, and \emph{partially contained} muons, when
the produced muon exits the detector. For fully contained events the
flavor, kinetic energy and direction of the charged lepton can be best
determined.  Higher energy muon neutrinos and antineutrinos can also
be detected indirectly by observing the muons produced in their
charged current interactions in the vicinity of the detector. These
are the so called \emph{upgoing muons}. Should the muon stop inside
the detector, it is classified as a \emph{stopping muon} while if the
muon track crosses the full detector the event is classified as a
\emph{through-going muon}. Downgoing muons from $\nu_\mu$ interactions
above the detector cannot be distinguished from the background of
cosmic ray muons. Higher energy $\nu_e$'s cannot be detected this way
as the produced $e$ showers immediately in the rock.

Atmospheric neutrinos were first detected in the 1960's by the
underground experiments in South Africa~\cite{africa} and the Kolar
Gold Field experiment in India~\cite{india}. These experiments
measured the flux of horizontal muons (they could not discriminate
between downgoing and upgoing directions) and although the observed
total rate was not in full agreement with theoretical predictions the
effect was not statistically significant. 

\begin{figure}\centering
    \includegraphics[width=0.99\textwidth]{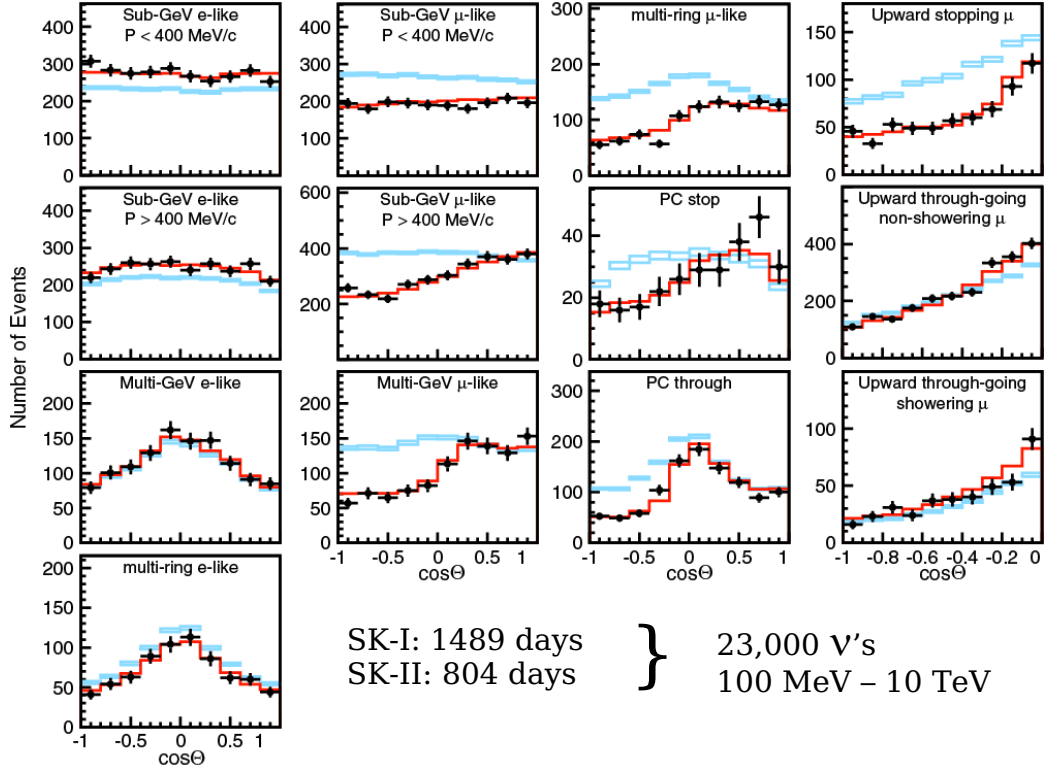}
    \caption{\label{fig:skatm}%
      The zenith angle distribution for fully-contained 1-ring events,
      multi-ring events, partially-contained events and upward muons
      from SK experiment~\cite{skII-walter}.  The points show the
      data, box histograms show the non-oscillated Monte Carlo events
      and the lines show the best-fit expectations for oscillations.} 
\end{figure}

A set of modern experiments were proposed and built in the 1970's and 
1980's. The original purpose was to search for nucleon decay, for
which atmospheric neutrinos constitute background. Two different
detection techniques were employed. In water Cherenkov detectors the
target is a large volume of water surrounded by photomultipliers which
detect the Cherenkov-ring produced by the charged leptons. The event
is classified as an electron-like (muon-like) event if the ring is
diffuse (sharp). In iron calorimeters, the detector is composed of a 
set of alternating layers of iron which act as a target and some
tracking element (such as plastic drift tubes) which allows the
reconstruction of the shower produced by the electrons or the tracks
produced by muons. Both types of detectors allow for flavor
classification of the events as well as the measurement of the
scattering angle of the outgoing charged lepton and some determination
of its energy. 

Since $\nu_e$ is produced mainly from the decay chain $\pi \to \mu
\nu_\mu$ followed by $\mu \to e \nu_\mu\nu_e$, one naively expects a
$2:1$ ratio of $\nu_\mu$ to $\nu_e$.  For higher energy events the
expected ratio is larger because some of the muons arrive to Earth
before they had time to decay. In practice, however, the theoretical
calculation of the ratio of muon-like interactions to electron-like
interactions in each experiment is more complicated.  In different
atmospheric flux 
calculations~\cite{bartol,honda,others1,others2,others3} the predicted
absolute fluxes of neutrinos produced by cosmic-ray interactions in
the atmosphere can vary at the 20\% level while the ratios of
neutrinos of different flavor are expected to be accurate to better
than 5\%. For this reason most of these early experiments presented
their results in terms of the flavor ratio of their event rates
compared to the theoretical expectation $R_{\mu/e} / R^{MC}_{\mu/e}$. 

The two oldest iron calorimeter experiments, Frejus~\cite{frejus} and
NUSEX~\cite{nusex}, found atmospheric neutrino fluxes in agreement
with the theoretical predictions. On the other hand, two water
Cherenkov detectors, IMB~\cite{IMB} and Kamiokande, detected a ratio
of $\nu_\mu$-induced events to $\nu_e$-induced events smaller than the
expected one by a factor of about 0.6.  Kamiokande further divided
their contained data sample into sub-GeV and multi-GeV events and
performed separate analyses for both sub-GeV neutrinos and multi-GeV
neutrinos~\cite{Kammulti}, which showed the same deficit.  This was
the original formulation of the atmospheric neutrino anomaly.  Whether
$R_{\mu/e} / R^{MC}_{\mu/e}$ was small because $\nu_\mu$ disappeared
or $\nu_e$ appeared or a combination of both could not be determined.
Furthermore, the fact that the anomaly was present only in water
Cherenkov detectors and not in iron calorimeters left the window open
for the suspicion of a possible systematic problem as the origin of
the effect.

Kamiokande also presented the zenith angular dependence of the deficit
for the multi-GeV neutrinos. The zenith angle, parametrized in terms
of $\cos\theta$, measures the direction of the reconstructed charged
lepton with respect to the vertical of the detector. Vertically
downgoing (upgoing) particles correspond to $\cos\theta=+1(-1)$.
Horizontally arriving particles come at $\cos\theta=0$. Kamiokande
results seemed to indicate that the deficit was mainly due to the
neutrinos coming from below the horizon.  Atmospheric neutrinos are
produced isotropically at a distance of about 15 km above the surface
of the Earth. Therefore neutrinos coming from the top of the detector
have traveled approximately those 15 kilometers before interacting
while those coming from the bottom of the detector have traversed the
full diameter of the Earth, $\sim 10^4$ Km before reaching the
detector. The Kamiokande distribution suggested that the deficit
increased with the distance between the neutrino production and
interaction points.

In the last ten years, the case for the atmospheric neutrino anomaly
became much stronger with the high precision and large statistics data
from Super-Kamiokande~\cite{SKatm} and it has received important
confirmation from the iron calorimeter detectors Soudan2~\cite{soudan}
and MACRO~\cite{MACRO}.  In June 1998, in the Neutrino98 conference,
SK presented \emph{evidence} of $\nu_\mu$ oscillations~\cite{SKatm}
based on the angular distribution for their contained event data
sample. Since then SK accumulated much more statistics and has also
studied the angular dependence of the upgoing muon sample.  In their
latest analyses Super-Kamiokande~\cite{skatmlast,skatm3nu} divides the
contained data sample into several subsamples according to the visible
energy in the event. On average contained events arise from neutrinos
with energies between several hundreds of MeV and several GeV. Upgoing
muons are divided in stopping muons (which arises from neutrinos
$E_\nu\sim 10$ GeV), and through-going muons (which are originated by
neutrinos with energies of the order of hundreds of GeV). 

The first run of Super-Kamiokande, usually referred as SK-I,
accumulated data during the period May 1996 to July 2001, and
corresponds to 1489 day exposure~\cite{skatmlast,skatm3nu}. After the
accident of 2001, the experiment resumed operation with a partial
coverage and during the so-called SK-II period (804 day exposure)
accumulated $\sim$ 50\% more statistics~\cite{skII-litch,skII-walter}.
In Fig.~\ref{fig:skatm} we show the data accumulated during both
periods~\cite{skII-walter}. Comparing the observed and the expected
(MC) distributions, we can make the following statements:
\begin{itemize}
  \item $\nu_e$ distributions are well described by the MC while
    $\nu_\mu$ presents a deficit. Thus the atmospheric neutrino
    deficit is mainly due to disappearance of $\nu_\mu$ and not the
    appearance of $\nu_e$.
    
  \item The suppression of contained $\mu$-like events is stronger for
    larger $\cos\theta$, which implies that the deficit grows with the
    distance traveled by the neutrino from its production point to the
    detector. This effect is more obvious for multi-GeV events because
    at higher energy the direction of the charged lepton is more
    aligned with the direction of the neutrino. It can also be
    described in terms of an up-down asymmetry which for the SK-I data
    is:
    \begin{equation} 
	\label{eq:aud}
	A_\mu \equiv \frac{U-D}{U+D}= -0.29 \pm 0.03
    \end{equation}
    where $U$ $(D)$ are the multi-GeV $\mu$-like events with zenith
    angle in the range $-1<\cos\theta<-0.2$ ($0.2<\cos\theta<1$).  It
    deviates from the SM value, $A_\mu=0$, by $\sim$ 10 standard
    deviations.
    
  \item The deficit on the number of through-going muons is smaller
    which implies that at larger energy the neutrino is less likely to
    disappear.  This is also parametrized in terms of the double ratio
    of the observed number versus expected number of through-going
    over stopping muons
    \begin{equation}
	\label{eq:rup}
	\frac{{N_\text{ST}} / {N_\text{TH}} |_\text{obs}}
	{{N_\text{ST}} / {N_\text{TH}} |_\text{MC}}
	= 0.53 \pm 0.16
    \end{equation}
    which deviates from the SM value of 1 by about 3 standard
    deviations.
\end{itemize}

These effects have been confirmed by the results of the iron
calorimeters Soudan2 and MACRO which removed the suspicion that the
atmospheric neutrino anomaly is simply a systematic effect in the
water detectors. The simplest and most direct interpretation of the
atmospheric neutrino anomaly is that of muon neutrino oscillations as 
we will discussed in Sec.~\ref{sec:3nu} and in the 
Appendix~\ref{sec:appatm}.


\subsection{Reactor Neutrinos}
\label{sec:expreac}

Neutrino oscillations are also searched for using neutrino beams from
nuclear reactors. Nuclear reactors produce $\bar\nu_e$ beams with
$E_\nu\sim$ MeV. Due to the low energy, $e$'s are the only charged
leptons which can be produced in the neutrino CC interaction. If the
$\bar\nu_e$ oscillated to another flavor, its CC interaction could not
be observed. Therefore oscillation experiments performed at reactors
are disappearance experiments. They have the advantage that smaller
values of $\Delta m^2$ can be accessed due to the lower neutrino beam
energy. 

There is a set of reactor experiments performed at relatively short or
intermediate baselines which did not find any positive evidence of
flavor mixing: Gosgen~\cite{gosgen}, Krasnoyarsk~\cite{krasnoyarsk},
Bugey~\cite{bugey}, CHOOZ~\cite{CHOOZ} and Palo
Verde~\cite{paloverde}. 

In particular CHOOZ searched for disappearance of $\bar{\nu}_e$'s
produced in a power station with two pressurized-water nuclear
reactors with a total thermal power of $8.5$ GW. At the detector,
located at $L\simeq 1$ km from the reactors, the $\bar{\nu}_e$
reaction signature is the delayed coincidence between the prompt $e^+$
signal and the signal due to the neutron capture in the Gd-loaded
scintillator.  The ratio between the measured and expected fluxes
averaged over the neutrino energy spectrum is given by
\begin{equation}
    \label{eq:rchooz}
    R = 1.01 \pm 2.8\% (\text{stat}) \pm 2.7\% (\text{syst}).
\end{equation}
Thus no evidence was found for a deficit in the flux. Furthermore
CHOOZ also presented their results in the form of the antineutrino
energy spectrum which showed no distortion. 

\begin{figure}\centering
    \includegraphics[width=2.5in]{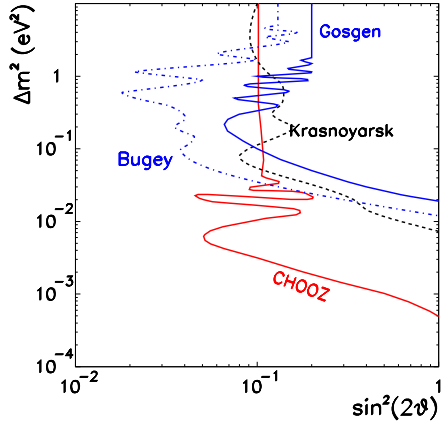}
    \caption{\label{fig:chooz2nu}%
      Excluded regions at 90\% for $\nu_e$ oscillations from searches
      in reactors experiments at short baselines.}
\end{figure}

In Fig.~\ref{fig:chooz2nu} we show the excluded regions in the
parameter space for two neutrino oscillations from these negative
results.  As we will see in Sec.~\ref{sec:3nu}, CHOOZ exclusion 
region extends to values of $\Delta m^2$ which are relevant for the
interpretation of atmospheric neutrino data. Consequently its results
play an important role in the global interpretation of the solar and
atmospheric neutrino data in the framework of three-neutrino mixing. 

Smaller values of $\Delta m^2$ can be accessed in a reactor experiment
using a longer baseline.  Pursuing this idea, the KamLAND 
experiment~\cite{klandprop}, a 1000 ton liquid scintillation detector,
is currently in operation in the Kamioka mine in Japan. This
underground site is located at an average distance of 150-210 km from
several Japanese nuclear power stations. The measurement of the flux
and energy spectrum of the $\bar\nu_e$'s emitted by these reactors 
provides a test of neutrino oscillations with $\Delta m^2\gtrsim
10^{-5}~\eVq$.

\begin{figure}\centering
    \includegraphics[width=8cm]{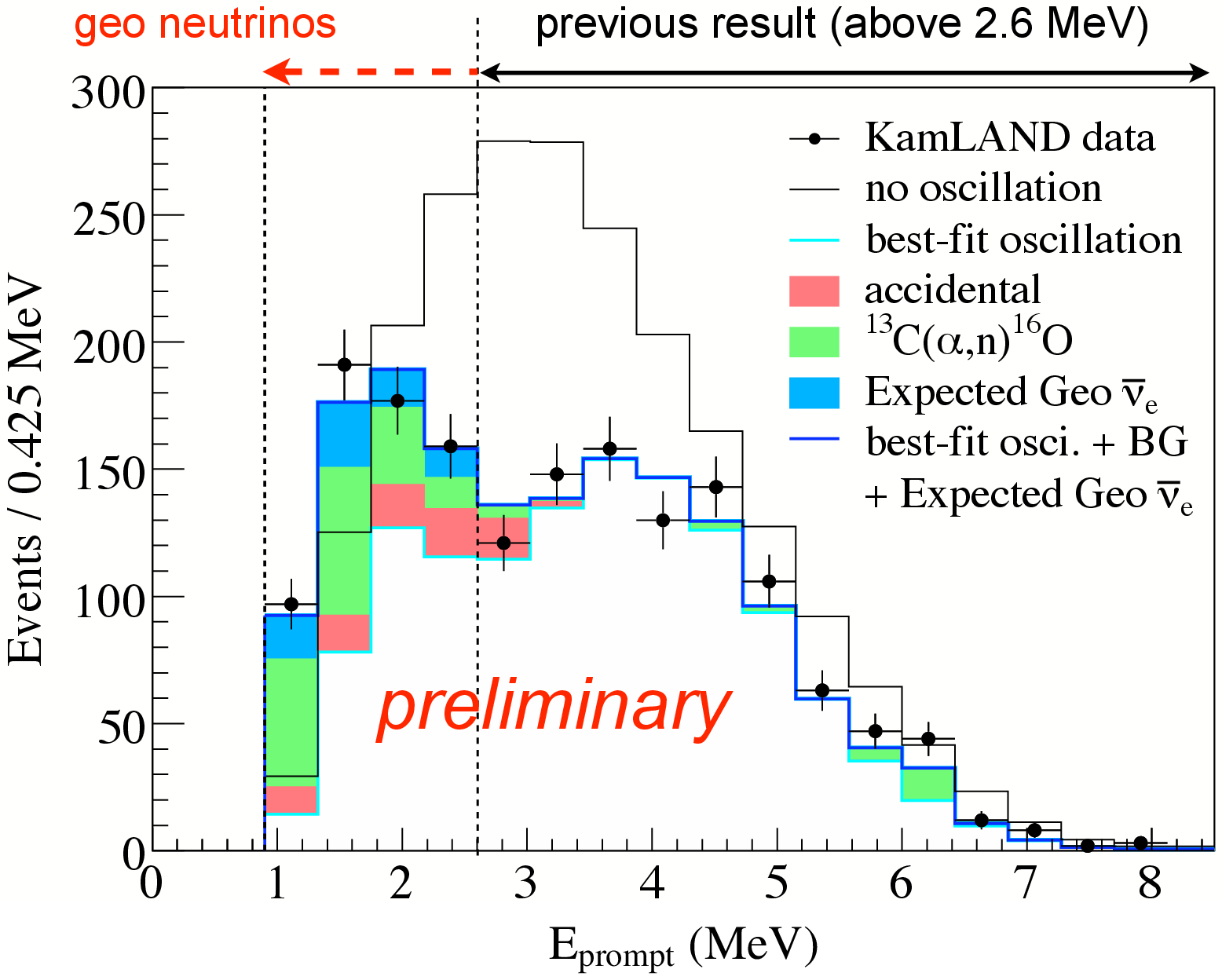}
    \caption{\label{fig:klandspec}%
      Prompt event energy spectrum of $\bar\nu_e$ events at KamLAND
      from Ref.~\cite{kland07}.}
\end{figure}

In their first result corresponding to an exposure of 162 ton-yr
(145.1 days), the ratio of the number of observed inverse
$\beta$-decay events to the number of events expected without
oscillations is $R_\text{KamLAND} ~=0.611 \pm 0.094$ for
$E_{\bar{\nu}_e}>$ 3.4 MeV~\cite{kland02}.  This deficit is
inconsistent with the expected rate for massless $\bar{\nu}_e$'s at
the 99.95\% confidence level. In June 2004 KamLAND also presented the
energy dependence of their events in the form of the prompt energy 
($E_\text{prompt}\simeq E_{\bar\nu_e}+m_p-m_n$)
spectrum~\cite{kland04}.  Finally, in September 2007 the collaboration
released a new analysis with increased statistics and a lower energy
threshold~\cite{kland07}. In Fig.~\ref{fig:klandspec} we show their
observed spectrum (from Ref.~\cite{kland07}) which clearly shows that
the deficit is energy dependent as expected from neutrino
oscillations. 


\subsection{Accelerator Neutrinos at Long Baselines}
\label{sec:explbl}

Conventional neutrino beams from accelerators are mostly produced by
$\pi$ decays (and some $K$ decays), with the pions produced by the
scattering of the accelerated protons on a fixed target:
\begin{equation}
    \begin{array}{lll}
	p + \text{target} \to & \multicolumn{2}{l}{\pi^\pm + X}
	\\
	& \pi^\pm \to & \mu^\pm + \nu_\mu (\bar \nu_\mu)
	\\
	& & \mu^\pm\to e^\pm + \nu_e (\bar\nu_e)
	+ \bar \nu_\mu (\nu_\mu)
    \end{array}
\end{equation}
Thus the beam can contain both $\mu$- and $e$-neutrinos and
antineutrinos. The final composition and energy spectrum of the
neutrino beam is determined by selecting the sign of the decaying
$\pi$ and by stopping the produced $\mu$ in the beam line. There is an
additional contribution to the electron neutrino and antineutrino 
flux from kaon decay.

Indeed the accelerator neutrino beams are very similar in nature to
the atmospheric neutrinos and they can be used to test the observed
oscillation signal with a controlled beam. Given the characteristic
$\Delta m^2$ involved in the interpretation of the atmospheric
neutrino signal, the intense neutrino beam from the accelerator must
be aimed at a detector located underground at a distance of several
hundred kilometers.

The first of these long baseline experiments with accelerator beams 
has been K2K~\cite{K2K} which run with a baseline of about 235 km from
KEK to SK.  MINOS~\cite{MINOS} is currently running with a baseline of
730 km from Fermilab, where the near detector is placed, to the Soudan
mine where the far detector is located.  

The results from both K2K and MINOS~\cite{K2K1,K2K2,MINOSdat}, both in
the observed deficit of events and in their energy dependence confirm
that accelerator $\nu_\mu$ oscillate over distances of several hundred
kilometers as expected from oscillations with the parameters
compatible with those inferred from the atmospheric neutrino data. 

\begin{figure}\centering
    \includegraphics[width=0.48\textwidth]{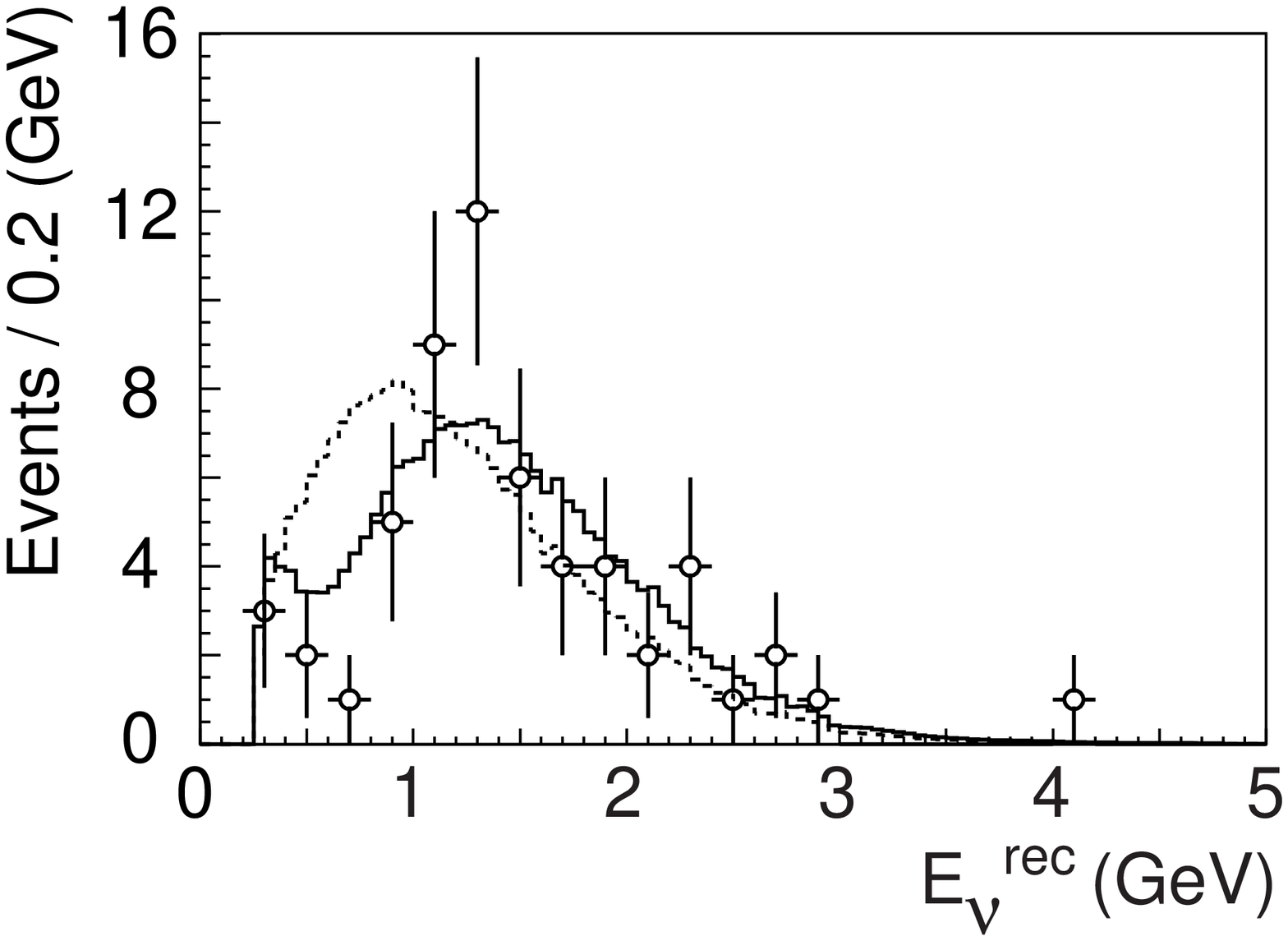}
    \hfill
    \includegraphics[width=0.48\textwidth]{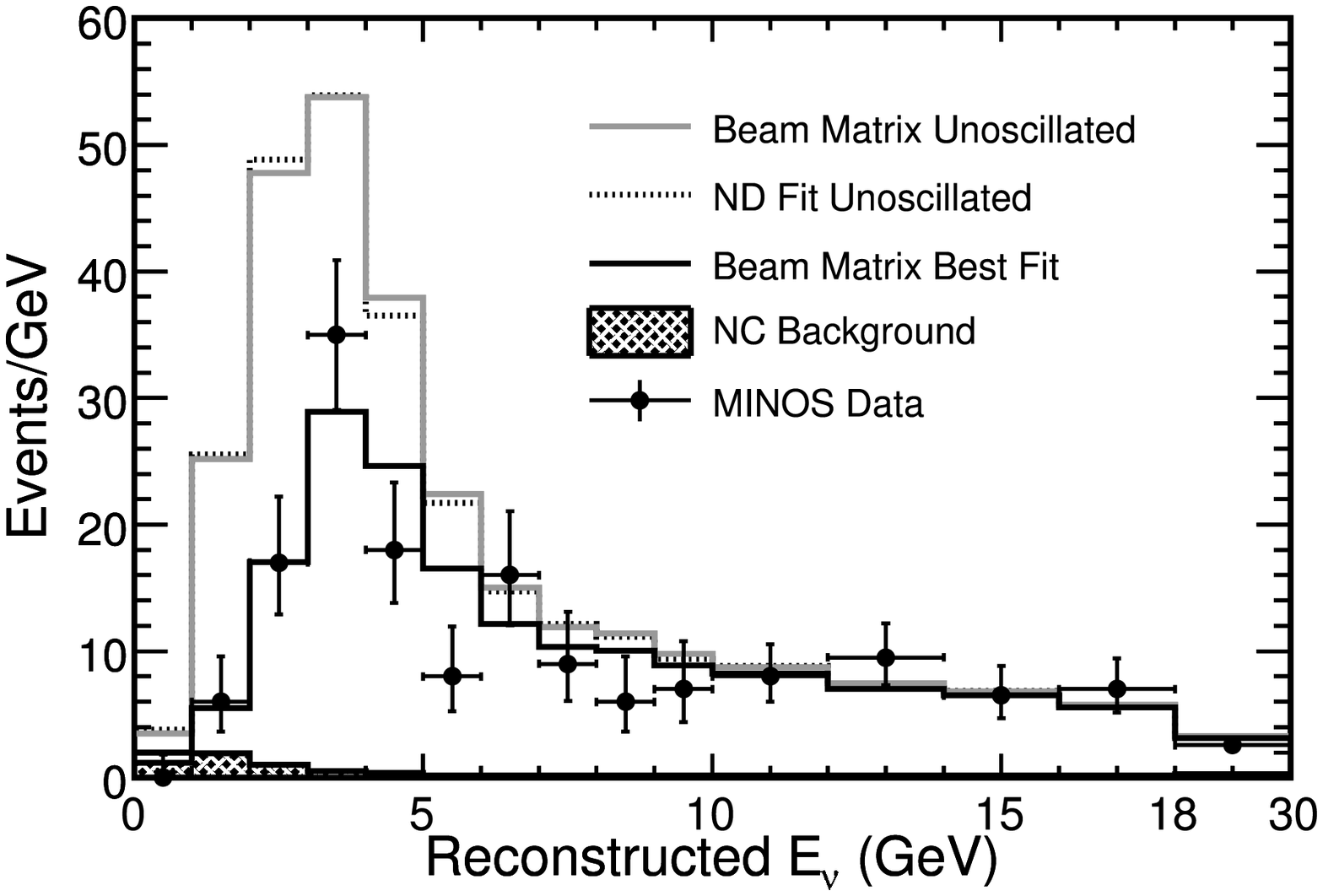}
    \caption{\label{fig:k2kminos}%
      (Left) The reconstructed $E_\nu$ distribution of the 1-ring
      $\mu$-like sample in K2K (from Ref.~\cite{K2K2}).  The dashed
      line is the expected spectrum without oscillation normalized to
      the observed number of events. The solid line is the best fit
      spectrum in presence of oscillations with $\Delta m^2 = 2.8
      \times 10^{-3}~\eVq$ and $\sin^2 2\theta=1$. (Right)
      Reconstructed energy spectra of the observed CC events in MINOS
      (from Ref.~\cite{MINOSdat}).  and comparison with the
      unoscillated and oscillated spectrum for $\Delta m^2 = 2.74
      \times 10^{-3}~\eVq$ and $\sin^2 2\theta = 1$.}
\end{figure}

In their last analysis K2K reported the observation of 107 fully
contained events while the expectation in the absence of oscillations
is $151^{+12}_{-10}$. In the left panel of Fig.~\ref{fig:k2kminos} we
show their latest data on the observed energy spectrum compared with
the expectations in the absence of oscillations as well as the best
fit in the presence of oscillations (from Ref.~\cite{K2K2}). 

The complete 5.4 kton MINOS far detector has been taking data since
the beginning of August 2003 and in March 2006 presented their first
results on the comparison of the rate and energy spectra of the
charged current neutrino interactions between the two detectors based
on a luminosity of $1.27\times 10^{20}$ protons on target.  They 
observed a total of 122 events below 10 GeV while the expectation
without oscillations is $238.7\pm 10.7$. In the right panel of
Fig.~\ref{fig:k2kminos} we show their published data on the observed
energy spectrum compared with the expectations in the absence of
oscillations as well as their best fit in the presence of oscillations
(from Ref.~\cite{MINOSdat}). In summer 2007 MINOS presented an updated
analysis based on a total integrated luminosity of $\sim$ $2.5\times
10^{20}$ protons on target~\cite{MINOSlast}.  In their preliminary
analysis they report a slightly lower value of $\Delta m^2 =
2.38^{+0.20}_{-0.16}~\eVq$ than in their previously published result.

The OPERA~\cite{OPERAprop,OPERA} neutrino detector at the underground
Gran Sasso Laboratory (LNGS) was designed to perform the first
detection of neutrino oscillations in appearance mode, through the
observation of $\nu_\tau$ appearance. It is placed in the high-energy,
long-baseline CERN to LNGS beam (CNGS) 730 km away from the neutrino
source. In August 2006 a first run with CNGS neutrinos was
successfully conducted. A first sample of neutrino events was
collected, statistically consistent with the integrated beam
intensity~\cite{OPERAdata}. 


\subsection{Accelerator Neutrinos at Short Baselines}

Most oscillation experiments performed with neutrino beams from
accelerators have characteristic distances of the order of hundreds of
meters.  We call them \emph{short baseline (SBL) experiments}.  With
the exception of the LSND experiment, which we discuss below, all
searches have been negative.  In Table~\ref{tab:sbl} we show the
limits on the various transition probabilities from the negative
results of the most restricting SBL experiments. Due to the short path
length, these experiments are not sensitive to the low values of
$\Delta m^2$ which we find when trying to explain either the solar or
the atmospheric neutrino data. 

\begin{table}\centering
    \catcode`?=\active \def?{\hphantom{0}}
    \caption{\label{tab:sbl}%
      90\% CL limit on the neutrino oscillation probabilities from the
      negative searches at short baseline experiments.} 
    \vspace{1mm}
    \begin{tabular}{|c|c|c|l|c|c|}
	\hline
	Experiment& Beam & Channel & \multicolumn{1}{c|}{Limit (90\%)}
	& $\Delta m^2_\text{min}$ (\eVq) & Ref.
	\\
	\hline
	CDHSW & CERN & $\nu_\mu\rightarrow \nu_\mu$
	& $P_{\mu\mu}>0.95$ & ?0.25? & \cite{CDHS}
	\\
	E776 & BNL  & $\nu_\mu\rightarrow \nu_e$
	& $P_{e\mu}<1.5\times 10^{-3}$ & ?0.075 & \cite{E776}
	\\
	E734 & BNL  & $\nu_\mu\rightarrow \nu_e$
	& $P_{e\mu}<1.6\times 10^{-3}$ & ?0.4?? & \cite{E734}
	\\
	KARMEN2 & Rutherford & $\bar\nu_\mu\rightarrow \bar\nu_e$ 
	& $P_{e\mu} <6.5\times 10^{-4}$ & ?0.05? & \cite{KARMEN}
	\\
	E531 & FNAL & $\nu_\mu\rightarrow \nu_\tau$
	& $P_{\mu\tau}<2.5\times 10^{-3}$ & ?0.9?? & \cite{E531}
	\\
	CCFR/ & FNAL 
	& $\nu_\mu\rightarrow \nu_e$
	& $P_{\mu e}<8\times 10^{-4}$ & ?1.6?? & \cite{CCFRmue,NUTEV}
	\\
	NUTEV & 
        & $\bar\nu_\mu\rightarrow \bar\nu_e$
	& $P_{\mu e}<5.5\times 10^{-4}$ & ?2.4?? & \cite{NUTEV}
        \\
        & & $\nu_\mu\rightarrow \nu_\tau$
	& $P_{\mu \tau}<4\times 10^{-3}$ & ?1.6?? & \cite{CCFRmt}
	\\ 
	& & $\nu_e\rightarrow \nu_\tau$
	& $P_{e\tau}<0.1$ & 20.0?? &\cite{CCFRet}
	\\
	Chorus& CERN & $\nu_\mu\rightarrow \nu_\tau$
	& $P_{\mu\tau}<3.4\times 10^{-4}$ & ?0.6?? & \cite{CHORUS}
	\\
	& & $\nu_e\rightarrow \nu_\tau$
	& $P_{e\tau}<2.6\times 10^{-2}$ & ?7.5?? & \cite{CHORUS}
	\\
	Nomad & CERN & $\nu_\mu\rightarrow \nu_\tau$
	& $P_{\mu\tau}<1.7\times 10^{-4}$ & ?0.7?? & \cite{NOMAD}
	\\
	& & $\nu_e\rightarrow \nu_\tau$
	& $P_{e\tau}<7.5\times 10^{-3}$ & ?5.9?? & \cite{NOMAD}
	\\
	& & $\nu_\mu\rightarrow \nu_e$
	& $P_{\mu e}<6\times 10^{-4}$ & ?0.4?? & \cite{NOMAD}
	\\
	\hline
    \end{tabular}
\end{table}

\begin{figure}\centering
    \raisebox{1.2mm}{\includegraphics[height=66mm]{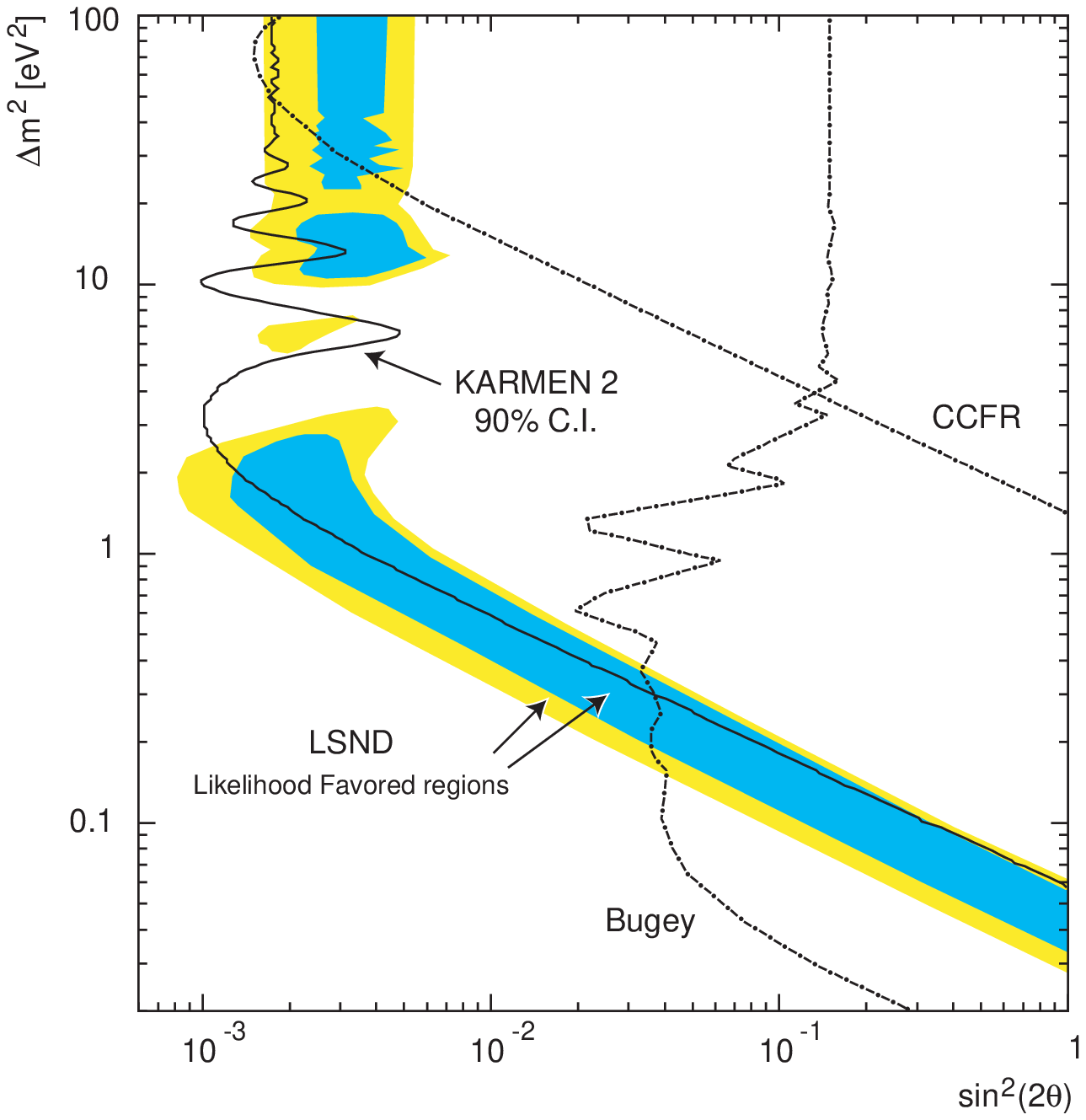}}
    \hfill
    \includegraphics[height=67.7mm]{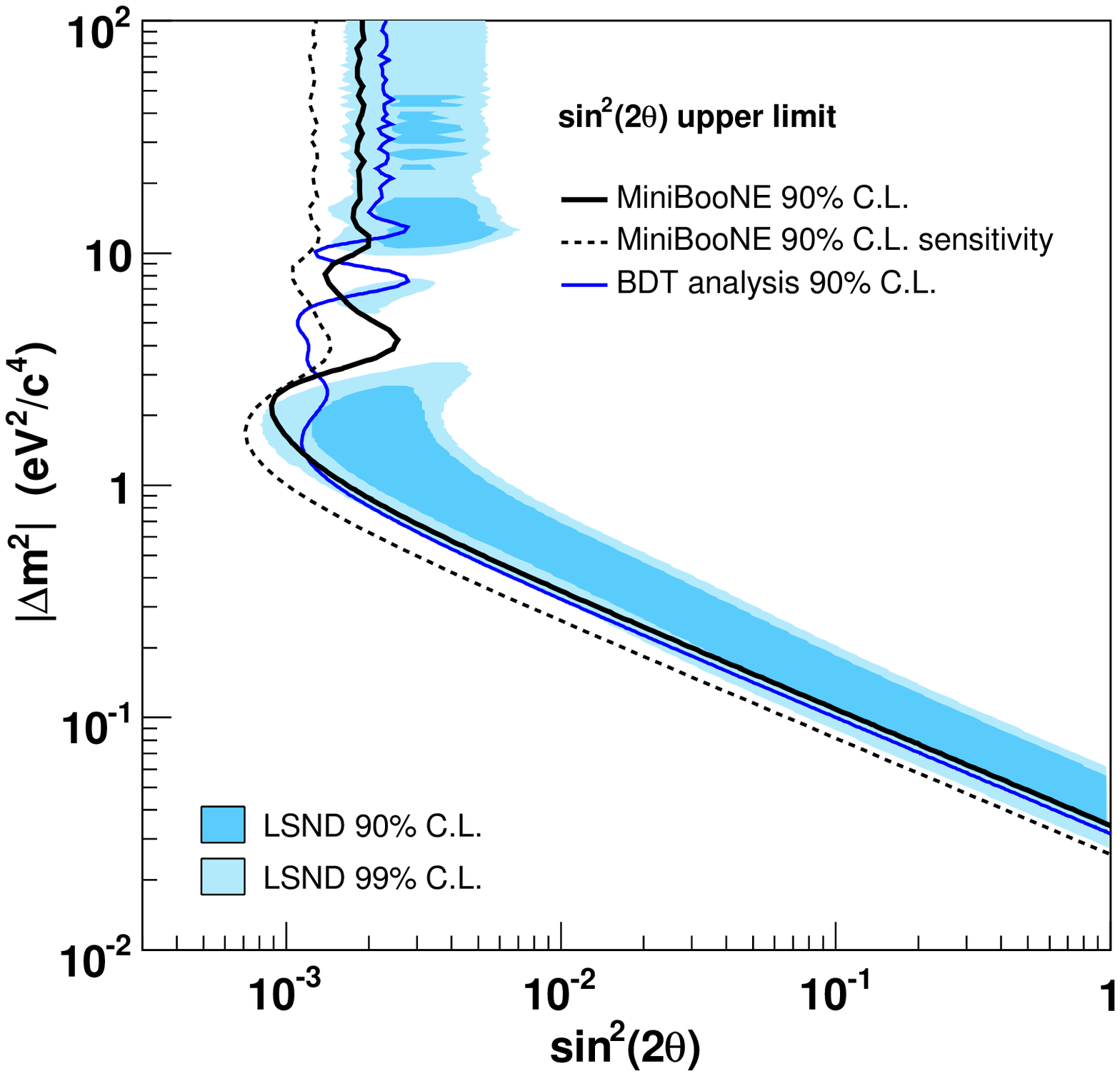}
    \caption{\label{fig:lsnd}%
      (Left) Allowed regions (at 90\% and 99\% CL) for
      $\nu_e\to\nu_\mu$ oscillations from the LSND experiment compared
      with the exclusion regions (at 90\% CL) from KARMEN2 and other
      experiments (from Ref.~\cite{KARMEN}).  (Right) 
      $\nu_\mu\rightarrow\nu_e$ excluded regions from MiniBooNE 
      compared to the LSND allowed region (from
      Ref.~\cite{miniboonelast}).}
\end{figure}

The only positive signature of oscillations at a short baseline 
laboratory experiment comes from the Liquid Scintillator Neutrino
Detector (LSND)~\cite{LSND} running at Los Alamos Meson Physics
Facility.  Its primary neutrino flux came from $\pi^+$'s produced in a
30-cm-long water target when hit by protons from the LAMPF linac with
800 MeV kinetic energy.  The detector was a tank filled with 167
metric tons of dilute liquid scintillator, located about 30 m from the
neutrino source.  The experiment observed an excess of events as
compared to the expected background while the excess was consistent
with $\bar\nu_\mu\to \bar\nu_e$ oscillations.  In the latest results
the total fitted excess was of $87.9 \pm 22.4 \pm 6$ events,
corresponding to an oscillation probability of $(2.64 \pm 0.67 \pm
0.45) \times 10^{-3}$. For oscillations between two neutrino states
these results lead to the oscillation parameters shown in
Fig.~\ref{fig:lsnd}.  The shaded regions are the 90\% and 99\%
likelihood regions from LSND. 

The region of parameter space which is favored by the LSND
observations has been partly tested by other experiments like the
KARMEN~\cite{KARMEN} experiment and very recently by 
MiniBooNE~\cite{miniboonelast}.

The KARMEN experiment was performed at the neutron spallation facility
ISIS of the Rutherford Appleton Laboratory.  They found a number of
events in good agreement with the total background expectation. The
corresponding exclusion curve from the final data set recorded with 
the full experimental set up of KARMEN2 in the two-neutrino parameter
space is given in Fig.~\ref{fig:lsnd} together with the favored
region for the LSND experiment. At large $\Delta m^2$, KARMEN2 results
exclude the region favored by LSND. At low $\Delta m^2$, KARMEN2
leaves some allowed space, but the reactor experiments at Bugey and
CHOOZ add stringent limits for the larger mixing angles. 

In Ref.~\cite{Church:2002tc} a combined statistical analysis of the
experimental results of the LSND and KARMEN search was performed. At a
combined confidence level of 36\,\%, they found no area of oscillation
parameters compatible with both experiments. For the complementary
confidence of $1-0.36=64$\,\%, they found two well defined regions of
oscillation parameters with either $\Delta m^2 \approx 7~\eVq$ or
$\Delta m^2 < 1~\eVq$ compatible with both experiments.

\subsubsection{MiniBooNE}

The MiniBooNE experiment~\cite{miniboone}, currently running at
Fermilab, searches for $\nu_\mu \to \nu_e$ oscillations and was
specially designed to make a conclusive statement about the LSND's
neutrino oscillation evidence.  In their 2002-2005 run they used a
$\nu_\mu$ beam of energy $0.5 - 1.0$~GeV initiated by a primary beam
of 8.89~GeV protons from the Fermilab Booster impinging on a 71~cm
long and 1~cm diameter beryllium target. The target is located inside
a magnet focusing horn.  The beam contains only a small intrinsic
$\nu_e$ component.  In January 2006, MiniBooNE switched the polarity
of the horn to select negative sign mesons and since then the
experiment has been collecting data using a beam of antineutrinos.

The MiniBooNE detector is a 12.2~m diameter sphere filled with 800
tons of pure mineral oil. The center of the detector is positioned at
a distance of 541~m from the front of the Beryllium target. The vessel
consists of two optically isolated regions separated by a support
structure. The inner region of 5.5~m radius is the neutrino target
region, while the outer volume forms the veto region. 
 
\begin{figure}\centering
    \includegraphics[width=0.6\textwidth]{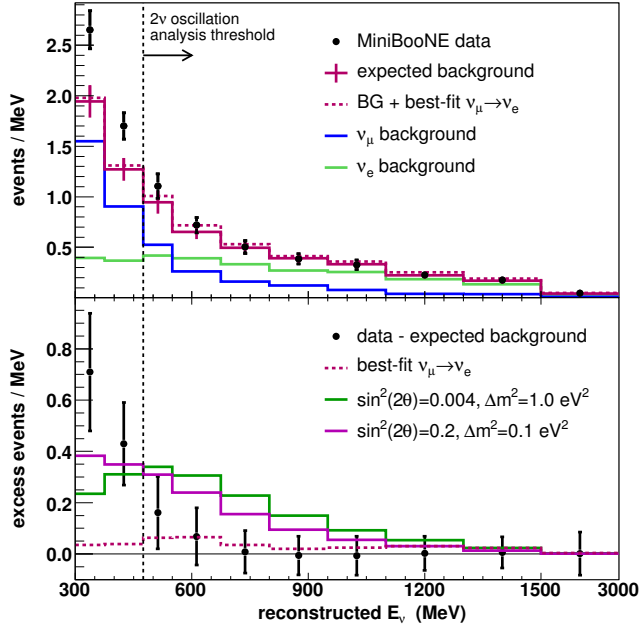}
    \caption{\label{fig:mbspec}%
      Reconstructed $E_\nu$ of the observed $\nu_e$ events recorded by
      MiniBooNE compared with the expected background and the
      expectations in the presence of oscillations with oscillation
      parameters characteristic of the LSND signal (from
      Ref.~\cite{miniboonelast}).}
\end{figure}

In their analysis of the neutrino data released on April
2007~\cite{miniboonelast} they studied the events with reconstructed
neutrino energy $300 < E_\nu < 3000$~MeV. Their observed spectrum of
$\nu_e$ events is shown in Fig.~\ref{fig:mbspec}. As seen in the
figure the spectrum presents and excess of $96\pm 17\pm 20$ events for
$E_\nu <475$~MeV while there is no significant excess of events for
$E_\nu >475$ ($22\pm 19\pm 35$ events above expectation).  In
Ref.~\cite{miniboonelast} MiniBooNE claims that the low-energy excess
cannot be explained by a $2\nu$-oscillation model and consequently the
collaboration performed their oscillation analysis using only the
events with $E_\nu >475$~MeV.  With this data they find a $\chi^2$
probability of 93\% for the null oscillation hypothesis and at 90\% CL
their single-sided roster scan excluded region shows no overlap with 
the 90\% allowed region of the LSND evidence for $2\nu$ oscillations
as seen in the right panel of Fig.~\ref{fig:lsnd} (from
Ref.~\cite{miniboonelast}). In Ref.~\cite{miniboonelast} they also
performed a joint analysis of LSND and MiniBooNE which excludes the
$2\nu$ oscillation hypothesis as an explanation of the LSND anomaly at
98\% CL.

%% file: sec.3nu.tex
\section{3-$\nu$ Mixing}
\label{sec:3nu}


\subsection{Dominant 2-$\nu$ Oscillations for Solar Neutrinos and
  KamLAND}
\label{sec:2nusolar}

The simplest explanation of the solar neutrino data described in
Sec.~\ref{sec:expsolar} is the oscillations of $\nu_e$ into an active
($\nu_\mu$ and/or $\nu_\tau$) or a sterile ($\nu_s$) neutrino.
Oscillations into pure sterile neutrinos are strongly disfavored by
the SNO data because if the beam comprises of only $\nu_e$'s and
$\nu_s$'s, the three observed CC, ES and NC fluxes should be equal (up
to effects due to spectral distortions), an hypothesis which is now
ruled out at more than $7\sigma$ by the SNO data (see
Eq.~\eqref{eq:snoev}).

The goal of the analysis of the solar neutrino data in terms of
neutrino oscillations is to determine which range of mass-squared
difference and mixing angle can be responsible for the observed
deficit. In order to answer this question in a statistically
meaningful way one must compare the predictions in the different
oscillation regimes with the observations, including all the sources
of uncertainties and their correlations. In the present analysis the
main sources of uncertainty are the theoretical errors in the 
prediction of the solar neutrino fluxes for the different reactions.
These errors are due to uncertainties in the twelve basic ingredients
of the solar model, which include the nuclear reaction rates
(parametrized in terms of the astrophysical factors $S_{11}$,
$S_{33}$, $S_{34}$, $S_{1,14}$ and $S_{17}$), the solar luminosity,
the metallicity $Z/X$, the Sun age, the opacity, the diffusion, and the
electronic capture of \Nuc{7}{Be}, $C_\text{Be}$. Another source of
theoretical error arises from the uncertainties in the neutrino
interaction cross section for the different detection processes. A
detailed description of the way to include all these uncertainties and
correlations can be found in Ref.~\cite{review} and references
therein.

\begin{figure}\centering
    \includegraphics[height=0.4\textheight]{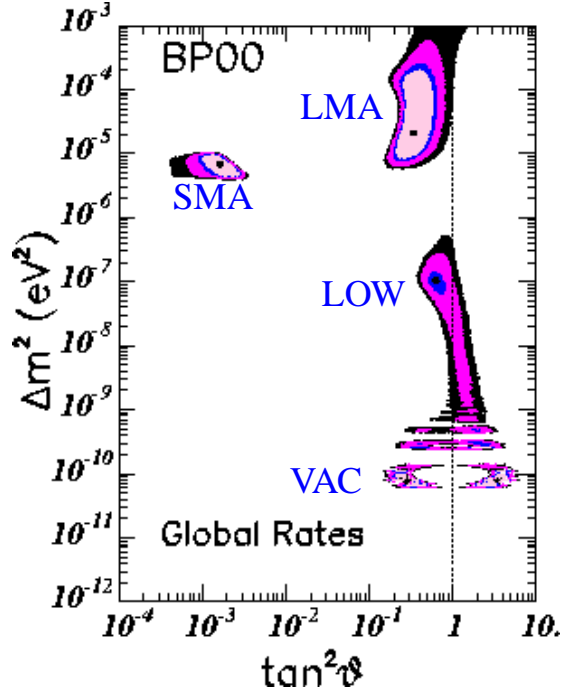}
    \caption{\label{fig:solrates}%
      Allowed oscillation parameters (at 90\%, 95\%, 99\% and 99.73\%
      CL) from the analysis of the total event rates of the Chlorine,
      Gallium, SK and the first SNO CC experiments (adapted from
      Ref.~\cite{oursnocc}).}
\end{figure}

As illustration we show in Fig.~\ref{fig:solrates} the results of the
analysis of the total event rates as it was in the summer of 2001
including the total rates from Chlorine, Gallium, SK and the first
determination of the CC event rates at SNO.  In the figure we plot the
allowed regions which correspond to 90\%, 95\%, 99\% and 99.73\%
($3\sigma$) CL for $\nu_e$ oscillations into active neutrinos
(2~d.o.f.). As seen in the figure, there were several oscillation
regimes compatible within errors with the experimental data. These
allowed parameter regions are denoted as \emph{MSW small mixing angle}
(SMA), \emph{MSW large mixing angle} (LMA), \emph{MSW low mass} (LOW)
and \emph{vacuum oscillations} (VAC).

For the LMA solution, oscillations for the \Nuc{8}{B} neutrinos occur
in the adiabatic regime and the survival probability is higher for
lower energy neutrinos. This situation fits well the higher rate
observed at gallium experiments. For the LOW solution, the situation
is opposite but matter effects in the Earth for pp and \Nuc{7}{Be}
neutrinos enhance the average annual survival probability for these
lower energy neutrinos. The combination of these effects still allows
a reasonable description of the Gallium rate. For the SMA solution the
oscillations for the \Nuc{8}{B} neutrinos occur in the non-adiabatic
regime while for the VAC solution the oscillation wavelength is of the
order of the Sun-Earth distance for \Nuc{8}{B} neutrinos.  

Further information on the different oscillation regimes can be
obtained from the analysis of the energy and time dependence data from
SK and SNO. For example, for LMA and LOW, the expected energy spectrum
at these experiments is very little distorted.  Also in the lower part
of the LMA region and in the upper part of the LOW region matter
effects in the Earth are important and some day-night variation is
expected.  For SMA, a positive slope of the energy spectrum is
predicted, with larger slope for larger mixing angle within SMA. For
VAC, large distortions of the energy spectrum are expected as imprints
of the $L/E$ dependence of the survival probability. The
quantification of these effects depends on the precise values of the
oscillation parameters.

The observed day-night spectrum in SK and SNO are essentially
undistorted in comparison to the SSM expectation and shows no
significant differences between the day and the night periods as
commented in Sec.~\ref{sec:expsolar}.  Consequently, a large region of
the oscillation parameter space where these variations are expected to
be large can be excluded. In particular:
\begin{itemize}
  \item SMA: within this region, the part with larger mixing angle 
    fails to comply with the observed energy spectrum, while the part
    with smaller mixing angles gives a not good enough fit to the
    total rates. 
    
  \item VAC: the observed undistorted energy spectrum cannot be
    accommodated.
    
  \item LMA and LOW: the small $\Delta m^2$ part of LMA and the LOW
    solution are eliminated because they predict a day-night variation
    that is larger than observed.
\end{itemize}

\begin{figure}\centering
    \includegraphics[height=0.35\textheight]{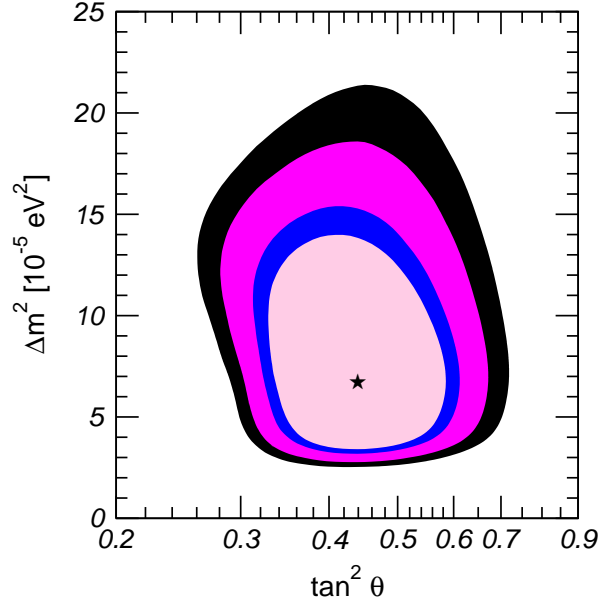}
    \caption{\label{fig:solar2nu}%
      Allowed oscillation parameters (at 90\%, 95\%, 99\% and 99.73\%
      CL) from the global analysis of the solar neutrino data.}
\end{figure}

Thus with the inclusion of the time and energy dependence of the
\Nuc{8}{B} neutrino fluxes at SK and SNO it was possible to select the
LMA as the most favored solution to the solar neutrino problem.  We
show in Fig.~\ref{fig:solar2nu} the allowed region of parameters which
correspond to 90\%, 95\%, 99\% and 99.73\% ($3\sigma$) CL for $\nu_e$
oscillations from the global analysis of the latest solar neutrino
data. The Borexino results are in perfect agreement with the 
expectations within the LMA region, but they are not included in the 
analysis because they are still not precise enough to have an impact
on the determination of the oscillation parameters.  

\begin{figure}\centering
    \includegraphics[height=0.35\textheight]{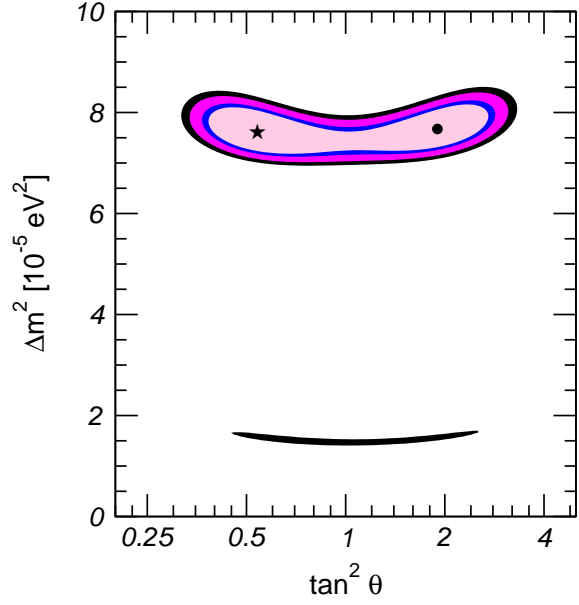}
    \caption{\label{fig:kland2nu}%
      Allowed oscillation parameters (at 90\%, 95\%, 99\% and 99.73\%
      CL) from the analysis of KamLAND data. Based on calculations by
      T.~Schwetz.}
\end{figure}

As mentioned in Sec.~\ref{sec:expe} these small values of $\Delta m^2$
could also be accessed the terrestrial experiment KamLAND using as
beam the $\bar\nu_e$'s from nuclear reactors located over distances of
the order of hundred kilometers.  Indeed the KamLAND results can be
interpreted in terms of $\bar\nu_e$ oscillations with parameters shown
in Fig.~\ref{fig:kland2nu}.\footnote{The analysis of the KamLAND
experiment presented here and in Sec.~\ref{sec:cptviol} is based on
the calculations performed by T.~Schwetz.}

\begin{figure}\centering
    \includegraphics[height=0.35\textheight]{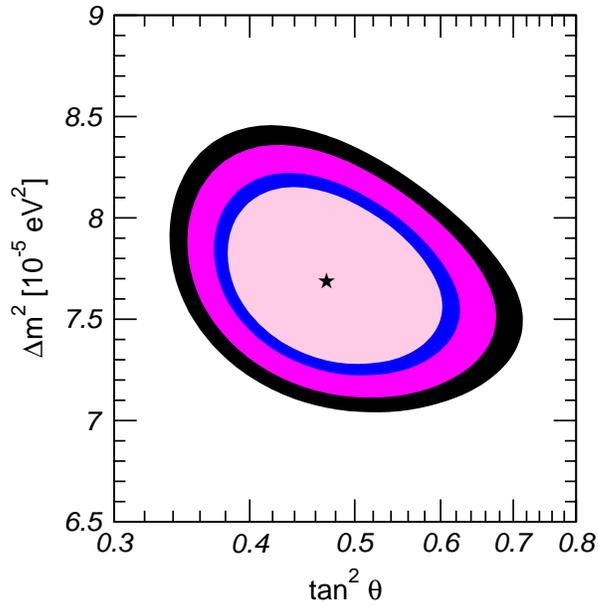}
    \caption{\label{fig:solkland2nu}%
      Allowed oscillation parameters (at 90\%, 95\%, 99\% and 99.73\%
      CL) from the combined analysis of solar and KamLAND data.}
\end{figure}

The most important aspect of Fig.~\ref{fig:kland2nu} is the
demonstration by KamLAND that anti-neutrinos oscillate with parameters
that are consistent with the LMA solution of the solar neutrino
problem. Under the assumption that CPT is satisfied, the anti-neutrino
measurements by KamLAND apply directly to the neutrino sector and the
two sets of data can be combined to obtain the globally allowed
oscillation parameters. The results of such an analysis are shown in
Fig.~\ref{fig:solkland2nu}.

\subsection{Dominant 2-$\nu$ Oscillations for Atmospheric 
  and LBL Neutrinos}
\label{sec:2nuatm}

The simplest and most direct interpretation of the atmospheric
neutrino data described in Sec.~\ref{sec:expatmos} is that of muon
neutrino oscillations. The required value of the oscillation
parameters can be easily estimated from the following observations:
\begin{itemize}
  \item The angular distribution of contained events shows that, for
    $E\sim 1$ GeV, the deficit comes mainly from $L\sim 10^2-10^4$ km.
    The corresponding oscillation phase must be maximal, $\frac{\Delta
    m^2 (\eVq) L(\text{km})}{2E(\text{GeV})}\sim 1$, which requires
    $\Delta m^2\sim 10^{-4}-10^{-2}~\eVq$.
    
  \item Assuming that all upgoing $\nu_\mu$'s which would lead to
    multi-GeV events oscillate into a different flavor while none of
    the downgoing ones do, the up-down asymmetry is given by $|A_\mu|
    = \sin^2 2\theta / (4-\sin^2 2\theta)$. The present one sigma
    bound reads $|A_\mu|>0.27$ which requires that the mixing angle is
    close to maximal, $\sin^2 2\theta > 0.85$.
\end{itemize}
In order to go beyond these rough estimates, one must compare in a 
statistically meaningful way the experimental data with the 
theoretical expectations. The most up-to-date details on our
determination of the expected rates in SK-I and SK-II and the
corresponding statistical analysis can be found in
Appendix~\ref{sec:appatm}.

\begin{figure}\centering
    \includegraphics[height=0.3\textheight]{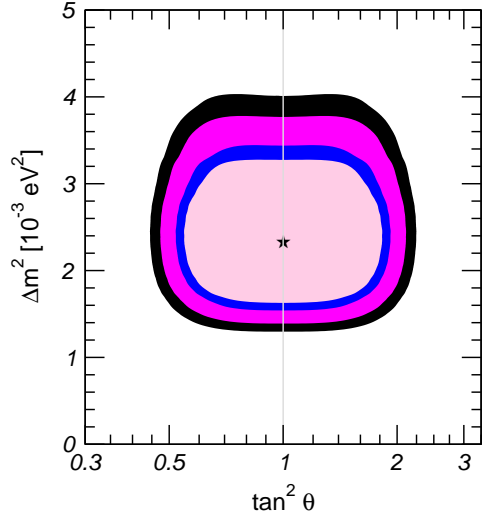}
    \caption{\label{fig:atmos2nu}%
      Allowed regions from the analysis of atmospheric neutrino data.
      The different contours correspond to at 90\%, 95\%, 99\% and
      $3\sigma$ CL.}
\end{figure}

Altogether the best interpretation of the atmospheric neutrino data is
the oscillation of $\nu_\mu$ into $\nu_\tau$.  In
Fig.~\ref{fig:atmos2nu} we plot the allowed regions from the global
analysis of atmospheric data.

Other oscillation channels are presently ruled out.  
$\nu_\mu\to\nu_e$ is excluded with high CL as the explanation to the
atmospheric neutrino anomaly for two different reasons:
\begin{itemize}
  \item SK high precision data show that the $\nu_e$ contained events
    are very well described by the SM prediction both in normalization
    and in their zenith angular dependence;
    
  \item Explaining the atmospheric data with $\nu_\mu\to\nu_e$
    transition has direct implications for the
    $\bar\nu_e\to\bar\nu_\mu$ transition.  In particular, there should
    be a $\bar\nu_e$ deficit in the CHOOZ reactor experiment which was
    not observed.
\end{itemize}
$\nu_\mu\rightarrow\nu_s$ is also ruled out as a possible explanation
of the atmospheric neutrino anomaly because the presence of matter
effects in this channel predict a flatter-than-observed angular
distribution of thru-upgoing muon events~\cite{Lipari:1998rf}.  Also
if $\nu_\mu$ oscillates into sterile neutrinos one expects a relative
suppression of the NC signal which has not been
observed~\cite{skatmsterile}. Furthermore recently Super-Kamiokande has
performed a dedicated analysis for the search for the effects of
appearance of tau neutrinos which disfavors the hypothesis of no
$\nu_\tau$ appearance at $2.4\sigma$~\cite{skatmtauap}.

\begin{figure}\centering
    \includegraphics[height=0.3\textheight]{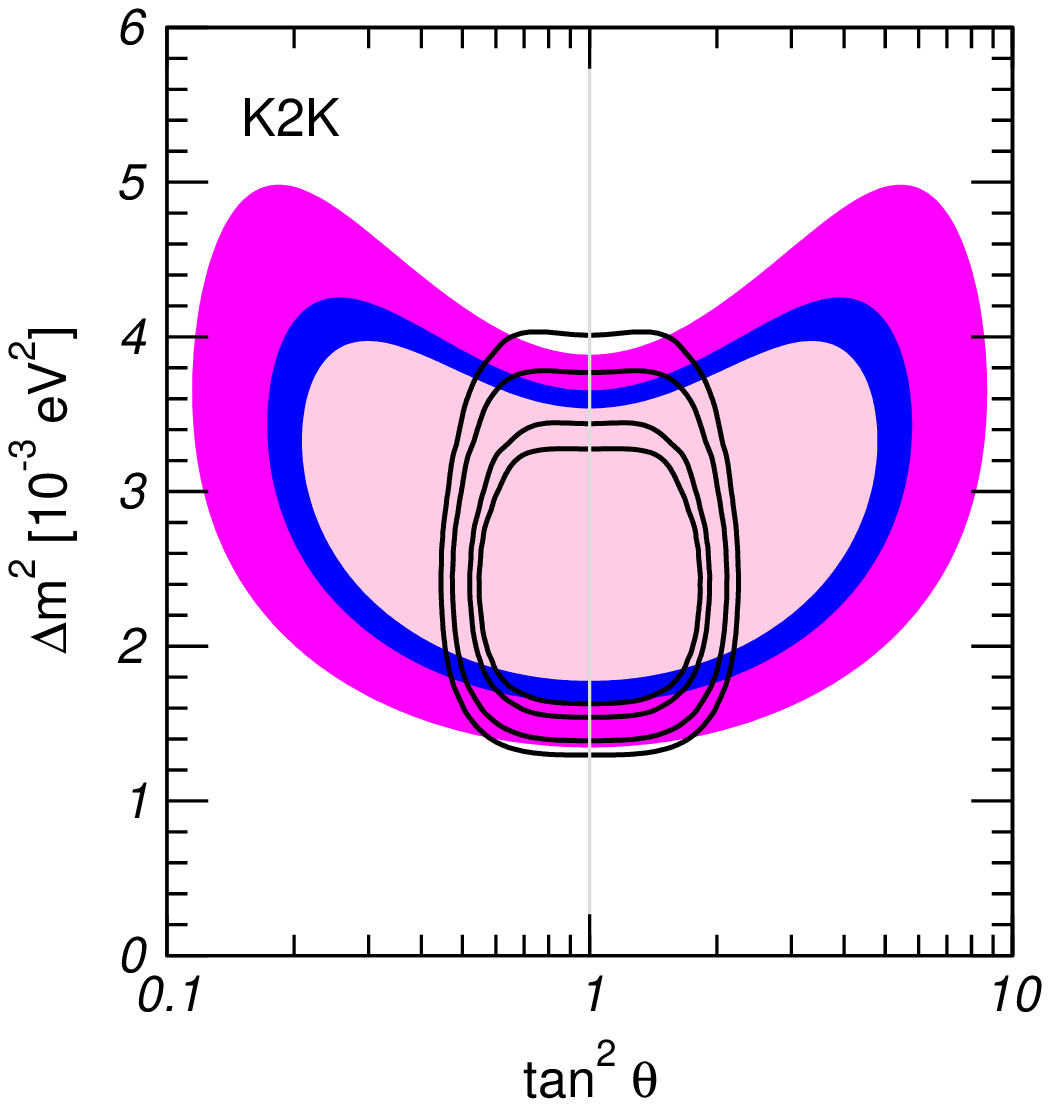}
    \includegraphics[height=0.3\textheight]{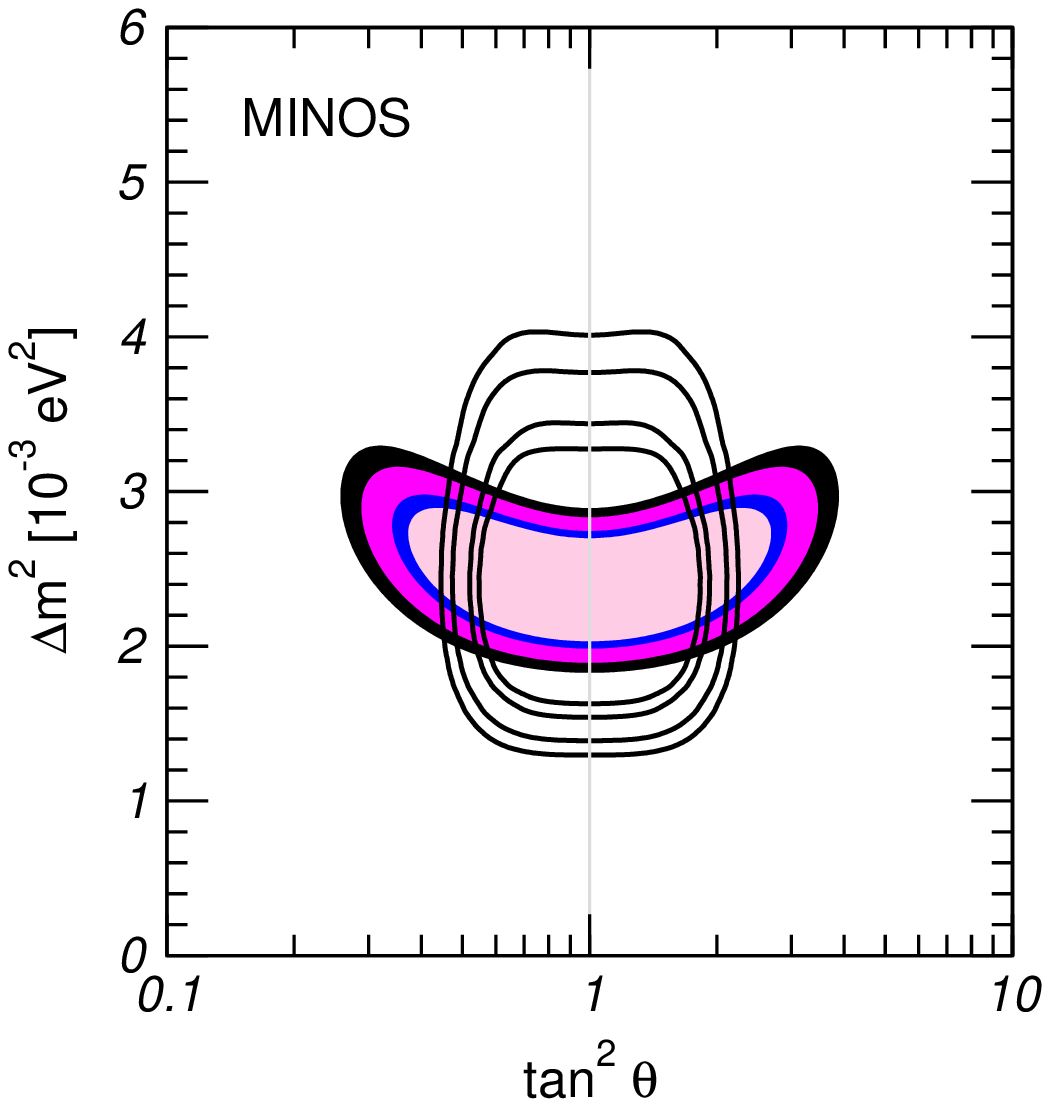}
    \caption{\label{fig:lbl2nu}%
      Allowed regions from the analysis of K2K (left) and MINOS
      (right) (full regions). For K2K only the 90\%, 95\% and and 99\%
      CL regions are shown.  For comparison we also show the
      corresponding allowed regions from ATM neutrinos at the same CL
      (lines).}
\end{figure}

As we have described in Sec.~\ref{sec:explbl}, the results of the LBL
experiments K2K and MINOS confirm, both in the observed deficit of
events and in their energy dependence, that accelerator $\nu_\mu$
oscillate over distances of several hundred kilometers as expected
from oscillations with the parameters previously inferred from the
atmospheric neutrino data. This is quantitatively illustrated in
Fig.~\ref{fig:lbl2nu} where we show the results of our analysis of the
K2K and MINOS data respectively.  As seen in the figure, both K2K and
MINOS provide an independent determination of the relevant $\Delta
m^2$ and, in particular, they favor the upper part of the atmospheric
mass-splitting while they have very limited sensitivity to the mixing
angle which is still dominantly determined by the atmospheric neutrino
data.

\subsection{Subdominant 3-$\nu$ Oscillation Effects}
\label{sec:3nuprobs}

From the results previously described it is obvious that the minimum
joint description of solar and atmospheric evidences requires that all
three known neutrinos take part in the oscillations.  In this case,
the mixing parameters are encoded in the $3 \times 3$ lepton mixing
matrix~\cite{MNS,ckm} which can be conveniently parametrized in the
standard form of Eq.~\eqref{eq:U3d}, since the two Majorana phases in
Eq.~\eqref{eq:U3m} do not affect neutrino
oscillations~\cite{BHP,majosc2}.

The determination of the oscillation probabilities for both solar and 
atmospheric neutrinos requires that one solves the evolution equation
of the neutrino system, Eq.~\eqref{eq:evol.5}, in the matter
background of the Sun and the Earth.  In the three-flavor framework,
the equation in the flavor basis can be written as: 
\begin{equation}
    \label{eq:evol.31} 
    i \frac{d\vec\nu}{dx} = H \, \vec\nu, \qquad
    H = U \cdot H_0^d \cdot U^\dagger + V \,,
\end{equation}
where $U$ is the lepton mixing matrix, $\vec\nu \equiv (\nu_e,\,
\nu_\mu,\, \nu_\tau)^T$ and
\begin{equation}
    \label{eq:evol.33}
    H_0^d =H_m-\frac{m_1}{2E}=
\frac{1}{2E}
    \diag\left( 0,\, \Delta m^2_{21},\, \Delta m^2_{31}\right) \,.
\end{equation}
$V$ is the effective potential that describes CC forward interactions
in matter, Eq.~\eqref{eq:evol.4}.

In total the three-neutrino oscillation analysis involves six
parameters: two mass differences (including two possible signs for one
of them), three mixing angles and one CP phase.  

\begin{figure}\centering
    \includegraphics[scale=0.4]{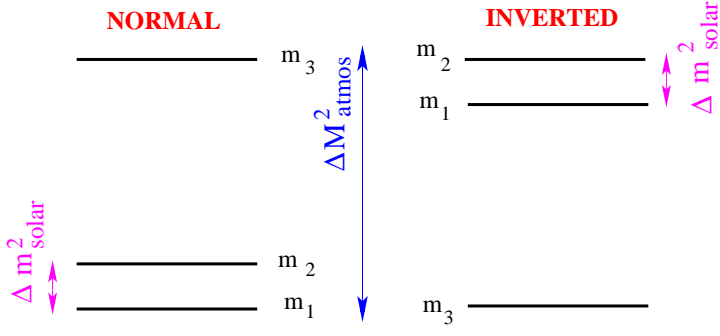}
    \caption{\label{fig:schemes}%
      Mass schemes for $3\nu$ oscillations}
\end{figure}

Without loss of generality one can chose the mass differences as shown
in Fig.~\ref{fig:schemes} so that $\Delta m^2_{21}$ is always
positive and there are two possible mass orderings which we denote as
\emph{normal} and \emph{inverted} and which correspond to the two
possible choices of the sign of $\Delta m^2_{31}$.  In this convention
the angles $\theta_{ij}$ can be taken without loss of generality to
lie in the first quadrant, $\theta_{ij} \in [0,\pi/2]$ and the phases
$\delta_\text{CP},\; \eta_i\in [0,2\pi]$.

The normal ordering is naturally related to hierarchical masses,
$m_1\ll m_2\ll m_3$, for which $m_2\simeq\sqrt{\Delta m^2_{21}}$ and 
$m_3\simeq\sqrt{\Delta m^2_{32}}$, or to quasi-degenerate masses,
$m_1\simeq m_2\simeq m_3\gg \Delta m^2_{21}, \Delta m^2_{32}$. On the
other hand, the inverted ordering implies that $m_3< m_1\simeq m_2$.  

With this assignment (see more below) $\Delta m^2_{21}$ and the mixing
angle $\theta_{12}$ have been chosen to be those that give the
dominant oscillations for solar neutrinos while $\Delta m^2_{31}$,
$\Delta m^2_{32}$ and $\theta_{23}$ give the dominant oscillation for
atmospheric neutrinos. 

Generic three-neutrino oscillation effects are:
\begin{itemize}
  \item Mixing effects because of the additional angle $\theta_{13}$
  \item Difference between Normal and Inverted schemes,
  \item Coupled oscillations with two different oscillation lengths,
  \item CP violating effects.
\end{itemize}
The strength of these effects is controlled by the values of the ratio
of mass differences, the mixing angle $\theta_{13}$ and the CP phase
$\delta_\text{CP}$.

In this respect, as we have seen in the previous sections, the
parameter space of solutions for solar and atmospheric oscillations in
Figs.~\ref{fig:atmos2nu} and~\ref{fig:solkland2nu} satisfy
\begin{equation}
    \label{eq:deltahier}
    \Delta m^2_{21}=\Delta m^2_\odot \ll \Delta m^2_\text{atm}=
|\Delta m^2_{31}|\simeq |\Delta m^2_{32}|.
\end{equation}
This hierarchy implies that even though in general the transition 
probabilities present an oscillatory behavior with two oscillation
lengths, in present experiments, such interference effects are not
very visible.

In this notation, the survival probability of reactor antineutrinos at
CHOOZ takes the form: 
\begin{equation} \begin{split}
    \label{eq:pchooz}
    P_{ee}^\text{CHOOZ}
    &= 1 - \cos^4\theta_{13}\sin^22\theta_{12}
    \sin^2\left(\frac{\Delta m^2_{21} L}{4 E} \right)
    \\
    & \hspace{2mm}
    - \sin^22\theta_{13} \left[ \cos^2\theta_{12}\sin^2\left(
    \frac{\Delta m^2_{31} L}{4 E} \right)
    + \sin^2\theta_{12}\sin^2 \left(
    \frac{\Delta m^2_{32} L}{4 E} \right) \right]
    \\
    & \simeq 1 - \sin^22\theta_{13}\sin^2\left(
    \frac{\Delta m^2_{31} L}{4 E} \right) \,,
\end{split} \end{equation}
where we have used that for the relevant values of energy and
distance, one can safely neglect Earth matter effects.  The second
equality holds under the approximation $\Delta m^2_{21}\ll E/L$ which
can be safely made for $\Delta m^2_{21}\leq 3\times 10^{-4}~\eVq$ Thus
effectively the analysis of the CHOOZ reactor data involves two
oscillation parameters the mass difference which drives the dominant
atmospheric and K2K oscillations, $\Delta m^2_{31}$, and the angle
$\theta_{13}$ which is severely constrained~\cite{Bilenky:2001jq}.

\subsubsection{Effects due to $\theta_{13}$ in Solar Neutrinos and
  KamLAND}

Let us first consider the analysis of solar and KamLAND neutrinos.  A
first simplification occurs because $L^\text{osc}_{31} = 4\pi E/\Delta
m^2_{31}$ is much shorter than the distance between the Sun and the
Earth for solar neutrinos or between the reactors and the detectors in
KamLAND.  Consequently, the oscillations related to
${L_{0,31}^\text{osc}}$ are averaged and the vacuum survival
probability can be obtained from Eq.~\eqref{eq:pab} with $\sin^2
X_{31} = \sin^2 X_{32} = 1/2$. It is trivial to show that it can be
written in the following form:
\begin{equation}
    \label{eq:p3}
    P^{3\nu}_{ee} = \sin^4\theta_{13}+\cos^4\theta_{13}
    P^{2\nu}_{ee}(\Delta m^2_{21},\theta_{12}) \,,
\end{equation}
For solar neutrinos one must also take into account the three-neutrino
mixing effects in the evolution in matter. In this case a second
simplification occurs since, for the evolution in both the Sun and the
Earth, $\Delta m^2_{31} \gg 2\sqrt{2} G_F n_e E \sin^22\theta_{13}$.
Consequently, matter effects on the evolution of $\nu_3$ can be
neglected. The net result is that for solar neutrinos the survival
probability can also be written as Eq.~\eqref{eq:p3} with
$P^{2\nu}_{ee}$ obtained taking into account evolution in the
effective density~\cite{3solap1,3solap2}:
\begin{equation}
    \label{eq:pote}
    n_{e}\Rightarrow n_e \cos^2 \theta_{13} \,.
\end{equation}
We conclude that the analysis of the solar and KamLAND data constrains
three of the six independent oscillation parameters: $\Delta m^2_{21},
\theta_{12}$ and $\theta_{13}$. 

Eq.~\eqref{eq:p3} reveals what is the dominant effect of a
non-vanishing $\theta_{13}$ in the solar and KamLAND neutrino survival
probability: the survival probability, $P^{2\nu}_{ee}$, gets reduced
by the factor $\cos^4\theta_{13}$, while an energy independent term,
$\sin^4\theta_{13}$, is added. Within the present allowed values of
$\theta_{13}$ the first effect is the most relevant.
\subsubsection{Effects due to $\theta_{13}$ in Atmospheric and LBL
  Neutrinos}

We discuss first the sub-leading effect due to the mixing angle
$\theta_{13}$ which is particularly easy to treat in the
\emph{hierarchical approximation} in which $\Delta m^2_{21}$-induced
oscillations are neglected in the atmospheric neutrino analysis. In
this approximation one can rotate away the angle $\theta_{12}$. Thus
the resulting survival probabilities do not depend on $\Delta
m^2_{21}$ and $\theta_{12}$. For instance for constant Earth matter
density, the various $P_{\alpha\beta}$ can be written as follows:
\begin{align}
    \label{eq:P3atmee}
    P_{ee} &= 1 - 4s^2_{13,m} c^2_{13,m} \, S_{31} \,,
    \\
    \label{eq:P3atmmm}
    P_{\mu\mu} &= 1 - 4s^2_{13,m} c^2_{13,m} s^4_{23} \, S_{31}
    -4 s^2_{13,m} s^2_{23} c^2_{23} \, S_{21}
    -4 c^2_{13,m} s^2_{23} c^2_{23} \, S_{32} \,,
    \\
    \label{eq:P3atmem} 
    P_{e\mu} &= 4 s^2_{13,m} c^2_{13,m} s^2_{23} \, S_{31} \,,
\end{align}
Here $\theta^m_{13}$ is the effective mixing angle in matter: 
\begin{equation}
    \label{eq:Phi}
    \sin 2\theta^m_{13} =
    \frac{\sin2\theta_{13}}{\sqrt{(\cos2\theta_{13}
	- 2 E V_e/\Delta m^2_{31})^2 + (\sin2\theta_{13})^2}}
\end{equation}
and $S_{ij}$ are the oscillating factors in matter:
\begin{equation}
    \label{eq:defSij}
    S_{ij} = \sin^2\left(\frac{\Delta\mu^2_{ij}}{4E}L\right).
\end{equation}
In Eq.~\eqref{eq:defSij}, $\Delta\mu^2_{ij}$ are the effective
mass-squared differences in matter:
\begin{align}
    \Delta\mu^2_{21} &= \frac{\Delta m^2_{31}}{2}\left(
    \frac{\sin 2\theta_{13}}{\sin 2\theta^m_{13}}-1\right)
    - E V_e \,,
    \\
    \Delta\mu^2_{32} &= \frac{\Delta m^2_{31}}{2}\left(
    \frac{\sin 2\theta_{13}}{\sin 2\theta^m_{13}}+1\right)
    + E V_e \,,
    \\
    \Delta\mu^2_{31} &= \Delta m^2_{31}
    \frac{\sin 2\theta_{13}}{\sin 2\theta^m_{13}} \,.
\end{align}
and $L$ is the path length of the neutrino within the Earth.

The main effect of $\theta_{13}$ is that now atmospheric neutrinos can
oscillate simultaneously in both the $\nu_\mu\to\nu_\tau$ and
$\nu_\mu\to\nu_e$ (and, similarly, $\nu_e\to\nu_\tau$ and
$\nu_e\to\nu_\mu$) channels. The oscillation amplitudes for channels
involving $\nu_e$ are controlled by the size of
$\sin^2\theta_{13}=|U_{e3}|^2$. Furthermore because of matter effects
the size of the effect is different for normal and inverted
hierarchies~\cite{Petcov:1998su,Akhmedov:1998ui,Akhmedov:1998xq,chizhov,Bernabeu:2001xn,Bernabeu:2003yp,PalPet,Choubey:2005zy,Akhmedov:2006hb}.

In Fig.~\ref{fig:zenelec} we show the expected zenith angular
distribution of contained e-like events (normalized to the
no-oscillation expectation) for $\sin^2\theta_{13}=0.04$.  From the
figure we see that the effect is most relevant for multi-GeV neutrinos
and larger for the normal-hierarchy than inverted orderings. Also the
effect can be a decrease or increase of the expected number of events
with respect to the $\theta_{13}=0$ prediction depending on whether
$\theta_{23}$ is in the first or second octant.  These results can be
understood as follows. From
Eqs.~\eqref{eq:P3atmee}--\eqref{eq:P3atmem} it is easy to show that 
for the case of constant matter density the expected flux of $\nu_e$
events in the hierarchical approximation can be written
as~\cite{Akhmedov:1998xq,chizhov}:
\begin{equation}
    \frac{N_e}{N_{e0}} - 1 =
    \langle P_{e\mu} \rangle \,
    \bar{r} (s_{23}^2 - \frac{1}{\bar{r}}) \,,
\end{equation}
where $\langle P_{e\mu} \rangle$ is the corresponding probability, 
Eq.~\eqref{eq:P3atmem}, averaged over energy and zenith angle, and
$\bar{r} = \Phi_{\mu0} / \Phi_{e0}$ is the ratio of the electron and
muon neutrino fluxes in the absence of oscillations in the relevant
energy and angular bin.

For instance, for sub-GeV events $\bar{r}\sim 2$.  So the effect
cancels for maximal $\theta_{23}$. For $\theta_{23}$ in the first
octant ($s_{23}^2<0.5$) there is a decrease in the number of electron
events as compared to the $\theta_{13}$ case while the opposite holds
for $\theta_{23}$ in the second octant. Thus the effect is suppressed
for maximal $\theta_{23}$ mixing. 

For multi-GeV events matter effects lead to an enhancement of the
effect which is slightly larger for the normal ordering where the
matter enhancement is in the neutrino channel. For sub-GeV events, the
matter term can be neglected and the effect of a non-vanishing
$\theta_{13}$ is smaller and it is the same for normal and inverted
ordering.

For K2K and MINOS, matter effects can be neglected and the relevant
survival probability takes the form
\begin{equation} \begin{split}
    \label{eq:pk2k}
    P^\text{K2K, MINOS}_{\mu\mu}
    &= 1 - 4 \left(s^4_{23} s^2_{13} c^2_{13}
    + c^2_{13} s^2_{23} c^2_{23}\right)
    \sin^2\left(\frac{\Delta m^2_{31}}{4E}L\right)
    \\
    & \simeq s_{13}^2\frac{\cos2\theta_{23}}{c_{23}^2}
    + \left( 1 - s_{13}^2\frac{\cos2\theta_{23}}{c_{23}^2} \right)
    P^\text{K2K,$2\nu$}_{\mu\mu}(\Delta m^2_{31},\theta_{23})
    + \mathcal{O}(s_{13}^4) \,.
\end{split} \end{equation}
So we find that in the approximation of Eq.~\eqref{eq:deltahier} the
analysis of the atmospheric and K2K+MINOS data constrains three of the
six independent oscillation parameters: $\Delta m^2_{31}$,
$\theta_{23}$ and $\theta_{13}$ and for atmospheric neutrinos also the
sign of $\Delta m^2_{31}$ is relevant. Consequently in this
approximation the mixing angle $\theta_{13}$ is the only parameter
common to both solar+KamLAND and atmospheric+K2K neutrino oscillations
and which may potentially allow for some mutual influence. 

\subsubsection{Effects due to $\Delta m^2_{21}$ in Atmospheric
  Neutrinos}

We next discuss the sub-leading effects due to $\Delta m^2_{21}$
oscillations for vanishing small value of
$\theta_{13}$~\cite{Kim:1998bv,yasuda1,yasuda2,sakai99,strumia,antonio,peresdm12,nohierold,nohiernew}.
In this approximation and for constant Earth matter density the
relevant oscillation probabilities can be written as: 
\begin{align}
    P_{ee} &= 1 - P_{e2} \,,
    \\
    P_{e\mu} &= c_{23}^2 P_{e2} \,,
    \\
    P_{\mu\mu} &= 1 - c_{23}^4 P_{e2} - 2s_{23}^2 c_{23}^2
    \, \left[1 - \sqrt{1 - P_{e2}\cos\phi} \right] \,,
\end{align}
where
\begin{equation}
    P_{e2} = \sin^2 2\theta_{12,m}
    \sin^2\left(\frac{\Delta m^2_{21} \, L}{4E}
    \frac{\sin 2\theta_{12}}{\sin 2\theta_{12,m}} \right) \,, 
\end{equation}
with
\begin{align}
    \label{eq:12mat}
    \sin 2\theta_{12,m} &=
    \dfrac{\sin 2\theta_{12}}{\sqrt{\left(
	\cos 2\theta_{12} \mp
	\dfrac{2 E V_e}{\Delta m^2_{21}} \right)^2
	+ \sin^2 2\theta_{12}}} \,,
    \\[1mm]
    \phi &\approx (\Delta m^2_{31} + s_{12}^2 \,
    \Delta m_{21}^2) \frac{L}{2E} \,.
\end{align}

\begin{figure}\centering
    \includegraphics[width=0.98\textwidth]{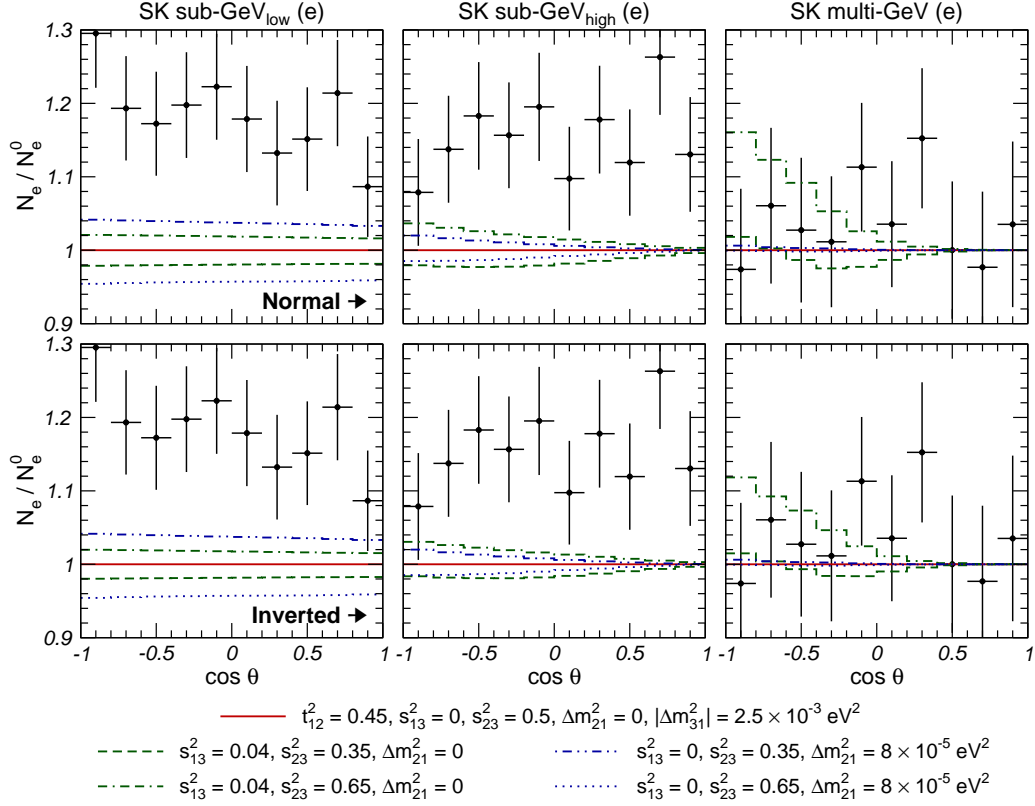}
      \caption{\label{fig:zenelec}%
	Comparison of the sub-leading effects due to $\Delta
	m^2_{21}$- and $\theta_{13}$-induced $\nu_e$ oscillations in
	the expected zenith angular distribution of e-like events.} 
\end{figure}

In Fig.~\ref{fig:zenelec} we show the angular distribution of
atmospheric $\nu_e$ for non-vanishing values of $\Delta m^2_{21}$ or
$\theta_{13}$.  As seen in these figures, unlike for $\theta_{13}$, 
the main effect of a small but non-vanishing $\Delta m^2_{21}$ is
mostly observable for sub-GeV electrons, and it can result either in
an increase or in a decrease of the expected number of events with
respect to the $\Delta m^2_{21}=0$ prediction depending on whether
$\theta_{23}$ is in the first or second octant. This behavior can be
understood in terms of the approximate analytical expressions: 
\begin{align}
    \label{eq:neal} 
    \frac{N_e}{N_{e0}}-1 
    &= \langle P_{e2} \rangle \,
    \bar{r} \left( c_{23}^2 - \frac{1}{\bar{r}} \right)
    \\
    \label{eq:nmual}
    \frac{N_\mu - N_{\mu}(\Delta m^2_{21} = 0)}{N_{\mu 0}}
    &= - \langle P_{e2} \rangle \,
    c_{23}^2 \left( c_{23}^2 - \frac{1}{\bar{r}} \right)
\end{align}
where $N_{e0}$ and $N_{\mu0}$ are the expected number of electron and
muon-like events in the absence of oscillations in the relevant energy
and angular bin and $N_{\mu}(\Delta m^2_{21} = 0)$ is the expected
number of muon-like events for $\Delta m^2_{21} = 0$. For sub-GeV
events, $\Delta m^2\ll 2 E V_e$ so 
\begin{equation}
    \label{eq:pe2alfa}
    P_{e2} = \sin^22\theta_{12} \left(
    \frac{\Delta m^2_{21}}{2E V_e} \right)^2\sin^2\frac{V_e L}{2} \,.
\end{equation}
According to Eqs.~\eqref{eq:neal} and~\eqref{eq:nmual} the sign of the
shift in the number of predicted events is opposite for electron and
muon-like events and it depends on the factor $c_{23}^2 -
\frac{1}{\bar{r}} \sim c_{23}^2-0.5$. So the effect cancels for
maximal $\theta_{23}$. For $\theta_{23}$ in the first octant,
$c_{23}^2>0.5$, there is an increase (decrease) in the number of
electron (muon) events as compared to the $\Delta m^2_{21} = 0$ case.
For $\theta_{23}$ in the second octant the opposite holds ( this is
the opposite behavior than the one due to $\theta_{13} \neq 0$
previously discussed).  We also see that the net shift is larger for
electron events than for muon events by a factor $c_{23}^2/\bar{r}$.
In summary for sub-GeV electrons, the shift in the expected number of
events is proportional to the deviation of $\theta_{23}$ from maximal
mixing and to $(\Delta m^2_{21})^2$, it is very weakly dependent on
the zenith angle, and it decreases with the energy.  

The present data may already give some hint of deviation of the 2-3
mixing from maximal as seen in Fig.~\ref{fig:skatm}. Indeed as the
figure illustrates, there is some excess of the $e-$like events in the
sub-GeV range. The excess increases with decrease of energy within the
sample as expected from a $\Delta m^2_{21}$ effect.

\subsubsection{Interference of $\theta_{13}$ and $\Delta m^2_{21}$ 
effects}

\begin{figure}\centering
    \includegraphics[width=0.98\textwidth]{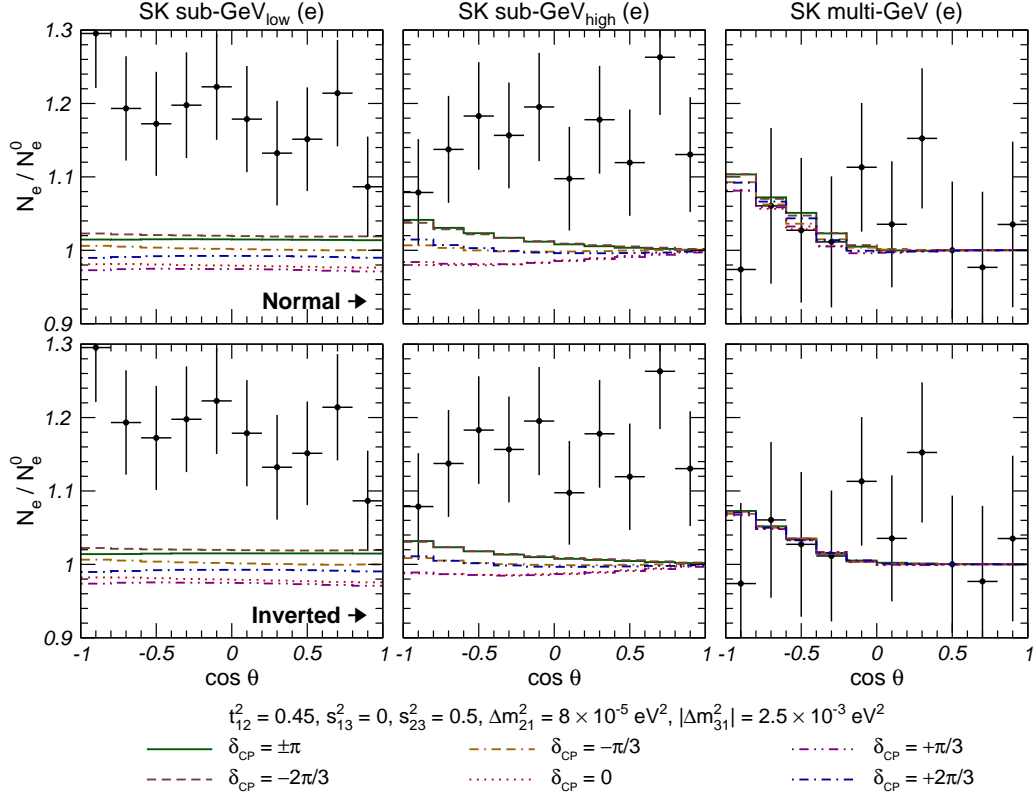}
    \caption{\label{fig:interf}%
      Sub-leading effect due to the interference of $\Delta m^2_{21}$-
      and $\theta_{13}$-induced $\nu_e$ oscillations in the expected
      zenith angular distribution of e-like events.} 
\end{figure}

Finally we comment on the possible effects due to the interference
between $\theta_{13}$- and $\Delta m^2_{21}$-induced
oscillations~\cite{nohierold,peresCPa,peresCPb} which could give
sensitivity to the CP violating phase $\delta_\text{CP}$. This effect
is most important for sub-GeV energies for which one can
write~\cite{peresCPb}:
\begin{multline}
    \label{eq:m12t13int}
    \frac{N_e}{N_e^0} -  1 \simeq
    \langle P_{e2} \rangle \,
    \bar{r} \left( c_{23}^2 - \frac{1}{\bar{r}} \right)
    + 2 \tilde{s}_{13}^2 \,
    \bar{r} \left( s_{23}^2 - \frac{1}{\bar{r}} \right)
    \\
    -\bar{r}
    \tilde{s}_{13} \tilde{c}_{13}^2 \sin 2\theta_{23} \left(
    \cos\delta_\text{CP} \langle R_2 \rangle - \sin\delta_\text{CP}
    \langle I_2 \rangle \right)
\end{multline}
where
\begin{align}
    P_{e2} &= \sin^2 2\theta_{12,m} \, \sin^2 \frac{\phi_m}{2} \,,
    \\
    R_2 &= - \sin 2\theta_{12,m} \cos 2\theta_{12,m} \,
    \sin^2 \frac{\phi_m}{2} \,,
    \\
    I_2 &= -\frac{1}{2} \sin 2\theta_{12,m} \, \sin\phi_m \,,
    \\ 
    \tilde\theta_{13} &\approx \theta_{13} \left(1 
    + \frac{2E \, V_e}{\Delta m^2_{31}} \right) \,.
\end{align}
Here $\phi_m$ is the phase oscillation in matter and $\theta_{12,m}$
is 12 the mixing angle in matter (Eq.~\eqref{eq:12mat}). As seen from
Eq.~\eqref{eq:m12t13int} the interference term (third term in the
equation) is not suppressed for maximal $\theta_{23}$ so it can
dominate for $\theta_{23}$ near maximal. Also it is proportional to
$\sin 2\theta_{23}$ and therefore it is not sensitive to the octant
of $\theta_{23}$.  Fig.~\ref{fig:interf} illustrates the possible size
of this effect.


\subsection{Global $3\nu$ Analysis of Oscillation Data}

\begin{figure}\centering
    \includegraphics[width=0.96\textwidth]{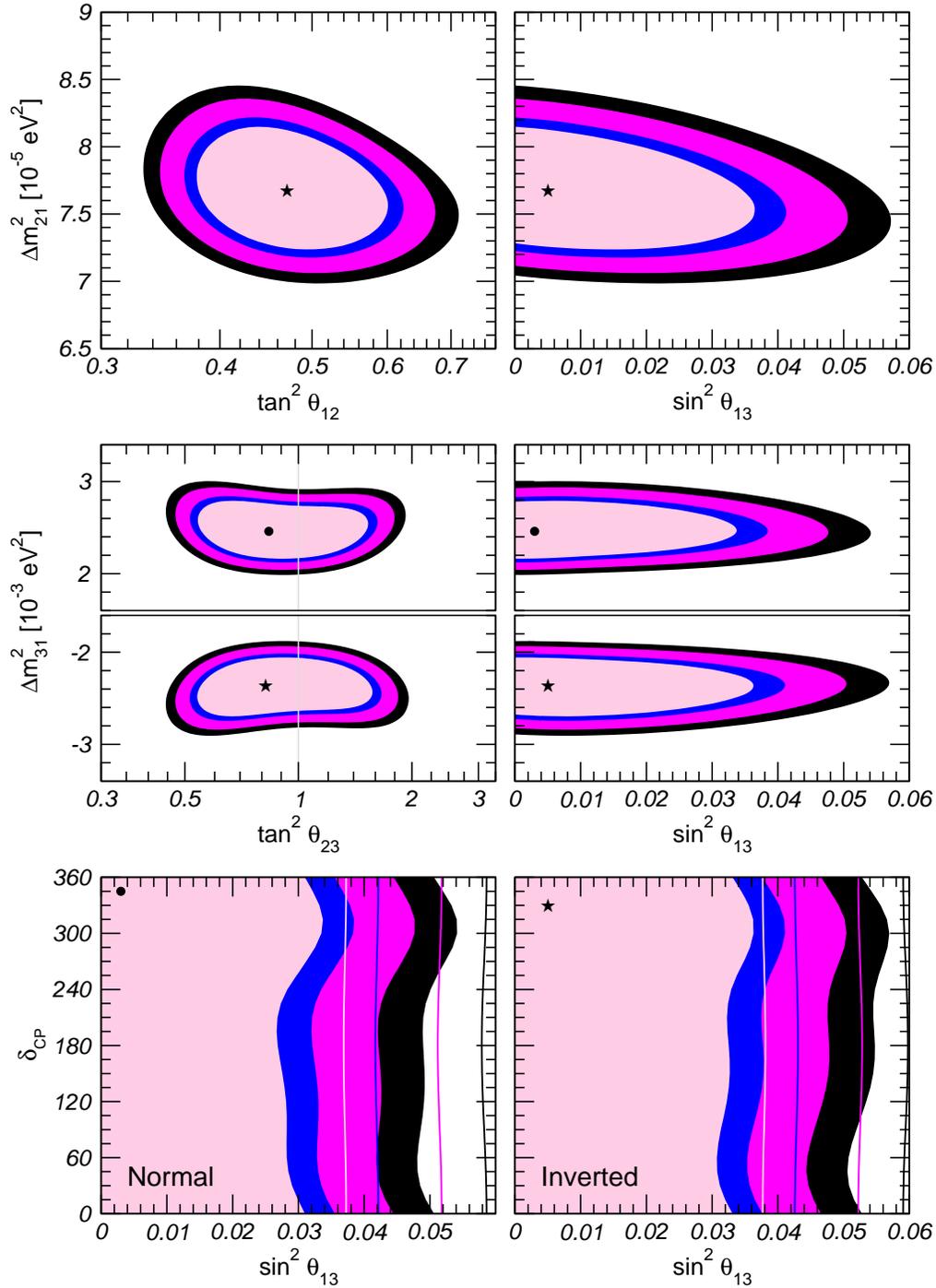}
    \caption{\label{fig:3nucontours}%
      Global $3\nu$ oscillation analysis. Each panels shows
      2-dimensional projection of the allowed 5-dimensional region
      after marginalization with respect to the undisplayed
      parameters.  The different contours correspond to the
      two-dimensional allowed regions at 90\%, 95\%, 99\% and
      $3\sigma$ CL. In the lowest panel the vertical lines correspond
      to the regions without inclusion of the atmospheric neutrino
      data.}
\end{figure}

\begin{figure}\centering
    \includegraphics[width=0.96\textwidth]{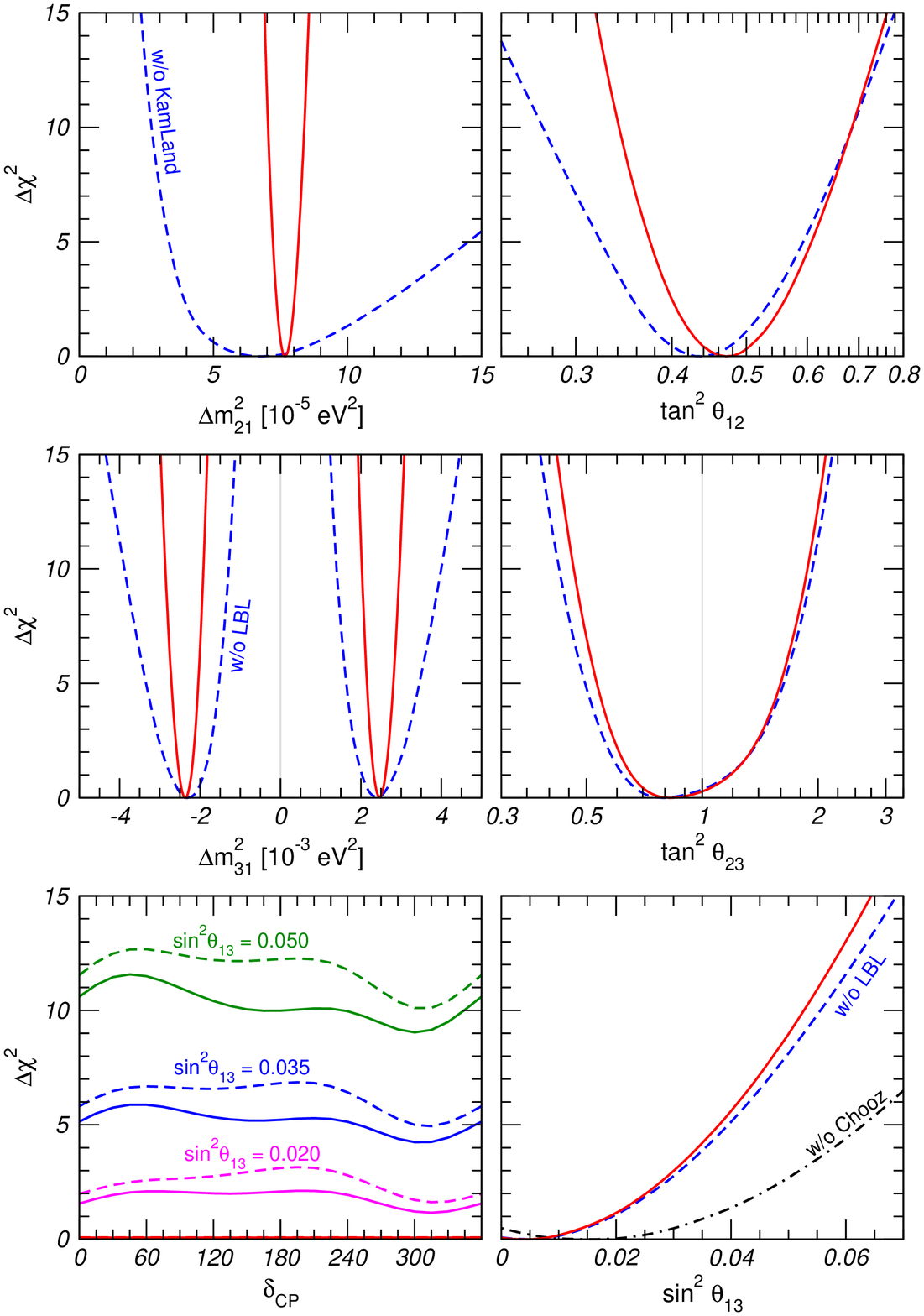}
    \caption{\label{fig:3nuchi}%
      Global $3\nu$ oscillation analysis. Each panels shows the 
      dependence of $\Delta\chi^2$ on each of the parameters from the
      global analysis (full line) compared to the bound prior to the 
      KamLAND data (dashed lines in the first row) and LBL data
      (dashed lines in the second rows). In the lowest panel we show
      the dependence of $\Delta\chi^2$ on $\theta_{13}$ for different
      sets of data as labeled in the curves. The individual $1\sigma$
      ($3\sigma$) bounds in Eqs.~\eqref{eq:3nuranges} can be read from
      the corresponding panel with the condition $\Delta\chi^2\leq 1$
      ($9$).}
\end{figure}

The results of the global combined analysis including all dominant and
subdominant oscillation effects are summarized in
Fig.~\ref{fig:3nucontours} and Fig.~\ref{fig:3nuchi} in which we show
different projections of the allowed 6-dimensional parameter space.
New to previous analysis is the inclusion of the latest MINOS and of
the SK-II atmospheric data and the inclusion in the analysis of the
effect of $\delta_\text{CP}$.

In Fig.~\ref{fig:3nucontours} we plot the correlated bounds from the
global analysis several pairs of parameters. The regions in each panel
are obtained after marginalization of $\chi^2_\text{global}$ with
respect to the three undisplayed parameters. The different contours
correspond to regions defined at 90\%, 95\%, 99\% and $3\sigma$ CL for
2~d.o.f.\ ($\Delta\chi^2 = 4.61$, $5.99$, $9.21$, $11.83$)
respectively. From the figure we see that the stronger correlation
appears between $\theta_{13}$ and $\Delta m^2_{31}$ as a reflection of
the CHOOZ bound. In the lower panels we show the allowed regions in
the ($\sin^2\theta_{13}$, $\delta_\text{CP}$) plane. As seen in the
figure, the sensitivity to the CP phase at present is marginal but we
find that the present bound on $\sin^2\theta_{13}$ can vary by about
$\sim$ 30\% depending on the exact value of $\delta_\text{CP}$. This
arises from the interference of $\theta_{13}$ and $\Delta m^2_{21}$ 
effects in the atmospheric neutrino observables as described above. To
illustrate this point we plot in the same panel the bounds on
$\sin^2\theta_{13}$ at 90\%, 95\%, 99\% and $3\sigma$ CL if the
atmospheric data is not included in the analysis (the vertical lines).

In Fig.~\ref{fig:3nuchi} we plot the individual bounds on each of the
six relevant parameters derived from the global analysis (full line).
To illustrate the impact of the LBL and KamLAND data we also show the
corresponding bounds when KamLAND and the LBL data are not included in
the analysis respectively.  In each panel, except the lower left one, 
the displayed $\chi^2$ has been marginalized with respect to the other
five parameters.  The lower left panel shows the $\chi^2$
(marginalized over all parameters but $\theta_{13}$) dependence of
$\delta_\text{CP}$ for fixed values of $\theta_{13}$.

The derived ranges for the six parameters at $1\sigma$ ($3\sigma$)
are:
\begin{equation} \begin{aligned}
    \label{eq:3nuranges}
    \Delta m^2_{21}
    &= 7.67 \,_{-0.21}^{+0.22} \,\left(_{-0.61}^{+0.67}\right)
    \times 10^{-5}~\eVq \,,
    \\
    \Delta m^2_{31} &=
    \begin{cases}
	-2.37 \pm 0.15 \,\left(_{-0.46}^{+0.43}\right)
	\times 10^{-3}~\eVq & \text{(inverted hierarchy)} \,,
	\\[1mm]
	+2.46 \pm 0.15 \,\left(_{-0.42}^{+0.47}\right)
	\times 10^{-3}~\eVq & \text{(normal hierarchy)} \,,
    \end{cases}
    \\
    \theta_{12} &=
    34.5 \pm 1.4 \,\left(_{-4.0}^{+4.8}\right) \,,
    \\
    \theta_{23} &=
    42.3 \,_{-3.3}^{+5.1} \,\left(_{\hphantom{0}-7.7}^{+11.3}\right) \,,
    \\
    \theta_{13} &=
    0.0 \,^{+7.9}_{-0.0} \,\left(^{+12.9}_{\hphantom{0}-0.0}\right) \,,
    \\
    \delta_\text{CP} &\in [0,\, 360] \,.
\end{aligned} \end{equation}

Figure~\ref{fig:3nuchi} illustrates that the dominant effect of the
inclusion of the laboratory experiments KamLAND, K2K and MINOS is the
better determination of the corresponding mass differences $\Delta
m^2_{21}$ and $\Delta m^2_{31}$ while the mixing angle $\theta_{12}$
is dominantly determined by the solar data and the mixing angle
$\theta_{23}$ is still most precisely measured in the atmospheric
neutrino experiments.  The non-maximality of the best $\theta_{23}$
observed in Eq.~\eqref{eq:3nuranges} is a pure 3-$\nu$ oscillation
effect associated to the inclusion of the $\Delta m^2_{21}$ effects in
the atmospheric neutrino analysis. As discussed above, although
statistically not very significant, this preference for non-maximal
2-3 mixing is a physical effect on the present neutrino data, induced
by the fact than an excess of events is observed in sub-GeV electrons
but not in sub-GeV muons nor, in the same amount, in the multi-GeV
electrons.  As a consequence, this excess cannot be fully explained by
a combination of a global rescaling and a ``tilt'' of the fluxes
within the assumed uncertainties. 

Finally, let's notice that the ranges in Eq.~\eqref{eq:3nuranges} are
not independent but they are correlated so they cannot be directly 
used to determine the leptonic mixing matrix at a given CL.  As
described in Ref.~\cite{ourthree} the leptonic mixing matrix can be
consistently extracted as follows. Starting by the
$\chi^2_\text{global}$ which is a function of the six parameters, we
can define the mass-marginalized $\chi^2$ function:
\begin{multline}
    \chi^2_\text{mix,global}(\theta_{12}, \theta_{23}, \theta_{13},
    \delta_\text{CP}) =
    \\
    \text{min}_{(\Delta m^2_{21}, \Delta m^2_{31})} \,
    \chi^2_\text{global}(\Delta m^2_{21}, \Delta m^2_{31},
    \theta_{12}, \theta_{23}, \theta_{13}, \delta_\text{CP}) \,.
\end{multline}
We study the variation of $\chi^2_\text{mix,global}$ as function of
each of the mixing combinations in $U$ as follows. For a given
magnitude $\bar{U}_{ij}$ of the entry $U(i,j)$ we define
$\chi^2(\bar{U}_{ij})$ as the minimum value of $\chi^2_\text{mix,
global}$ with the condition $|U(i,j)(\theta_{12}, \theta_{23},
\theta_{13}, \delta_\text{CP})| = \bar{U}_{ij}$.  The allowed range of
the magnitude of the entry $ij$ at a given CL is then defined as the
values $\bar{U}_{ij}$ verifying
\begin{equation}
    \chi^2(\bar{U}_{ij})-\chi^2_\text{global,min} \leq
    \Delta\chi^2(\text{CL, 1~d.o.f.}) \,.
\end{equation}
With this procedure we derive the following values on the magnitude of
the elements of the complete matrix, at 90\% CL:
\begin{equation}
    |U|_\text{90\%} =
    \begin{pmatrix}
	0.80 \to 0.84 & ~\quad~ 0.53 \to 0.60 & ~\quad~ 0.00 \to 0.17 \\
	0.29 \to 0.52 & ~\quad~ 0.51 \to 0.69 & ~\quad~ 0.61 \to 0.76 \\
	0.26 \to 0.50 & ~\quad~ 0.46 \to 0.66 & ~\quad~ 0.64 \to 0.79
    \end{pmatrix}
\end{equation}
and at the $3\sigma$ level:
\begin{equation}
    |U|_{3\sigma} =
    \begin{pmatrix} 
	0.77 \to 0.86 & ~\quad~ 0.50 \to 0.63 & ~\quad~ 0.00 \to 0.22 \\
	0.22 \to 0.56 & ~\quad~ 0.44 \to 0.73 & ~\quad~ 0.57 \to 0.80 \\
	0.21 \to 0.55 & ~\quad~ 0.40 \to 0.71 & ~\quad~ 0.59 \to 0.82
    \end{pmatrix} \,.
\end{equation}
By construction these limits are obtained under the assumption that
$U$ is unitary. In other words, the ranges in the different entries of
the matrix are correlated due to the fact that, in general, the result
of a given experiment restricts a combination of several entries of
the matrix, as well as to the constraints imposed by unitarity.  As a
consequence choosing a specific value for one element further
restricts the range of the others. Effects in the determination of the
leptonic mixing matrix due to violations of unitarity have been
considered in Ref.~\cite{belen}.

%% file: sec.fluxes.tex
\section{Learning about Solar and Atmospheric Neutrino Fluxes}
\label{sec:fluxes}


\subsection{Motivation}
\label{sec:fluxmot}

As we have described in previous sections the flavor oscillation 
hypothesis has been supported by an impressive wealth of neutrino 
experimental data. Originally, the two most important pieces of
evidence came from solar and atmospheric neutrino experiments.

The expected number of solar and atmospheric neutrino events in an
experiment depends on a variety of components: the neutrino fluxes, 
the neutrino oscillation parameters and the neutrino interaction cross
section in the detector. Although the original goal of the experiments
was the understanding of the solar and atmospheric neutrino fluxes,
once it was found that the observed anomalies seemed to indicate that
neutrinos oscillated, the main focus of the experiments changed to 
the determination of the neutrino masses and mixing.  Consequently, in
the standard analysis, such as the ones described in 
Sec.~\ref{sec:3nu}, the remaining components of the event rate
computation are inputs taken from other sources. In particular, as
discussed in Sec.~\ref{sec:expsolar}, the fluxes of solar neutrinos
are taken from the results of Solar Model simulations~\cite{bs05} 
which describe the properties of the Sun and its evolution based on a
set of observational parameters and basic assumptions such as
spherical symmetry, thermal equilibrium, etc\dots. Similarly, the
fluxes of atmospheric neutrinos are taken from the results of
numerical calculations, such as those of
Refs.~\cite{honda,bartol,others1,others2,others3}, which are based on
the convolution of the primary cosmic ray spectrum with the expected
yield of neutrinos per incident cosmic ray.

The oscillation of $\nu_e$ and $\nu_\mu$'s can also be tested at
terrestrial facilities. In particular, the disappearance of reactor
antineutrinos can be used to detect $\bar\nu_e$ oscillations as
described in Sec.~\ref{sec:expreac}, and $\nu_\mu$ oscillations can be
studied in Long Baseline experiments (see Sec.~\ref{sec:explbl}) using
as neutrino source a controlled beam of accelerator neutrinos. As we 
have seen, the results of KamLAND~\cite{kland04}, K2K~\cite{K2K2} and
MINOS~\cite{MINOSdat} have confirmed both in the observed deficit of
events and in their energy dependence, that reactor $\bar\nu_e$ and
accelerator $\nu_\mu$ oscillate as expected from oscillations with the
parameters inferred from the solar and atmospheric neutrino data 
respectively. Furthermore they already provide a competitive
independent determination of the relevant $\Delta m^2$.

The attainable accuracy in the independent determination of the
relevant neutrino oscillation parameters from non-solar and
non-atmospheric neutrino experiments makes it possible to attempt an
inversion of the strategy: to use the oscillation parameters
(independently determined in reactor and LBL neutrino experiments) as
inputs in the solar and atmospheric neutrino analysis in order to
extract the solar and atmospheric neutrino fluxes directly from the
data.  Alternatively one can perform global fits to natural (solar and
atmospheric) and terrestrial (reactor and accelerator) neutrino data
in which both the oscillation parameters and the natural fluxes are
extracted simultaneously. 

There are several motivations for such direct determination of the
solar and atmospheric neutrino fluxes.  First of all it would provide
a cross-check of the standard flux calculations as well as of the size
of the associated uncertainties (which, being mostly theoretical, are
difficult to quantify).  Also, such program quantitatively expands the
physics potential of future solar and atmospheric neutrino
experiments.  


\subsection{Learning How the Sun Shines}
\label{sec:fluxsol}

The idea that the Sun generates power through nuclear fusion in its
core was first suggested in 1919 by Sir Arthur Eddington, who pointed
out that the nuclear energy stored in the Sun could explain the
apparent age of the Solar System.  

In 1939, Hans Bethe described in an epochal paper~\cite{bethe39} two
nuclear fusion mechanisms by which main sequence stars like the Sun
could produce the energy corresponding to their observed luminosities.
The two mechanisms have become known as the $pp$ chain and the CNO
cycle~\cite{bahcall}. For both the $pp$ chain and the CNO cycle the
basic energy source is the burning of four protons to form an alpha
particle,two positrons, and two neutrinos.  In the $pp$ chain, fusion
reactions among elements lighter than $A = 8$ produce a characteristic
set of neutrino fluxes, whose spectral energy shapes are known but
whose fluxes must be calculated with a detailed solar model.  In the
CNO chain, with \Nuc{12}{C} as a catalyst, \Nuc{13}{N} and \Nuc{15}{O}
beta decays are the primary source of neutrinos.

The first sentence in Bethe's paper reads: ``It is shown that the most
important source of energy in ordinary stars is the reactions of
carbon and nitrogen with protons." Bethe's conclusion about the
dominant role of the CNO cycle relied upon a crude model of the Sun.
Over the next two and a half decades, the results of increasingly more
accurate laboratory measurements of nuclear fusion reactions and more
detailed solar model calculations led to the theoretical inference
that the Sun shines primarily by the $pp$ chain rather than the CNO
cycle. Currently, solar model calculations imply~\cite{bs05} that
98.5\% of the solar luminosity is provided by the $pp$ chain and only
1.5\% is provided by CNO reactions.

Despite the obvious appeal of the theory, simple observations of the
solar luminosity are not enough to demonstrate that nuclear fusion is,
in fact, the solar energy source neither the role of the CNO versus
the $pp$ chain in the energy generation.  Only neutrinos, with their
extremely small interaction cross sections, can enable us to see into
the interior of a star and thus verify directly the hypothesis of
nuclear energy generation in stars~\cite{bahcall64}. 

Indeed from the earliest days of solar neutrino research, a primary
goal of the field was to test the energy generation model of the Sun
and in particular the solar model prediction that the Sun shines by
the $pp$ chain and not by the CNO cycle~\cite{bahcall69a,bahcall69b}.
However this task was made difficult by the fact that neutrino
oscillations occur and they change in an energy dependent way the
probability that electron type neutrinos created in the Sun reach the
Earth as electron type neutrinos. This affects both the overall number
of events in the solar neutrino experiments and the relative
contribution expected from the different components of the solar
neutrino spectrum. Because of these complications, the extraction of
the fluxes from the solar neutrino data was not possible, and, for
example, one could find neutrino oscillation solutions in which
99.95\% of the Sun's luminosity is supplied by the CNO
cycle~\cite{cnoold}. 

As a consequence until very recently, it was necessary to assume the
standard solar model predictions for all the solar neutrino fluxes and
their uncertainties in order to determine reasonably constrained
values for neutrino oscillation parameters. Only the upcome of the
real time experiments Super-Kamiokande and SNO and the independent
determination of the oscillation parameters using reactor
antineutrinos at KamLAND allowed for the attempt to extract the solar
neutrino fluxes and their uncertainties directly from the 
data~\cite{ourbobe,ourcno,ourpostnu04,roadmap,Bandyopadhyay:2006jn} as
we describe next.

\subsubsection{Solar Neutrino Fluxes from Neutrino Data}

There are eight thermonuclear reactions which can produce neutrinos 
in the Sun. Neutrino fluxes are named by the corresponding source
reaction.  Five reactions produce $\nu_e$ in the $pp$ chain ($pp$,
$pep$, $hep$, \Nuc{7}{Be}, and \Nuc{8}{B}) and three in the CNO cycle
(\Nuc{13}{N}, \Nuc{15}{O}, and \Nuc{17}{F}).  Most of these reactions
produce a neutrino spectrum characteristic of $\beta$ decay while in
some cases, like the \Nuc{7}{Be} neutrinos, the spectrum is almost
monochromatic, with an energy width of about 2 keV which is
characteristic of the temperature in the core of the Sun. In general,
the physics which determines the neutrino energy spectrum in each of
the reactions is well understood. Thus in the extraction of the solar
neutrino fluxes from the neutrino data one assumes that the energy
dependence of the fluxes is well determined by nuclear physics and
only the rates of the different reactions is to be tested. Under this
assumption, the empirical determination of the solar fluxes reduces to
extracting from the data the value of the eight normalization
constants giving the rate of each of the contributions. For
convenience these are usually parametrized in terms of some factors
$f_i$ giving the ratios of the ``true'' solar neutrino fluxes and the
fluxes predicted by some solar model:
\begin{equation}
    \label{eq:fidefinition}
    f_i \equiv \frac{\phi_i}{\phi_i^\text{SSM}} \,,
\end{equation}
with $i=pp$, $pep$, $hep$, \Nuc{7}{Be}, \Nuc{8}{B}, \Nuc{13}{N},
\Nuc{15}{O}, \Nuc{17}{F}.

The most directly determined flux is the \Nuc{8}{B} flux because both 
Super-Kamiokande and SNO detect the interaction of \Nuc{8}{B}
neutrinos. Under the hypothesis of no sterile neutrino mixing this
flux is exactly given by the NC rate observed at SNO
\begin{equation}
    f_B = \frac{R^\text{NC,exp}_\text{SNO}}{R^\text{NC,SSM}_\text{SNO}}
    = 0.87\pm 0.08 \,.
\end{equation}
where $R^\text{SSM}_\text{SNO}$ is the NC rate for the SNO experiment 
that is predicted by the standard solar model in the absence of
oscillations:
\begin{equation}
    \label{eq:defrsno}
    R^\text{NC,SSM}_\text{SNO}=
    \int \phi^\text{SSM}(\Nuc{8}{B}, E_\nu) \, \sigma^{NC}(E_\nu) \, 
    R^\text{SSM}_\text{SNO} \, dE_\nu \,.
\end{equation}
$E_\nu$ is the neutrino energy and $\sigma^{NC}$ is the weighted
average $\nu_e$-d NC interaction cross-section, including the
experimental energy resolution function.

Independently of the presence of sterile neutrinos the \Nuc{8}{B} flux
can be extracted from the CC rate at SNO~\cite{ourbobe} once the
oscillation parameters have been independently determined, for
example, at KamLAND:
\begin{equation}
    \label{eq:fbodef}
    f_\text{B} =
    \frac{R^\text{CC,exp}_\text{CC,SNO}}{R^\text{SSM}_\text{SNO}} \times
    \frac{1}{\langle P_{ee}(\Delta m^2,\theta) \rangle_\text{SNO}}, 
\end{equation}
where ${\langle P_{ee}(\Delta m^2,\tan^2\theta)\rangle_\text{SNO}}$ is
the average survival probability for electron-flavor neutrinos 
detected at SNO in the CC interactions for a given value of the
oscillation parameters:
\begin{equation}
    \label{eq:defnpee}
    \langle P_{ee} (\Delta m^2,\theta)\rangle_\text{SNO} =
    \frac{1}{R^\text{CC,SSM}_\text{SNO}}
    \int \phi^\text{SSM}(\Nuc{8}{B}, E_\nu) \,
    \sigma^{CC}(E_\nu) \, P_{ee}(E_\nu,\Delta m^2,\theta) \,
    dE_\nu \,.
\end{equation}

Similarly the lower energy fluxes can be extracted from the 
radiochemical experiments. For example, the expected event rate in the
gallium experiments is a sum of the contributions from the different
neutrino fluxes:
\begin{multline}
    \label{eq:rga}
    R_\text{Ga} =
    f_\text{B} \, R^{\Nuc{8}{B},\text{SSM}}_\text{Ga}
    \langle P_{ee}(\Delta m^2,\theta) \rangle^{\Nuc{8}{B}}_\text{Ga} \,
    \\[2mm]
    + f_\text{Be} \, R^{\Nuc{7}{Be},\text{SSM}}_\text{Ga}
    \langle P_{ee}(\Delta  m^2,\theta)\rangle^{\Nuc{7}{Be}}_\text{Ga}
    + \sum_i f_i \, R^{\phi_i,SSM}_\text{Ga}
    \langle P_{ee} (\Delta m^2,\theta) \rangle^{\phi_i}_\text{Ga} \,.
\end{multline}
where the average survival probabilities are obtained similarly to 
Eq.~\eqref{eq:defnpee} with the corresponding fluxes and cross
sections. The last term in Eq.~\eqref{eq:rga} contains the
contributions from $hep$, $pep$, CNO and $pp$ neutrinos. Substituting
the value of $f_\text{B}$ determined from the KamLAND and SNO CC
measurements, Eq.~\eqref{eq:fbodef}, into Eq.~\eqref{eq:rga}, one can
solve Eq.~\eqref{eq:rga} for $f_\text{Be}$ by equating $R_\text{Ga} =
R^\text{exp}_\text{Ga}$. In order to do so one has to to assume that
all the solar neutrino fluxes but the \Nuc{8}{B} and \Nuc{7}{Be}
fluxes are equal to the values predicted by the SSM. 

Alternatively, instead of trying to determine one flux at a time, one
can perform a global fit to all solar and reactor neutrino data in
which both the oscillation parameters and the flux normalization
constants $f_i$ are determined.  A key ingredient in these type of
analysis is the imposition of the luminosity
constraint~\cite{luminosity}. The luminosity constraint implements, in
a global way for the Sun, the constraint of conservation of energy for
nuclear fusion among light elements. Each neutrino flux is associated
with a specific amount of energy released to the star and therefore a
particular linear combination of the solar neutrino fluxes is equal to
the solar luminosity (in appropriate units). One can write the
luminosity constraint as
\begin{equation}
    \label{eq:genconstraint}
    \frac{L_\odot}{4\pi \text{(A.U.)}^2} =
    \sum\limits_i \alpha_i \phi_i \,,
\end{equation}
where $L_\odot$ is the solar luminosity measured at the Earth's
surface, 1 $A.U.$ is the average Earth-Sun distance, and the
coefficient $\alpha_i$ is the amount of energy provided to the star by
nuclear fusion reactions associated with each of the important solar
neutrino fluxes, $\phi_i$. The coefficients $\alpha_i$ are calculated
accurately in Ref.~\cite{luminosity}.  An additional simplification
comes from the fact that the ratio of the $pep$ neutrino flux to the
$pp$ neutrino flux is fixed to high accuracy because they have the
same nuclear matrix element. 

At present, this strategy yields the following value for the flux 
normalization constants
\begin{equation}
    \label{eq:solfluxfit}
    \begin{aligned}
	f_{pp} = f_{pep} &= 1.0\pm 0.02 \\
	f_{\Nuc{8}{B}} &= 0.88\pm 0.04 \\
	f_{\Nuc{7}{Be}} &= 1.03^{+0.24}_{-1.03} \\
	f_{\Nuc{13}{N}} &= 0.0^{+7.6}_{-0.0} \\
	f_{\Nuc{15}{O}} &= 0.0^{+5.0}_{-0.0} \\
	f_{\Nuc{17}{F}} &= 0.0^{+2.1}_{-0.0}\\
    \end{aligned}
\end{equation}
at 1$\sigma$. Concerning the very small $hep$ flux, the global fit has
very little sensitivity to this flux.  At present the best
determination of this flux is an upper bound, $f_{hep}\lesssim 5$,
which has been derived from the search of events with $E_\nu>16$ MeV
at the SNO experiment~\cite{hepbound}.
 
From the results in Eq.~\eqref{eq:solfluxfit} it is possible, for
example, to extract the allowed range of the fraction of the Sun's
luminosity that arises from CNO reactions
as~\cite{ourcno,ourpostnu04}:
\begin{equation}
    \label{eq:cnosum}
    \frac{L_{CNO}}{L_\odot} =
    \sum\limits_{i=\text{N,O,F}} \left(
    \frac{\alpha_i}{10~\text{MeV}} \right) a_i \, f_i \,,
\end{equation}
where $a_i$ is the ratio of the neutrino flux $i$ predicted by the
standard solar model to the characteristic solar photon flux defined
by $L_\odot/[4\pi\text{(A.U.)}^2(\text{10 MeV})]$. At present this
determination gives a bound 
\begin{equation}
    \label{eq:limcno}
    \frac{L_\text{CNO}}{L_\odot} =
    0.0^{+2.7}_{-0.0}\, (^{+7.3}_{-0.0})\,\%
\end{equation}
at $1\sigma$ ($3\sigma$).  So the global fit to all the data yields a
constraint which is consistent with the solar model prediction at
$1\sigma$ which is an important empirical confirmation of the SSM and,
in general, of our understanding of the Sun. 

As mentioned above a very important ingredient on the determination of
the solar fluxes given in Eq.~\eqref{eq:solfluxfit} is the imposition
of the luminosity constraint. For example, it is this constraint that
implies that the $pp$ flux is known with a precision, $\pm 2\%$,
comparable to the theoretical uncertainty, $\pm 1\%$ in the SSM
prediction.  On the contrary if one does not impose the luminosity
constraint on the extraction of the solar neutrino fluxes one finds
that both the $pp$ and \Nuc{7}{Be} fluxes as well as the CNO
luminosity fraction are very poorly known.

Alternatively one may try to test the luminosity constraint itself, by
comparing the inferred luminosity based on the neutrino fluxes
extracted from the fit without imposing that condition, with the
observed photon luminosity.  Such a test can tell us whether there are
any energy generation mechanisms beyond nuclear fusion.  In addition,
we can learn whether the Sun is in a steady state, because the
neutrino luminosity tells us how it burns today, while the photons
tell us how it burned over 40,000 years ago.  With the existing data,
however, the experimental precision is not enough to allow for such a
test. In other words the current comparison of these luminosities is
not very precise~\cite{roadmap}. Quantitatively we obtain that the
inferred luminosity based on the neutrino fluxes is:
\begin{equation}
    \label{eq:lnuoverlphoton}
    \frac{L_\odot\text{(neutrino-inferred)}}{L_\odot}
    = 1.4^{+0.2}_{-0.3} \,(^{+0.7}_{-0.6}) \,.
\end{equation}
We see that, at $3\sigma$, the inferred luminosity can be 2.1 times
larger than the measured photon luminosity, or 0.8 times smaller.  The
fact that the solar neutrino flux is overwhelmingly $pp$ neutrinos
means that the precision of this comparison approximately scales with
the precision of a measurement of the $pp$ flux. Therefore a future
low energy solar neutrino experiment has the potential to perform such
a fit as we describe in Sec.~\ref{sec:futsolar}.


\subsection{General Strategy for Atmospheric Flux Determination}
\label{sec:fluxatm1}

The determination of atmospheric neutrino fluxes directly from the
atmospheric neutrino data is technically more involved than for solar
neutrinos because there are four different fluxes to be determined:
$\nu_e$, $\nu_\mu$, $\bar\nu_e$, and $\bar\nu_\mu$ which, after
integration over the azimuthal angle, are a function of two variables:
the zenith angle and the energy of the neutrino.  Unlike for solar
neutrinos, there is no simple physics which can determine the angular
and energy dependence of the fluxes thus not only the normalization of
the fluxes but also their functional dependence has to be extracted
from the data. Consequently the fully empirically determination of the
atmospheric fluxes from atmospheric neutrino data requires a generic
parametrization of the energy and angular functional dependence of the
fluxes which is valid in all the range of energies where there is
available data. Such parametrization does not exist.

The problem of the unknown functional form for the neutrino flux can
be bypassed by the use of neural networks as interpolants.  Artificial
neural networks have long been used in different fields, from biology
to high energy physics, and from pattern recognition to business
intelligence applications. In this context artificial neural networks
allow the parametrization of the atmospheric neutrino flux without
having to assume any functional behavior.  Indeed, the problem of the
\emph{deconvolution} of the atmospheric flux from experimental data on
event rates is rather close in spirit to the determination of parton
distribution functions in deep-inelastic scattering from
experimentally measured structure functions~\cite{nnpdf}. 
Consequently a similar strategy can be applied to determine the
atmospheric fluxes. This approached was followed in
Ref.~\cite{ourneutnet} and can be summarized as follows:
\begin{itemize}
  \item In the first stage, a Monte Carlo sample of replicas of the
    experimental data on neutrino event rates (``artificial data'') is
    generated. These can be viewed as a sampling of the probability
    measure on the space of physical observables at the discrete
    points where data exist. 
    
  \item In the second stage one uses neural networks to interpolate
    between these points.  In order to do so one has to first
    determine the atmospheric event rates for a given atmospheric
    flux, and second to compare these rates to the data in order to
    tune the best-fit form of input neural flux distribution. This
    process is called the ``training of the neural network''.
\end{itemize}
Combining these two steps, the space of physical observables is mapped
onto the space of fluxes, so the experimental information on the
former can be interpolated by neural networks in the latter.

We describe briefly now how this procedure can be applied to the data
from Super-Kamiokande. The starting point, of course, is the observed
event rates:
\begin{equation}
    R_i^\text{(exp)}, \qquad i=1, \ldots, N_\text{dat} \,, 
\end{equation}
which contain information on the value of the atmospheric fluxes in
the range of 0.1 GeV $\lesssim E_\nu\lesssim$ few TeV and on their
flavor and angular dependence.  

For this data the experimental correlation matrix can be constructed
as:
\begin{equation} \label{eq:cormat}
    \rho^\text{(exp)}_{ij} =
    \frac{(\sigma^\text{stat}_i)^2 \delta_{ij}
      + \displaystyle\sum_{n=1}^{N_\text{cor}}
      \sigma^\text{cor}_{ni} \, \sigma^\text{cor}_{nj}}
    {\sigma^\text{tot}_{i} \, \sigma^\text{tot}_{j}} \,,
\end{equation}
where the statistical uncertainty is given by
\begin{equation}
    \sigma^\text{stat}_i=\sqrt{R_i^\text{(exp)}} \,.
\end{equation}
The $N_\text{cor}$ correlated uncertainties are computed from the
couplings factors, $\pi_i^n$ to the corresponding \emph{pull}
$\xi_{n}$~\cite{Fogli:2002pt} (see Appendix~\ref{sec:appatm}):
\begin{equation}
    \sigma^\text{cor}_{ni} \equiv R_i^\text{(exp)} \pi_i^n \,, 
\end{equation}
and the total error is computed adding the statistical and correlated
errors in quadrature.

The purpose of the artificial data generation is to produce a Monte
Carlo set of `pseudo--data', \textit{i.e.} $N_\text{rep}$ replicas of
the original set of $N_\text{dat}$ data points:
\begin{equation} \label{eq:replicas}
    R^{(\text{art})(k)}_i \,, \qquad k=1,\dots,N_\text{rep} \,,
    \qquad i=1,\dots,N_\text{dat} \,,
\end{equation}
such that the $N_\text{rep}$ sets of $N_\text{dat}$ points are
distributed according to an $N_\text{dat}$--dimensional multi-gaussian
distribution around the original points, with expectation values equal
to the central experimental values, and error and covariance equal to
the corresponding experimental quantities.

This is achieved by defining
\begin{equation} \label{eq:gen}
    R_i^{(\text{art})(k)} =
    R_i^\text{(exp)} + r_i^{(k)} \sigma_i^\text{tot},
    \qquad i=1,\ldots,N_\text{dat} \,,
    \qquad k=1,\ldots,N_\text{rep} \,,
\end{equation}
where $N_\text{rep}$ is the number of generated replicas of the
experimental data, and where $r_i^{(k)}$ are univariate gaussian
random numbers with the same correlation matrix as experimental data,
that is they satisfy
\begin{equation}
    \langle r_i^{(k)}r_j^{(k)}\rangle_\text{rep}
    = \rho_{ij}^\text{(exp)}
    + \mathcal{O} \left(\frac{1}{N_\text{rep}} \right) \,.
\end{equation}
Because the distribution of the experimental data coincides (for a
flat prior) with the probability distribution of the value of the
event rate $R_i$ at the points where it has been measured, this Monte
Carlo set gives a sampling of the probability measure at those
points~\cite{cowanstat}.

The second step consists of training $N_\text{rep}$ neural networks.
Each neural network parametrizes a differential flux,
$\Phi^{(\text{net})}(E_\nu, c_\nu, t, \vec\omega)$ which in principle
depends on the neutrino energy $E_\nu$, the zenith angle
$\cos\theta_{\nu}\equiv c_\nu$ and the neutrino \emph{type} $t$
($t=1,\ldots,4$ labels the neutrino flavor: electron neutrinos and
antineutrinos, and muon neutrinos and antineutrinos), as well as on
the parameters $\vec\omega$ of the neural network, and it is based on
all the data in one single replica of the original data set.

The process which determines the function $\Phi^{(\text{net})(k)}$
which better describes each of the $k=1, \dots, N_\text{rep}$ sets of
artificial data, $\{R_i^{(\text{art})(k)}\}$, is what it is called
training of the neural network. It involves two substeps. First for a
given $\Phi^{(\text{net})}(E_\nu, c_\nu, t, \vec\omega)$ the expected
atmospheric event rates have to be computed. Second the neural network
parameters $\vec\omega$ have to be determined by minimizing some error
function.  Usually the determination of the parameters that define the
neural network, its weights, is performed by maximum likelihood and
the minimization of the corresponding error function is performed with
the use of genetic algorithms.  In such algorithms, the minimization
is ended after a number of iterations large enough so that the error
function stops decreasing, that is, when the fit has converged, but
not too large to prevent overlearning of the net.

Nothing has to be assumed about the functional form of
$\Phi^{(\text{net})}(E_\nu, c_\nu, t, \vec\omega)$ whose value is only
known after the full procedure of training is finished.  There are,
however, some requirements about the choice of the architecture of the
neural network.  As discussed in Ref.~\cite{f2ns} such choice cannot
be derived from general rules and it must be tailored to each specific
problem. The main requirements for an optimal architecture are, first
of all, that the net is large enough so that the results are stable
with respect small variations of the number of neurons (in this case
the neural net is called \emph{redundant}) and, second, that this net
is not so large than the training times become prohibitive.

Thus at the end of the procedure, one ends up with $N_\text{rep}$
fluxes, with each flux $\Phi^{(\text{net})(k)}$ given by a neural net.
The set of $N_\text{rep}$ fluxes provide the best representation of
the corresponding probability density in the space of atmospheric
neutrino fluxes: for example, the mean value of the flux at a given
value of $E_\nu$ is found by averaging over the replicas, and the
uncertainty on this value is the variance of the values given by the
replicas.  

In particular, for any given value of the energy $E_\nu$, the zenith
angle $\cos\theta$ and the neutrino type $t$, one can compute the
average atmospheric neutrino flux as
\begin{equation}
    \langle \Phi^{(\text{net})}\rangle_\text{rep}(E_\nu,c_\nu,t) =
    \frac{1}{N_\text{rep}} \sum_{k=1}^{N_\text{rep}}
    \Phi^{(\text{net})(k)}(E_\nu,\cos\theta,t)
\end{equation}
and the standard deviation as
\begin{equation}
    \sigma_{\Phi}^2(E_\nu,c_\nu,t) = 
    \frac{1}{N_\text{rep}} \sum_{k=1}^{N_\text{rep}}
    \left( \Phi^{(\text{net})(k)}(E_\nu,c_\nu,t) \right)^2
    - \langle \Phi^{(\text{net})}\rangle_\text{rep}^2 (E_\nu,c_\nu,t) \,.
\end{equation}


\subsection{Energy Dependence of Atmospheric Fluxes from Neutrino Data}
\label{sec:fluxatm2}

The procedure sketched above can be applied to determine the
atmospheric neutrino and antineutrino fluxes of $\nu_e$ and $\nu_\mu$
flavors as a function of the neutrino energy and zenith angle. 
However, the precision of the available experimental data is not
enough to allow for a separate determination of the energy, zenith
angle and type dependence of the atmospheric fluxes. Consequently in
Ref.~\cite{ourneutnet} it was assumed that the zenith and type
dependence of the flux is known with some precision and only its
energy dependence was extracted from the data. 

In this case the neural flux parametrization is:
\begin{equation}
    \label{eq:fluxnn}
    \Phi^{(\text{net})}(E_\nu, c_\nu, t) \equiv 
    \frac{d^2\Phi^{(\text{net})}_t}{dE_\nu dc_\nu}
    = \text{NN}(E_\nu)
    \frac{d^2\Phi^{(\text{ref})}_t}{dE_\nu\, dc_\nu}
\end{equation}
where $\text{NN}(E_\nu)$ is the neural network output when the input
is the neutrino energy $E_\nu$ 
\begin{equation}
    \text{NN}(E_\nu) \equiv \text{NN}(E_\nu, \vec\omega) \,.
\end{equation}
and it depends on the neutrino energy, as well as on the parameters
$\vec\omega$ of the neural network. For convenience in
Eq.~\eqref{eq:fluxnn} the neural network flux has been normalized to a
reference differential flux, $\Phi^{(\text{ref})}$, which can be taken
to be, for example, the most recent computations of either the
Honda~\cite{honda} or the Bartol~\cite{bartol} collaborations,
extended to cover also the high-energy region by consistent matching
with the Volkova fluxes~\cite{volkova}.

Notice that in what respects the normalization and energy dependence
of the fluxes, the choice of reference flux is irrelevant.  Any
variation on the normalization or on the energy dependence of the
reference flux can be compensated by the corresponding variation of
$\text{NN}(E_\nu)$ so that the output flux $\Phi^{(\text{net})}$ will
be the same. The dependence of the results of the analysis on the
reference flux comes because of the differences among the different
flux calculations in angular and flavor dependence.

Nothing further has to be assumed about the function
$\text{NN}(E_\nu)$ other than the requirements on the optimal neural
network architecture.  For the particular problem of determining only
the energy dependence of the flux the neural network must have a
single input neuron (whose value is $\log(E_\nu$)) and a final output
neuron (whose value is the $\text{NN}(E_\nu)$ ) and a number of hidden
layers with several neurons each.  For example an architecture with
two hidden layers with 5 neurons each satisfies the above requirements
in the present case.

\begin{figure}\centering
    \includegraphics[width=0.70\textwidth]{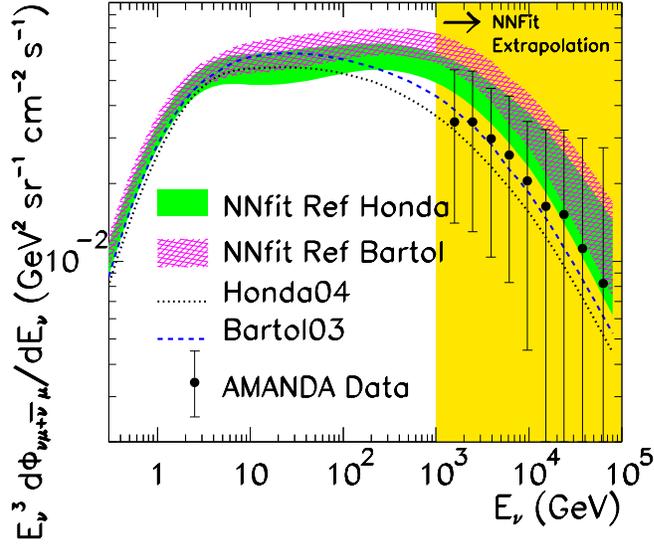}
    \caption{\label{fig:atmflux}%
      Results for the neural network fit for the angular averaged muon
      neutrino plus antineutrino flux and comparison with numerical 
      computations. The fluxes are also shown extrapolated 
      extrapolated to the high energy region and compared to the
      corresponding data from AMANDA~\cite{icecubedata}. The neural
      network fit was performed assuming oscillations with $\Delta
      m_\text{atm}^2 = 2.2\times 10^{-3}~\eVq$ and
      $\tan^2\theta_\text{atm} = 1$}
\end{figure}

The results from the fit are shown in Fig.~\ref{fig:atmflux}.  In the
figure we show the neural network angular averaged muon neutrino and
antineutrino fluxes obtained using as reference the fluxes of
Honda~\cite{honda} and Bartol~\cite{bartol} groups respectively.  In
the figure it is also shown the results of the computations of these 
two groups. The results of the neural network fit are shown as the
$\langle \Phi^{(\text{net})}\rangle_\text{rep}\pm \sigma_{\Phi}$ band
as a function of the neutrino energy. From the figure we see that:
\begin{itemize}
  \item The neural network fluxes are independent of the reference
    flux choice for $E_\nu \lesssim 10$ GeV as expected since both
    Honda and Bartol calculations give very similar angular and
    flavor ratios at those energies.  
    
  \item At those energies the present uncertainty in the extracted
    fluxes is larger than the range of variations between
    calculations.
    
  \item For $E_\nu\lesssim 1$ TeV, which is the energy range for which
    Super-Kamiokande data is available, the fluxes obtained from the 
    neural network fits are in reasonable agreement with the results
    from the calculations of Honda and Bartol groups. 
    
  \item The fits prefer a slightly higher flux than the Honda and
    Bartol calculations at higher energies.
\end{itemize}
All this indicates that until about $E_\nu\sim 1$ TeV we have a good
understanding of the normalization of the fluxes and the present
accuracy from Super-Kamiokande neutrino data is comparable with the
theoretical uncertainties from the numerical calculations.  If one
assumes that the present uncertainties of the angular dependence have
been properly estimated, it turns out that they have very little
effect on the determination of the energy dependence of the fluxes.

In Fig.~\ref{fig:atmflux} we also show the extrapolation of the
results of the fit to the high energy region compared to the data from
AMANDA~\cite{icecubedata}. It is important to recall that the behavior
of neural networks in the extrapolation region is not determined by
its behavior where data is available, as it would happen in fits with
usual functional forms.  As a consequence the values of the extracted
fluxes in the extrapolation region can be extremely unphysical.
Nevertheless the figure illustrates the reach of the presently
available data at higher energies.

%% file: sec.npatm.tex
\section{Testing New Physics in Atmospheric and LBL $\nu$ Oscillations}
\label{sec:npatm}

Oscillations are not the only possible mechanism for flavor
transitions. These can also be generated by a variety of forms of
nonstandard neutrino interactions or properties. In general these
alternative mechanisms share a common feature: they require the
existence of an interaction (other than the neutrino mass terms) that
can mix neutrino flavors.  

In this section we will describe the phenomenology associated with
some scenarios which affect mostly $\nu_\mu\rightarrow\nu_\tau$
atmospheric and LBL neutrino oscillations.  Among others we will
describe effects due to violations of the equivalence principle
(VEP)~\cite{VEP1,VEP2,VEP3,qVEP}, non-standard neutrino interactions
with matter~\cite{wolf}, neutrino couplings to space-time torsion
fields~\cite{torsion}, violations of Lorentz invariance
(VLI)~\cite{VLI1,VLI2,Choubey:2003ke} and of CPT
symmetry~\cite{VLICPT1,VLICPT2,VLICPT3}, neutrino
decay~\cite{bardec1,bardec2}, and of non-standard decoherence
effects~\cite{lisideco1}. The discussion of some scenarios which could
mostly affect solar $\nu_e$ and reactor $\bar\nu_e$ oscillations will
be postponed to Section~\ref{sec:npsolar}. 

From the point of view of atmospheric and LBL neutrino oscillation
phenomenology, the most relevant feature of the scenarios that we will
describe is that, in general, they imply a departure from the $E^{-1}$
energy dependence of the oscillation
wavelength~\cite{yasuda3,flanagan}.  Prior to the highest-statistics
SK data, some of these scenarios could provide a good description
--~alternative to $\Delta m^2$ neutrino oscillations~-- of the
atmospheric neutrino phenomenology~\cite{oldatmfitnp,NSI2,bardec1}.
However, with more precise data, and in particular with the expansion
of the energy range covered by atmospheric neutrino data due to the
inclusion of the upward-going muons, these alternative scenarios
became disfavored as leading mechanism to explain the
observations~\cite{fogli1,lipari,NSI3,lisidec1}.  The results from LBL
experiments further singled out oscillations as the dominant mechanism
of $\nu_\mu\leftrightarrow\nu_\tau$ transitions~\cite{fogli2}. Indeed,
the present experimental precision is such that the presence of these
form of new physics could be observed in the data even if they are
subdominant to oscillations~\cite{ouratmnp,fogli2} as we describe
next.


\subsection{New Physics in $\nu_\mu\rightarrow \nu_\tau$ Oscillations:
No-damping Effects}
\label{sec:npatmforma}

Generically all the new physics scenarios which we are going to 
consider in this section induce new sources of lepton flavor mixing 
in addition to the ``standard'' $\Delta m^2$ oscillations ($\Delta
m^2$-OSC) whose effect in the evolution of the three-neutrino ($+$)
and antineutrino ($-$) system can be determined by solving:
\begin{equation}
    i\, \frac{d\vec\nu}{dx} = H_\pm \, \vec\nu
\end{equation}
where $H_\pm$ is the Hamiltonian in the flavor basis:
\begin{equation}
    \label{eq:hnpgen}
    H_\pm = \frac{1}{2E} U M^2 U^\dagger 
+ V
    + \sum_n \sigma_n^\pm \, E^n {U}_{n,\text{NP}} \, \delta_n \,
    {U}^\dagger_{n,\text{NP}} \,,
\end{equation}
$U$ is the mass-flavor mixing matrix, Eq.~\eqref{eq:U3d}, $M$ is the
neutrino mass matrix, $V$ is the effective potential describing the
standard coherent forward interactions of the neutrinos with matter, 
Eq.~\eqref{eq:evol.4}. $\sigma_n^\pm$ accounts for a possible relative
sign of the new physics effects between neutrinos and antineutrinos, 
$\delta_n(r) $ is a diagonal matrix which parametrizes the size of the
energy differences due to the new physics and ${U}_{n,\text{NP}}$ is
the flavor mixing matrix induced by the new physics effects. 

As discussed in Sec.~\ref{sec:3nu}, as consequence of the fact that
$\Delta m^2_{21}/\vert \Delta m^2_{31}\vert \approx 0.03$ and the
smallness of $\theta_{13}$ we found that for the existing data the
3-$\nu$ oscillations effectively factorize into 2-$\nu$ oscillations
of the two different subsystems: solar plus KamLAND, and atmospheric
plus LBL.

In principle, with the inclusion of the new physics terms it is not
warranted that such factorization will hold.  Nevertheless in what
follows we describe some new physics effects in atmospheric and LBL
neutrino oscillations associated to the dominant $\nu_\mu\rightarrow
\nu_\tau$ oscillations and we will do so under the assumption that the
new effects still allow for an effective 2$\nu$ mixing analysis (a
specific example of possible effects due to departures from this
approximation will be discussed in Sec.~\ref{sec:nseatm}). In all
cases:
\begin{equation}
    \label{eq:deltas}
    M^2=
\frac{\Delta m_{31}^2}{2}
    \begin{pmatrix}
	-1 &  0 \\	
	\hphantom{-}0 & ~1
    \end{pmatrix} \,,
    \qquad
    \delta_n = \frac{\Delta \delta_n}{2}
    \begin{pmatrix}
	-1 & ~0 \\
	\hphantom{-}0 & ~1
    \end{pmatrix}
\end{equation}
and 
\begin{equation} 
    \label{eq:rotat}
    U = U_{23} =
    \begin{pmatrix}
	\hphantom{-}\cos\theta_{23} & ~\sin\theta_{23} \\
	-\sin\theta_{23} & ~\cos\theta_{23}
    \end{pmatrix},
    \quad
    U_{n,\text{NP}} =
    \begin{pmatrix}
	\hphantom{-}\cos\xi_n\hphantom{e^{-i\eta_n}} 
	& ~\sin\xi_n e^{\pm i\eta_n} 
	\\
	-\sin\xi_n e^{\mp i\eta_n} 
	& ~\cos\xi_n\hphantom{e^{-i\eta_n}}
    \end{pmatrix},
\end{equation}
where one must account for possible non-vanishing relative phases
$\eta_n$ between the standard $\Delta m^2$ terms and $\Delta \delta_n$
one. 

Eq.~\eqref{eq:hnpgen} can describe, for example, the evolution of
$\nu_\mu$ and $\nu_\tau$'s in the presence of VEP due to non universal
coupling of the neutrinos, $\gamma_1\neq \gamma_2$ ($\nu_1$ and
$\nu_2$ being related to $\nu_\mu$ and $\nu_\tau$ by a rotation
$\theta_G$), to the local gravitational potential
$\phi$~\cite{VEP1,VEP2,VEP3} with 
\begin{equation}
    \label{eq:veq}
    \Delta\delta_1 =
    2 |\phi|(\gamma_1- \gamma_2) \equiv 2 |\phi| \Delta\gamma \,,
    \qquad \xi_1 = \theta_G \,,
    \qquad \sigma_1^+ = \sigma_1^- \,.
\end{equation}
For constant potential $\phi$, this mechanism is phenomenologically
equivalent to the VLI induced by different asymptotic values of the
velocity of the neutrinos, $c_1\neq c_2$, with $\nu_1$ and $\nu_2$
being related to $\nu_\mu$ and $\nu_\tau$ by a rotation
$\theta_v$~\cite{VLI1,VLI2}. In this case
\begin{equation}
    \label{eq:vli}
    \Delta\delta_1 = (c_1- c_2)=\Delta c \,,
    \qquad \xi_1 = \theta_v \,,
    \qquad \sigma_1^+ = \sigma_1^- \,.
\end{equation}
VEP for massive neutrinos due to quantum effects discussed in
Ref.~\cite{qVEP} can also be parametrized as Eq.~\eqref{eq:hnpgen}
with $n=2$. 

Non-universal coupling of the neutrinos, $k_1\neq k_2$ ($\nu_1$ and
$\nu_2$ being related to the $\nu_\mu$ and $\nu_\tau$ by a rotation
$\theta_Q$), to a space-time torsion field $Q$~\cite{torsion} lead to
an energy independent contribution to the oscillation wavelength with 
\begin{equation}
    \label{eq:torsion}
    \Delta\delta_0= Q (k_1- k_2)\equiv Q \,\delta k \,,
    \qquad \xi_0 = \theta_Q \,,
    \qquad \sigma_0^+ = \sigma_0^- \,.
\end{equation}
Violation of CPT due to Lorentz-violating effects also lead to an
energy independent contribution to the oscillation
wavelength~\cite{VLICPT1,VLICPT2,VLICPT3} with
\begin{equation}
    \label{eq:cpt}
    \Delta\delta_0 = b_1-b_2 \equiv \delta b \,,
    \qquad \xi_0 = \theta_{{CPT}} \,,
    \qquad \sigma_0^+ = -\sigma_0^-
\end{equation}
where $b_i$ are the eigenvalues of the Lorentz violating CPT-odd
operator $\bar{\nu}_L^\alpha b_\mu^{\alpha\beta} \gamma_\mu
\nu_L^\beta$ and $\theta_v$ is the rotation angle between the
corresponding neutrino eigenstates and the flavor
eigenstates~\cite{VLICPT2}.

In all these scenarios, if $\Delta \delta_n$ is constant along the
neutrino trajectory, the expression of $P_{\nu_\mu \to\nu_\mu}$ takes
the form~\cite{VLICPT2}:
\begin{equation}
    \label{eq:prob}
    P_{\nu_\mu \to \nu_\mu} = 1 - P_{\nu_\mu \to \nu_\tau} =
    1 - \sin^2 2\Theta \, \sin^2 \left( 
    \frac{\Delta m^2_{31} L}{4E} \, \mathcal{R} \right) \,.
\end{equation}
where the correction to the $\Delta m^2$-OSC wavelength,
$\mathcal{R}$, and to the global mixing angle, $\Theta$, verify
\begin{align}
    \mathcal{R} \cos 2\Theta \,
    & = \cos 2\theta_{23} + \sum_n\, R_n\, \cos 2\xi_n \,,
    \\
    \mathcal{R} \sin 2\Theta \
    & = \lvert \sin 2\theta_{23} 
    + \sum_n\, R_n \,\sin 2\xi_n \, e^{i\eta_n} \rvert \,,
\end{align}
with $R_n$ being the ratio between $\Delta m^2$--induced and the
$\Delta\delta_n$--induced contributions to the oscillation wavelength
\begin{equation}
    R_n = \sigma_n^+ \frac{\Delta\delta_n E^n}{2} \, \frac{4E}{\Delta m^2_{31}} \,.
\end{equation}
For $P_{\bar{\nu}_\mu \to \bar{\nu}_\mu}$ the same expressions hold
with the exchange $\sigma_n^+ \to \sigma_n^-$ and $\eta_n \to
-\eta_n$.

For scenarios with one new physics source characterized by a unique
$n$:
\begin{align}
    \label{eq:Theta}
    \sin^2 2\Theta &= \frac{1}{\mathcal{R}^2} \left(
    \sin^2 2\theta_{23} + R_n^2 \sin^2 2\xi_n
    + 2 R_n \sin 2\theta_{23} \sin 2\xi_n \cos\eta_n \right) \,,
    \\[2mm]
    \label{eq:Xi}
    \mathcal{R}
    &= \sqrt{1 + R_n^2 + 2 R_n \left( \cos 2\theta_{23} \cos 2\xi_n
      + \sin 2\theta_{23} \sin 2\xi_n \cos\eta_n \right)} \,.
\end{align}
In this case the physical intervals of variation of the five
parameters $\Delta m^2$, $\theta$, $\Delta\delta_n$, $\xi_n$, $\eta_n$
can be easily found from the symmetries of the Hamiltonian and the
oscillation probabilities.  In particular for a given value of
$\sigma_n^+$ the Hamiltonian is invariant under the following
transformations:
\begin{itemize}
  \item $\theta_{23} \to \theta_{23} + \pi$,
  \item $\xi_n \to \xi_n + \pi$,
  \item $\eta_n \to \eta_n + 2\pi$,
  \item $\Delta m^2_{31} \to -\Delta m^2_{31} \quad \text{and} \quad
    \theta_{23} \to \theta_{23} + \pi/2$,
  \item $\Delta\delta_n \to -\Delta\delta_n \quad \text{and} \quad
    \xi_n \to \xi_n + \pi/2$,
  \item $\xi_n \to -\xi_n \quad \text{and} \quad \eta_n \to \eta_n + \pi$.
\end{itemize}
Furthermore, the relevant survival probabilities $P_{\nu_\mu \to
\nu_\mu}$ and $P_{\bar{\nu}_\mu \to \bar{\nu}_\mu}$ are not affected
by a change in the overall sign of the Hamiltonian, as well as change
in the global phase of its non-diagonal components. Therefore, we also
have:
\begin{itemize}
  \item $\theta_{23} \to \theta_{23} + \pi/2 \quad \text{and} \quad
    \xi_n \to \xi_n + \pi/2$,
  \item $\theta \to -\theta \quad \text{and} \quad
    \xi_n \to - \xi_n$,
  \item $\eta_n \to -\eta_n$.
\end{itemize}
The above set of symmetries allows us to define the ranges of
variation of the five parameters as follows:
\begin{equation} \begin{aligned}
    \label{eq:intervals}
    (a) &~ \Delta m^2_{31} \geq 0 \,, & \qquad
    (c) &~ 0 \leq \theta_{23} \leq \pi/2 \,,
    \\
    (b) &~ \Delta\delta_n \geq 0 \,, & \qquad
    (d) &~ 0 \leq \xi_n \leq \pi/4 \,, 
    \\
    & \qquad & (e) &~ 0 \leq \eta_n \leq \pi \,.
\end{aligned} \end{equation}
Thus in the general case one can cover the mixing parameter space by
using, for instance, $0 \leq \sin^2\theta_{23}\leq 1$ and $0 \leq
\sin^22\xi_n \leq 1$.

For the case of real relative phase, $\eta_n \in \{0,\,\pi\}$, it is
possible to absorb the two values of $\eta_n$ into the sign of
$\xi_n$. In this case one can drop $(e)$ and replace $(d)$ by:
\begin{equation}
    \label{eq:xietazer0}
    (d')\; -\pi/4 \leq \xi_n \leq \pi/4
\end{equation}
and use instead $-1\leq\sin 2\xi_n\leq 1$.

Finally we notice that the above derivation is valid for a given sign
of $\sigma_n^+$. Keeping the convention of $\Delta m^2_{31} > 0$ and 
$\Delta\delta_n > 0$ the survival probability for the opposite sign
can be obtained by the exchange
\begin{equation}
    \label{eq:sign}
    \sin^2\theta_{23} \to 1-\sin^2\theta_{23}
    \quad \text{and} \quad
    \eta_n \to \pi - \eta_n \,.
\end{equation}

Similarly the effect of non-standard neutrino-matter interactions
(NSI) which can be cast as a neutral or charged vector
current~\cite{wolf,NSI2} in the evolution of the $\nu_\mu$--$\nu_\tau$
system can be parametrized in terms of the effective 
Hamiltonian~\cite{NSI2,Fornengo:2001pm}:
\begin{equation} 
    \label{eq:halnsi}
    H_\pm =
    \frac{\Delta m^2_{31}}{4 E} {U}_{23}
    \begin{pmatrix}
	-1 & ~0 \\
	\hphantom{-}0 & ~1
    \end{pmatrix}
    {U}_{23}^\dagger
    \pm \sqrt{2} G_F \sum_f N_f(r)
    \begin{pmatrix}
	\varepsilon_{\mu\mu}^{f\pm} & \varepsilon_{\mu\tau}^{f\pm} \\
	{\varepsilon_{\mu\tau}^{f\pm}}^\star & \varepsilon_{\tau\tau}^{f\pm}
    \end{pmatrix}
\end{equation}
where the coefficients $\varepsilon_{\alpha\beta}^{f\pm}$ parametrize
the deviation from standard neutrino interactions: $\sqrt{2} \, G_F
N_f(r) \varepsilon_{\alpha\beta}^{f+}$ is the forward scattering
amplitude of the process $\nu_\alpha + f \to \nu_\beta + f$, and
$\varepsilon_{\alpha\beta}^{f-}$ gives the corresponding amplitude for
antineutrinos. Here $N_f(r)$ is the number density of the fermion $f$
along the path $\vec{r}$ of the neutrinos propagating in the Earth 
which can be obtained, for example, from the PREM~\cite{PREM} matter
density profile and the standard chemical composition with
proton-nucleon ratio $Y_p = 0.497$ in the mantle and $0.468$ in the
core.

If one assumes that the new interactions for neutrinos and
antineutrinos are the same so that $\varepsilon_{\alpha\beta}^{f+} =
(\varepsilon_{\alpha\beta}^{f-})^\star \equiv
\varepsilon_{\alpha\beta}^f$, the Hamiltonian contain three new real
parameters, which can be chosen to be $(\varepsilon_{\tau\tau}^f -
\varepsilon_{\mu\mu}^f)$, $|\varepsilon_{\mu\tau}^f|$ and
$\arg(\varepsilon_{\mu\tau}^f)$. 

For scenarios where neutrinos have non-standard interactions with only
one fermion type $f$, Eq.~\eqref{eq:halnsi} can be seen as a special
case of Eq.~\eqref{eq:hnpgen} with $n=0$, $\sigma_0^- = -\sigma_0^+$,
and
\begin{equation} \begin{gathered}
    \label{eq:parnsi}
    \Delta\delta_0 = 2\sqrt{2}\, G_F \, N_f(r)\, F
    \equiv 4.58 \times 10^{-22}\, Y_f(r) \,
    \frac{\rho(r)_\text{Earth}}{3~\text{g} / \text{cm}^3} \,
    F \, \text{GeV} \,,
    \\[2mm]
    \sin(2\xi) = \frac{|\varepsilon_{\mu\tau}^f|}{F} \,,
    \qquad
    \eta = \arg(\varepsilon_{\mu\tau}^f) \,,
    \qquad
    F = \sqrt{|\varepsilon_{\mu\tau}^f|^2 +
      \frac{(\varepsilon_{\tau\tau}^f - \varepsilon_{\mu\mu}^f)^2}{4}} \,, 
    \\[2mm]
    Y_e = Y_p \,,
    \qquad
    Y_u = 1+Y_p \,,
    \qquad
    Y_d = 2-Y_p \,.
\end{gathered} \end{equation}
The main difference with the previous scenarios is that NSI only 
affect the evolution of neutrinos when crossing the Earth, and their
strength changes along the neutrino trajectory. Consequently the
flavor transition probability cannot be simply read from
Eq.~\eqref{eq:prob} and its evaluation requires the numerical solution
of the neutrino evolution in the Earth matter.  


\subsection{Sensitivity at Current Experiments}
\label{sec:noatmresults}

\begin{figure}\centering
    \includegraphics[width=5.3in]{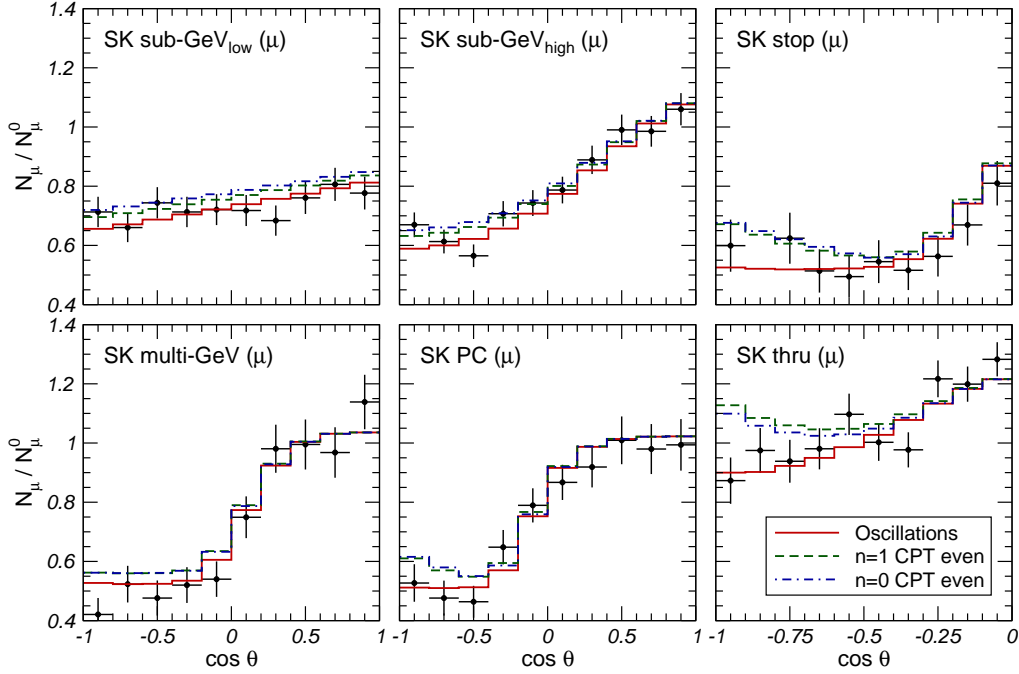}
    \caption{\label{fig:nph-zenith}%
      Zenith-angle distributions (normalized to the no-oscillation
      prediction) for the Super--Kamiokande $\mu$-like events. The
      full line gives the distribution for the best fit of $\Delta
      m^2$-OSC, $\Delta m^2 = 2.4\times 10^{-3}~\eVq$ and
      $\sin^2\theta = 0.5$. The dashed and dotted lines give the
      distributions for $\Delta m^2$-OSC plus new physics scenarios
      for $n=1$ and $n=0$ with $\Delta\delta_1 = 2.4\times 10^{-24}$
      and $\Delta\delta_0 = 6.6\times 10^{-23}$~GeV respectively. In
      both cases $\eta=\xi=0$ and the oscillation parameters have been
      set to their best fit values.}
\end{figure}

In Fig.~\ref{fig:nph-zenith} we illustrate the effect in the
atmospheric neutrino events distributions for $\Delta m^2$-OSC plus
sub-dominant CPT-even tensor-like and vector-like new physics 
effects, for some characteristic values of the parameters. In both
cases $R_n$ is a growing function of $E$ and the new effects become
most relevant in the higher energy samples, in particular for upward
going muons.

\begin{figure}\centering
    \includegraphics[width=4.5in]{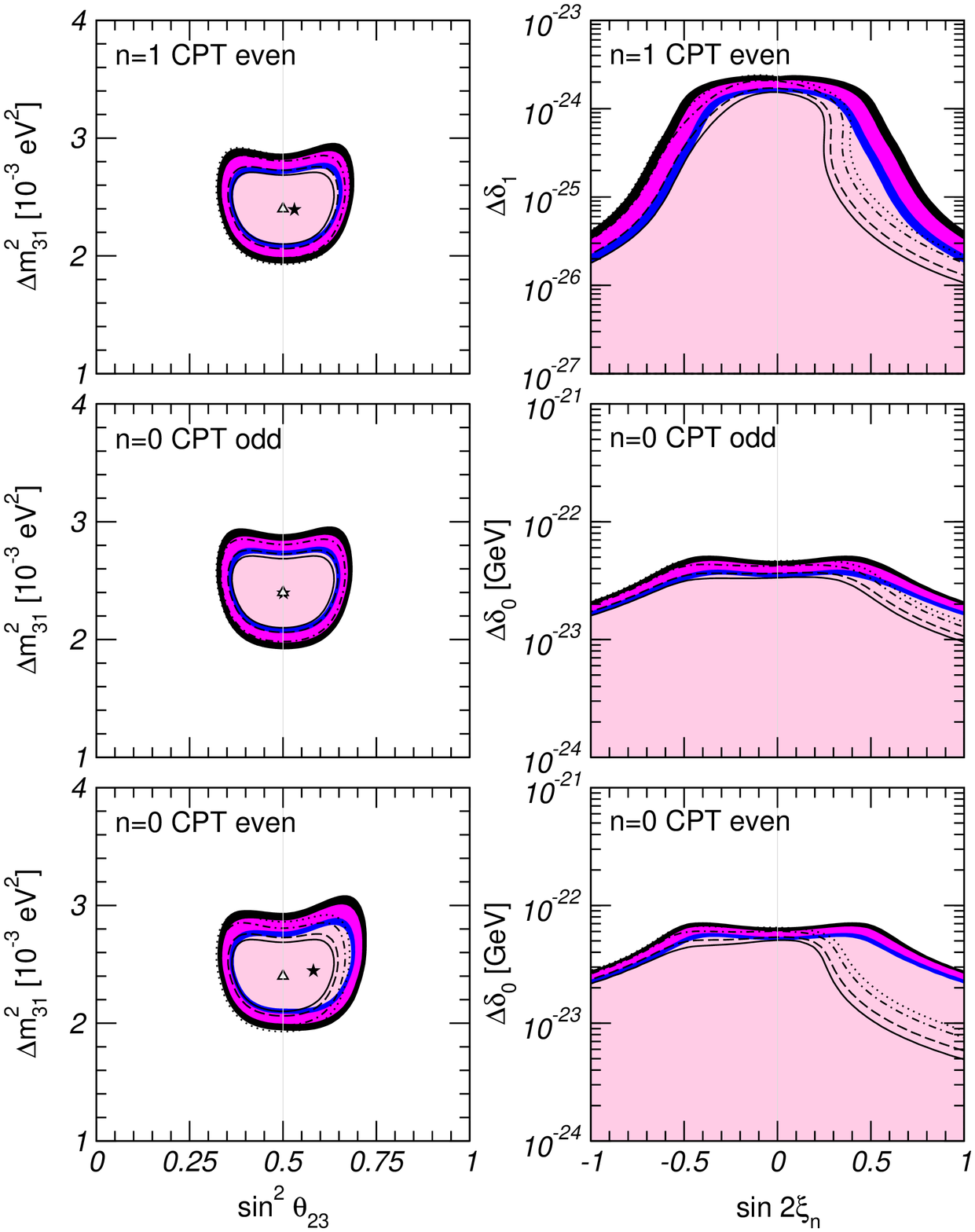}
    \caption{\label{fig:nph-funny}%
      Allowed parameter regions for the analysis of atmospheric and
      LBL data in presence of $\nu_\mu \to \nu_\tau$ oscillations and
      different new physics effects as labeled in the figure. Each
      panel shows a two-dimensional projection of the allowed
      five-dimensional region after marginalization with respect to
      the three undisplayed parameters. The different contours
      correspond to the two-dimensional allowed regions at 90\%, 95\%,
      99\% and $3\sigma$ CL. The filled areas in the left panels show
      the projected two-dimensional allowed region on the oscillation
      parameters $\Delta m^2$--$\sin^2\theta$ plane. The best fit
      point is marked with a star. For the sake of comparison we also
      show the lines corresponding to the contours in the absence of
      new physics and mark with a triangle the position of the best
      fit point. The results are shown for the chosen relative sign
      $\sigma_n^+ = +1$; for $\sigma_n^+ = -1$ the corresponding
      region would be obtained by $\sin^2\theta \to 1-\sin^2\theta$.
      The regions on the right panels show the allowed values for the
      parameters characterizing the strength and mixing of the new
      physics. The full regions corresponds to arbitrary values of the
      phase $\eta_n$ while the lines correspond to the case $\eta_n
      \in \{0,\,\pi\}$.}
\end{figure}

\begin{figure}\centering
    \includegraphics[width=4.5in]{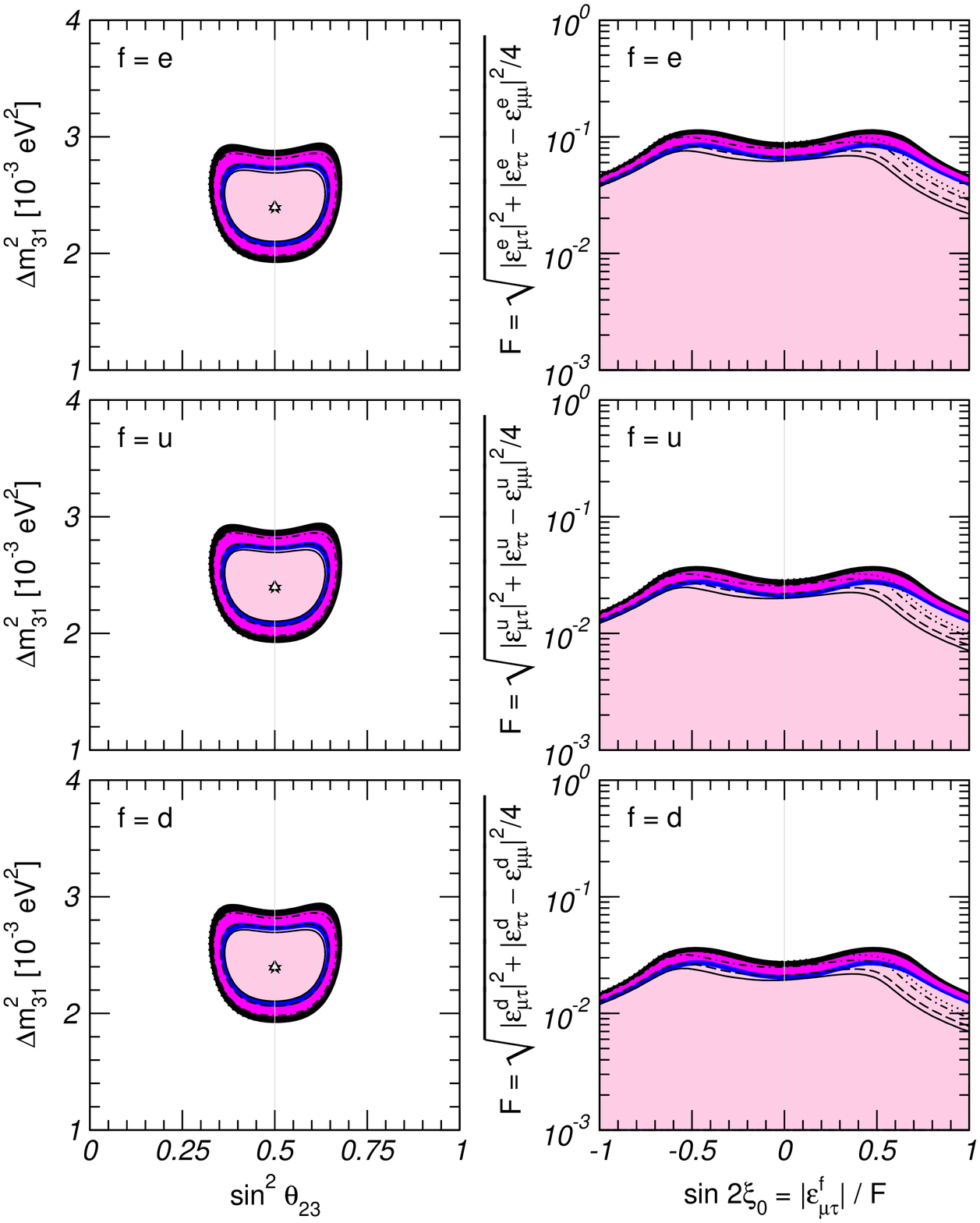}
    \caption{\label{fig:nph-nsni}%
      Same as Fig.~\ref{fig:nph-funny} for the case of $\text{$\Delta
      m^2$-OSC} + \text{NSI}$.} 
\end{figure}

Figs.~\ref{fig:nph-funny} and~\ref{fig:nph-nsni} show the
two-dimensional projections of the allowed parameter region for the
analysis of atmospheric and LBL data in presence of $\nu_\mu \to
\nu_\tau$ oscillations and different new effects.  The corresponding
results for the case of NSI are presented in Fig.~\ref{fig:nph-nsni}.

From the figures we see that the best fit point is always very near
the best fit point of pure $\Delta m^2$-OSC.  In other words, the data
does not show any evidence of presence of new physics even as a
sub-dominant effect.  Consequently the analysis allow us to derive
well-defined upper bounds on the new physics parameters. The figures
also illustrate that generically the bounds on $\Delta\delta_n$
tightens for larger values of $\xi_n$, being this effect stronger for
effects leading to sub-dominant oscillations with stronger energy
dependence. Imposing that the Hamiltonian is real does not
substantially affect these conclusions.

Altogether the analysis of atmospheric and LBL neutrino data allows to
impose the following at 90\% ($3\sigma$) bounds on possible VLI in the
$\nu_\mu$--$\nu_\tau$ sector via CPT-even effects: 
\begin{equation}
    \label{eq:VLIlim}
    \frac{|\Delta c|}{c} \leq 1.3\times 10^{-24}
    ~ (2.1\times 10^{-24}) \,,
\end{equation}
and on the possible VEP:
\begin{equation}
    \label{eq:VEPlim}
    |\phi\, \Delta \gamma| \leq 6.3\times 10^{-25}
    ~ (1.0\times 10^{-24}) \,.
\end{equation}
$\nu_\mu$--$\nu_\tau$ the VLI via CPT-odd effects is constrained to
\begin{equation}
    \label{eq:CPTlim}
    |\delta b| \leq 2.9\times 10^{-23}
    ~ (4.6\times 10^{-23})~\text{GeV} \,,
\end{equation}
and non-universality of the neutrino couplings to a torsion field
verify
\begin{equation}
    \label{eq:torsionlim}
    |Q \,\delta k| \leq 4.7\times 10^{-23}
    ~ (6.6\times 10^{-23})~\text{GeV}.
\end{equation}

For the case of non-standard neutrino interactions the corresponding
90\% ($3\sigma$) bounds read:
\begin{equation} \begin{aligned}
    \label{eq:NSIreallim}
    (-0.058) \; {-0.038} \leq \varepsilon_{\mu\tau}^e & \leq 0.024 \; (0.043) \,,
    & \qquad |\varepsilon_{\tau\tau}^e - \varepsilon_{\mu\mu}^e| &\leq 0.11 \; (0.17) \,,
    \\[1mm]
    (-0.019) \; {-0.012} \leq \varepsilon_{\mu\tau}^u & \leq 0.008 \; (0.014) \,,
    & \qquad |\varepsilon_{\tau\tau}^u - \varepsilon_{\mu\mu}^u| &\leq 0.036 \; (0.056) \,,
    \\[1mm]
    (-0.019) \; {-0.012} \leq \varepsilon_{\mu\tau}^d & \leq 0.008 \; (0.014) \,,
    & \qquad |\varepsilon_{\tau\tau}^d - \varepsilon_{\mu\mu}^d| &\leq 0.035 \; (0.054)
\end{aligned} \end{equation}
for real NSI, and
\begin{equation} \begin{aligned}
    \label{eq:NSIcplxlim}
    |\varepsilon_{\mu\tau}^e| &\leq 0.038 \; (0.058) \,,
    & \qquad |\varepsilon_{\tau\tau}^e - \varepsilon_{\mu\mu}^e| &\leq 0.12 \; (0.19) \,,
    \\[1mm]
    |\varepsilon_{\mu\tau}^u| &\leq 0.012 \; (0.019) \,,
    & \qquad |\varepsilon_{\tau\tau}^u - \varepsilon_{\mu\mu}^u| &\leq 0.039 \; (0.061) \,,
    \\[1mm]
    |\varepsilon_{\mu\tau}^d| &\leq 0.012 \; (0.019) \,,
    & \qquad |\varepsilon_{\tau\tau}^d - \varepsilon_{\mu\mu}^d| &\leq 0.038 \; (0.060)
\end{aligned} \end{equation}
for the general case of complex $\varepsilon_{\mu\tau}^f$. Note that
the different chemical composition of the Earth mantle and core have
very little impact on NSI: from Fig.~\ref{fig:nph-nsni} and
Eqs.~\eqref{eq:NSIreallim} and~\eqref{eq:NSIcplxlim} we see
immediately that the bounds on $\varepsilon_{\alpha\beta}^u$ are
practically identical to those on $\varepsilon_{\alpha\beta}^d$, and
are three times stronger than those on $\varepsilon_{\alpha\beta}^e$.
It is therefore convenient to define:
\begin{equation}
    \label{eq:NSIeffeps}
    \varepsilon_{\alpha\beta} \equiv
    \sum_f \left< \frac{Y_f}{Y_e} \right> \varepsilon_{\alpha\beta}^f
    \approx \varepsilon_{\alpha\beta}^e
    + 3 \varepsilon_{\alpha\beta}^u
    + 3 \varepsilon_{\alpha\beta}^d
\end{equation}
which allows to generalize the former results to the case where
non-standard interactions with electrons, up-type quarks and down-type
quarks are present \emph{simultaneously}. By construction, the bounds
and allowed regions for the effective parameters
$\varepsilon_{\alpha\beta}$ coincide with those for
$\varepsilon_{\alpha\beta}^e$ shown in Eqs.~\eqref{eq:NSIreallim}
and~\eqref{eq:NSIcplxlim} and in Fig.~\ref{fig:nph-nsni}.


\subsection{Non-standard Neutrino Interactions in the
  $\nu_e\leftrightarrow \nu_\tau$ Channel}
\label{sec:nseatm}

The case of non-standard neutrino interactions in the
$\nu_\mu\leftrightarrow\nu_\tau$ channel considered in the previous
section is only a special case of the most general scenario where
\emph{all} the known neutrino flavors have non-standard interactions
with matter. Neglecting the solar mass splitting, $\Delta m_{21}^2$,
and setting $\theta_{13}$ to zero, the evolution of the neutrino
system is described by the Hamiltonian:
\begin{equation}
    \label{eq:nseHamil}
    H_\pm = \frac{\Delta m^2_{31}}{4E}
    \begin{pmatrix}
	-1 & 0 & 0 \\
	0 & -\cos 2\theta_{23} & \sin 2\theta_{23} \\
	0 & \hphantom{-}\sin 2\theta_{23} & \cos 2\theta_{23}
    \end{pmatrix}
    \pm \sqrt{2} G_F N_e(r)
    \begin{pmatrix}
	1 + \varepsilon_{ee}^\pm
	& \varepsilon_{e\mu}^\pm
	& \varepsilon_{e\tau}^\pm
	\\
	\varepsilon_{e\mu}^{\pm\star}
	& \varepsilon_{\mu\mu}^\pm
	& \varepsilon_{\mu\tau}^\pm
	\\
	\varepsilon_{e\tau}^{\pm\star}
	& \varepsilon_{\mu\tau}^{\pm\star}
	& \varepsilon_{\tau\tau}^\pm
    \end{pmatrix}
\end{equation}
Neutrino scattering tests, like those of NuTeV~\cite{nutev} and
CHARM~\cite{charm}, mainly constrain the NSI couplings of the muon
neutrino: $|\varepsilon_{e\mu}| \lesssim 10^{-3}$ and
$|\varepsilon_{\mu\mu}|\lesssim 10^{-3}-10^{-2}$. The limits they
place on $\varepsilon_{ee}$, $\varepsilon_{e\tau}$, and
$\varepsilon_{\tau\tau}$ are rather loose, \textit{e.g.},
$|\varepsilon_{\tau\tau}^{uR}| < 3$, $-0.4 < \varepsilon_{ee}^{uR} <
0.7$, $|\varepsilon_{\tau e}^u| < 0.5$, $|\varepsilon_{\tau e}^d| <
0.5$~\cite{Davidson:2003ha}. 

Given the above bounds one can safely set $\varepsilon_{e\mu}$ and
$\varepsilon_{\mu\mu}$ to zero in the analysis. Furthermore, motivated
by the results of the previous section also $\varepsilon_{\mu\tau}$
can be neglected. In this approximation, the Hamiltonian contains five
parameters: $\Delta m^2_{31}$ and $\theta_{23}$ and three NSI
couplings $\varepsilon_{\tau\tau}$, $\varepsilon_{ee}$, and
$\varepsilon_{e\tau}$ these last two giving the new three-neutrino
mixing effects. 

\begin{figure}\centering
    \includegraphics[width=4.5in]{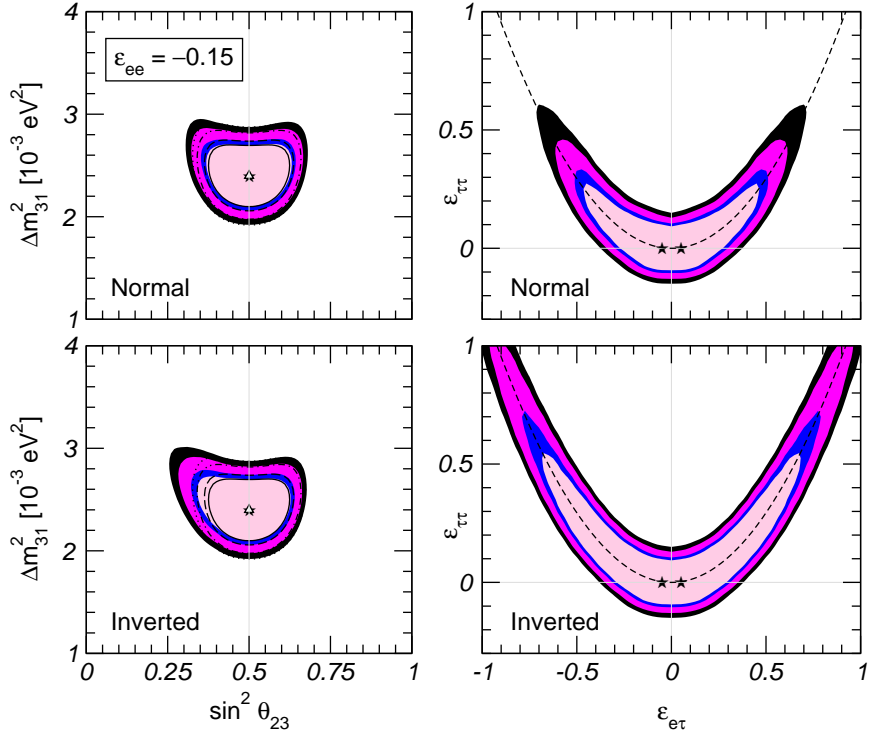}
    \caption{\label{fig:nph-nsne}%
      Same as Fig.~\ref{fig:nph-funny} for the case of non-standard
      neutrino interactions in the $\nu_e\leftrightarrow \nu_\tau$
      channel. Upper (lower) panels correspond to the case of normal
      (inverted) mass hierarchy. The black dashed lines in the right
      panels correspond to the parabola $(1 + \varepsilon_{ee}) \,
      \varepsilon_{\tau\tau} = |\varepsilon_{e\tau}|^2$ defined in
      Eq.~\eqref{eq:nseParab}.}
\end{figure}

As illustration of the effects due to the presence of these two new
couplings we show in Fig.~\ref{fig:nph-nsne} the results of the
atmospheric and LBL neutrino analysis of Ref.~\cite{michelalex} in
which for simplicity a fixed value $\varepsilon_{ee} =-0.15$ was set.
The complete analysis of the full parameter space, as well as a
discussion of subleading effects due to non-zero $\theta_{13}$ and
$\Delta m^2_{21}$, can be found in~\cite{cecialex1,cecialex2}.
Comparing with Fig.~\ref{fig:nph-nsni} we see that the presence of 
NSI in the $\nu_e\leftrightarrow \nu_\tau$ channel allows for the
relaxation of the lower bound on $\theta_{23}$, especially for the
case of inverted hierarchy. On the other hand, the determination of
$\Delta m^2_{31}$ is robust. However, the most differentiating feature
is the appearance of a line in the $(\varepsilon_{e\tau},\,
\varepsilon_{\tau\tau})$ plane where the effects of NSI cancel to a
great extent, so that the fit is quite good even for large values of
$\varepsilon_{e\tau}$ and $\varepsilon_{\tau\tau}$.

These features can be understood as follows. In the high energy limit,
$E \gtrsim 10$~GeV, $\sqrt{2} G_F N_e \gg \Delta m_{31}^2/4E$, so for
large values of the NSI parameters the matter term in the
Hamiltonian~\eqref{eq:nseHamil} dominates over the vacuum term. In
general, the presence of large matter effects produces deformations of
the energy spectrum and therefore spoils the accurate description of
the data provided by the pure vacuum solution. This is the reason why
most of the points in the $(\varepsilon_{\mu\tau},\,
\varepsilon_{\tau\tau})$ plane are excluded. However, along the
parabola:
\begin{equation}
    \label{eq:nseParab}
    (1 + \varepsilon_{ee}) \, \varepsilon_{\tau\tau} =
    |\varepsilon_{e\tau}|^2 \,,
\end{equation}
the matter term in $H_\pm$ has two degenerate eigenvalues, meaning
that there is a two-dimensional subspace where the NSI effects
vanishes. In this subspace, neutrino oscillations exhibit the $1/E$
behavior typical of vacuum oscillations, and a good fit of the data
is possible at the price of a shift in the oscillation parameters to
lower values of $\theta_{23}$ and larger values of $\Delta m_{31}^2$.
Such a shift is responsible for the extension to lower mixing angles
of the allowed region in the $(\Delta m_{31}^2,\, \theta_{23})$ plane,
and introduce a tension with low-energy data (which are only
marginally affected by NSI, and therefore favor the ``standard''
values of $\Delta m_{31}^2$ and $\theta_{23}$) which eventually cut
the otherwise infinite parabola of degenerate solutions. 

The main conclusion is that in this case it is possible to find
regions of parameters in which the new physics effects cancel in the
existing observables. As a consequence, while giving no positive
evidence for non-standard physics the data cannot exclude large NSI.
For example the $3\sigma$ bound $|\varepsilon_{\tau\tau}|< 0.17$
described in the previous section can be significantly relaxed by the
addition of nonzero $\varepsilon_{e\tau}$ and $\varepsilon_{ee}$, and
values as large as $\varepsilon_{\tau\tau} \sim 1$ are allowed.


\subsection{Damping Effects in $\nu_\mu\rightarrow \nu_\tau$ Oscillations:
  Decay and Decoherence}

We describe in this section the phenomenology of atmospheric neutrino
flavor mixing in the presence of new physics effects which, unlike
those described in Sec.~\ref{sec:npatmforma}, lead to a decrease with
time of the oscillation amplitude. In particular we will concentrate
on the possibility of neutrino decay and of non-standard physics
leading to quantum decoherence.

For the sake of concreteness, when discussing neutrino decay, we will
focus on scenarios in which only the heaviest neutrino ($\nu_3$ by
convention) decays and its decay is invisible so that the decay
products are either outside of the $\nu_\mu$ $\nu_\tau$ neutrino
ensemble or their are so degraded in energy that they do not
contribute to the observed event rates and they are fast enough to
have interesting effects in atmospheric neutrino phenomenology

Fast invisible neutrino decay for $\nu_3$ to $\nu_{i}$ Majorana
neutrinos can be induced by the effective interaction:
\begin{equation}
    \label{eq:majdec}
    \mathcal{L}_I = g_{i3} \, \bar{\nu}_{iL}^c \, \nu_{3L} \ ,J
    + \text{h.c.} \,,
\end{equation}
where $J$ is the Majoron field~\cite{maj1,maj2,maj3,maj4,maj5,maj6}, 
which has to be dominantly singlet with only a small triplet
admixture, in order to satisfy the constraints from the invisible
decay width of $Z$~\cite{yosi}. 

A decay model for Dirac neutrinos can also be constructed via the
coupling between the right--handed $SU(2)$--singlet states given
by~\cite{acker}
\begin{equation} 
    \label{eq:dirdec}
    \mathcal{L}_I = g_{i3} \, \bar{\nu}_{iR}^c \, \nu_{3R} \, \chi
    + \text{h.c.} \,,
\end{equation}
where $\chi$ is a complex scalar field with lepton number $-2$, $I_W =
0$, and hypercharge zero and which should be light compared to the
neutrino masses.

With the interaction of Eq.~\eqref{eq:dirdec} or
Eq.~\eqref{eq:majdec}, the rest-frame lifetime of $\nu_3$ is given by
\begin{equation}
    \tau_3 = \frac{16\pi}{g_{i3}^2}
    \frac{m_3^3}{\Delta m^2_{3i} (m_3 + m_i)^2} \,,
\end{equation}
Furthermore it can be shown that the mass difference between the
decaying neutrino and the daughter one has to verify that
\begin{equation}
    \label{eq:declimit}
    \Delta m^2_{3j} > 0.73~\eVq \,,  
\end{equation}
to satisfy constraints from $K\to\mu + \text{neutrals}$
decay~\cite{bardec1}.

With these assumptions neutrino decay can be accounted for in the
evolution equation by introducing an imaginary part in the Hamiltonian
which is proportional to the decay width:
\begin{equation}
    \label{eq:nudec}
    i\, \frac{d\vec\nu}{dx} = U \left[
    \frac{\Delta m_{31}^2}{4E}
    \begin{pmatrix}
	-1 & 0 \\	
	\hphantom{-}0 & ~1
    \end{pmatrix} 
    - i \frac{m_3}{2 \, \tau_3 \, E}
    \begin{pmatrix}
	0 & 0 \\
	0 & 1
    \end{pmatrix} \right] U^\dagger \, \vec\nu \,,
\end{equation}
where $\tau_3$ is the $\nu_3$ lifetime.

Solving Eq.~\eqref{eq:nudec} one gets the survival probability of
$\nu_\mu$ as given by:
\begin{multline}
    \label{eq:pmumudec}
    P(\nu_\mu\to \nu_\mu) = \cos^4 \theta_{23} + \sin^4 \theta_{23} 
    e^{-\frac{m_3\, L}{\tau_3 \, E}}
    \\
    + 2 \sin^2\theta_{23} \cos^2 \theta_{23} 
    e^{-\frac{m_3\, L}{2 \tau_3\, E}} 
    \cos \left( \frac{\Delta m_{31}^2 L}{2E} \right) \,.
\end{multline}

As mentioned above, Eq.~\eqref{eq:pmumudec} can describe for example
the decay $\nu_3 \rightarrow \bar{\nu}_{1,2} + J (\chi)$. In this case
the mass difference between the decaying state and the daughter state
is the same as in Eq.~\eqref{eq:pmumudec}. Therefore the constraint
Eq.~\eqref{eq:declimit} implies that the oscillating term in
Eq.~\eqref{eq:pmumudec} averages to zero and $P(\nu_\mu \to \nu_\mu)$
simplifies to
\begin{equation}
    P(\nu_\mu\to \nu_\mu) = \cos^4 \theta_{23} + \sin^4 \theta_{23} 
    e^{-\frac{m_3\, L}{\tau_3\, E}}  .
\end{equation}

This decay scenario was proposed in Ref.~\cite{bardec1} where it was 
found that it could describe the $L/E_\nu$ dependence and the
asymmetry of the contained events in Super-Kamiokande if $\sin^2
\theta_{23} \sim 0.87$ and $m_3 / \tau_3 \sim 1~\text{GeV} / D_E$
(where $D_E$ is the diameter of the Earth). 

However, when studied in detail, it was shown that the description of 
the global contained event sample in this scenario is worse than in
the case of oscillations. This arises from the strong energy
dependence of the survival probability while the contained data both
in the sub-GeV and multi-GeV samples present a similar deficit. As a
consequence the allowed decay lifetimes which give a good description
to the sub-GeV and multi-GeV data are not the same and the decay
hypothesis cannot produce enough up-down asymmetry for the multi-GeV
sample without conflicting with the sub-GeV data.

This behavior becomes particularly lethal when trying to describe the
upward going muon data since for lifetimes favored by the contained
event data very little muon conversion is expected already for
stopping muons in contradiction with observation.  Based on this fact
this mechanism was ruled out in its simpler form in
Refs.~\cite{lipari,lisidec1}.

Another possibility, mentioned in Ref.~\cite{bardec1} and developed in
Ref.~\cite{bardec2} was the decay of $\nu_3$ is into a state $\nu_j$
with which it does not mix. In this case, the $\Delta m_{31}^2$ in
Eq.~\eqref{eq:pmumudec} is not directly related to the mass difference
between the decaying state and the daughter neutrino. Consequently
$\Delta m_{31}^2$ can be taken to be very small for all energy ranges
relevant for atmospheric neutrinos ($\Delta m^2_{31}<10^{-4}~\eVq$) so
that the the oscillating term in Eq.~\eqref{eq:pmumudec} is one and
$P(\nu_\mu\to \nu_\mu)$ becomes
\begin{equation}
    \label{eq:pmumudec2}
    P(\nu_\mu \to \nu_\mu) = \left( \cos^2 \theta_{23}
    + \sin^2 \theta_{23} e^{-\frac{m_3\, L}{2 \tau_3\, E}} \right)^2 \,.
\end{equation}
This scenario was explored in detail in Ref.~\cite{bardec2} where it
was found that a good fit to the contained and upgoing muon 
atmospheric data at that time could be obtained for the choice of
parameters
\begin{equation}
    \tau_3/m_3 = 63~\text{km/GeV}, \qquad \sin^2 \theta_{23} = 0.30 \,.
\end{equation}
In Ref.~\cite{atmle} Super-Kamiokande collaboration presented the
study of the muon neutrino disappearance probability as a function of
neutrino flight length L over neutrino energy E. They found a dip in
the L/E distribution, as predicted from the sinusoidal flavor
transition probability of neutrino oscillation.  This analysis found
that the decay probability in Eq.~\eqref{eq:pmumudec2} provided a
worse fit (by about $3.4\sigma$) than the pure oscillation hypothesis
to the observed distribution.  

Another mechanism that was proposed as a modification to the
oscillation pattern was the possibility of new sources of quantum
decoherence. Quantum gravity, as suggested by Hawking in the context
of black-hole thermodynamics~\cite{Hawking} could provide a source for
this effect.  In this approach any physical system is inherently
``open'', due to its unavoidable, decohering interactions with a
pervasive ``environment'' (the spacetime and its Planck-scale
dynamics~\cite{Giddins,Ellis1}).  In Ref.~\cite{lisideco1} this
mechanism was studied as an alternative interpretation of the
atmospheric neutrino data as we briefly summarize here.

In order to include decoherence effects in the neutrino flavor
oscillation equations it is convenient to use the density matrix
formalism. In this formalism the evolution of the neutrino ensemble 
is determined by the Liouville equation for the density matrix 
$\rho(t) = \nu(t)\otimes \nu(t)^\dagger$
\begin{equation}
    \label{eq:eqdensity}
    \frac{d\rho}{dx} = -i [H,\, \rho] \,,
\end{equation}
where $H$ is given by Eq.~\eqref{eq:hnpgen}. The survival probability
in Eq.~\eqref{eq:prob} is given by $P_{\mu\mu}(t) = \Tr[\Pi_{\nu_\mu}
\, \rho(t)]$, where $\Pi_{\nu_\mu} = \nu_\mu \otimes \nu_\mu$ is the
$\nu_\mu$ state projector, and with initial condition $\rho(0) =
\Pi_{\nu_\mu}$. An equivalent equation can be written for the
antineutrino density matrix.

For the case of oscillations between two neutrino states the hermitian
operators $\rho$, $H$ and the flavor projectors $\Pi_{\nu_\mu}$ and
$\Pi_{\nu_\tau}$ can be expanded in the basis formed by the unit
matrix and the three Pauli matrices $\sigma_i$.  In particular we can
write
\begin{equation}
    \label{eq:rhoHpauli}
    \rho(x) = \frac{1}{2} \big[I
    + \vec{p}(x) \cdot \vec\sigma \big] \,,
    \qquad
    H = \frac{1}{2} \vec{h} \cdot \vec\sigma \,,
\end{equation}
and the evolution of the neutrino ensemble is determined by a
precession-like equation of the three-vector $\vec p(t)$
\begin{equation}
    \frac{d\vec{p}}{dx} = \vec{p}(x) \times \vec{h} \,.
\end{equation}

In this language modifications of Eq.~\eqref{eq:eqdensity} which
emerge from dissipative interactions with an environment can be
parametrized by introducing an extra term $\mathcal {D}[\rho]$,
\begin{equation}
    \label{eq:eqdensitydeco}
    \frac{d\rho}{dx} = -i [H,\, \rho] - \mathcal{D}[\rho] \,,
\end{equation}
which violates the conservation of $\Tr(\rho^2)$ and allows
transitions from pure to mixed states.  The operator $\mathcal {D}$
has the dimension of an energy, and its inverse defines the typical
(coherence) length after which the system gets mixed. 

In Ref.~\cite{lisideco1} it was assumed that the decoherence term took
the Lindblad form~\cite{Lindblad} 
\begin{equation}
    \label{eq:lindblad}
    \mathcal{D}[\rho] = \sum_n 
    \lbrace \rho,\,  D_n \, D_n^\dagger \rbrace 
    -2 D_n\, \rho\, D^\dagger_n \,,
\end{equation}
which arises from the requirement of complete positivity~\cite{CPos}
as implied by a a linear, Markovian, and trace-preserving map
$\rho(0)\to\rho(t)$. 

In Eq.~\eqref{eq:lindblad} the operators $D_n$ arise from tracing away
the environment dynamics and here they parametrize the new physics
effects. for which there are no first-principle calculations.
Consequently, in Ref.~\cite{lisideco1} a phenomenological approach was
taken in which it was simply required that the $\nu$ system obeys the
laws of thermodynamics: the increase of the entropy and the
conservation of the average energy. These two requirements imply that:
\begin{equation}
    D_n = D^\dagger_n \qquad \text{and}\qquad [H,\, D_n] = 0 \,,
\end{equation}
so that Eq.~\eqref{eq:lindblad} becomes
\begin{equation}
    \mathcal{D}[\rho] = \sum_n \, [D_n,\, [D_n,\, \rho] ] \,.
\end{equation}
As it was discussed above for the case of oscillations, the hermitian
operators $\rho$, $\Pi_{\nu_\mu}$, $H$, and $D_n$, can be expanded
onto the basis formed by the unit matrix and the three Pauli matrices
so that in addition to Eq.~\eqref{eq:rhoHpauli} we can write
\begin{equation}
    D_n = \frac{1}{2} \vec{d}_n \cdot \vec{\sigma} \,,
\end{equation}
with this Eq.~\eqref{eq:eqdensitydeco} can be then written as a Bloch
equation,
\begin{equation}
    \frac{d\vec{p}}{dx} = \vec{p}(x) \times \vec{h} -G \cdot \vec p \,,
\end{equation}
where the tensor ${G} = \sum_n |\vec{d}_n|^2 \, I - \vec{d}_n \otimes
\vec{d}_n^T$. This equation has a simple physical interpretation: the
first term induces flavor oscillations, as discussed above, while the
decoherence term ${G}\cdot\vec{p}$ is responsible for their damping.

The requirement of conservation of the average energy, $[H,\, D_n] =
0$, implies that each vector $\vec{d}_n$ is parallel to $\vec{h}$.
Therefore, the tensor ${G}$ takes the form
\begin{equation}
    G = \diag(\gamma,\, \gamma,\, 0)
    \qquad\text{with}\qquad
    \gamma = \sum_n |\vec{d}_n|^2 \,.
\end{equation}
With this the $\nu_\mu$ survival probability takes the form
\begin{equation}
    \label{eq:pmumudeco}
    P_{\mu\mu} = 1 - \frac{1}{2} \sin^2 2\theta_{23} \left[
    1 - e^{-\gamma L} \cos\left(
    \frac{\Delta m^2_{31} L}{2 E}\right) \right] \,.
\end{equation}
The L/E analysis of SK~\cite{atmle} disfavored the pure decoherence
scenario (corresponding to the limit $\Delta m^2_{31} = 0$ in 
Eq.~\eqref{eq:pmumudeco}) by about $3.8\sigma$ when compared to the
pure oscillation case.  In Ref.~\cite{fogli2} a global analysis to the
Super-Kamiokande and first K2K data was performed allowing for
simultaneous oscillations and decoherence effects as described by
Eq.~\eqref{eq:pmumudeco}. They found that the best fit was obtained by
the pure oscillation scenario so that no evidence for decoherence
effects was present in the data. Conversely the analysis allowed to
set an upper bound on the decoherence parameter
\begin{equation}
    \gamma^2 \lesssim 3 \times 10^{-3}~\eVq \,,
\end{equation}
at $3\sigma$.

It is interesting to notice that both neutrino decay and decoherence
lead to an exponentially decreasing $P_{\mu\mu}$ which is
qualitatively similar as can be seen by comparing
Eq.~\eqref{eq:pmumudec} and Eq.~\eqref{eq:pmumudeco}.  Consequently
the decay and decoherence scenarios cannot be easily distinguished in
$\nu_\mu$ disappearance.  However, they can be distinguished through
the appearance modes, \textit{i.e.}, through neutral current (NC)
events and $\tau$ appearance events in SK. This is so because in the
in the decoherence case the total number of active neutrinos is
conserved while in the decay scenario it decreases with L/E. 


\subsection{Effects at Neutrino Telescopes: Propagation in Matter of 
High Energy Oscillating Neutrinos}

In contrast to the $E$ energy dependence of the conventional
oscillation length, the new physics scenarios that we discussed in
Sec.~\ref{sec:npatmforma} predict neutrino oscillations with
wavelengths that are constant or decrease with energy. Therefore the
effects are more visible at higher energies.

Neutrino telescopes~\cite{icecubelect} are underwater or under-ice
devices aimed at detecting high energy neutrinos from distant sources.
In neutrino telescopes, neutrinos are detected through the observation
of Cherenkov light emitted by charged particles produced in neutrino
interactions. The neutrino induced events can be categorized as either
muon tracks or showers. Cosmic ray muons and muons from CC 
$\nu_{\mu}$ interactions are the origin of tracks. Showers results
from neutrino interactions --~ $\nu_e$ or $\nu_\tau$ CC interactions,
and NC interactions initiated by all three flavors~-- inside or near
the detector. Because of the large range of muons, kilometers to tens
of kilometers for the energies considered, the effective volume of the
detector for muon neutrinos is significantly larger than the
instrumented volume. Furthermore, the angular resolution for muon
tracks is superior to that for showers.  

Despite their main goal is the observation of neutrinos of
extraterrestrial origin, a neutrino telescope can also detect
atmospheric neutrinos which, as a matter of fact, constitute the main
background in their searches for astrophysical sources.  However, as
it is many times the case in particle physics experiments, the study
of the background can still leave an important room for discovery.
Indeed neutrino telescopes, with an energy reach in the $0.1 \sim
10^4$~TeV range for atmospheric neutrinos, are also ideal instruments
to search for the new physics effects discussed above. For most of
this energy interval standard $\Delta m^2$ oscillations are suppressed
and therefore the observation of an angular distortion of the
atmospheric neutrino flux or its energy dependence provide a clear
signature for the presence of new physics mixing neutrino flavors.

The Hamiltonian of Eq.~\eqref{eq:hnpgen} describes the coherent
evolution of the $\nu_\mu$--$\nu_\tau$ ensemble for any neutrino
energy.  But high-energy neutrinos propagating in the Earth can also
interact inelastically with the Earth matter either by CC and NC and
as a consequence the neutrino flux is attenuated.  This attenuation is
qualitatively and quantitatively different for $\nu_\tau$'s and
$\nu_\mu$'s. Muon neutrinos are absorbed by CC interactions while tau
neutrinos are regenerated because they produce a $\tau$ that decays
into another tau neutrino before losing energy~\cite{hs}. As a
consequence, for each $\nu_\tau$ lost in CC interactions, another
$\nu_\tau$ appears (degraded in energy) from the $\tau$ decay and the
Earth never becomes opaque to $\nu_\tau's$.  Furthermore, as pointed
out in Ref.~\cite{kolb}, a new secondary flux of $\bar\nu_\mu$'s is
also generated in the leptonic decay $\tau\rightarrow
\mu\bar\nu_\mu\nu_\tau$.

Attenuation and regeneration effects of incoherent neutrino fluxes can
be consistently described by a set of coupled partial
integro-differential cascade equations (see for
example~\cite{reno1,reno2,reno3,reno4} and references therein).  In
this way, for example, the observed $\nu_\mu$ and oscillation-induced
$\nu_\tau$ fluxes (and the associated event rates in a high energy
neutrino telescope) from astrophysical sources has been evaluated.
Alternatively, these effects can be accounted for in a Monte Carlo
simulation of the neutrino propagation in
matter~\cite{hs,kolb,crotty}.  Whatever the technique used, because of
the long distance traveled by the neutrinos from the source, the
oscillations average out and the neutrinos arriving at the Earth can
be treated as an incoherent superposition of mass eigenstates.

For atmospheric neutrinos this is not the case because oscillation,
attenuation, and regeneration effects occur simultaneously when the
neutrino beam travels across the Earth's matter. For the
phenomenological analysis of conventional neutrino oscillations this
fact can be ignored because the neutrino energies covered by current
experiments are low enough for attenuation and regeneration effects to
be negligible. Especially for non-standard scenario oscillations,
future experiments probe high-energy neutrinos for which the
attenuation and regeneration effects have to be accounted for
simultaneously. 

In order to include all these effects in the neutrino flavor
oscillation equations it is convenient to use the density matrix
formalism as introduced in the previous section. In this formalism
attenuation effects due to CC and NC interactions can be introduced by
relaxing the condition $\Tr(\rho)=1$.  In this case 
\begin{equation}
    \rho(x) = \frac{1}{2} \big[p_0(x)
    + \vec{p}(x) \cdot \vec\sigma \big] \,,
\end{equation}
and
\begin{equation}
    \frac{d\rho(E,x)}{dx} = -i\big[ H(E),\; \rho(E,x) \big]
    - \sum_\alpha \frac{1}{2\lambda^\alpha_\text{int}(E,x)}
    \big\lbrace \Pi_\alpha, \, \rho(E,x) \big\rbrace \,,
\end{equation}
where we have explicitly exhibited the energy dependence and 
\begin{equation} \begin{aligned}[]
    [\lambda^\alpha_\text{int}(E,x)]^{-1} &\equiv
    [\lambda^\alpha_\text{CC}(E,x)]^{-1} +
    [\lambda_\text{NC}(E,x)]^{-1} \,,
    \\
    [\lambda^\alpha_\text{CC}(E,x)]^{-1} &=
    n_T(x)\, \sigma^{\alpha}_\text{CC}(E) \,,
    \\
    [\lambda_\text{NC}(E,x)]^{-1} &=
    n_T(x)\, \sigma_\text{NC}(E) \,.
\end{aligned} \end{equation}
$n_T(x)$ is the number density of nucleons at the point $x=ct$. 
$\sigma^{\alpha}_\text{CC}(E)$ is the cross section for CC
interaction, $\nu_\alpha + N \rightarrow l_\alpha + X$, and
$\sigma_\text{NC}(E)$ is the cross section for $\nu_\alpha + N
\rightarrow \nu_\alpha + X$ which is flavor independent. Thus we
obtain four equations that describe the evolution of the neutrino
system because one has to take into account both the flavor
precession of the vector $\vec p(E,t)$ as well as the neutrino
intensity attenuation encrypted in the evolution of $p_0(E,t)$.

$\nu_\tau$ regeneration and neutrino energy degradation can be
accounted for by coupling these equations to the shower equations for
the $\tau$ flux, $F_\tau(E_\tau,t)$ (we denote by $F$ the differential
fluxes $d\phi / (dE \, d\cos\theta)$).  Muons decouple from the
evolution equations because they range out electromagnetically in the
Earth matter before they can produce a $\nu_\mu$ by decay or CC
interaction. It is convenient to define the \emph{neutrino flux
density matrix} $F_\nu(E,x) = F_{\nu_\mu}(E,x_0) \, \rho(E,x)$ where
$F_{\nu_\mu}(E,x_0)$ is the initial neutrino flux. The equations can
be written as:
\begin{align}
    \label{eq:nshower}
    \begin{split}
	\frac{d F_\nu(E_\nu,x)}{dx}
	&= -i\big[ H,\, F_\nu(E_\nu,x) \big]
	-\sum_\alpha \frac{1}{2\lambda^\alpha_\text{int}(E_\nu,x)}
	\big\lbrace \Pi_\alpha,\, F_\nu(E_\nu,x) \big\rbrace
	\\
	& \hspace{5mm} +
	\int_{E_\nu}^\infty \frac{1}{\lambda_\text{NC}(E'_\nu,x)}
	F_\nu(E'_\nu,x)
	\frac{d N_\text{NC}(E'_\nu,E_\nu)}{d E_\nu} dE'_\nu
	\\
	& \hspace{5mm} +
	\int_{E_\nu}^\infty \frac{1}{\lambda^\tau_\text{dec}(E_\tau,x)}
	F_\tau(E_\tau,x) 
	\frac{d N_\text{dec}(E_\tau,E_\nu)}{d E_\nu} dE_\tau \, \Pi_\tau
	\\
	& \hspace{5mm} +
	\text{Br}_\mu \,
	\int_{E_\nu}^\infty
	\frac{1}{\lambda^\tau_\text{dec}(E_\tau,x)}\, 
	\bar{F}_\tau(\bar E_\tau,x)
	\frac{d \bar{N}_\text{dec}(\bar{E}_\tau,E_\nu)}
	{d E_\nu} d\bar{E}_\tau\, \Pi_\tau \,,
    \end{split}
    \\
    \label{eq:tshower}
    \begin{split}
	\frac{d F_\tau(E_\tau,t)}{d\,x}
	&= -\frac{1}{\lambda^\tau_\text{dec}(E_\tau,x)} F_\tau(E_\tau,x)
	\\
	& \hspace{5mm} +
	\int_{E_\tau}^\infty
	\frac{1}{\lambda^\tau_\text{CC}(E_\nu,t)}
	\Tr \big[ \Pi_\tau \, {F_\nu}(E_\nu,t) \big]
	\frac{d N_\text{CC}(E_\nu,E_\tau)}{d E_\tau} d E_\nu \,.
    \end{split}
\end{align}
$\lambda^\tau_\text{dec}(E_\tau,x)=\gamma_\tau\, c\, \tau_\tau$. 
$\tau_\tau$ is the $\tau$ lifetime and $\gamma_\tau = E_\tau/m_\tau$
is its gamma factor.

The CC and NC distributions are defined as:
\begin{equation} \begin{aligned}
    \frac{d N_\text{NC}(E'_\nu,E_\nu)}{d E_\nu}
    &\equiv \frac{1}{\sigma_\text{NC}(E'_\nu)}
    \frac{d\sigma_\text{NC}(E'_\nu,E_\nu)}{dE_\nu} \,,
    \\[1mm]
    \frac{d N_\text{CC}(E_\nu,E_\tau)}{d E_\tau} 
    &\equiv \frac{1}{\sigma^{\tau}_\text{CC}(E_\nu)} 
    \frac{d\sigma^\tau_\text{CC}(E_\nu,E_\tau)}{dE_\tau} \,,
\end{aligned} \end{equation}
The $\tau$ decay distribution $\frac{d N_\text{dec} (E_\tau,E_\nu)}{d
E_\nu}$ can be found, for example, in Ref.~\cite{reno1} and $\frac{d
\bar N_\text{dec} (\bar E_\tau,E_\nu)} {d E_\nu}$ in
Ref.~\cite{gaisser}.
 
The third term in Eq.~\eqref{eq:nshower} represents the neutrino
regeneration by NC interactions and the fourth term represents the
contribution from $\nu_\tau$ regeneration, $\nu_\tau \rightarrow
\tau^- \rightarrow \nu_\tau$, describing the energy degradation in the
process.  The secondary $\nu_\mu$ flux from $\bar\nu_\tau$
regeneration, $\bar\nu_\tau \rightarrow \tau^+ \rightarrow
\bar\nu_\tau + \mu^+ + \nu_\mu$, is described by the last term where
we denote by over-bar the energies and fluxes of the $\tau^+$.
$\text{Br}_\mu = 0.18$ is the branching ratio for this decay. In
Eq.~\eqref{eq:tshower} the first term gives the loss of taus due to
decay and the last term gives the $\tau$ generation due to CC
$\nu_\tau$ interactions. In writing these equations the tau energy
loss has been neglected because it is only relevant at much higher
energies.

An equivalent set of equations can be written for the antineutrino
flux density matrix and the for the $\tau^+$ flux. Both sets of
equations are coupled due to the secondary neutrino flux term.

\begin{figure}\centering
    \includegraphics[width=5in]{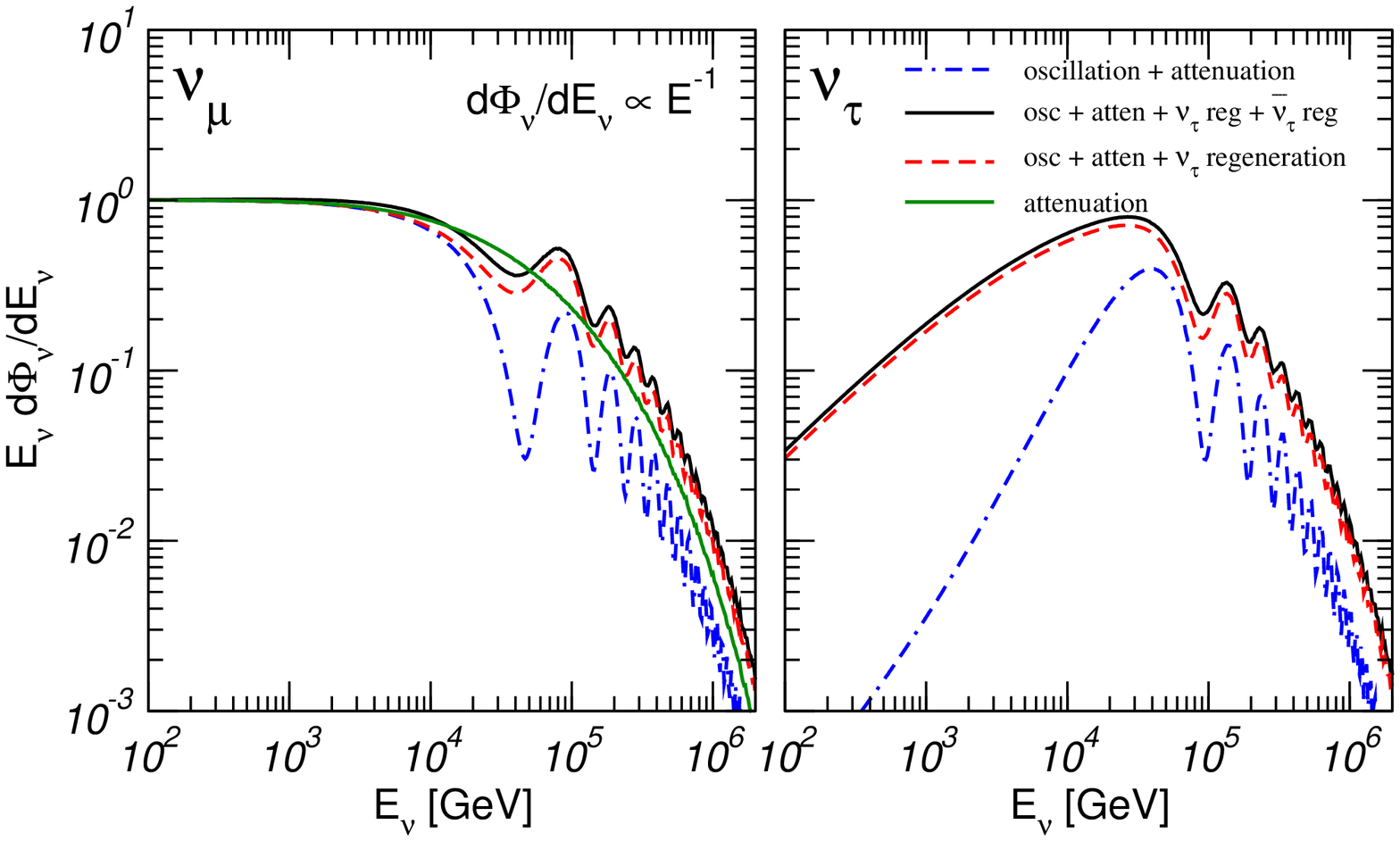}
    \includegraphics[width=5in]{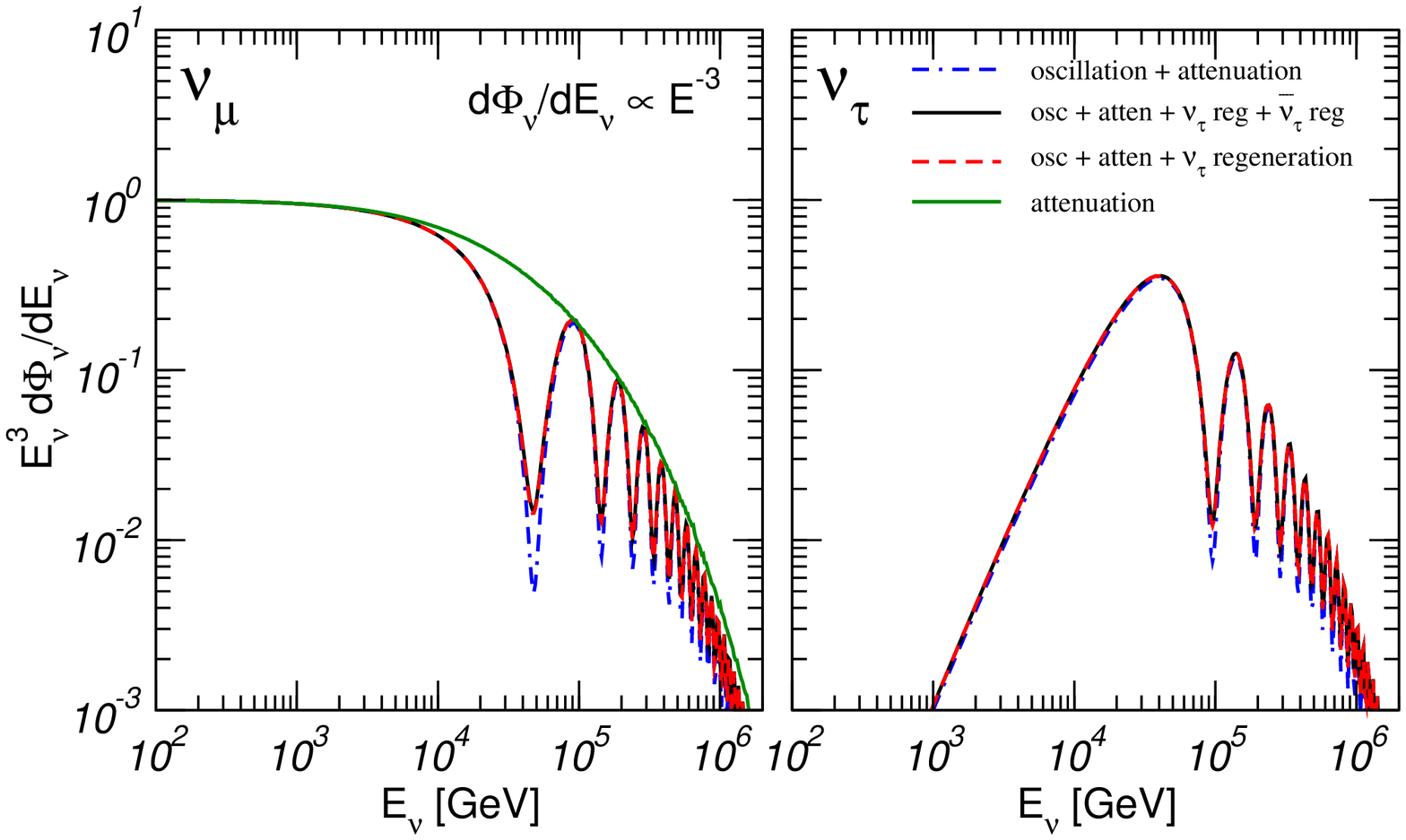}
    \caption{\label{fig:ice-cascade}%
      Vertically upgoing neutrinos after traveling the full length of
      the Earth taking into account the effects due to VLI
      oscillations, attenuation in the Earth, $\nu_\tau$ regeneration
      and secondary $\bar\nu_\tau$ regeneration (see text for
      details).} 
\end{figure}

After solving this set of ten coupled evolution equations that
describe propagation through the Earth one obtains the neutrino fluxes
in the vicinity of the detector from
\begin{equation}
    \label{eq:fluxdet}
    \frac{d\phi_{\nu_\alpha}(E,\theta)}
    {dE\, d\cos\theta} = \Tr\big[ F_\nu(E,L=2R\cos\theta) \,
    \Pi_\alpha \big] \,.
\end{equation}

Figure.~\ref{fig:ice-cascade} illustrates the interplay between the
different terms in Eqs.~\eqref{eq:nshower} and~\eqref{eq:tshower}. The
figure covers the example of VLI-induced oscillations with $\delta
c/c=10^{-27}$ and maximal $\xi_{vli}$ mixing.  The upper panels show
the final $\nu_\mu$ and $\nu_\tau$ fluxes for vertically upgoing
neutrinos after traveling the full length of the Earth for the initial
conditions $d\Phi(\nu_\mu)_0 / dE_\nu = d\Phi(\bar\nu_\mu)_0 / dE_\nu
\propto E^{-1}$ and $d \Phi(\nu_\tau)_0 / dE_\nu = d
\Phi(\bar\nu_\tau)_0 / dE_\nu = 0$.

The figure illustrates that the attenuation in the Earth suppresses
the neutrino fluxes at higher energies. The effect of the attenuation
in the absence of oscillations is given by the dotted thin line in the
left panel. Even in the presence of oscillations this effect can be
well described by an overall exponential suppression~\cite{gaisser,ls}
both for $\nu_\mu$'s and the oscillated $\nu_\tau$'s. In other words,
the curve for ``oscillation + attenuation" can be reproduced simply by
multiplying the initial flux by the oscillation probability and an
exponential damping factor: 
\begin{multline}
    \label{eq:fluxapp}
    \frac{d\phi_{\nu_\alpha}(E,\theta,L=2R \cos\theta)}{dE d\cos\theta}
    = \frac{d\phi_{\nu_\mu,0}(E,\theta)}{dE d\cos\theta} \,
    P_{\mu\alpha}(E,L=2R \cos\theta)
    \\
    \times \exp\big\lbrace -X(\theta)[\sigma_\text{NC}(E)
    + \sigma_\text{CC}^\alpha(E)] \big\rbrace \,,
\end{multline}
where $X(\theta)$ is the column density of the Earth.

The main effect of energy degradation by NC interactions (the third
term in Eq.~\eqref{eq:nshower}) that is not accounted for in the
approximation of Eq.~\eqref{eq:fluxapp} is the increase of the flux in
the oscillation minima (the flux does not vanish in the minimum)
because higher energy neutrinos end up with lower energy as a
consequence of the NC interactions.  The difference between the
dash-dotted line and the dashed line is due to the interplay between
the $\nu_\tau$ regeneration effect (fourth term in
Eq.~\eqref{eq:nshower}) and the flavor oscillations.  As a
consequence of the first effect, we see in the right upper panel that
the $\nu_\tau$ flux is enhanced because of the regeneration of higher
energy $\nu_\tau$'s, $\nu_\tau(E) \rightarrow \tau^- \rightarrow
\nu_\tau(E'<E)$, that originated from the oscillation of higher
energies $\nu_\mu$'s.  In turn this excess of $\nu_\tau$'s produces an
excess of $\nu_\mu$'s after oscillation which is seen as the
difference between the dashed curve and the dash-dotted curve in the
left upper panel.  Finally the secondary effect of $\bar\nu_\tau$
regeneration (last term in Eq.~\eqref{eq:nshower}), $\bar\nu_\tau(E)
\rightarrow \tau^+ \rightarrow \mu^+ \, \bar\nu_\tau \, \nu_\mu
(E'<E)$, results into the larger $\nu_\mu$ flux (seen in the left
upper panel as the difference between the dashed and the thick full
lines). This, in turn, gives an enhancement in the $\nu_\tau$ flux
after oscillations as seen in the right upper panel.

The lower panels show the final $\nu_\mu$ and $\nu_\tau$ fluxes for an
atmospheric-like energy spectrum $d{\Phi(\nu_\mu)_0} / dE_\nu =
d\Phi(\bar\nu_\mu)_0 / dE_\nu \propto E^{-3}$ and $d\Phi(\nu_\tau)_0 /
dE_\nu = d{\Phi(\bar\nu_\tau)_0} / dE_\nu = 0$. In this case all
regeneration effects are suppressed.  Regeneration effects result in
the degradation of the neutrino energy and the more steeply falling
the neutrino energy spectrum, the smaller the contribution to the
total flux. Therefore, in this case, the final fluxes can be
relatively well described by the approximation in
Eq.~\eqref{eq:fluxapp}.

\subsubsection{Example: VLI-induced Oscillations at ICECUBE}
\label{sec:results}

\begin{figure}\centering
    \includegraphics[width=5in]{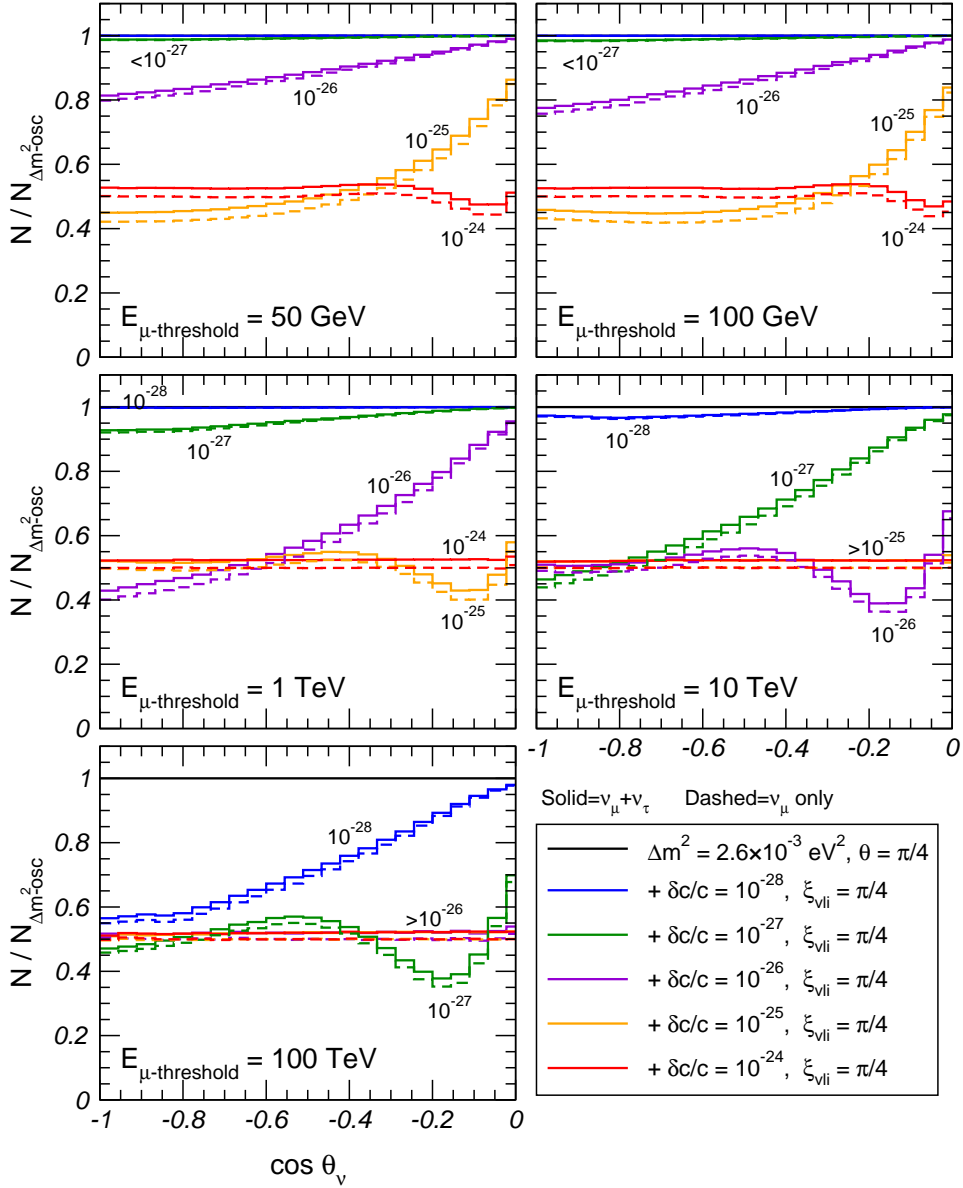}
    \caption{\label{fig:ice-zenith}%
      Zenith angle distributions for muon induced events for different
      values of the VLI parameter $\Delta c/c$ and maximal mixing
      $\xi_{vli}=\pi/4$ for different threshold energy 
      $E_\mu^{fin}>E_\text{threshold}$ normalized to the expectations
      for pure $\Delta m^2$ oscillations. The dashed line includes
      only the $\nu_\mu$-induced muon events and the full line
      includes both the $\nu_\mu$-induced and $\nu_\tau$-induced muon
      events.}
\end{figure}

\begin{figure}\centering
    \includegraphics[width=3.5in]{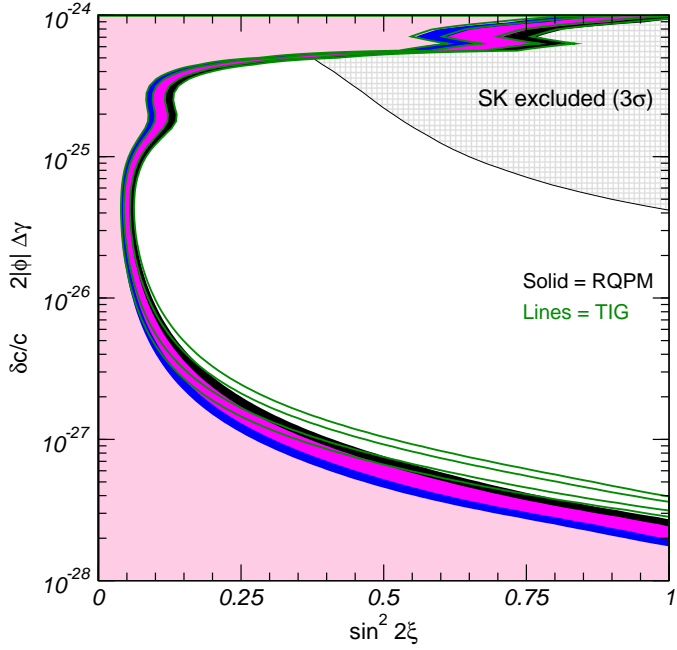}
    \caption{\label{fig:ice-chisq}%
      Sensitivity limits in the $\Delta c/c,\, \xi_\text{vli}$ at
      90\%, 95\%, 99\% and $3\sigma$ CL.  The hatched area in the
      upper right corner is the present $3\sigma$ bound from the
      analysis of SK+K2K+Minos data in Sec.~\ref{sec:noatmresults}. In
      order to estimate the uncertainty associated with the poorly
      known charm meson production cross sections at the relevant
      energies, the results are presented for two different models of
      charm production: the recombination quark parton model (RQPM)
      developed by Bugaev \textit{et al.}~\cite{rqpm} (filled regions)
      and the model of Thunman \textit{et al.}\ (TIG)~\cite{tig} (full
      lines) that predicts a smaller rate.}
\end{figure}

For illustration we present here the results of Ref.~\cite{ouricecube}
on the physics potential of ICECUBE~\cite{Icecube}, now under
construction to unravel these new physics effects.  ICECUBE will
consist of 80 kilometer-length strings, each instrumented with 60
10-inch photomultipliers spaced by 17~m.  The deepest module is 2.4~km
below the surface. The strings are arranged at the apexes of
equilateral triangles 125~m on a side. The instrumented detector
volume is a cubic kilometer.

The most obvious effect of neutrino oscillations induced by the new
physics at ICECUBE will be an energy dependent distortion of the
zenith angle distribution of atmospheric muon neutrino events.  Also
one has to take into account that together with $\nu_\mu$-induced muon
events, oscillations also generate $\mu$ events from the CC
interactions of the $\nu_\tau$ flux which reaches the detector
producing a $\tau$ that subsequently decays as $\tau \rightarrow \mu
\, \bar\nu_\mu \, \nu_\tau$ and produces a $\mu$ in the detector. 

We show in Fig.~\ref{fig:ice-zenith} the zenith angle distributions
for muon induced events for different values of the VLI parameter
$\Delta c/c$ and maximal mixing $\xi_{vli} = \pi/4$ for different
threshold energy $E_\mu^{fin}>E_\text{threshold}$ normalized to the
expectations for pure $\Delta m^2$ oscillations.  The full lines
include both the $\nu_\mu$-induced events and $\nu_\tau$-induced
events while the last ones are not included in the dashed curves.  As
seen in the figure, for a given value of $\Delta c/c$ there is a range
of energy for which the angular distortion is maximal. Above that
energy, the oscillations average out and result in a constant
suppression of the number of events.  Inclusion of the
$\nu_\tau$-induced events events leads to an overall increase of the
event rate but slightly reduces the angular distortion as a
consequence of the ``anti-oscillations'' of the $\nu_\tau$'s as
compared to the $\nu_\mu$'s.  The expected sensitivity bounds which
were be obtained in Ref.~\cite{ouricecube} from the statistical
analysis of this distortion are shown in Fig.~\ref{fig:ice-chisq}.

The figure illustrates the physics potential for discovery beyond the
present bounds by more than two orders of magnitude even within the
context of this very conservative analysis. 

%% file: sec.npsolar.tex
\section{Non Standard Medium Effects in Solar and Reactor Neutrinos}
\label{sec:npsolar}

As discussed in Sec.~\ref{sec:npatm} oscillations are not the only
possible mechanism for flavor transitions. They can also be generated
by a variety of forms of nonstandard neutrino interactions or
properties.  In this section we describe some phenomenology associated
with models which affect $\nu_e$ oscillations and which can be
observed or constrained by the combined analysis of solar and KamLAND
experiments.

At present solar neutrino data is the most sensitive probe that we
have on flavor effects in neutrino propagation in a dense medium. On
the other hand, because of the smaller density of the Earth
background, such ``environmental'' effects are expected to be much
less relevant in the oscillation of reactor $\bar\nu_e$'s observed in
KamLAND. Consequently the requirement of simultaneously describing
solar and KamLAND data is an important test on new physics scenarios
which induce non-universal effective couplings of $\nu_e's$ to the
particles in the background.

Generically in most neutrino mass models, new sources of environmental
dependence (ED) of the effective neutrino mass arise as a natural
feature due to the presence of non-standard neutrino interactions. The
form in which the presence of the non-standard interactions affect the
neutrino propagation depends on the tensor structure of the new force.
If the new interaction can be cast as a neutral or charged vector
current, it will contribute as an energy independent potential to the
neutrino evolution equation in addition to the MSW potential of the
Standard Model. On the other hand tensor neutral forces lead to an
energy decreasing potential while new physics in the form of Yukawa
interactions of neutrinos and matter with a neutral scalar particle
modify the kinetic part of the neutrino evolution equation.

Assuming that the oscillation of solar and KamLAND antineutrinos are
still dominated by a single mass scale even in the presence of these
effects one can write their evolution equation as: 
\begin{equation}
    \label{eq:evol}
    i \frac{d}{dx}
    \begin{pmatrix}
	\nu_e \\ \nu_a
    \end{pmatrix} = \left[
    \frac{1}{2E_\nu}
    U_{12} \, M^2(x) \, U_{12}^\dagger + V(x)
\right]
    \begin{pmatrix}
	\nu_e \\ \nu_a
    \end{pmatrix} \,,
\end{equation} 
where we have defined $M(x)$ as the modified neutrino mass matrix 
which will not be diagonal and will vary along the neutrino trajectory
but that is \emph{energy independent} and have the same sign for
neutrinos and antineutrinos. V(x) contains the standard MSW potential
as well as all other effects which are not included in $M(x)$. We
label $\nu_a = \cos\theta_{23}\nu_\mu + \sin\theta_{23}\nu_\tau$.

In what follows we illustrate the possible phenomenological
consequences of the presence of new interactions in the solar neutrino
oscillations in two different scenarios: mass varying neutrinos
(MaVaNs) models associated to the origin of the cosmic
acceleration~\cite{dark1,dark1b,dark2} and the possibility of long
range leptonic forces~\cite{lee-yang}.


\subsection{Mass Varying Neutrinos in the Sun: $\nu$ Density Effects}
\label{sec:mavas1}

One of the truly challenging and open questions in both cosmology and
particle physics is the nature of the dark energy in the Universe. In
parallel, one would like to explain also why, in the present epoch,
the energy density associated with dark energy and that of matter
happen to be approximately the same if these densities have different
temperature dependence.  This coincidence is resolved dynamically if
the dark energy density tracks (some component) of the matter
density~\cite{tracking1,tracking2,tracking3}. 

In Refs.~\cite{dark1,dark1b} it was proposed that the dark energy
density tracks the energy density in neutrinos.  This implies that the
energy density associated with dark energy depends on the neutrino
masses and in turn, neutrino masses are not fixed but variable with
their magnitude being a function of the neutrino density.

Besides the possible interesting cosmological
effects~\cite{dark1,dark1b,cosmocons1,cosmocons2,cosmocons3,cosmocons4},
from the point of view of neutrino oscillation phenomenology the
unavoidable consequence of these scenarios is that the neutrino mass
depends on the local neutrino density and therefore can be different
in media with high neutrino densities such as the Sun~\cite{ourmavas}.

In the simplest realization of this scenario neutrino mass arises from
the interaction with a scalar field, the acceleron, whose effective
potential changes as a function of the neutrino density.  For most
purposes, the derivation of the effective neutrino mass in the
presence of the neutrino background can be made in a model independent
way using the neutrino mass $m_\nu$ as the dynamical field (without
making explicit use of the dependence of $m_\nu$ on the acceleron
field). In this approach at low energies the effective Lagrangian for
$m_\nu$ is 
\begin{equation} 
    \mathcal{L} = m_\nu \bar\nu^c \nu + V_{tot}(m_\nu) \,,
\end{equation}
where $V_{tot}(m_\nu) = V_\nu(m_\nu) + V_0(m_\nu)$ contains the
contribution to the energy density both from the neutrinos as well as
from the scalar potential.  The condition of minimization of $V_{tot}$
determines the physical neutrino mass.

The contribution of a neutrino background to the energy density is
given by
\begin{equation} 
    \label{eq:deltav}
    V_\nu = \int \frac{d^3 k}{(2\pi)^3} \,
    {\sqrt{k^2 + m_\nu^2}} \, f(k) \,,
\end{equation}
where $f(k)$ is the sum of the neutrino and antineutrino occupation
numbers for momentum $k$. $V_\nu$ receives contribution from the
cosmological Big Bang remnant neutrinos as well as from any other
neutrinos that might be present in the medium. Thus in general
\begin{equation}
    \label{eq:Vnu}
    V_{\nu}(m_\nu) = V_\text{C$\nu$B} + V_{\nu,\text{medium}}
    = m_\nu \, n^\text{C$\nu$B} + V_{\nu,\text{medium}} \,,
\end{equation}
where we have used that in the present epoch relic neutrinos are non
relativistic. $n^\text{C$\nu$B} = 112~\text{cm}^{-3}$ for each
neutrino species. In a medium like the Sun, which contains an
additional background of relativistic neutrinos,
$V_{\nu,\text{medium}}$ is given by Eq.~\eqref{eq:deltav}.

Thus in the Sun, the condition of minimum of the effective potential 
reads
\begin{equation}
    \label{eq:minimization}
    \left. \frac{\partial V_\text{tot}(m_\nu)}{\partial m_\nu}
    \right|_{m_\nu} = 0
    \quad \Rightarrow 
    \quad  V'_0(m_\nu) + n^\text{C$\nu$B}(1 + m_\nu \, \mathcal{A}) = 0 \,,
\end{equation}
where $\mathcal{A}$ is the average inverse energy parameter normalized
to the CMB neutrino density
\begin{equation}
    \label{eq:a}
    \mathcal{A} \equiv \frac{1}{n^\text{C$\nu$B}}
    \int \frac{d^3 k}{(2\pi)^3} \,
    \frac{1}{\sqrt{k^2 + m_\nu^2}} \, f_\text{Sun}(k) \,.
\end{equation}

In the SSM the distribution of relativistic electron neutrino sources
in the Sun is assumed to be spherically symmetric and it is described
in terms of radial distributions $p_i(r)$ for $i=pp$, \Nuc{7}{Be}, N,
O, $pep$, F, and \Nuc{8}{B} fluxes.  As a consequence, the density of
neutrinos in the Sun is only a function of the distance from the
center of the Sun, $x$.  It can be obtained integrating over the
contributions at point $x$ due to the neutrinos isotropically emitted
by each point source, as:
\begin{equation}
    \label{eq:density2}
    n^\text{Sun}(x) = 4.6 \times 10^4~\text{cm}^{-3} \,
    \frac{1}{x} \sum_i \alpha_i
    \int 4\pi r \, \log\frac{x+ r}{|x- r|}\, p_i( r)\, dr \,,
\end{equation}
where both $r$ and $z$ are given in units of $R_\odot$ and $\alpha_i =
\Phi_{\nu,i} / \Phi_{\nu,pp}$. 

\begin{figure}\centering
    \includegraphics[width=2.5in]{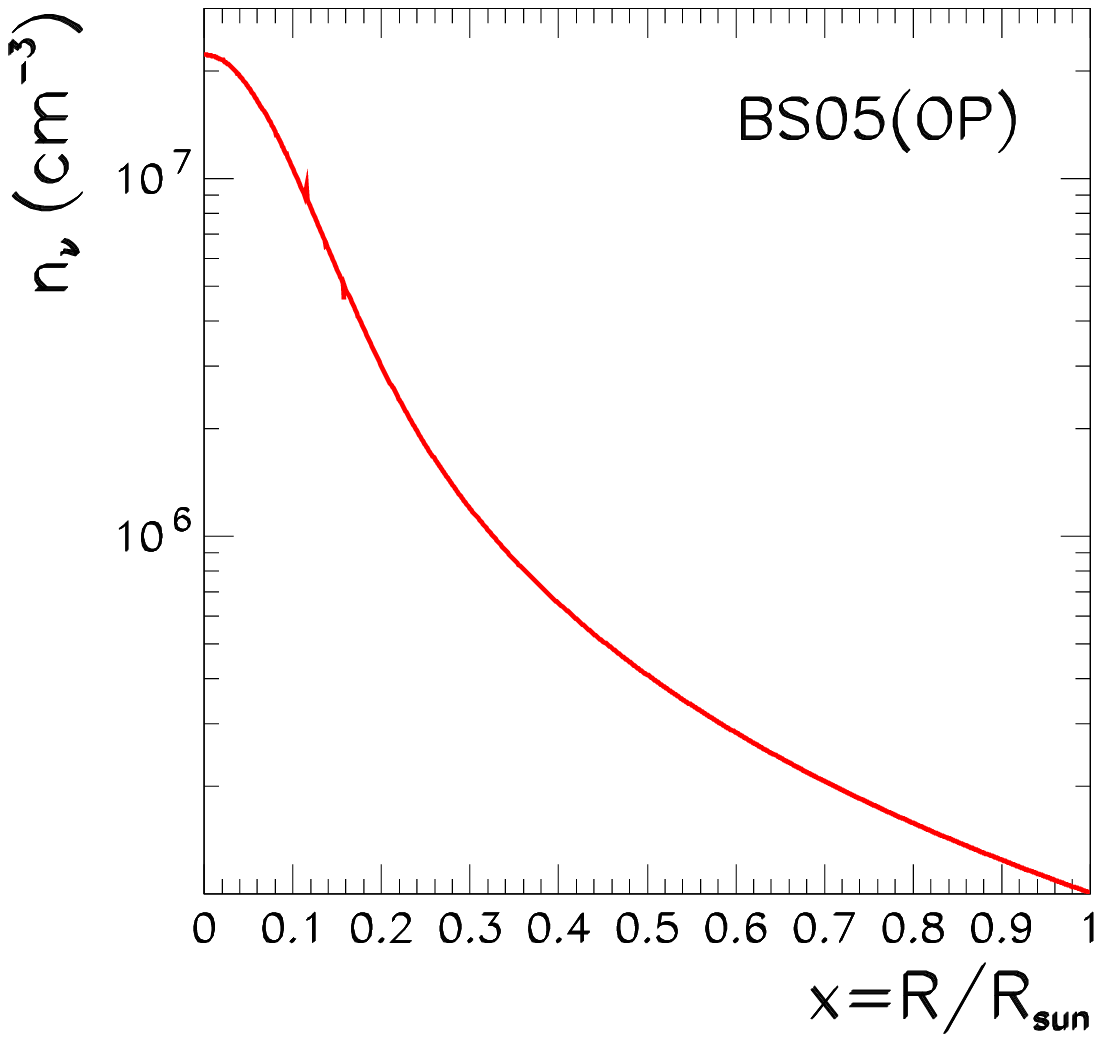}
    \includegraphics[width=2.5in]{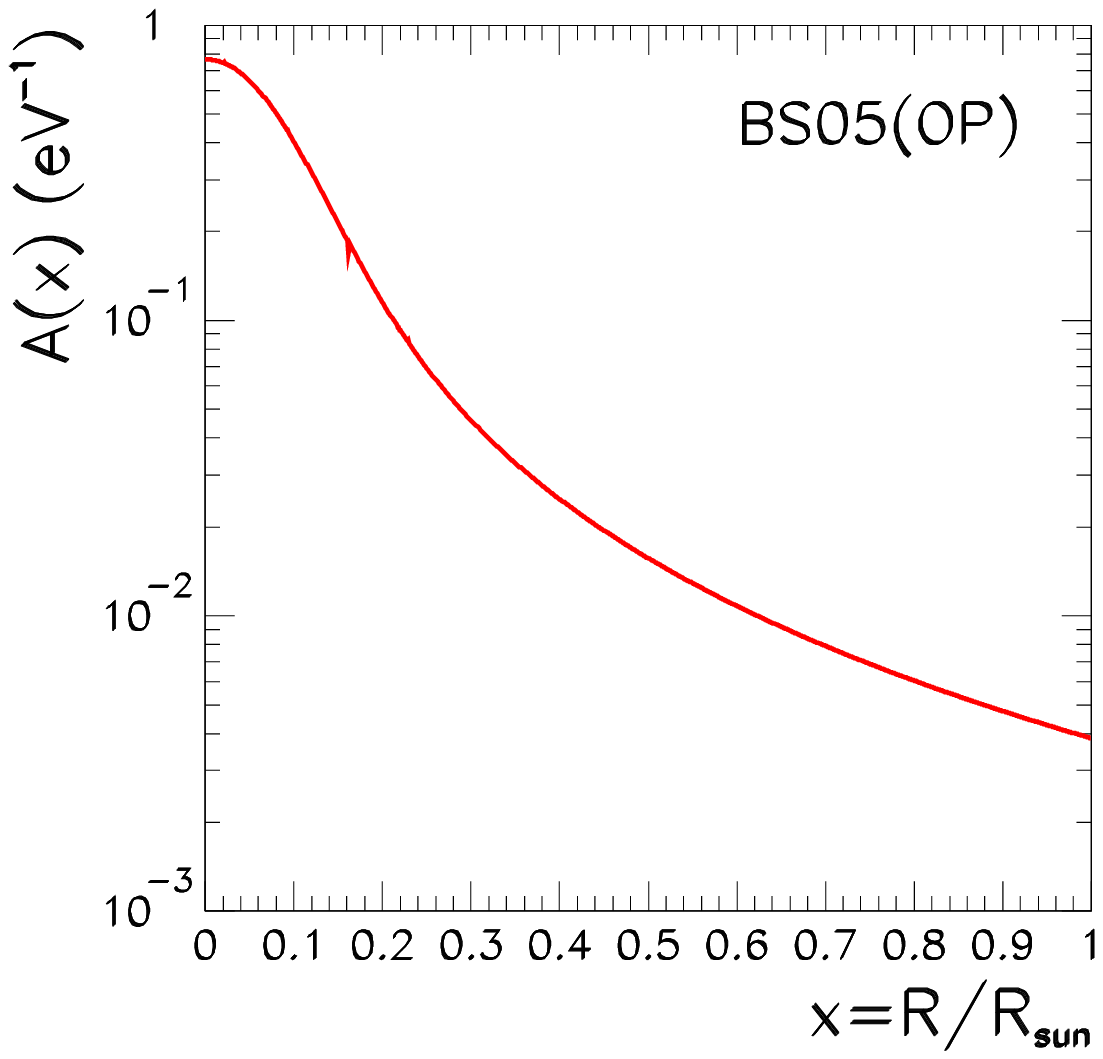}
    \caption{\label{fig:nudens}%
      Density of relativistic neutrinos in the Sun and the
      corresponding $\mathcal{A}$ factor as a function of the distance
      from the center of the Sun.} 
\end{figure}

Altogether one gets the density of relativistic neutrinos in the Sun
shown in Fig.~\ref{fig:nudens}.  As seen in the figure the neutrino
density is maximum at the center of the Sun where it reaches $2.2
\times 10^7~\text{cm}^{-3}$. It decreases by over two orders of
magnitude at the edge of the Sun.

Correspondingly the average inverse energy of the most abundant $pp$
flux is: 
\begin{gather}
    \int \frac{1}{E} \, \frac{d\Phi_{pp}}{dE}(E) \, dE = 
    2.7 \times 10^{5} \, \text{cm}^{-2} \, \text{s}^{-1} \,
    \text{eV}^{-1} \,,
\end{gather}
so the average inverse energy parameter normalized to the CMB neutrino
density~\eqref{eq:a} is:
\begin{equation}
    \label{eq:asun}
    \mathcal{A}(x) = 0.00186~\text{eV}^{-1}\, \frac{1}{x} 
    \sum f_i \int 4\pi r \, \log\frac{x+r}{|x-r|} \, p_i(r) \, dr \,,
\end{equation}
where $f_i = 2.3\times 10^{-2}$, $2\times 10^{-3}$, $1\times 10^{-3}$,
$3.6\times 10^{-4}$, $2.7\times 10^{-5}$, and $4\times 10^{-6}$ give
the small relative contribution from the \Nuc{7}{Be}, N, O, $pep$, F,
and \Nuc{8}{B} fluxes, respectively.  In deriving Eq.~\eqref{eq:asun}
the neutrino mass has been neglected with respect to its
characteristic energy in the Sun.

In Fig.~\ref{fig:nudens} we plot the factor $\mathcal{A}(x)$ in 
Eq.~\eqref{eq:asun}. As seen in the figure $\mathcal{A}(x)\sim
\mathcal{O}(1)$ eV$^{-1}$ in the region of maximum density, as
expected, since $\mathcal{A} \sim (n^\text{sun}/n^\text{C$\nu$B})
(1/\langle E\rangle)$ with $\langle E\rangle\sim 0.1$~MeV being the
characteristic $pp$ neutrino energy.  

Solving Eq.~\eqref{eq:minimization} with the $\mathcal{A}(x)$ term
above one finds the effective value of the neutrino mass as a function
of the solar neutrino density, while the vacuum neutrino mass
$m_\nu^0$ can be found from the corresponding condition outside of any
non-relic neutrino background
\begin{equation}
    \label{eq:minimization0}
    \left. \frac{\partial V_{tot}(m^0_\nu)}{\partial m_\nu}
    \right|_{m_\nu^0} = 0
    \quad\Rightarrow\quad
    V'_0(m^0_\nu) +n^\text{C$\nu$B} = 0 \,.
\end{equation}
It is clear from Eqs.~\eqref{eq:minimization}
and~\eqref{eq:minimization0} that the precise shift induced in the
neutrino mass by the presence of an additional neutrino density
depends on the exact form of the scalar potential $V_0(m_\nu)$. In
general one can parametrize the scalar potential as
\begin{equation}
    V_{0}(m_\nu) = \Lambda^4 \, f\left(\frac{m_\nu}{\mu} \right) \,,
\end{equation}
factoring out an overall scale $\Lambda^4$ which would set the scale
of the cosmological constant in a standard scenario and a function $f$
which depends on the dimensionless ratio $m_\nu/\mu$, where $\mu$ is
an accessory mass scale which will have no particular role for our
discussion.
 
The observation that the equation of state for the dark energy, 
\begin{equation}
    \omega + 1 =
    -\frac{m^0_\nu \, V'_0(m^0_\nu)}{V_\text{tot}(m^0_\nu)} \,,
\end{equation}
must have $\omega \approx - 1$ implies that the scalar potential must
be fairly flat
\begin{equation}
    \label{eq:flatness}
    \frac{dV_0(m_\nu)}{dm_\nu} \ll 1 \,.
\end{equation}
Furthermore Eq.~\eqref{eq:minimization0} implies 
\begin{equation}
    \label{eq:mono}
    \frac{dV_0(m_\nu)}{dm_\nu} < 0\,,
\end{equation}
this is, the potential must be a monotonically decreasing function of
$m_\nu$. 

Given the requirements~\eqref{eq:flatness} and~\eqref{eq:mono} 
several suitable forms of the function $f(m_\nu/\mu)$ can be 
proposed.  For example for a logarithmic form
\begin{equation} 
    \left(\frac{m_\nu}{\mu}
    \right) = \log\left(\frac{\mu}{m_\nu} \right).
\end{equation}
In this case from Eqs.~\eqref{eq:minimization}
and~\eqref{eq:minimization0} one gets the equation for the neutrino
mass shift
\begin{equation}
    \label{eq:logeq}
    m_\nu - m^0_\nu = -\mathcal{A} \, m^2_\nu 
\end{equation}
whose solution in the limit of small $\mathcal{A}$ is
\begin{equation}
    \label{eq:logsolution}
    m_\nu = m^0_\nu - \mathcal{A} (m^0_\nu)^2 + \ldots 
\end{equation}
Eq.~\eqref{eq:logeq} shows explicitly that the relative shift in the
neutrino mass due to the additional neutrino background 
$(m_\nu-m^0_\nu) / m_\nu$ grows in magnitude with the neutrino mass
scale.
Similar results can be found for other forms of the potential as long
as they verify the conditions of flatness and
monotony~\cite{ourmavas}.

\subsubsection{Effects on Solar Neutrino Oscillations}
\label{sec:osc}

In order to determine the effect of the scenario on the solar neutrino
oscillations one needs to extend the previous discussion to two or
more neutrinos.  In the simplest case one can assume that the coupling
to the dark sector leads to a shift of the neutrino masses but does
not alter the leptonic flavor structure which is determined either by
other non-dark contributions to the neutrino mass or from the charged
lepton sector of the theory. In this approximation the effective
Lagrangian for the neutrinos can be written as
\begin{equation} 
    \mathcal{L} = \sum_i m_i \bar\nu^c_i
    \nu_i + \sum_i \left[ m_i \, n^\text{C$\nu$B}_i 
    + V_{\nu_i,\text{medium}} + V_0(m_i)\right] \,, 
\end{equation}
and the condition of minimum of the effective potential implies: 
\begin{equation}
    \label{eq:min3}
    (m_i-m_{0i}) = -m_i^2 \, \mathcal{A}_i 
\end{equation}
where
\begin{equation}
    \mathcal{A}_i = \frac{1}{n^\text{C$\nu$B}_i} 
    \int \frac{d^3 k}{(2\pi)^3} \,
    \frac{1}{\sqrt{k^2 + m_i^2}} \, f_{\text{Sun},i}(k) \,.
\end{equation}
So even in this simple case of no leptonic mixing from the scalar
potential, there is a generation dependence of the $\mathcal{A}$
factor from the flavor dependence of the background neutrino density
which appears in the case of solar neutrinos because in the Sun only
$\nu_e$'s are produced. For $\theta_{13} = 0$ only the states $\nu_1$
and $\nu_2$ have their masses modified by the presence of the solar 
neutrino background as given in Eq.~\eqref{eq:min3} with 
\begin{equation} \begin{aligned}
    n_1(x) &= \cos^2\theta_{12} \, n_{\nu_e}(x)
    &~\Rightarrow\quad \mathcal{A}_1(x)
    &= \cos^2\theta_{12} \, \mathcal{A}(x) \,,
    \\
    n_2(x) &= \sin^2\theta_{12} \, n_{\nu_e}(x)
    &~\Rightarrow\quad \mathcal{A}_2(x) 
    &= \sin^2\theta_{12} \, \mathcal{A}(x) \,,
\end{aligned} \end{equation}
where $\theta_{12}$ is the vacuum mixing angle and $\mathcal{A}(x)$ is
given in Eq.~\eqref{eq:asun}. 

The evolution equation for the two neutrino state in the Sun can be
cast as Eq.~\eqref{eq:evol} with
\begin{equation}
    M^2(x) =
    \begin{pmatrix}
	m_1^2(x) &  0 \\
	0 & m_2^2(s) 
    \end{pmatrix}
    \quad\text{and}\quad
    V(x) =
    \begin{pmatrix}
	V_e(x) & ~0 \\
	0 & ~0 
    \end{pmatrix} \,,
\end{equation}
where $m_1(x)$ are the solutions of Eq.~\eqref{eq:min3} and $V_e$ is
the standard MSW potential Eq.~\eqref{eq:evol.4}.

Thus the effective ``kinetic'' (we label it kinetic to make it
explicit that it does not contain the MSW potential) mass difference
in the Sun is
\begin{equation} \begin{split}
    \label{eq:deltam}
    \Delta m^2_\text{kin}(x) &= m^2_2(x) - m_1^2(x)
    \\[1mm]
    &\simeq \Delta m^2_{21,0} \,
    [1 - 3\mathcal{A}_2(x) \, m_{01}] + 
    2 \, [\mathcal{A}_1(x) - \mathcal{A}_2(x)] m_{01}^3
    + \dots
\end{split} \end{equation}
$\Delta m^2_{21,0} = m_{02}^2 - m_{01}^2$ and $\theta_{12}$ are vacuum
mass difference and mixing angle.  Neglecting the vanishingly small
contribution from the terrestrial neutrino background, these are the
parameters measured with reactor antineutrinos at KamLAND.  We see
that a generic characteristic of these scenarios is that they
establish a connection between the effective $\Delta m^2$ in the Sun
and the absolute neutrino mass scale. 

More quantitatively from Eq.~\eqref{eq:deltam} we read that, as long
as the different massive neutrinos have different projections over
$\nu_e$ ($\mathcal{A}_1\neq \mathcal{A}_2$), $\Delta
m^2_\text{kin}(x)$ receives a contribution from the solar neutrino
background which rapidly grows with the neutrino mass scale $m_{01}$.
For the particular scenario that we are studying $\mathcal{A}_1(x)-
\mathcal{A}_2(x)=\cos 2\theta_{12}\, \mathcal{A}(x)>0$ so the
effective kinetic mass splitting is positive and larger than the
vacuum one in the resonant side for neutrinos.

\begin{figure}\centering
    \includegraphics[width=3.2in]{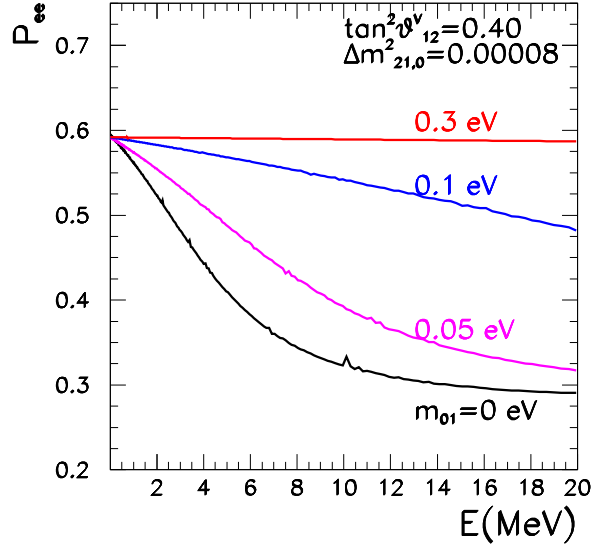}
    \caption{\label{fig:prob}%
      Survival probability of solar $\nu_e$'s as a function of the
      neutrino energy. This survival probability has been obtained for
      neutrinos produced around $x=0.05$ as it is characteristic of
      \Nuc{8}{B} neutrinos.}
\end{figure}

\begin{figure}\centering
    \includegraphics[width=4.5in]{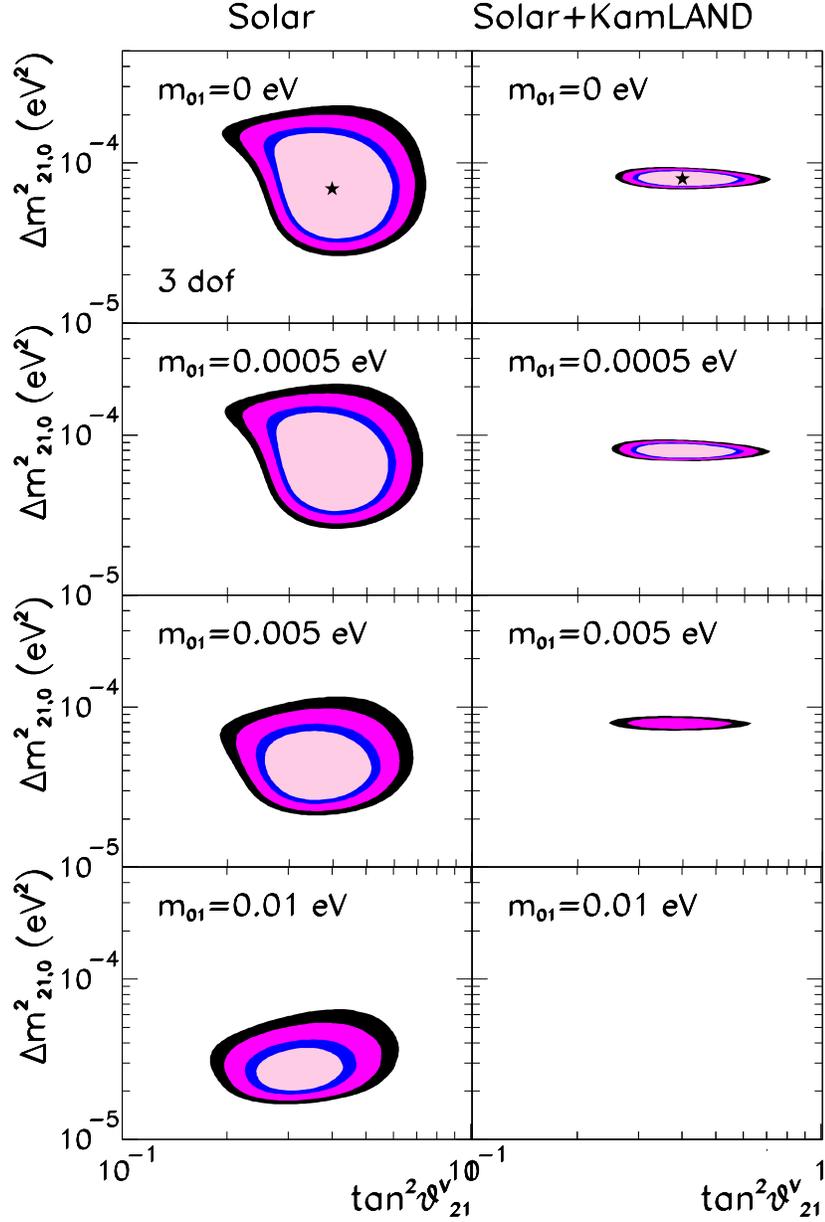}
    \caption{\label{fig:regions}%
      Allowed regions from the global analysis of solar and solar plus
      KamLAND data in the $(\Delta m^2_{21,0},\tan^2\theta_{12},
      m_{01})$ parameter space, shown for 4 sections at fixed values
      of $m_{01}$.  The different contours corresponds to 90\%, 95\%,
      99\%, and 3$\sigma$ CL for 3 d.o.f. The global minima are marked
      with a star.}
\end{figure}

The $\nu_e$ survival probability is obtained by solving numerically
the evolution equation.  In most of the parameter space the evolution
of the neutrino system is adiabatic and the survival probability is
well reproduced by the standard formula, Eq.~\eqref{eq:peeadiab}, with
an effective mixing angle in matter at the neutrino production point
$x_0$, $\theta^m_{12,0}$.  It includes both the effect of the point
dependent kinetic mass splitting as well as the effect of the MSW
potential $V_e(x)$ 
\begin{equation}
    \label{eq:mixing}
    \cos2\theta^m_{12,0}=
    \frac{\Delta m^2_\text{kin}(x_0)\cos2\theta_{12}-A(x_0)}
    {\sqrt{[\Delta m^2_\text{kin}(x_0)\cos2\theta_{12}-A(x_0)]^2
	+ [\Delta m^2_\text{kin}(x_0)\sin2\theta_{12}]^2}}
\end{equation}
where $A(x_0) = 2 E V_e(x_0)$.  In Fig.~\ref{fig:prob} the survival
probability is plotted for different values of the neutrino mass scale
$m_{01}$.  As can be seen in the figure, due to the different
contributions of the solar neutrino background to the two mass
eigenstates, the energy dependence of the survival probability is
rapidly damped even for mildly degenerated neutrinos.  As a
consequence, in these cases, it is not possible to simultaneously
accommodate the observed event rates in solar neutrino experiments and
in KamLAND as shown in Fig.~\ref{fig:regions}.

\subsection{Mass Varying Neutrinos in the Sun: Matter Density Effects}
\label{sec:mavas2}

In principle it is possible that the acceleron couples not only to the
neutrinos but also to the visible matter.  Such coupling would be
induced by non-renormalizable operators and it would imply that
neutrino masses depend on the visible matter background density as
well. This matter background dependence could lead to interesting
phenomenological consequences for neutrino
oscillations~\cite{dark2,Zurek,Barger,ourdark2}.
 
Generically one can parametrize these effects in an effective low
energy model containing the Standard Model particles plus a light
scalar ($\phi$) of mass $m_S$ which couples very weakly both to
neutrinos ($\nu_i$) and the matter fields $f=e,n,p$:
\begin{multline}
    \label{eq:lag}
    \mathcal{L} = 
    \sum_i \bar\nu_i \left(i \slash\!\!\!\partial - m_{0i} \right) \nu_i
    + \sum_f \bar f\left(i \slash\!\!\!\partial - m^0_f\right) f
    + \frac{1}{2} \left[\phi \left(\partial^2 - m_S^2 \right)
    \phi\right]
    \\
    + \sum_{ij} \lambda^\nu_{ij} \bar\nu_i \nu_j \phi
    + \sum_{f} \lambda^f \bar f f \phi \,,
\end{multline}
$\lambda^{\nu}_{ij}$ and $\lambda^f$ are, respectively, the effective
neutrino-scalar and matter-scalar couplings.  

In the context of the dark energy-related MaVaNs models of
Ref~\cite{dark1,dark1b,dark2} discussed in the previous section the
scalar $\phi$ would be the acceleron --~with mass in the range
$m_S\sim 10^{-6}$--$ 10^{-8}$ eV~-- which, when acquiring a
non-vanishing expectation value, $\langle\phi\rangle$, gives a
contribution to the neutrino mass. This in turn implies that the
acceleron effective potential receives a contribution which changes as
a function of the neutrino density, so that 
\begin{equation}
    \label{eq:lnu}
    \lambda^\nu = \left.
    \frac{\partial m_\nu}{\partial\phi}\right|_{\langle\phi\rangle} \,.
\end{equation}
$\lambda^f$, the effective low energy couplings of the acceleron to
visible matter, come from non-renormalizable operators which couple 
the acceleron to the visible matter, such as might arise from quantum
gravity.  They are constrained by tests of the gravitational inverse
square law (ISL) which require~\cite{Adelberger}
\begin{equation}
    \label{eq:isl}
    \lambda^n, \lambda^p \lesssim 10^{-21}
\end{equation}
for any scalar with $m_S\gtrsim 10^{-11}$~eV.

Eq.~\eqref{eq:lag} implies that in a medium with some additional 
neutrino background (either relativistic or non-relativistic) as well 
as non-relativistic matter (electrons, protons and neutrons),
neutrinos acquire masses which obey the following set of integral
equations
\begin{equation} \begin{aligned}
    \label{eq:numass}
    m_{ij}(x) &= m_{0i} \delta_{ij} - M_{ij}(x),
    \\
    M_{ij}(x) &= \frac{\lambda^\nu_{ij}}{m_S^2} \left(\sum_f
    \lambda^f n_f(x) + \sum_a \lambda^\nu_{aa} 
    \int \frac{d^3 k}{(2\pi)^3} \,
    \frac{M_{aa}}{\sqrt{k^2 + M_{aa}^2}} \,f_a(r,k) \right).
\end{aligned} \end{equation}
$n_f(x)$ is the number density for the fermion $f$, and $f_a(x,k)$ is
the sum of neutrino and antineutrino ``$a$'' occupation numbers for
momentum $k$ in addition to the cosmic background neutrinos.

The results in Eq.~\eqref{eq:lnu} and Eq.~\eqref{eq:numass} 
correspond to the first order term in the Taylor expansion around the
present epoch background value of $\phi$. In general, for the required
flat potentials in these models, one needs to go beyond first order
and the neutrino mass is not linearly proportional to the number
density of the particles in the background as shown in
Sec.~\ref{sec:mavas1}. The exact dependence on the background
densities is function of the specific form assumed for the scalar
potential. This is mostly relevant for the neutrino density
contribution to the neutrino mass, while for small enough couplings to
the matter potential one expects the linear approximation to hold
better.

As seen in the previous section for solar neutrinos the neutrino
background contribution is only relevant as long as the neutrinos are
not very hierarchical.  On the other hand for solar neutrinos of
hierarchical masses ($m_{01}\simeq 0$) the dominant contribution to
the neutrino mass is due to the matter background
density.\footnote{Also it has been argued~\cite{zaldarriaga} that,
generically, these models contain a catastrophic instability which
occurs when neutrinos become non-relativistic. As a consequence the
acceleron coupled neutrinos must be extremely light which implies that
the neutrino spectrum must indeed be hierarchical. Recently there have
been some discussion on the conditions required to evade this
constraint~\cite{mota1,mota2,mota3}.} In this section we will describe
the phenomenology of solar and KamLAND neutrinos in this scenario for
which 
\begin{equation}
    \label{eq:mrmat}
    M_{ij}(x) = \frac{\lambda^\nu_{ij}}{m_S^2}
    \sum_f \lambda^f n_f(x) \,.
\end{equation}

\subsubsection{Effects on Solar Neutrinos and KamLAND}

Assuming that the oscillation of solar and reactor antineutrinos are
still dominated by a single mass scale even in the presence of these
effects one can parametrize their evolution equation
Eq.~\eqref{eq:evol} with 
\begin{equation}
    M^2(x) =
    \begin{pmatrix}
	M_1^2(x) &  M_3^2(x) \\
	M_3^2(x) & [m_{02}-M_2(x)]^2
    \end{pmatrix}
    \quad\text{and}\quad
V(x)=    \begin{pmatrix}
	V_e(x) & ~0 \\
	0 & ~0 
    \end{pmatrix} \,,
\end{equation}
where $M_i(r)$ are the ED contributions to the neutrino masses 
Eq.~\eqref{eq:mrmat}.

In general, for given matter density profiles, Eq.~\eqref{eq:evol} has
to be solved numerically. But in most of the parameter space allowed
by KamLAND and solar data the transition is adiabatic and the survival
probability is well reproduced by the standard formula,
Eq.~\eqref{eq:peeadiab}, with an effective mixing angle in matter at
the neutrino production point $x_0$, $\theta^m_{12,0}$: 
\begin{equation}
    \label{eq:costm}
    \cos 2\theta^m_{12,0}=
    \frac{(\Delta \tilde{M}_{21}^2(x_0)\cos 2\tilde{\theta}_{12,0}
      -2 E V_e(x_0))}
    {\sqrt{[\Delta \tilde{M}_{21}^2(x_0) 
	\cos 2\tilde{\theta}_{12,0} - 2 E V_e(x_0)]^2 +
	[\Delta \tilde{M}_{21}^2(x_0) \sin 2\tilde{\theta}_{12,0}]^2}} \,,
\end{equation}
with
\begin{gather}
    \label{eq:deltatilde}
    \Delta \tilde{M}_{21}^2(x_0) = 2\sqrt{M_3^4(x_0)
      + \left(\frac{\Delta M^2_{21}(x_0)}{2}\right)^2}
    \\
    \label{eq:costhetatilde}
    \cos 2\tilde{\theta}_{12,0} =
    \frac{\dfrac{\Delta M_{21}^2(x_0)}{2}\cos 2\theta_{21} -
      M_3^2(x_0)\sin 2\theta_{12}}{\sqrt{M_3^4(x_0)
	+ \left(\dfrac{\Delta M^2_{21}(x_0)}{2}\right)^2}}
\end{gather}
and where
\begin{equation}
    \label{eq:M21}
    \Delta M^2_{21}(x_0) = [m_{02} - M_2(x_0)]^2 - M_1^2(x_0) \,.
\end{equation}

As discussed above, in general, $M_i(r)$ can be an arbitrary function
of the background matter density.  In the linear approximation given
in Eq.~\eqref{eq:mrmat} and for $\lambda^e\ll\lambda^n=\lambda^p
\equiv \lambda^N$, these terms can be parametrized as:
\begin{equation}
    \label{eq:mral}
    M_i(x) = \alpha_i \,
    \left[\frac{\rho(x)}{\text{gr}/\text{cm}^3}
    \right] \,,
\end{equation}
where $\rho$ is the matter density, and the characteristic value of
the $\alpha$ coefficients is
\begin{equation}
    \label{eq:alfa}
    \alpha \sim 4.8 \times 10^{23} \,
    \lambda^\nu \, \lambda^N \, \left(
    \frac{10^{-7} \text{eV}}{m_S} \right)^2 \, \text{eV} \,.
\end{equation}

One must notice, however, that, as long as the transition is
adiabatic, the survival probability only depends on the value of
$M_i(x)$ at the neutrino production point.  Therefore it only depends
on the exact functional form of $M_i(x)$ via the averaging over the
neutrino production point distributions.

The survival probability for anti-neutrinos, $P_{\bar{e}\bar{e}}$,
which is relevant for KamLAND, takes the form 
\begin{equation}
    \label{eq:peekam}
    P_{\bar{e}\bar{e}} = 1 -
    \sin^2 2\theta^m_{12,0} \, \sin^2 \left(
    \frac{\Delta m^2_{KL} L}{2E} \right) \,,
\end{equation}
where $\cos 2\theta^m_{12,0}$ is defined as in Eq.~\eqref{eq:costm}
and $\Delta m^2_{KL}$ is the denominator of this equation but
replacing $V_e$ by $- V_e$ and assuming a constant matter density
$\rho \sim 3$ gr/cm$^3$, typical of the Earth's crust.

\begin{figure}\centering
    \includegraphics[width=4.0in]{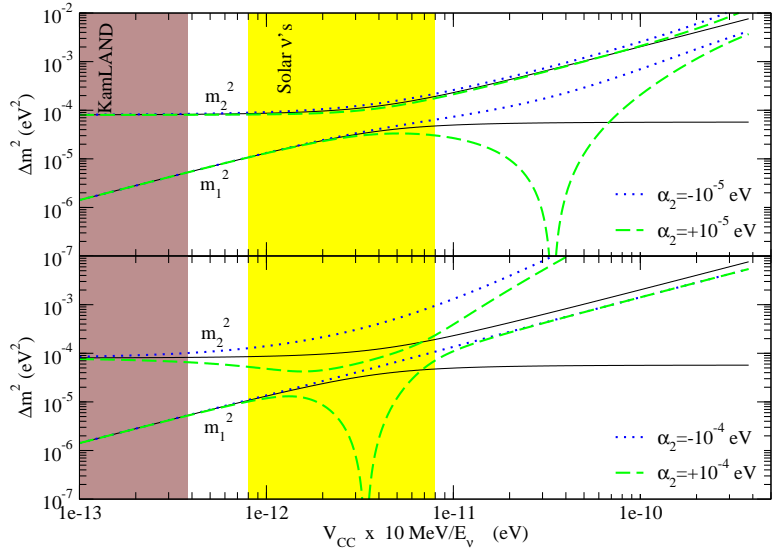}
    \caption{\label{fig:mass}%
      Evolution of the neutrino mass eigenstates in matter with
      $m_{02}^2 = 8 \times 10^{-5}~\eVq$, $\tan^2\theta_{12} = 0.4$
      and different values of the ED parameters as labeled. The solid
      lines represent the standard MSW evolution. The shaded regions
      correspond to typical values of $V_CC \equiv V_e$ in neutrino
      production region in the center of the Sun for the Solar $\nu$'s
      region, and a constant Earth crust density of 3 g/cm$^3$, with a
      proton density fraction of $Y = 0.5$ and neutrino energies
      varying from 3 to 10 MeV for the KamLAND region.}
\end{figure}

To illustrate the expected size of the effect we show in
Fig.~\ref{fig:mass} the evolution of the mass eigenstates $m_1$ and
$m_2$ in matter as a function of $V_e E$ for different values of
$\alpha_2$ (keeping $\alpha_1 = \alpha_3 = 0$).  As a reference, we
also show in this figure the standard MSW evolution curve (solid line)
for the oscillation parameters at $\Delta m^2_{0,21} = (m_{02})^2 =
8\times 10^{-5}~\eVq$ and $\tan^2\theta_{12}=0.4$. From this plot we
can appreciate that in the region relevant to solar neutrino
experiments the evolution of the mass eigenstates is not significantly
different from the MSW one if $|\alpha_2| \lesssim 10^{-5}$~eV.  For
larger values of $\alpha_2$, such as $|\alpha_2| = 10^{-4}$~eV, we
expect solar neutrinos to be affected.  On the other hand, KamLAND
data is very little affected by the ED terms in this range of
$\alpha_2$.

Fig.~\ref{fig:mass} also illustrates a curious feature of these
scenarios: the fact that it is possible to find a value of the matter
dependence term which exactly cancels $\Delta m^2_{0,21}$. It can be
seen, directly from Eqs.~\eqref{eq:evol}, that if for a particular
point, $r_0$, in the medium, $m_{02} = M_2(r_0)$ and $M_1 = M_3 = 0$
($\alpha_1 = \alpha_3 = 0$) the lower mass eigenstate will be zero
while the higher one will be at the corresponding value of $2V_e(r_0)
E$.

Non-adiabatic effects in the Sun can also occur. In the region of
relatively small $\alpha$ parameters, non-adiabaticity occurs when the
parameters are ``tuned'' to give a vanishing effective $\Delta
m^2_{21}$ (the denominator of Eq.~\eqref{eq:costm}).  This can be
achieved, for example, with $\alpha_1=\alpha_2=0$ by solving the
following set of equations inside the Sun:
\begin{align}
    (m_{02})^2 \cos 2 \theta - 2 M^2_3(x) \sin 2 \theta 
    &=  2 E V_e(x),
    \\[1mm]
    (m_{02})^2 \sin 2 \theta + 2 M^2_3(x) \cos 2 \theta
    &= 0.
\end{align}
It can be shown that for $\alpha_3 = i\, 5.5 \times 10^{-5}$~eV,
$\tan^2\theta = 0.3$ and $E = 10$~MeV this set of equations are
fulfilled at $r / R_\odot \sim 0.027$, and the neutrinos would suffer
a non-adiabatic transition on their way out of the Sun. However, in
general for the small values of the $\alpha$ parameters discussed
here, these non-adiabatic effects do not lead to a better description
of the solar neutrino data.

More generically, non-adiabatic effects occur for sufficiently large
values of the $\alpha$ parameters so that one can disregard the
standard MSW potential $V_e$ and the vacuum mass $m_{02}$ with respect
to the matter density mass dependent terms.  In this case, as seen
from Eq.~\eqref{eq:costm}, the mixing angle inside the Sun is constant
and controlled by the $\alpha's$. At the border of the Sun, as the
density goes to zero, the mixing angle is driven to its vacuum value
in a strongly non-adiabatic transition. This scenario would be
equivalent to a vacuum-like oscillation for solar neutrinos with the
ED of neutrino mass having to play a leading role in the
interpretation of terrestrial neutrino experiments. Given the strong
constraints from atmospheric neutrino experiments on new physics
scenarios described in Sec.~\ref{sec:npatm}, it is difficult to 
foresee that this scenario could lead to a successful global
description of the oscillation data. 

\begin{figure}\centering
    \includegraphics[width=3.5in]{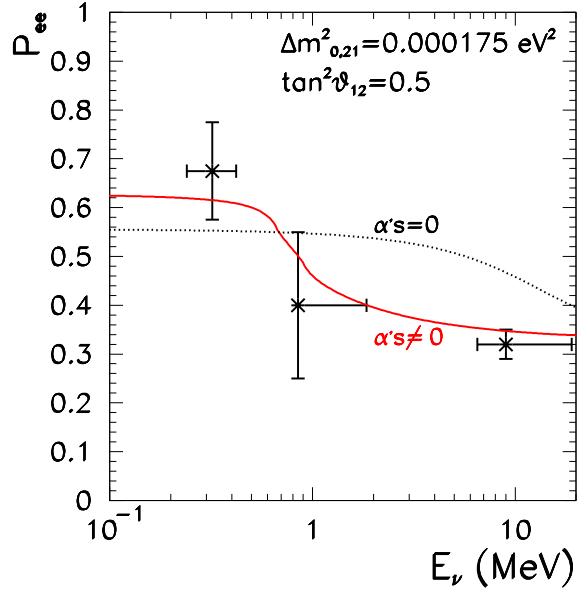}
    \caption{\label{fig:probhigh}%
      $\nu_e$ survival probability in the Sun versus neutrino energy
      for the best fit point in the high-$\Delta m^2$ region in the
      presence of ED effects. The dotted line is the survival
      probability for conventional oscillations ($\alpha_i = 0$) with
      the same values of $\Delta m^2_{21,0}$ and $\theta_{12}$. These
      survival probabilities have been obtained for neutrinos produced
      around $x = 0.05$ as it is characteristic of \Nuc{8}{B} and
      \Nuc{7}{Be} neutrinos. The data points are the extracted average
      survival probabilities for the low energy ($pp$), intermediate
      energy (\Nuc{7}{Be}, $pep$ and CNO) and high energy solar
      neutrinos (\Nuc{8}{B} and $hep$) from Ref.~\cite{Barger}.}
\end{figure}

Altogether it is found that for specific values of the ED couplings 
some modification of the allowed $\Delta m^2$ regions are possible. 
For example in Ref.~\cite{Barger} it was shown that the presence of 
these effects can improve the agreement with solar neutrino data
within the LMA region while being perfectly consistent with KamLAND
data.  In Ref.~\cite{ourdark2} it was also found that the description
of the solar data in the high-$\Delta m^2$ KamLAND region can be
significantly improved and there is a new allowed solution in the
global solar plus KamLAND analysis at the 98.9\% CL around
$\tan^2\theta_{12} = 0.5$ and $\Delta m^2_{0,21} = 1.75\times
10^{-4}~\eVq$.  We show in Fig.~\ref{fig:probhigh} the survival
probability for this best fit point in that high-$\Delta m^2$ region
in the presence of ED effects.  This solution will be further tested
by a more precise determination of the antineutrino spectrum in
KamLAND. In particular with an improvement of the systematic error
down to 4\% and an accumulated statistics of 3Kt-years, the LMAI
solution could be ruled out beyond 3$\sigma$~\cite{klandnu06}.

\begin{figure}\centering
    \includegraphics[width=4.5in]{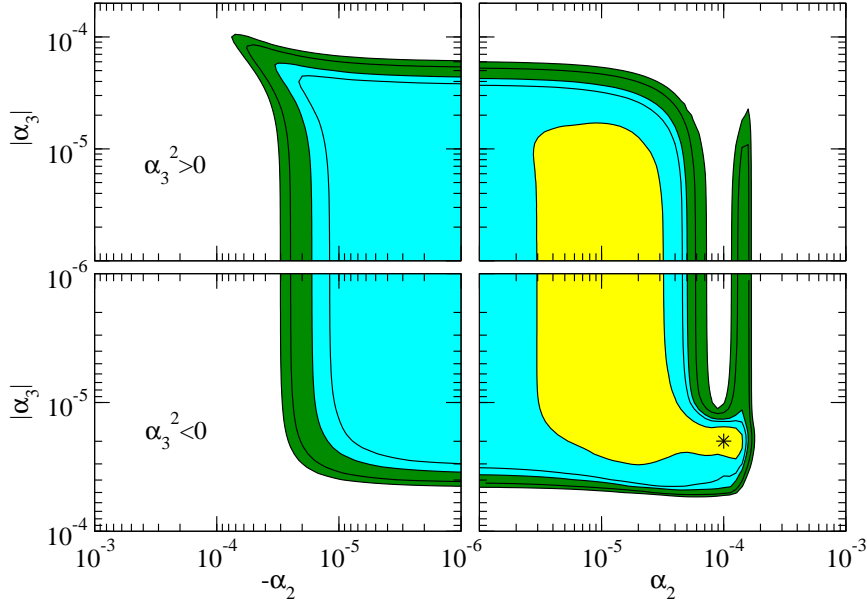}
    \caption{\label{fig:alfaregions}%
      Allowed regions from the global analysis of solar and solar plus
      KamLAND data in the $(\alpha_2,\alpha_3)$ parameter space. The
      curves correspond to 1$\sigma$, 90\%, 95\%, 99\% and $3\sigma$
      CL (2 d.o.f.).  The best fit point at $\alpha_2 = 10^{-4}$~eV
      and $\alpha_3 = i\, 2.0\times 10^{-5}$~eV, represented by a
      star, is also shown.}
\end{figure}

More generically, the global analysis of solar and KamLAND data can be
used to constraint the possible size of the ED contribution to the
neutrino mass and correspondingly of the possible size of the
acceleron couplings to matter and neutrinos.  This is illustrated in
Fig.~\ref{fig:alfaregions} where we show the result of such global
analysis in the form of the allowed two-dimensional regions in the
$(\alpha_2, \alpha_3)$ (for $\alpha_1 = 0$) parameter space after
marginalization over $\Delta m^2_{0,21}$ and $\tan^2\theta_{12}$.  

From the combined analysis one can derive the following 3 $\sigma$, 
bounds (with 1 d.o.f.)
\begin{equation} \begin{gathered}
    \label{eq:limits}
    -5.6\times 10^{-5} \leq \alpha_2 / \text{eV}
    \leq 1.7 \times 10^{-4} \,,
    \\
    |\alpha_3| / \text{eV} \leq
    \begin{cases}
	8 \times 10^{-5} & \text{for}\quad \alpha_3^2 > 0 \,,
	\\
	5 \times 10^{-5} & \text{for}\quad \alpha_3^2 < 0 \,.
    \end{cases}
\end{gathered} \end{equation}
These bounds can be converted into a limit on the product of the 
characteristic effective neutrino-scalar and matter-scalar couplings. 
For example, at 90\% CL,
\begin{equation}
    \label{eq:copl} 
    |\lambda^\nu \, \lambda^N | \, \left(
    \frac{10^{-7} \, \text{eV}}{m_S} \right)^2
    \leq 3.0 \times 10^{-28} \,.
\end{equation}
This bound can be compared to those derived from tests of the ISL,
Eq.~\eqref{eq:isl}.  We conclude that if the scalar also couples to
neutrinos with coupling
\begin{equation}
    \lambda^\nu \gtrsim 3.0 \times 10^{-7} \left(
    \frac{m_S}{10^{-7} \, \text{eV}} \right)^2 \,,
\end{equation}
the analysis of solar and KamLAND data yields a more restrictive
constraint on the matter-scalar couplings than ISL tests.


\subsection{Leptonic Long Range Forces}

We now turn to the possible effects arising from long range forces
coupling to flavor symmetries.  Since the seminal work of Lee and
Yang~\cite{lee-yang} it has become standard to consider long-range
forces coupling to baryon and/or lepton number. These forces violate
the universality of free fall and thus they can be tested by
E\"otv\"os-type experiments, as pointed out
in~\cite{lee-yang,okun1,okun2,dolgov2}. At present such tests yield a
bound the ``fine structure'' constant of a vector force associated to
$L$:
\begin{equation}
    \label{eq:okun}
    k_V < 10^{-49} \,.
\end{equation}

Since neutrinos are massive the lepton flavor symmetries $L_i$
($i=e,\, \mu,\, \tau$) cannot be exact in nature. Thus if an
electronic (or muonic, taunic) force exists, we may expect it to be of
arbitrary but finite range. When the range of the force is less than
the Earth-Sun distance, the bound~\eqref{eq:okun} is no longer valid.
Other experiments~\cite{fischbach,Adelberger} using the Earth as
electronic source instead of using the Sun place bounds, which however
are much less strict than~\eqref{eq:okun}.

Neutrino oscillations are sensitive probes to such forces. If there is
a new force coupled to the leptonic flavor numbers, its presence will
affect neutrino oscillations when neutrinos travel through regions
where a flavor dependent density of leptons is present.  As discussed
in Ref.~\cite{ourlepto} the effect of the new interaction on the
oscillation pattern depends on its Lorentz structure.  Nevertheless,
as we will show below for scalar, vector or tensor interactions of
large enough range the modification of the evolution equation can
always be casted in terms of a unique function which solely depends on
the background density of leptons --~the source of the force~-- and
the range of the interaction.

In this form, in Ref.~\cite{mohanty} it was discussed that atmospheric
neutrino data can constrain the strength of vector forces coupled to
$L_e-L_\mu$, $k_V (e\mu)\leq 5.5 \times 10^{-52}$, and to
$L_e-L_\tau$, $k_V (e\tau)\leq 6.4 \times 10^{-52}$ (at 90\% CL) when
the range of the force is the Earth-Sun distance. Also, new vector
forces coupling individually to muonic or tauonic number can be
constrained, giving however worse bounds.  For example, primordial
nucleosynthesis considerations provide the bound $k_V(\mu,\tau) < 1.8
\times 10^{-11}$~\cite{gm1}.

\begin{figure}\centering
    \includegraphics[width=3.5in]{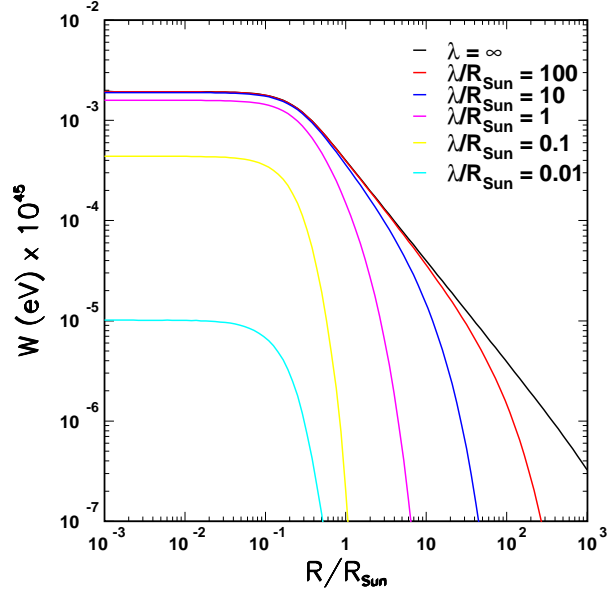}
    \caption{\label{fig:pot}%
      Leptonic \emph{potential} function $W(r)$ due to the density of
      electron in the Sun as a function of the distance from the solar
      center in units of $R_{\odot}$ for various ranges $\lambda$.} 
\end{figure}

Because of the larger electron density in the solar environment
stronger effects are expected in solar neutrino
observables~\cite{gm2,ourlepto,joshipura}.  If the new interaction is
flavor diagonal its effect in the evolution of atmospheric neutrinos
does not modify the hierarchy~\eqref{eq:deltahier} and the 2$\nu$
oscillation factorization still holds. In this case in general, one
can write the solar neutrino evolution equation in the presence of the
new force as Eq.~\eqref{eq:evol} where both ${M}(x)$ and $V(x)$ depend
on the Lorentz structure of the leptonic interaction as follows:
\begin{itemize}
  \item Let's start studying the case in which the new force is
    mediated by a neutral vector boson $A_\alpha$ with a small finite
    mass $m$ so that the new contribution to the Lagrangian is:
    \begin{equation}
	\label{eq:leffv}
	\mathcal{L} = - 
	g_1 A_\alpha \bar\psi_\nu \gamma^\alpha \psi_\nu
	-  g_1 A_\alpha \bar\psi_e \gamma^\alpha \psi_e \,.
    \end{equation}
    The electrons in the Sun can be thought as the source of a solar
    leptonic field, described by the external static classical vector
    potential
    \begin{equation}
	A_\mu^{ext}(\vec r) 
	\equiv \frac{1}{\sqrt{(2 \pi})^3}
	\int d^3k\, e^{i \vec k \vec r}\, \tilde A_\mu^{ext}(\vec k) \,,
    \end{equation}
    with
    \begin{equation}
	\tilde A_\mu^{ext}(\vec k)= -\frac{1}{k^2-m^2}\, g_1\,
	\tilde{j}^e_\mu(k) \,,
    \end{equation}
    where $k^2=-|\vec k |^2$ and
    \begin{equation}
	\tilde j^e_\mu(k) = \frac{1}{\sqrt{(2 \pi})^3} \,
	\int d^3\rho \, e^{-i \vec k \vec\rho} \,
	\bar\psi_e(\vec\rho) \gamma_\mu \psi_e(\vec\rho) \,,
    \end{equation}
    In the NR limit one can neglect the spatial components of the
    current and the only non-vanishing piece is:
    \begin{equation}
	\tilde j^e_0(k) = \frac{1}{\sqrt{(2 \pi})^3}
	\int d^3\rho \, e^{-i \vec k \vec\rho} \,
	\bar\psi_e(\vec\rho) \gamma_0 \psi_e(\vec\rho) =
	\frac{1}{\sqrt{(2 \pi})^3}
	\int d^3\rho \, e^{-i \vec k \vec\rho} \, n_e(\vec\rho) \,.
    \end{equation}
    $n_e(\vec\rho)$ is the electron number density.  Thus the only
    non-vanishing component of the vector potential is
    \begin{equation} \begin{aligned}
	A_0^{ext}(\vec r)
	&= \frac{g_1}{(2 \pi)^3} \int \d^3\rho \, n_e(\rho) \int d^3k \, 
	e^{i\vec k (\vec r - \vec\rho)} \, \frac{1}{|\vec k|^2+m^2}
	\\
	&= \frac{g_1}{4 \pi} \,	\int \d^3\rho \, n_e(\vec\rho) \,
	\frac{e^{-|\vec\rho - \vec r| / \lambda}}{|\vec\rho - \vec r|}
    \end{aligned} \end{equation}
    where $\lambda=1/m$. Introducing this result in
    Eq.~\eqref{eq:leffv} we get a contribution to the Lagrangian for
    the neutrinos of the form
    \begin{equation}
	-\frac{g_1^2}{4\pi} \int \d^3\rho \, n_e(\vec\rho) \,
	\frac{e^{-|\vec\rho - \vec r| / \lambda}}{|\vec\rho - \vec r|} \,
	\bar\psi_\nu \gamma^0 \psi_\nu \,,
    \end{equation}
    which can be interpreted as a contribution to the potential energy
    for the neutrinos $V(r) = k_V(e) W(r)$ with $k_V(e) =
    \dfrac{g_1^2}{4\pi}$ and
    \begin{equation} \begin{split}
	\label{eq:W}
	W(r) &=	\int
	\frac{e^{-|\vec\rho - \vec r| / \lambda}}{|\vec\rho - \vec r|}
	\, d^3\rho
	\\
	&= \frac{2\pi\lambda}{r} \int_0^{R_\text{Sun}} n_e(\rho) \, \rho
	\left[ e^{-|\rho - r| / \lambda} - e^{-(\rho + r) / \lambda}
	\right] \, d\rho
    \end{split} \end{equation}
    where $R_{\odot}$ is the radius of the Sun, $n_e(r)$ is the
    electron number density in the medium (assumed here to be
    spherically symmetric) and $\lambda$ is the range of the
    interaction.  
    In summary we can write the solar neutrino evolution equation in
    the presence of the new force as Eq.~\eqref{eq:evol} where 
    \begin{equation}
	\label{eq:vector}
	M_{ij} = m_{0i} \, \delta_{ij} 
	\quad \text{and} \quad V(x) = V_e(x) + k_V(e) \, W(r)
    \end{equation}
    with $k_V(e) = \dfrac{g_1^2}{4\pi}$. As a consequence of the
    vector structure of the force, the new leptonic potential adds to
    the MSW potential with the same energy dependence and sign. It
    will accordingly flip sign when describing antineutrino
    oscillations.
    
  \item In the same fashion one can show that if the new force is
    mediated by a neutral spin $J=0$ particle, $\phi$, so that the new
    contribution to the Lagrangian reads:
    \begin{equation}
	\mathcal{L} = 
	-g_0 \phi \bar \psi_\nu \psi_\nu - g_0 \phi \bar \psi_e \psi_e
    \end{equation}
    the corresponding external scalar ``potential'' due to the solar
    electrons is:
    \begin{equation}
	\phi^{ext}(r)= -\frac{g_0}{4 \pi}\, W(r) \,,
    \end{equation}
    which gives a contribution to the Lagrangian for the neutrinos of
    the form
    \begin{equation}
	\frac{g_0^2}{4\pi}\, W(r) \, \bar\psi_\nu \psi_\nu \,,
    \end{equation}
    which can be interpreted as a contribution to the effective $\nu_e$
    mass. So the solar neutrino evolution equation in the presence of
    the new force takes the form Eq.~\eqref{eq:evol} where now
    \begin{equation}
	\label{eq:scalar}
	M =
	\begin{pmatrix}
	    m_{01} & 0 \\
	    0 & m_{02}
	\end{pmatrix}
	- U_{12}^\dagger \,
	\begin{pmatrix}
	    k_S(e) \,W(x) & ~0 \\
	    0 & ~0
	\end{pmatrix} \, U_{12}
	\quad \text{and} \quad V(r) = V_e(x) \,.
    \end{equation}
    $k_S(e) = \dfrac{g_0^2}{4\pi}$ and the $k_S(e) \, W(x)$ term has
    the same sign for neutrinos and antineutrinos. $m_{0i}$ are the
    neutrino masses in vacuum.
    
  \item Similarly for a $L_e$-coupled force mediated by a tensor field
    of spin $J=2$, $\chi_{\alpha\beta}$:
    \begin{equation}
	\mathcal{L}= 
	-g_2 \,
	\chi_{\alpha\beta} T^{\alpha\beta}_\nu
	-g_2 \,
	\chi_{\alpha\beta} T^{\alpha\beta}_e
    \end{equation}
    where $T^{\alpha\beta}$ is the energy-momentum tensor for either
    electrons or neutrinos:
    \begin{equation}
	T^{\alpha\beta} = \left[ \bar \psi \gamma^\alpha i
	\partial^\beta \psi - i \partial^\alpha \bar \psi
	\gamma^\beta \psi \right] \,.
    \end{equation}
    The solar electrons can be the source of an static tensor field
    whose only non-vanishing contributions are:
    \begin{equation}
	\chi^{ext}_{00}(r) = \chi^{ext}_{ii}(r)=
	-g_2 \, m_e \frac{1}{4\pi} \, W(r) \,,
    \end{equation}
    where $m_e$ is the mass of the electron.  The action of this
    potential on the neutrinos is
    \begin{equation}
	- m_e\, \frac{g_e^2}{4\pi}\, W(r)\,
	(T^{00}_nu\, +\, \sum_i T^{ii}_\nu) =
	-m_e\, E\, \frac{g_e^2}{4\pi}\, W(r)\,
	\bar\psi_\nu \gamma^0 \psi_\nu \,,
    \end{equation}
    where in the last line we have used the Dirac equation for a
    massless neutrino of energy $E$ (introducing the neutrino mass
    here leads to a higher order correction on the neutrino evolution
    equation). This can be interpreted as a contribution to the
    neutrino potential so that the neutrino evolution equation takes 
    the form Eq.~\eqref{eq:evol} with
    \begin{equation}
	\label{eq:tensor}
	M_{ij} = m_{0i} \, \delta_{ij} 
	\quad\text{and}\quad
	V(x) = V_e(x)- E k_T(e) W(x) \,.
    \end{equation}
    $k_T(e) = m_e \dfrac{g_2^2}{4\pi}$ (notice that the coupling
    constant $g_2$ has dimensions $1/E$).  As for the case of a scalar
    leptonic force, the tensor force is always symmetric when changing
    from neutrinos to antineutrinos. This is so because what couples
    to the tensor field $\chi$ is in fact the energy momentum tensor
    of the leptons which has to be symmetric under the exchange of
    particles and antiparticles.
\end{itemize}
We see that for all cases one can define the \emph{universal} function
$W$ as Eq.\eqref{eq:W} determining the effect of the force coupled to
$L_e$ at a point $r$ from the center of the Sun. In Fig.~\ref{fig:pot}
we show the function $W(x)$ in the Sun as a function of the distance
in units of $R_{\odot}$ for various ranges $\lambda$.

\begin{figure}\centering
    \includegraphics[width=68mm,height=90mm]{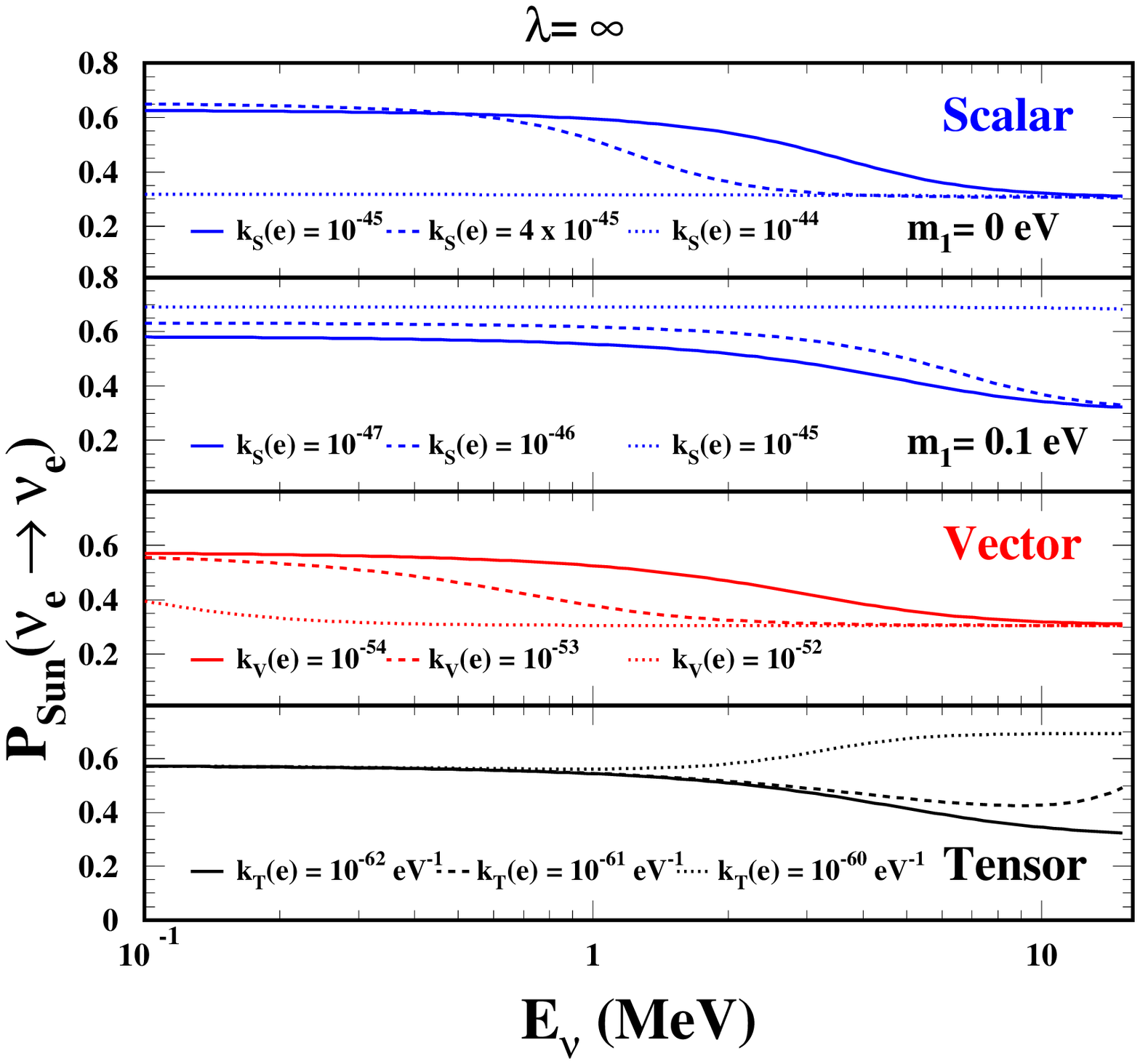}\hfill
    \includegraphics[width=68mm,height=90mm]{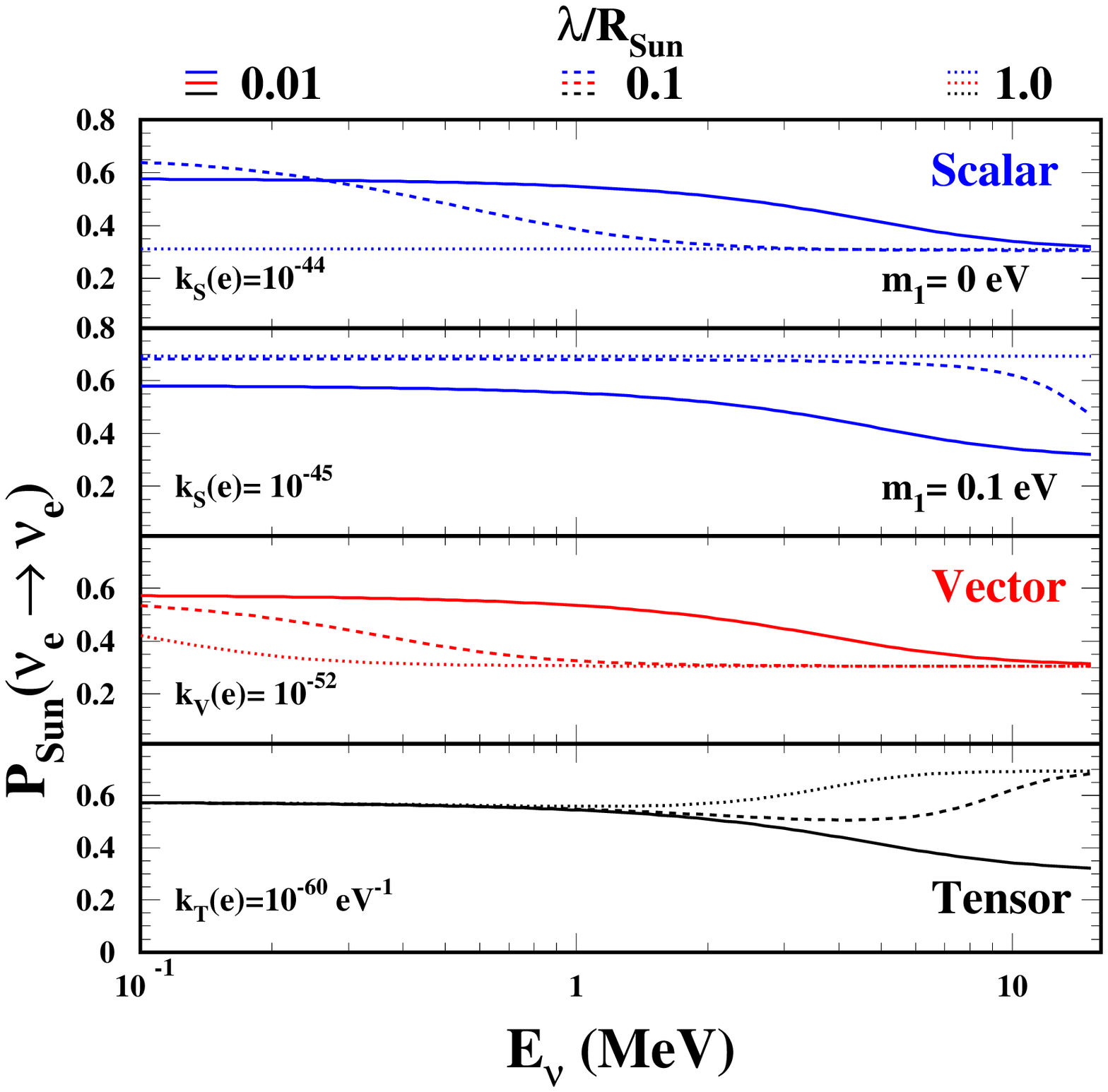}
    \caption{\label{fig:infty}%
      Survival probability of $\nu_e$ in the Sun as a function of the
      neutrino energy $E$ for an infinite range scalar (first two
      panels), vector (third panel) and tensor (lower panel) leptonic
      force, for various values of the strength and range. For all
      curves $\tan^2\theta_{12}=0.44$ and $\Delta m^2_{21} = 7.9
      \times 10^{-5}~\eVq$.} 
\end{figure}

For the range of parameters of interest the evolution in the Sun and 
from the Sun to the Earth is always adiabatic.  The energy dependence
of the resulting survival probability of solar $\nu_e$ at the sunny
face of the Earth is shown in Fig.~\ref{fig:infty}.  From this figure
we see that for $m_1=0$, values of $k_S(e) \gtrsim
10^{-45}$--$10^{-44}$ will conflict with the existing solar neutrino
data while for $m_1 = 0.1$ eV even smaller values of the coupling,
$k_S(e) \lesssim 10^{-46}$--$10^{-45}$, will be rule out.  In the
third panel we show the vector case for some values of $k_V(e)$.  One
expects from this that our analysis will lead to bounds $k_V(e)
\lesssim 10^{-54}$--$10^{-53}$.  Finally, in the lower panel the
tensor case is displayed.  In this case one expects that the data will
constrain $k_T(e) \lesssim 10^{-61}$--$10^{-60}$ eV$^{-1}$.

\begin{figure}\centering
    \includegraphics[width=0.65\textwidth,clip]{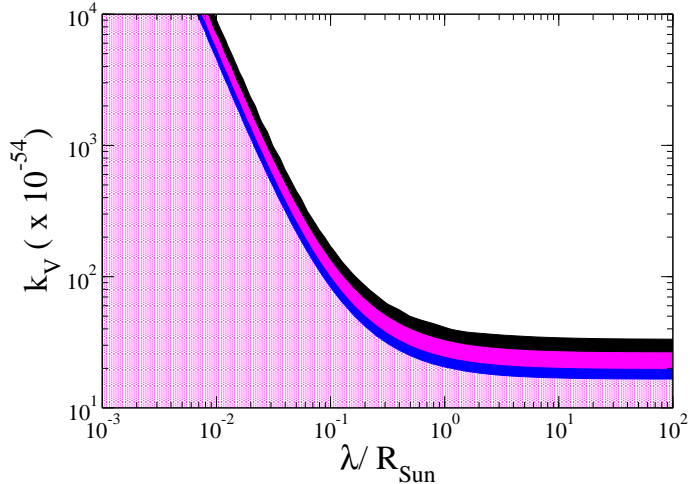}
    \caption{\label{fig:krvector}%
      Allowed regions from the global analysis of solar plus KamLAND
      data in the $(k_V, \lambda/R_\text{Sun})$ at 90\%, 95\%, 99\%
      and 3$\sigma$ CL (2 d.o.f.) for a leptonic vector force.} 
\end{figure}

More quantitatively in Ref.~\cite{ourlepto} it was shown that the
combined analysis of solar and KamLAND data provides the following
bounds for infinite range $\lambda = \infty$
\begin{align}
    \label{eq:scalar_conc}
    k_S & \leq 5 \times 10^{-45}
    \\
    \label{eq:vector_conc}
    k_V & \leq 2.5\times 10^{-53}
    \\
    \label{eq:tensor_conc}
    k_T & \leq 1.7 \times 10^{-60}~\text{eV}^{-1}
\end{align}
at $3\sigma$ (1 d.o.f.). These bounds are practically the same for any
$\lambda \gtrsim 10\, R_\text{Sun}$. For $\lambda \lesssim 10\,
R_\text{Sun}$, the bound slowly worsens, and for $\lambda \sim 0.1 \,
R_\text{Sun}$ we have that the bound on $k_i$ is a factor less than 10
worse than~\eqref{eq:scalar_conc}--\eqref{eq:tensor_conc} as
illustrated in Fig.~\ref{fig:krvector}.

In some cases these effects can also modify the allowed ranges of
$\Delta m^2$ for specific values of the couplings.  In particular in
the presence of scalar forces for highly hierarchical neutrinos and
$4.5 \leq k_S(e) / 10^{-45} \leq 8$ and/or vector forces with $2 \leq
k_V(e) / 10^{-54} \leq 30$ the description of the solar data in the
high-$\Delta m^2$ region of KamLAND can be significantly improved and
there is a new allowed solution at the 98.9\% CL around
$\tan^2\theta_{12} = 0.5$ and $\Delta m^2_{0,21} = 1.7 \times
10^{-4}~\eVq$~\cite{ourlepto}.


\subsection{Neutrino Magnetic Moment}
\label{sec:magnetic}

Non-zero neutrino masses can manifest themselves also through
non-standard neutrino electromagnetic properties. For example, if the
lepton sector in the Standard Model is minimally extended in analogy
with the quark sector, neutrinos get Dirac masses ($m_\nu$) and their
magnetic moments (MMs) are tiny~\cite{Fujikawa:yx},
\begin{equation}
    \label{eq:dirac}
    \mu_\nu \simeq 3 \times 10^{-19} ~
    \mu_B \left( \frac{m_\nu}{1~\text{eV}} \right) \,,
\end{equation}
where $\mu_B$ is the Bohr magneton. 

In general models, however, there is no such direct proportionality
between the neutrino mass and its coupling to the electromagnetic
interactions.  Consequently neutrinos can have sizable magnetic moment
and/or electric dipole moment (EDM) couplings which can show a
non-trivial flavor structure.  Thus the existence of any
electromagnetic neutrino moment well above the expectation in
Eq.~\eqref{eq:dirac} would signal that some special mechanism---which
goes beyond the SM---is at work.  Therefore, neutrino electromagnetic
properties are interesting probes of new physics.

In the flavor basis the interaction of Dirac neutrinos with a MM and
an EDM with the electromagnetic field is described by the Hamiltonian
\begin{equation}
    \label{eq:HD}
    H_\text{em}^D =
    \frac{1}{2} \bar{\nu}_R \lambda
    \sigma^{\alpha \beta} \nu_L F_{\alpha \beta} + \text{h.c.}
\end{equation}
Here $\nu^T_{L(R)} = (\nu_e,\, \nu_\mu,\, \nu_\tau,\, \nu_s,\,
\ldots)_{L(R)}$ is the vector of the left-handed (right-handed)
flavor eigenfields including an arbitrary number of sterile
neutrinos. The hermitian matrices $\mu$ of MMs and $d$ of EDMs are
condensed in the non-hermitian matrix
\begin{equation}
    \label{eq:lambda}
    \lambda = \mu - i d
    \quad\text{with}\quad
    \mu = \frac{ \lambda + \lambda^\dagger}{2} \,, \quad
    d = \frac{i( \lambda - \lambda^\dagger )}{2} \,.
\end{equation}
In the mass basis, Eq.\eqref{eq:diracmassdiag}, the matrix
\eqref{eq:lambda} transforms as
\begin{equation}
    \tilde \lambda = {V^\nu_R}^\dagger \lambda V^\nu \,.
\end{equation}
Similarly, for Majorana neutrinos:
\begin{equation}
    \label{eq:HM}
    H_\text{em}^M =
    - \frac{1}{4} \nu_L^T C^{-1} \lambda
    \sigma^{\alpha \beta} \nu_L F_{\alpha \beta} + \text{h.c.} \,,
\end{equation}
where $C$ is the charge conjugation matrix. For Majorana neutrinos CPT
conservation implies that the matrix $\lambda$, defined as in
Eq.~\eqref{eq:lambda}, is antisymmetric and correspondingly the MM and
EDM matrices are antisymmetric and hermitian~\cite{Schechter:cpphase}.
In the mass basis of the Majorana neutrino fields
\begin{equation}
    \tilde \lambda = {V^\nu}^T \lambda V^\nu \,.
\end{equation}

Neutrino MM and EDM's can be searched for by detecting their effect 
in the neutrino-electron scattering cross section. In the SM, neutrinos
interact with electrons only via weak currents. But a neutrino MM
and/or EDM adds an extra component due to photon
exchange~\cite{Bardin:wr, Kyuldjiev:1984kz}. For a neutrino produced
in a source with flavor $\beta$, $\nu_\beta$, this additional
contribution reads:
\begin{equation}
    \label{eq:cross}
    \frac{d\sigma_\text{em}}{dT}
    = \frac{\alpha^2 \pi}{m_e^2 \, \mu_B^2} \left(
    \frac{1}{T} - \frac{1}{E_\nu}\right) \mu^2_\text{eff}
    (\nu_\beta) \,.
\end{equation}
Here $T$ denotes the kinetic energy of the recoil electron, $E_\nu$ is
the energy of the incoming neutrino.  The effective MM square is given
by~\cite{Grimus:2000tq}
\begin{equation}\label
    {eq:mmeff}
    \mu^2_\text{eff} (\nu_\beta)=  
    a_{\beta -}^\dagger \lambda^\dagger \lambda a_{\beta -} +
    a_{\beta +}^\dagger \lambda \lambda^\dagger a_{\beta +} \,.
\end{equation}
The vectors $a_{\beta \pm}$ denote the neutrino amplitudes for
negative and positive helicities, respectively, at the detector after
propagation, so that the state arriving in the detector is:
\begin{equation}
    \label{eq:instate}
    |\nu\rangle_\text{det} =
    \sum_{\alpha=e,\mu,\tau,s,\ldots} \left(
    a_{\beta -}^\alpha | \nu^{(-)}_\alpha \rangle +
    a_{\beta +}^\alpha | \nu^{(+)}_\alpha \rangle \right) \,.
\end{equation}
In the massless limit the negative helicity states are left-handed
neutrinos whereas the positive helicity states are sterile
right-handed neutrinos in the Dirac case and right-handed
antineutrinos in the Majorana case. In general, for neutrinos coming
from distance sources, these amplitudes depend on the initial neutrino
flavor, its energy and the distance between source and detector and
are obtained by solving the neutrino evolution equation as described
below.  

The electromagnetic cross section adds to the weak cross section and
allows to extract information on the TM matrix $\lambda$.  In this
respect it is important to notice that the square of the effective MM
given in Eq.~\eqref{eq:mmeff} is independent of the basis 
chosen~\cite{Grimus:2000tq} and it only depends on the neutrino
flavor at the source and on its propagation.

At laboratory experiments with terrestrial baselines the vectors
$a_{\beta \pm}$ are simply unity vectors pointing on the corresponding
neutrino ($-$) (or antineutrino $(+)$) flavor direction, since in
those experiments the baseline is much too short for any oscillations
to develop and then $|\nu\rangle_\text{det} =|\nu_\beta\rangle$. For
example, for a reactor neutrino experiment the source is $\bar\nu_e$
and we have $a_{\beta-}=a_{e-}- = 0$ and $a_{\beta+} = a_{e+} = (1,\,
0,\, 0)^T$.

At present, laboratory experiments give 90\% CL bounds on the neutrino
magnetic moments of $1.9 \times 10^{-10}~\mu_B$~\cite{rovno}, $9
\times 10^{-11}~\mu_B$~\cite{munu}, $7.4 \times
10^{-11}~\mu_B$~\cite{texono} and $5.8 \times
10^{-11}~\mu_B$~\cite{gemma} for the electron neutrino, $6.8 \times
10^{-10}~\mu_B$ for the muon neutrino~\cite{LSND} and $3.9 \times
10^{-7}~\mu_B$ for the tau neutrino~\cite{DONUT} (see also
Ref.~\cite{pdg}). On the other hand, astrophysics and cosmology
provide limits of the order of $10^{-12}$ to $10^{-11}$ Bohr
magnetons~\cite{Raffelt:gv}. Improved sensitivity for the electron
neutrino MM from reactor neutrino experiments is expected, while
experiments with tritium sources aims to reach
$10^{-12}~\mu_B$~\cite{MAMONT, McLaughlin:2003yg}.

For solar neutrino phenomenology it was also important the realization
that the existence of magnetic transition (flavor non-diagonal)
moments (TM) leads to the phenomenon of spin-flavor precession
(SFP)~\cite{Schechter:cpphase,SFP}, when they move through a magnetic
field, as might happen in the Sun. In this case their evolution
equation is (we follow the notation of Ref.~\cite{Grimus:2000tq}):
\begin{equation}
    \label{eq:sfpevol}
    i\frac{d}{dz}
    \begin{pmatrix}
	\varphi_- \\ 
	\varphi_+
    \end{pmatrix} =
    \begin{pmatrix}
	V_L + \frac{1}{2 E} M_\nu^\dagger M_\nu
	& -B_+ \lambda^\dagger
	\\
	-B_- \lambda
	& V_R + \frac{1}{2 E} M_\nu M_\nu^\dagger
    \end{pmatrix}
    \begin{pmatrix}
	\varphi_- \\ 
	\varphi_+
    \end{pmatrix} \,.
\end{equation}
In this equation, $\varphi_-$ and $\varphi_+$ denote the vectors of
neutrino flavor wave functions corresponding to negative and positive
helicity, respectively, and $E$ denotes the neutrino energy.  $M_\nu$
denotes the neutrino mass matrix in the flavor basis.  If the
neutrino propagates along the $z$-axis the relevant magnetic field
components are:
\begin{equation}
    B_\pm = B_x \pm iB_y \,.
\end{equation}
The matter potential $V_L$ (see Sec.~\ref{sec:matterosc}) is given by
\begin{equation}
    V_L = \sqrt{2}\, G_F\, \diag(
    n_e - n_n/2,\, -n_n/2,\, -n_n/2,\, 0,\, \ldots) \,,
\end{equation}
where $n_e$ ($n_n$) is the electron (neutron) density in the sun.  For
Dirac neutrinos $V_R = 0$ while for Majorana neutrinos $V_R = -V_L$.

SFP was shown to affect the propagation of solar neutrinos in an
important way, due to the effects of matter~\cite{Akhmedov:uk,
Lim:1987tk} which could be resonant, or non-resonant~\cite{Miranda}
depending on the relative sign among the kinetic term, the matter
potential and the magnetic one in Eq.~\eqref{eq:sfpevol}.  

There is also another important difference between Majorana and Dirac
neutrinos for the SFP mechanism. For Dirac Neutrinos, the resulting
states are right-handed neutrinos which are sterile and therefore
undetectable while for Majorana neutrinos SFP converts left-handed
$\nu_e$ into right-handed $\bar{\nu}_\mu$ or $\bar{\nu}_\tau$, which
can be detected in accordance with the SNO NC measurement.

Within the expected magnitudes and profiles of the solar magnetic
field, SFP of Majorana neutrinos could be the dominant source of the
observed solar neutrino transitions provided that $10^{-9}~\eVq
\lesssim \Delta m^2_\odot \lesssim 10^{-7}~\eVq$ (see
Ref.~\cite{Miranda} for a recent fit).  Thus the results of KamLAND,
which are unaffected by this mechanism and point out to a much larger
$\Delta m^2$, rule out SFP as the dominant mechanism for solar
neutrino flavor conversion.

Conversely one can use the independent determination of the
oscillation parameters by KamLAND to set a bound on the size of the
subdominant effects associated to SFP.  Following this approach the
experimental limits on solar $\bar\nu_e$ fluxes~\cite{klandsolanue}
can be used to derive a constraint on the product of the neutrino
magnetic moment and the solar magnetic field. With the modeling of the
magnetic field in Refs.~\cite{pulido,Miranda2} this procedure resulted
into a bound
\begin{equation}
    \mu_\text{eff} (\nu_{\Nuc{8}{B}})
    \lesssim 10^{-10} {\rm -} 10^{-12}~\mu_B \,,
\end{equation}
where the range spans the uncertainty associated with the assumptions
made over the solar magnetic field.

\subsubsection{Bounds on neutrino magnetic moments from solar and
  reactor data}

A solar-model independent bound on solar neutrino MMs can be derived
from the observation that the presence of neutrino TMs also affect the
neutrino detection process via Eq.~\eqref{eq:cross}. As discussed in
Ref.~\cite{Grimus:2002vb}, sizable TMs would contribute to the elastic
neutrino--electron scattering in the Super-Kamiokande experiment, so
that data from this experiment can be used to constrain
electromagnetic neutrino properties~\cite{Beacom:1999wx,
joshipura:bp}.
Moreover, data from reactor neutrino experiments which use elastic
neutrino--electron scattering for neutrino detection are also
sensitive to TMs, so by combining both data sets one can improve the
bounds as we summarize next. In what follows we will assume that
Majorana neutrinos.  

The effect of neutrino TM's in solar neutrino detection can be
described by Eq.~\eqref{eq:cross} with
\begin{equation}
    \label{eq:mu2effLMA}
    \mu_\odot^2 =
    |\mathbf\Lambda|^2 - P^{3\nu}_{e3} |\Lambda_3|^2 -
    \sum_{j,k=1}^2 \left\langle (\tilde{a}^j_{e-})
    (\tilde{a}^k_{e-})^* \right\rangle \Lambda_j^* \Lambda_k \,.
\end{equation}
where we have introduced the vectors $\mathbf\Lambda =
(\Lambda_\alpha)$ and $\tilde{\mathbf\Lambda} = (\Lambda_j)$ in the
flavor and mass basis, respectively:
\begin{equation}
    \label{eq:defL}
    \lambda_{\alpha\beta} =
    \varepsilon_{\alpha\beta\gamma} \Lambda_\gamma
    \quad \text{and} \quad 
    \lambda_{jk} = \varepsilon_{jkl} \Lambda_l \,,
\end{equation}
and the amplitudes in the mass basis $\tilde{a}_{e-}={V^\nu}^\dagger
a_{e-}$. Thus, in the flavor basis $\lambda_{e \mu} = \Lambda_\tau$,
$\lambda_{\mu\tau} = \Lambda_e$ and $\lambda_{\tau e} = \Lambda_\mu$.
Note also that 
\begin{equation}
    \label{eq:basisindep}
    |\mathbf\Lambda|^2 = \frac{1}{2}
    \Tr \left( \lambda^\dagger \lambda \right)
    \quad \Rightarrow \quad
    |\mathbf\Lambda| = |\tilde{\mathbf\Lambda}| \,.
\end{equation}
The brackets $\left\langle \dots \right\rangle$ in the last term in
Eq.~\eqref{eq:mu2effLMA} denote the average over the production point,
Earth-Sun distance and zenith angle, and $P^{3\nu}_{ej} \equiv
\left\langle |\tilde{a}_{e-}^j|^2 \right\rangle$ is the probability
that the neutrino produced in the core of the Sun as a $\nu_e$ arrives
at the detector as a mass eigenstate $\nu_j$.

In order to derive Eq.~\eqref{eq:mu2effLMA} one has to solve the
evolution equation Eq.~\eqref{eq:sfpevol} with the initial condition
$\varphi_-^T(z_0) = (1,0,\ldots)$, $\varphi_+^T(z_0) = (0,\ldots)$
($\nu_e$ produced at $z_0$) and and final conditions $a_{e\mp} \equiv
\varphi_\mp(z_\text{det})$, where $z_\text{det}$ is the distance
between the neutrino production point in the Sun and its detection
point in the Earth. This task is considerably simplified by the fact
that, as shown in in Ref.~\cite{Miranda} for an effective MM of
$10^{-11}~\mu_B$ a characteristic solar magnetic field of the order of
80 KGauss has practically no effect on the LMA solution.  Hence,
helicity is conserved in solar neutrino propagation, so that $a_{e+} =
0$. Furthermore oscillations with $\Delta m^2_{31}$ can be averaged
out.

Eq.~\eqref{eq:mu2effLMA} can be further simplified by realizing that
due to fast vacuum oscillations on the way from the Sun to the Earth
the neutrino state arriving at the Earth are an incoherent mixture of
mass eigenstates. Also, in the LMA region Earth matter effects are
very small, so that can be neglected. Taking this into account and 
neglecting terms of order $\theta_{13}^2$ one gets
\begin{equation}
    \label{eq:finalLMA}
    \mu_\odot^2 = |\mathbf\Lambda|^2 - |\Lambda_2|^2 +
    P^{2\nu}_{e1} \left( |\Lambda_2|^2 - |\Lambda_1|^2 \right) \,.
\end{equation}
The probability $P^{2\nu}_{e1}$ is the usual $2\nu$ oscillation
probability which is a function of the ratio $\Delta m^2_{21}/E_\nu$
and the solar mixing angle $\theta_{12}$.  Notice that 
Eq.~\eqref{eq:finalLMA} naturally makes no distinction between MMs and
EDMs. Constraining $|\mathbf\Lambda|$, \emph{all} elements of the TM
matrix will be bounded at the same time.

\begin{figure}\centering
    \includegraphics[width=0.99\linewidth]{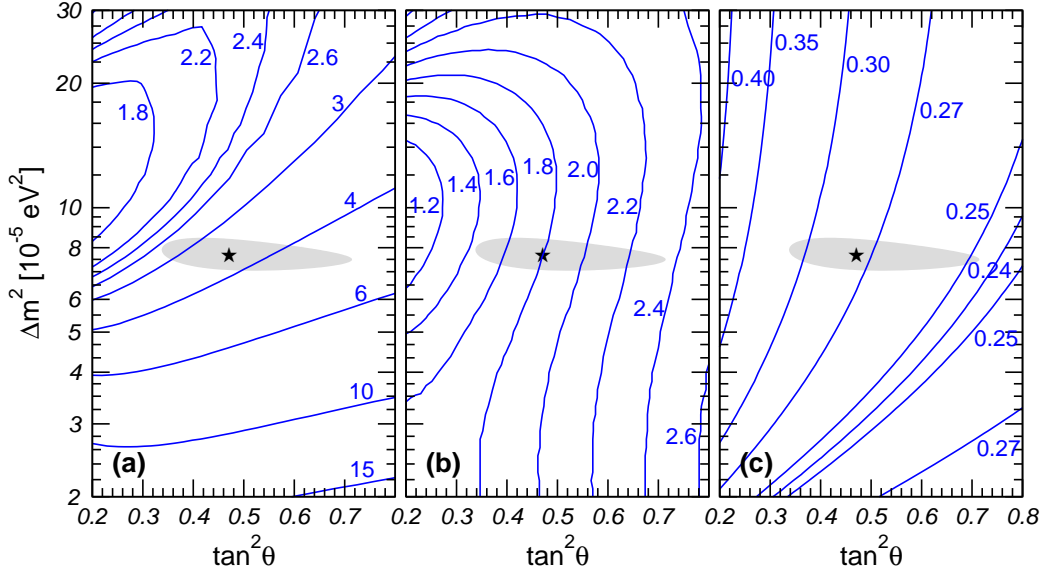}
    \caption{\label{fig:magnetic}%
      Contours of the 90\% CL bound on $|\mathbf\Lambda|$ in units of
      $10^{-10}~\mu_B$, from the analysis of solar data only (left
      panel), of solar and reactor data (central panel), and after 3
      years of Borexino data-taking (right panel). The gray area show
      the current $3\sigma$ allowed region; the best fit point is
      marked with a star. Figure adapted from
      Ref.~\cite{Tortola:2004vh}.}
\end{figure}

Fixing the oscillation parameters at the current best fit point given
in Sec.~\ref{sec:3nu}, one obtains the following 90\% CL bound:
\begin{equation}
    |\mathbf\Lambda| < 3.6 \times 10^{-10}~\mu_B
    \quad\text{(solar data only).}
\end{equation}
However, such a bound substantially depends on the values of the
neutrino oscillation parameters. In Fig.~\ref{fig:magnetic}(a) we show
contours of the 90\% CL bound on $|\mathbf\Lambda|$ in the
$(\tan^2\theta_{12},\, \Delta m^2_{21})$ plane. We see that the bound
gets stronger for smaller values of $\tan^2\theta$, whereas in
rightmost part of the LMA region the bound is only $5\times
10^{-10}~\mu_B$.

On the other hand for reactor neutrinos, as discussed above, $a_{e-} =
0$ and $a_{e+} = (1,\, 0,\, 0)^T$ and the resulting $\mu_\text{eff}^2$
relevant in reactor experiments is given as
\begin{equation} \begin{split}
    \label{eq:reactorMM}
    \mu_\text{R}^2 &= |\Lambda_\mu|^2 + |\Lambda_\tau|^2
    \\[1mm]
    &= |\mathbf\Lambda|^2 - c^2 |\Lambda_1|^2 - s^2 |\Lambda_2|^2 
    -2 s c |\Lambda_1| |\Lambda_2| \cos\delta \,,
\end{split} \end{equation}
where $c = \cos\theta_{12}$ and $s = \sin\theta_{12}$.  From this
relation it is clear that reactor data on its own \emph{cannot}
constrain all TMs contained in $\lambda$, since $\Lambda_e$ does not
enter in Eq.~\eqref{eq:reactorMM}.  Also, notice that the relative
phase $\delta = \arg(\Lambda_1^* \Lambda_2)$ between $\Lambda_1$ and
$\Lambda_2$ appears in addition to $|\mathbf\Lambda|$, $|\Lambda_1|$
and $|\Lambda_2|$. However it is clear that the effects on reactor 
antineutrinos do not depend on $\Delta m^2$. 

Thus combining solar and reactor data one can obtain considerably
stronger bounds. At the best fit point at 90\% CL
\begin{equation}
    |\mathbf\Lambda| < 1.8 \times 10^{-10}~\mu_B
    \quad\text{(solar + reactor data).}
\end{equation}
In Fig.~\ref{fig:magnetic}(b) we show the contours of the bound in the
$(\tan^2\theta_{12},\, \Delta m^2_{21})$ plane for the combination of
solar and reactor data. The bound becomes weaker as
$\tan^2\theta_{12}$ increases: for example, for $\tan^2\theta_{12} =
0.7$ we get $|\mathbf\Lambda| \le 2.4\times 10^{-10}~\mu_B$.

Substantial improvements on this bound are expected from
Borexino~\cite{borexino}. This experiment is mainly sensitive to the
solar \Nuc{7}{Be} neutrino flux, which will be measured by elastic
neutrino--electron scattering. Therefore, Borexino is similar to SK,
the main difference being that the energy of the monochromatic
\Nuc{7}{Be} neutrinos is roughly one order of magnitude smaller than
the average energy of the \Nuc{8}{B} neutrino flux relevant for SK.
The expected 90\% sensitivity to $|\mathbf\Lambda|$ as a function of
the neutrino oscillation parameters after three years of Borexino data
taking is shown in Fig.~\ref{fig:magnetic}(c).  Fixing the oscillation
parameters at the current best-fit point one expects:
\begin{equation}
    \label{eq:borexinobound}
    |\mathbf\Lambda| \le 2.8 \times 10^{-11}~\mu_B
    \quad\text{(after 3 years of Borexino).}
\end{equation}
Thus the expected sensitivity is about one order of magnitude stronger
than the bound from existing data.

%% file: sec.future.tex
\section{Future Facilities}
\label{sec:fut}


\subsection{Solar Neutrino Experiments: Motivations and Expectations}
\label{sec:futsolar}

The first forty years of solar neutrino research has demonstrated that
new physics may appear when we carry out neutrino experiments in a new
domain of sensitivity. Most of the new solar neutrino experiments 
under construction or consideration aim at measuring the energy of 
individual neutrino-induced events below or of the order of 1 MeV, a
domain in which solar neutrino energies could not previously be 
measured. Remember that more than 98\% of the predicted flux of solar
neutrinos lies below 1 MeV. In Table~\ref{tab:futsolar} (from
Ref.~\cite{nakamura}) we list the low energy solar neutrino
experiments either under construction or being proposed.

\begin{table}\centering
    \caption{\label{tab:futsolar}%
      Listing of upcoming solar neutrino experiments and projects
      under development. Taken from Ref.~\cite{nakamura}.}
    \vspace{1mm}
    \begin{tabular}{llll}
	\hline
	Experiment  & Flux & Technique & Size
	\\
	\hline
	BOREXINO~\cite{borexino}
	& \Nuc{7}{Be}  & ES & 100 ton liquid scintillator
	\\
	KamLAND~\cite{klandprop} 
	& \Nuc{7}{Be}  & ES & 1000 ton liquid scintillator
	\\
	LENS~\cite{lens}    & $pp$, \Nuc{7}{Be}
	& CC & 60 ton In-loaded scintillator
	\\ 
	MOON~\cite{moon}    & $pp$, \Nuc{7}{Be} 
	& CC & 3.3 ton \Nuc{100}{Mo} foil + plastic scint
	\\
	Lithium~\cite{lithium} & CNO & radiochem & 10 ton Li
	\\
	XMASS~\cite{xmass}  & $pp$, \Nuc{7}{Be} & ES  & 10 ton liquid Xe
	\\
	HERON~\cite{heron}  & $pp$, \Nuc{7}{Be}  & ES & 10 ton superfluid He
	\\
	CLEAN~\cite{clean}  & $pp$, \Nuc{7}{Be}  & ES  & 130 ton liquid Ne
	\\
	SNO+~\cite{sno+}    & $pep$, CNO & ES& 1000 tom liquid scintillator
	\\
	\hline
    \end{tabular}
\end{table}

There are four primary motivations for doing such low energy solar
neutrino experiments.

First, as we have seen in Sec.~\ref{sec:3nuprobs}, according to the
currently accepted LMA oscillation solution the daytime survival
probability for solar neutrinos can be written to a good approximation
in the following simple form 
\begin{equation}
    \label{eq:plmaday}
    P_{ee} = \cos^4\theta_{13} (\frac{1}{2} + \frac{1}{2}
    \cos2\theta_{12m,0} \cos2\theta_{12}) \,,
\end{equation}
where the mixing angle in matter is
\begin{equation}
    \label{eq:defthetaM}
    \cos2\theta_{12m,0} =
    \frac{\cos2\theta_{12} - \beta}
    {\sqrt{(\cos2\theta_{12}-\beta)^2 + \sin^22\theta_{12}}} \,.
\end{equation}
$\beta$ is the ratio between the oscillation length in matter and the
oscillation length in vacuum
\begin{equation} \begin{split}
    \label{eq:betaconvenient}
    \beta &=
    \frac{2 \sqrt2 G_F \cos^2\theta_{13} n_{e,0} E_\nu}{\Delta m^2}
    \\
    &= 0.22 \, \cos^2\theta_{13}
    \, \left[ \frac{E_\nu}{1~\text{MeV}} \right]
    \, \left[ \frac{\mu_e\,\rho}{100~\text{g~cm}^{-3}} \right]
    \, \left[ \frac{7 \times 10^{-5}~\eVq}{\Delta m^2} \right] \,.
\end{split} \end{equation}
$\mu_e$ is the electron mean molecular weight and $\rho$ is the total 
density both evaluated at the neutrino production point.\footnote{For
the electron density at the center of the standard solar model, $\beta
= 0.22$ for $ E = 1$~MeV, $\theta_{13} = 0$, and $\Delta m^2 = 7\times
10^{-5}~\eVq$.}

If $\beta < \cos 2\theta_{12} \sim 0.38$ (at the best fit point), the
survival probability corresponds to vacuum averaged oscillations,
\begin{equation}
    \label{eq:peevacuum}
    P_{ee} = \cos^4\theta_{13}
    \, \left( 1 - \frac{1}{2} \sin^22\theta_{12} \right)
    \quad [\beta < \cos 2\theta_{12}, ~\text{vacuum}] \,.
\end{equation}
If $\beta > 1$, the survival probability corresponds to matter
dominated oscillations,
\begin{equation}
    \label{eq:peeadiab13}
    P_{ee} = \cos^4\theta_{13} \, \sin^2\theta_{12}
    \quad [\beta > 1, ~\text{MSW}] \,.
\end{equation}
The critical energy at which $\beta = \cos 2\theta_{12}$ is different
for the different neutrino sources since the fraction of the neutrino
flux that is produced at a given radius (\textit{i.e.}, density and
$\mu_e$) differs from one neutrino source to another. In particular
\begin{equation}
    \label{eq:critical}
    E_\text{crit} \simeq
    \begin{cases}
	1.7~\text{MeV} & (\Nuc{8}{B});
	\\
	2.1~\text{MeV} & (\Nuc{7}{Be});
	\\
	3.1~\text{MeV} & (pp).
    \end{cases}
\end{equation}
which means that to a very good approximation, \Nuc{8}{B} neutrinos
are always in the MSW regime, Eq.~\eqref{eq:peeadiab13}, while $pp$
and \Nuc{7}{Be} neutrinos are in the vacuum averaged regime,
Eq.~\eqref{eq:peevacuum}. Clearly, this is a prediction which can only
be directly tested by measuring the energy spectrum of solar neutrino
fluxes at low energies.

Second, new solar neutrino experiments will provide accurate
measurements of the fluxes of the important $pp$ and \Nuc{7}{Be} solar
neutrino fluxes, which together amount to more than 98\% of the total
flux of solar neutrinos predicted by the standard solar model. These
measurements will test the solar model predictions for the main
energy-producing reactions, predictions that are more precise than for
the higher-energy neutrinos. Using only the measurements of the solar
neutrino fluxes, one can determine the current rate at which energy is
being produced in the solar interior and can compare that energy
generation rate with the observed photon luminosity emitted from the
solar surface. This comparison will constitute a direct and accurate
test of the fundamental idea that the Sun shines by nuclear reactions
among light elements. Moreover, the neutrino flux measurements will
test directly a general result of the standard solar model, namely,
that the Sun is in a quasi-steady state in which the interior energy
generation rate equals the surface radiation rate.

Third, as seen in Eq.~\eqref{eq:peevacuum}, the survival probability
at low energies depends mostly on the mixing angle $\theta_{12}$ (and
only very weakly on the mixing angle $\theta_{13}$ given its present
constraint). Thus the determination of the survival probability for
low energy solar neutrinos and its comparison with the one for
\Nuc{8}{B} neutrinos will make possible a precise measurement of the
vacuum mixing angle, $\theta_{12}$, as well as a slightly improved
constraint on $\theta_{13}$. The increased robustness in determining
mixing angles will be very useful in connection with searches for CP
violation. Uncertainties in the CP-conserving neutrino parameters
could compromise the determination of the CP violating phase. As we
will see in Sec.~\ref{sec:futlblmot} the size of the CP violating
effects is always proportional to the product of the three mixing
angles and the two mass differences. Therefore at present the 
dominant source of uncertainty for the possible determination of the
leptonic CP violation is the unknown value of $\theta_{13}$ and the
neutrino mass ordering. Nevertheless, once those parameters are known,
the ultimate sensitivity to the CP phase will be given by the 
precision with which all the three mixing angles and the two mass
differences are known. 

Fourth, there may be entirely new physical phenomena that show up only
at the low energies, the very long baseline, and the great sensitivity
to matter effects provided by solar neutrino experiments as discussed
in Sec.~\ref{sec:npsolar}.

Next we briefly describe the expected sensitivity which can be 
achieved at the different types of future solar neutrino experiments.
More details can be found in Refs.~\cite{roadmap,bpsolrev}.

\subsubsection{\Nuc{7}{Be} Experiments}
\label{sec:be7}

Before Borexino the solar plus reactor experiments provided only loose
constraints on the \Nuc{7}{Be} solar neutrino flux, corresponding to
approximately a $\pm 40\%$ uncertainty at $1\sigma$.  

Measuring the flux from the 0.86 MeV monoenergetic line of \Nuc{7}{Be}
solar neutrinos in real-time is the main goal of the the Borexino
experiment~\cite{borexino} which started in 2007 taking data in the
Laboratori Nazionali del Gran Sasso in Italy.  Several tens of events
are expected daily at Borexino. In their first results they have
provided a measurement of the $\nu-e$ scattering rate with a precision
of 30\% (15\% statistics and 25\% systematics). 

In Ref.~\cite{roadmap} it was shown that a measurement of the $\nu-e$
scattering rate accurate to $\pm 10\%$ or better will reduce by a
factor of four the uncertainty in the measured \Nuc{7}{Be} neutrino
flux. Moreover, the $10\%$ \Nuc{7}{Be} flux measurement will reduce
the uncertainty in the crucial $pp$ flux by a factor of about 2.5 if
the luminosity constraint is assumed. A \Nuc{7}{Be} measurement
accurate to $\pm 3\%$ would provide another factor of two improvement
in the accuracy of the \Nuc{7}{Be} and $pp$ solar neutrino fluxes.

Similarly the KamLAND~\cite{klandprop} experiment, once its phase as
an reactor antineutrino experiment ends, will proceed with the
measurement of the \Nuc{7}{Be} solar neutrino flux. In order to do so
they need to reduce their present level of radioactive backgrounds by
six orders of magnitude.

However, the \Nuc{7}{Be} solar neutrino experiments are not expected
to provide significantly more accurate values for the neutrino
oscillation parameters than what we expect to be available after the
termination of the reactor antineutrino program in KamLAND.

\subsubsection{$pp$ Experiments}
\label{sec:pp}

A new generation of experiments aiming at a high precision real time
measurement of the low energy solar neutrino spectrum is now under
study (see Table~\ref{tab:futsolar}.  They seek to measure the primary
$pp$, $pep$ and CNO neutrino fluxes in real time either via neutrino
capture or neutrino-electron scattering. The expected rates at these
experiments for the proposed detector sizes are of the order of $\sim
1-10$ $pp$ neutrinos a day. Consequently, with a running time of two
years, they can reach a sensitivity of a few percent in the total
neutrino rate at low energy, provided that they can achieve sufficient
background rejection.

As described above, an accurate measurement of the $pp$ solar neutrino
flux will provide a direct test of the fundamental ideas underlying
the standard solar model. The $pp$ measurement will make possible the
determination of the total solar luminosity from just neutrino
experiments alone.  The global combination of a \Nuc{7}{Be}
experiment, plus a $pp$ experiment, plus the existing solar and
KamLAND data, and would make possible a precise determination of the
solar neutrino luminosity~\cite{roadmap}. A $pp$ solar neutrino
experiment accurate to 5\% would make possible a measurement of the
solar neutrino luminosity to 4\% and a 1\% $pp$ experiment would
determine the solar luminosity to the accuracy implied below:
\begin{equation}
    \label{eq:lnuoverlphotonbepp}
    \frac{L_\odot~\text{(neutrino-inferred)}}{L_\odot}
    = 0.99 \pm 0.02 \,.
\end{equation}

\begin{figure}\centering
    \includegraphics[height=0.3\textheight]{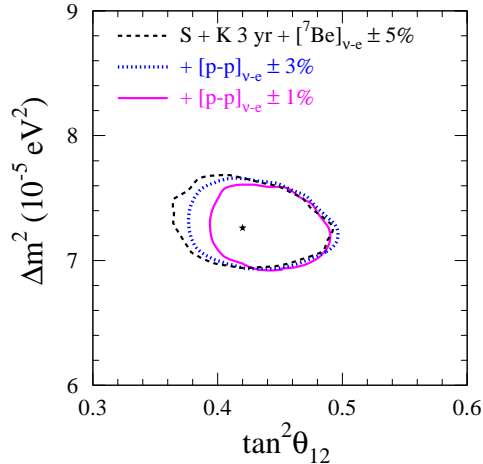}
    \caption{\label{fig:futsolarosc}%
      Improvement on the determination of the oscillation parameters
      from a precise determination of the $pp$ solar neutrino flux at
      5 and/or 1\%. Taken from Ref.~\cite{roadmap}.}
\end{figure}

Furthermore if the standard solar model is correct to the stated
accuracy ($\pm 1\%$ for the total $pp$ neutrino flux), and if there is
no new physics that shows up below $0.4$ MeV, then a measurement of
the $pp$ flux to an accuracy of better than $\pm 3\%$ can 
significantly the present experimental knowledge of $\tan^2
\theta_{12}$ illustrated in
Fig.~\ref{fig:futsolarosc}~\cite{roadmap,Bandyopadhyay:2004cp}.

\subsubsection{A $pep$ Experiment}
\label{sec:pep}

It is possible to fill the Sudbury Neutrino Observatory with liquid
scintillator after the physics program with heavy water is
completed~\cite{sno+}. This would enable the detection of electron
antineutrinos and, thanks to its deep location, the measurement of
solar $pep$ and CNO neutrinos with high statistics.

The ratio of the $pep$ to the $pp$ neutrino flux is robustly
determined by the standard solar model calculations.  The ratio is
determined more accurately than the individual fluxes because the
ratio only depends weakly on the solar model characteristics.  As a
consequence a measurement of the $\nu-e$ scattering rate by $pep$
solar neutrinos (a $1.4$ MeV neutrino line) would yield essentially
equivalent information about neutrino oscillation parameters and solar
neutrino fluxes as a measurement of the $\nu-e$ scattering rate by
$pp$ solar neutrinos~\cite{roadmap,bpsolrev}.

\subsubsection{Large Water Cherenkov Detectors}
\label{sec:protondecay}

Mega-ton class Water Cherenkov Detectors are being proposed as the
next generation of Long Baseline and Proton Decay
experiments~\cite{uno1,uno2,T2K,hyperk}.  Such detectors, can also
make a unique and important test of mater oscillations using
\Nuc{8}{B} solar neutrinos. Only a very large detector could have an
event rate sufficiently high to detect with some statistical
confidence the small day-night effect expected in the LMA solution due
to the conversion of solar neutrinos in the Earth matter. It would
also provide a much more precise measurement (much better than 1\%) of
the total event rate for the scattering of \Nuc{8}{B} solar neutrinos
by electrons.

Furthermore a first detection of the very rare but high energy $hep$
neutrinos should also be possible. In Ref.~\cite{bpsolrev} it was
estimated that assuming that the SSM predicted $hep$ flux is correct,
it could be determined with a $4\sigma$ or better accuracy over ten
years.


\subsection{Future LBL Experiments: Motivation and Challenges}
\label{sec:futlblmot}

At the end of the presently running neutrino experiments, many
questions will still remain open. Even after the non confirmation from
MiniBooNE of the LSND signal and the possibility of explaining all
data in terms of oscillations among the three known neutrinos, at the
end of the presently approved experimental program we will still be
ignorant about: (i) the value of $\theta_{13}$ (if not within the
limited reach of MINOS and CNGS), (ii) the sign of $\Delta m^2_{13}$,
and (iii) the possibility of CP violation in the lepton sector.

The generic requirements of an oscillation experiment to be able to
measure these parameters can be understood by examining the relevant
oscillation probabilities for neutrinos propagating in the constant
Earth matter potential, $V_E\sim 10^{-13}$~eV, and expand them in the
known-to-be-small parameters $\Delta m^2_{21} / \Delta m^2_{31}$,
$\Delta m^2_{21} L/E$, $\Delta m^2_{21}/ (E\, V_E)$, and
$\theta_{13}$: 
\begin{align}
    \label{eq:lblpem} 
    \begin{split}
	P(\nu_e \to \nu_\mu)
	& \simeq s_{23}^2\, \sin^2 2 \theta_{13}
	\, \left( \frac{\Delta_{31}} {B_\mp} \right)^2
	\, \sin^2 \left( \frac{B_\mp \, L}{2} \right)
	\\
	& + \tilde{J} \, \frac{\Delta_{12}}{V_E}
	\, \frac{\Delta_{31}}{B_\mp}
	\, \sin\left( \frac{V_E L}{2} \right) 
	\, \sin\left( \frac{B_\mp L}{2} \right) 
	\, \cos\delta_\text{CP}
	\, \cos\left( \frac{\Delta_{31} \, L}{2} \right)
	\\
	& \pm \tilde{J} \, \frac{\Delta_{21}}{V_E}
	\, \frac{\Delta_{31}}{B_\mp}
	\, \sin\left( \frac{V_E L}{2} \right) 
	\, \sin\left( \frac{B_\mp L}{2} \right) 
	\, \sin\delta_\text{CP}
	\, \sin\left( \frac{\Delta_{31} \, L}{2} \right)
	\\
	& + \left( \frac{\Delta_{21}}{V_E} \right)^2
	\, \sin^2 2\theta_{12} c^2_{23}
	\, \sin^2 \left( \frac{V_E L}{2} \right)
    \end{split}
    \\[2mm]
    \label{eq:lblpmm}
    \begin{split}
	P(\nu_\mu \to \nu_\mu)
	& \simeq 1 - c_{13}^2 \, \sin^2 2\theta_{23}
	\, \sin^2\left( \frac{\Delta_{31} \, L}{2} \right)
	\\
	& + (\Delta_{21} L) c^2_{13} \, c^2_{12} \, \sin^2 2\theta_{23}
	\, \cos\left( \frac{\Delta_{31}L}{2} \right)
    \end{split}
    \\[2mm]
    \label{eq:lblpee}
    P(\nu_e \to \nu_e)
    & \simeq 1 - \sin^22\theta_{31}
    \, \sin^2\left( \frac{\Delta_{31} \, L}{2} \right)
    - \left( \frac{\Delta_{21} L}{2} \right)^2 \, c_{13}^4
    \, \sin^2\theta_{12}
\end{align}
where 
\begin{align}
    \Delta_{ij} & =\frac{\Delta m^2_{ij}}{2E_\nu} \,,
    \\
    B_\pm &= \Delta_{31} \pm V_E \,,
    \\
    \tilde{J} &= c_{13} \, \sin^2 2\theta_{13}
    \, \sin^2 2\theta_{23} \, \sin^2 2\theta_{12} \,.
\end{align}
In the above expressions the upper (lower) sign applies to
oscillations of neutrinos (antineutrinos).

From these expressions we see that in order to measure the missing 
parameters, the following is required of future LBL experiments:
\begin{itemize}
  \item[(i)] To best discriminate normal and inverted mass orderings
    matter effects must be relevant so one can observe the dominant
    interference between the $\Delta m^2_{31}$ and $V_E$ terms in 
    $B_\pm$ in Eq.~\eqref{eq:lblpem}. This requires a very long
    baseline. With a shorter baseline sensitivity to the the ordering
    can be achieved if one has information on four oscillation
    channels $\nu_e\rightarrow \nu_\mu$ , $\nu_\mu\rightarrow \nu_e$,
    $\bar\nu_e\rightarrow \bar\nu_\mu$ and $\bar\nu_\mu\rightarrow
    \bar\nu_e$~\cite{thomas}.  In principle vacuum oscillations are
    also sensitive to the difference between normal and inverted
    ordering via the difference between $\Delta m^2_{31}$ and $\Delta
    m^2_{32}$ oscillation wavelengths. This effect is higher order
    than the ones shown in Eqs.~\eqref{eq:lblpem}--~\eqref{eq:lblpee}
    and it appears both in electron and muon
    disappearance~\cite{petpiai1,petpiai2,dgk,paren1,paren2}. 

  \item[(ii)] To measure $\theta_{13}$, since these effects are small
    because of the smallness of this mixing angle, one needs a very
    intense beam with very good background rejection and excellent
    systematics.  It can be performed by detection of $\nu_e$
    appearance in a $\nu_\mu$ beam (with maximal sensitivity given by
    the first term in Eq.~\eqref{eq:lblpem}) or by disappearance of
    reactor $\bar\nu_e$ as seen from Eq.~\eqref{eq:lblpee}.
    
  \item[(iii)] To better measure the exact value of $\theta_{23}$ and
    determine whether it is exactly equal to $\pi/4$ one needs an
    intense $\nu_\mu$ beam. This is so because the cleanest channel is
    $\nu_\mu$ disappearance but, as seen from Eq.~\eqref{eq:lblpmm},
    the dependence with the deviation from maximal mixing is only
    quadratic since $\sin^2 2\theta_{23} = 1 - 4(\sin^2\theta{23} -
    \frac{1}{2})^2$. Notice also, that to this order there is no
    sensitivity in this channel to the octant of $\theta_{23}$.  
    
  \item[(iv)] To detect CP violation the best option is to have
    intense beams with exchangeable initial state to compare the
    oscillations of neutrinos and antineutrinos which would allow to
    isolate the $\sin\delta_\text{CP}$ term in Eq.~\eqref{eq:lblpem}.
    In principle some sensitivity to $\delta_\text{CP}$ can also be
    achieved by using only neutrinos (or antineutrinos) from the
    different $L/E$ energy dependence of the $\cos\delta_\text{CP}$
    and $\sin\delta_\text{CP}$ pieces in Eq.~\eqref{eq:lblpem}. For
    this effect to be observable $\theta_{13}$ should be not too
    small. 
\end{itemize}

\begin{figure}\centering
    \includegraphics*[width=4.5in]{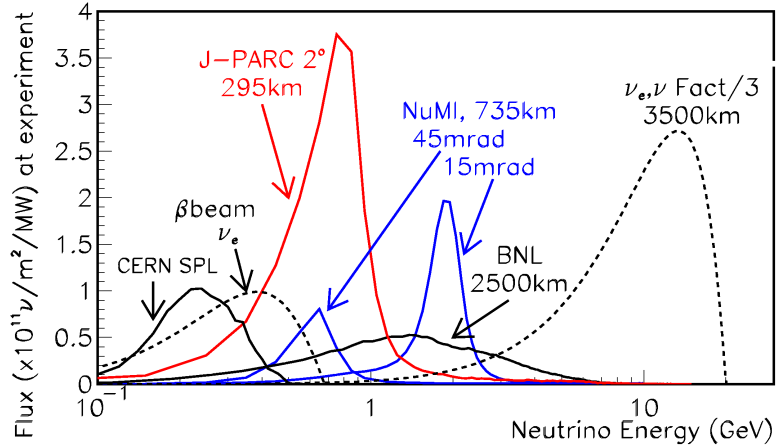}
    \caption{\label{fig:futfluxes}%
      Neutrino fluxes for several of the proposed experiments. Taken
      from Ref.~\cite{harris}.}
\end{figure}

New facilities and experiments are being proposed which can implement
some (or all) of these conditions. In particular, for future neutrino
oscillation experiments four type of facilities are under
consideration: 
\begin{itemize}
  \item[a)] Conventional neutrino
    superbeams~\cite{superbeams1,superbeams2,superbeams3,T2K,nova1,nova2}
    from the decay of pions generated from a proton beam dump with a
    detector either on or off axis. In these facilities the main beam
    consists of $\nu_\mu$'s and the experiments can search for both
    $\nu_e$ appearance and $\nu_\mu$ disappearance.
    
  \item[b)] Very intense medium baseline ($L\sim$ few km) $\bar\nu_e$
    reactor disappearance experiment~\cite{futreactors1,futreactors2}
    with two detectors to minimize the systematic uncertainties and
    allow for a precise determination of $\theta_{13}$.
    
  \item[c)] Neutrino Factories: These are neutrino beams from muon
    decay in muon storage rings~\cite{nufact1,nufact2,nufactrev}. 
    This provides a very clean $\nu_\mu$ and $\bar\nu_e$ beam (or
    vice-versa) with well known energy spectrum. The dominant search is
    the appearance of ``wrong sign'' muons from the oscillation of the
    $\bar\nu_e$ although all other oscillation channels can also be
    observed.

  \item[d)] Beta beam: A beam of pure $\nu_e$ or $\bar\nu_e$ from
    heavy ion decay~\cite{betabeams} with which both $\nu_e$
    disappearance and $\nu_\mu$ appearance is searched for. 
\end{itemize}
In Table~\ref{tab:futfacilities} we list the main features of the some
of the LBL experiments under consideration.  Fig.~\ref{fig:futfluxes}
(from Ref.~\cite{harris}) illustrates some of the characteristic
neutrino fluxes at the different experiments.  A detail recent study
of the comparison among different possible configurations in the US
can be found in Ref.~\cite{uslbl}. 

\begin{table}\centering
    \caption{\label{tab:futfacilities}%
      Characteristics of some of the future LBL experiments.}
    \vspace{1mm}
    \begin{tabular}{lccccc}
	\hline
	Experiment &  $L$~[km] & $\langle E_\nu \rangle$
	& Power (MW) & Mass (kton) & channel
	\\ 
	\hline
	\multicolumn{6}{l}{\bf First Generation Superbeams:}
	\\
	T2K     &  295 &  0.7 GeV & 0.8 & 22.5 & $\nu_\mu \to \nu_{e,\mu}$
	\\ 
	NuMI-OA & 700--900 & 2 GeV & 0.4 &  50 & $\nu_\mu \to \nu_{e,\mu}$
	\\
	\hline
	\multicolumn{6}{l}{\bf Reactor Experiments:}
	\\
	D-CHOOZ  & 1.05 & $\sim$ few MeV & $2\times 4250$ & 0.011 & $\nu_e \to \nu_e$
	\\
	\hline
	\multicolumn{6}{l}{\bf Next Generation Superbeams:}
	\\
	T2HK      &  295 &  0.7 GeV & 4 &  450  & $\nu_\mu \to \nu_{e,\mu}$
	\\
	SNuMI-OA & 700--900 &  2 GeV & 2 &  100 & $\nu_\mu \to \nu_{e,\mu}$
	\\
	BNL2NUSL & $>2500$  & 1  GeV & 1 &  500 & $\nu_\mu \to \nu_{e,\mu}$
	\\
	CERN SPL & 130 & 0.4 GeV & 4 &  400 & $\nu_\mu \to \nu_{e,\mu}$
	\\
	\hline
	{\bf $\beta$ beam} & 130--3000 
	& 0.2--5 GeV& 0.04 & 400
	& $\nu_e \to \nu_{e,\mu}$ 
	\\
	\hline
	{\bf $\nu$ factory} & 700--3000 & 7--40 GeV &  4 & 50 
	& $\nu_{e,\mu} \to \nu_{e,\mu,\tau}$
	\\
	\hline
    \end{tabular}
\end{table}

In general, the independent determination of the missing pieces of the
puzzle at these experiments becomes challenging because in the
relevant oscillation probabilities there appear three independent
two-fold parameter degeneracies~\cite{bargerdeg,burguet,mn1,mn2}.  In
brief: 
\begin{itemize}
  \item ($\theta_{23}$, $\pi/2 - \theta_{23}$)
    degeneracy~\cite{foglideg,bargerdeg} which means that the same 
    event rates are predicted for a value of $\theta_{23}$ mixing
    angle than for the value $\pi/2 - \theta_{23}$.  This is due to
    the fact that most sensitivity to the $\theta_{23}$ angle is
    achieved from precise measurement of $\nu_\mu$ disappearance and
    the dominant piece of the relevant probability
    Eq.~\eqref{eq:lblpmm}, is invariant under the exchange
    $\theta_{23} \to \pi/2 - \theta_{23}$.
    
  \item ($\delta_\text{CP}$, $\theta_{13}$) degeneracy~\cite{burguet}.
    This means that by simultaneously changing $(\delta_\text{CP}$ and
    $\theta_{13})$ it is possible to predict the same number of
    observed events. This is due to the fact that the appearance
    probability Eq.~\eqref{eq:lblpem} takes the same values for
    different pairs of parameters ($\delta_\text{CP}$, $\theta_{13}$).
    
  \item The third degeneracy arises because it is possible to 
    change simultaneously the sign of $\Delta m^2_{31}$ and the value
    of $\delta_\text{CP}$ without changing the predicted number of
    observed events.  It arises from the fact that in
    Eq.~\eqref{eq:lblpem} a change in $\sgn(\Delta m^2_{31})$ can be
    compensated by an offset in $\delta_\text{CP}$~\cite{mn1,mn2}. 
\end{itemize}
As a consequence if one only measures the total number of events in
the different channels at a given facility (fixed $L$) one can find
different sets of parameters which are able to fit the data.

The phenomenological efforts in this front concentrate on the study of
how the combination of data from experiments performed at different
baselines and/or with different beam
types~\cite{comb1,comb2,comb3,comb4} as well as the use of subdominant
oscillation channels involving $\nu_\tau$'s~\cite{silver} and the
measurement of the energy spectrum of the 
events~\cite{golden,bnlnuwg,lindspec} can help in resolving these
degeneracies. 

Next we briefly describe the expectations at some of these proposed
facilities.


\subsection{First Generation of Superbeam and Reactor Experiments}
\label{sec:futlblfirst}

There are two different type of experiments being built or proposed 
as a first step in the determination of the full leptonic mixing
matrix. They make use of either a neutrino \emph{superbeam} with an
detector \emph{off axis} or a reactor antineutrino beam with a near
and far detector.

Neutrino superbeams are defined as conventional neutrino beams
produced using megawatt-scale high-energy proton drivers.  The
conventional neutrino beam is produced using the very powerful primary
proton beam which after hitting a target, creates a secondary beam
composed mostly of charged pions and some kaons. They are then allowed
to decay to produce a tertiary neutrino beam.  The secondary beam are
focused and allowed to decay in a long decay channel and it can be
charge-sign selected to produce either a neutrino beam from positive
meson decays or an antineutrino beam from negative meson decays. 

The resulting neutrino beam consists mostly of muon neutrinos (or
antineutrinos) from $\pi^\pm \to \mu + \nu_\mu$ decays, with a small
``contamination'' of electron neutrinos, electron antineutrinos, and
muon antineutrinos from muon, kaon, and charmed meson decays.  The
fractions of $\nu_e$, $\bar\nu_e$ and $\bar\nu_\mu$ in the beam depend
critically on the beamline design.

The contamination from other neutrino species is a handicap for the
appearance experiments searching for $\nu_\mu \to \nu_e$ transition,
which as we have seen above, is the dominant channels in the
determination of $\theta_{13}$, the mass ordering and the CP phase.
However, the signal to noise ratio can be improved if one could design
a conventional superbeam which was nearly monochromatic. In order to
made such a \emph{narrow band} neutrino beam the
\emph{off-axis}~\cite{offaxis} technique is being put forward. The
basic point is that because the pion decay is a two-body decay, there
is an angle with respect to the focused pion direction where the broad
band of pion energies will produce a narrow band of neutrino energies.
In the near future two narrow band beams are being foreseen: one for
the T2K experiment in Japan~\cite{T2K} and another one using the
Fermilab NuMI beamline~\cite{nova1,nova2,flare}.

T2K makes use of the new 0.8 MW hadron facility being constructed at
JAERI, J-PARC with an initial proton energy of 40 GeV.  In combination
with that facility there will be a neutrino beam line which will be
aimed at the Super-Kamiokande detector. This off axis beam will allow
the T2K experiment to run at peak neutrino energies between 550 and
700 MeV. 

At Fermilab the NuMI beamline, while providing on axis neutrinos for
the MINOS experiment, is also producing off axis neutrinos. For
example a detector place at about 20mrad from the NuMI beamline would
see a very narrow neutrino beam peaked at 2 GeV. Using this idea the
NO$\nu$a concept~\cite{nova1,nova2} was proposed to used a fine
grained calorimeter based on liquid scintillator as a detector.  Also,
a proposal to use a Liquid Argon Time Projection Chamber as a
detector, FLARE~\cite{flare}, has been put forward, 

In addition to the intrinsic background due to the beam contamination,
the appearance experiments are also challenged by the background due
to neutral current $\pi^0$ production in the detector which mimics the
electron neutrino signal. The relevance of this background is
different for different detection techniques and it seems better
eliminated with a liquid argon type detector.

Recently, there has been a lot of activity to investigate the
potential of new reactor neutrino experiments~\cite{futreactors1}
based on the observation that the performance of previous experiments,
such as CHOOZ~\cite{CHOOZ}, can be significantly improved if a near
detector is used to control systematics and if the statistics is
increased. The main advantage of these experiments is that they are
free of degeneracies as they only measure the disappearance
probability Eq.~\eqref{eq:lblpee} which gives a direct determination
of $\theta_{13}$.  

A number of possible sites are discussed, including reactors in
Brasil, China, France, Japan, Russia, Taiwan, and the US (see
Ref~\cite{futreactors1,futreactors2} for an extensive list).  Of these
projects the Double-Chooz experiment~\cite{dchooz} has the opportunity
to obtain results first. The experiment will employ two almost
identical detectors of medium size. The far detector will be situated
in the same cavern as CHOOZ but will be able detect ~50000 electron
antineutrinos in three years of operation with a systematic
uncertainty of 0.6\%.  From the second generation reactor experiments,
a more advance project is the experiment proposed to be located in
Daya Bay~\cite{dayabay}, China. Its basic layout consists of three
underground experimental halls, one far, two near where eight
identical cylindrical detectors will be located (four of them at the
far site). The goal of the Daya Bay experiment is to reach a
sensitivity of 0.01 or better in $\sin^2 2\theta_{13}$ at 90\% CL.

\begin{figure}\centering
    \includegraphics[width=0.8\textwidth]{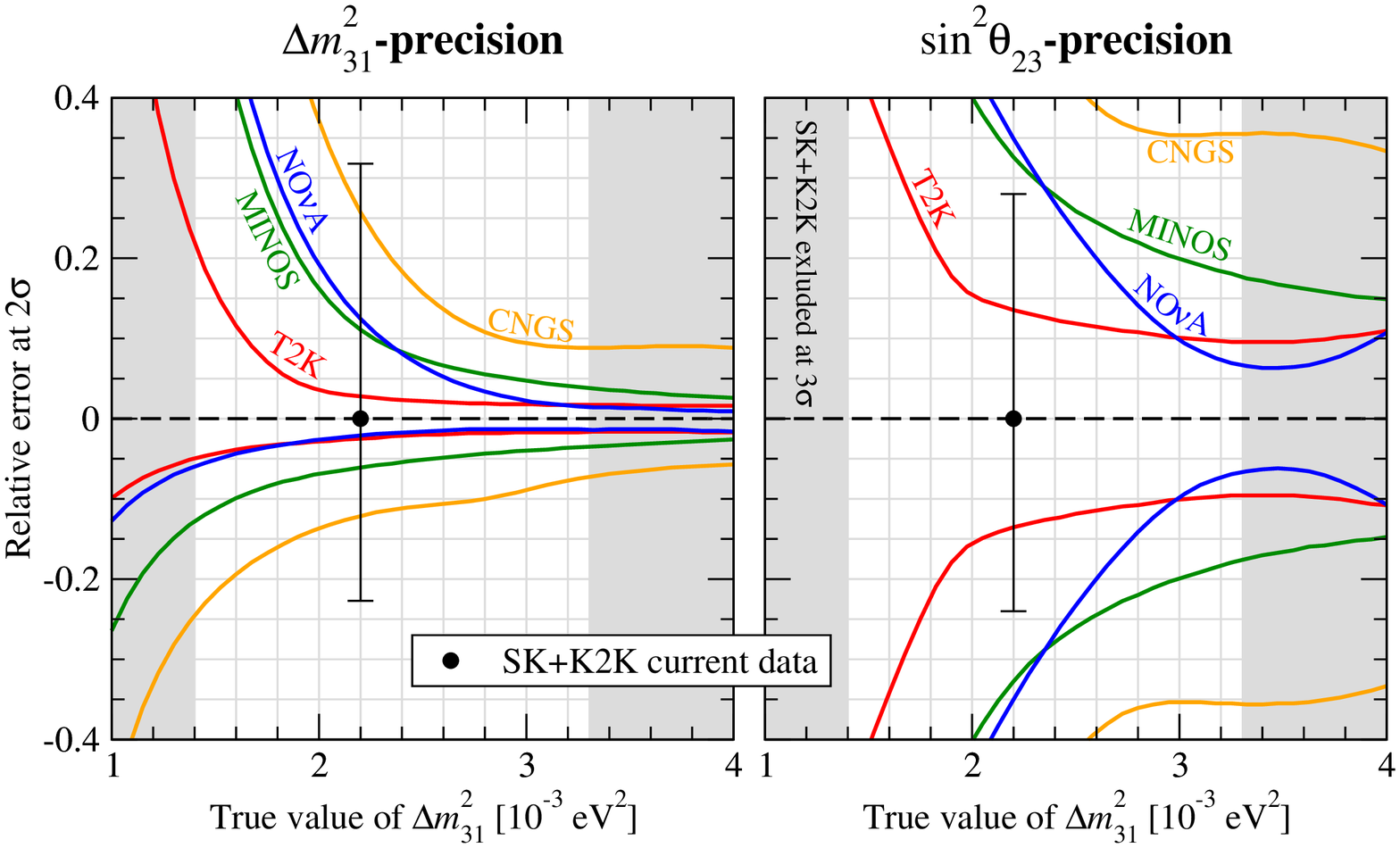}
    \includegraphics[width=0.6\textwidth]{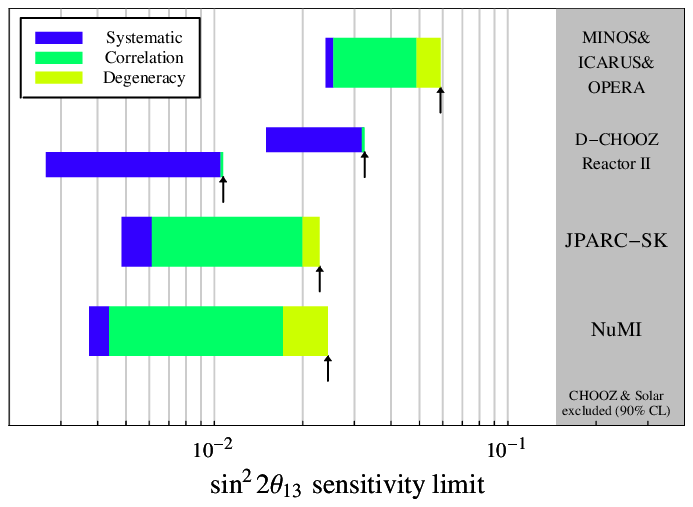}
    \caption{\label{fig:nfutteta13}%
      Expected sensitivity at near future oscillation experiments. 
      Precision of $\Delta m^2_{31}$ (left), and $\sin^2\theta_{23}$
      (center) as a function of the true value of $\Delta m^2_{31}$
      ($\theta_{23}^\text{true} = \pi/4$), and sensitivity to $\sin^2
      2\theta_{13}$ (right) for $\Delta m^2_{31} = 2\times
      10^{-3}~\eVq$. See Ref.~\cite{lindner10} for details.}
\end{figure}

As illustration of the expectations in these experiments we show in
Fig.~\ref{fig:nfutteta13} the results from Ref.~\cite{lindner10} on
the expected precision in the determination of $\Delta m^2_{31}$,
$\theta_{23}$ and their sensitivity to $\theta_{13}$. From these
results we read that the largest improvement is expected in the
determination of $|\Delta m^2_{32}|$ whose uncertainty can be reduced
to about a 10\% while only a mild improvement is expected in the
determination of $\theta_{23}$ because these experiments are mainly
sensitive only to $\sin^2 2 \theta_{23}$.  

The expected sensitivity to $\theta_{13}$ is given in the lower panel
of Fig.~\ref{fig:nfutteta13}. In this figure, from
Ref.~\cite{lindner10}, systematics refers to the usual experimental
and theoretical systematic errors such as errors in the overall signal
and background normalization as well as in their energy dependence, 
uncertainties in the reconstruction efficiencies etc\dots. Correlation
and degeneracy uncertainties refer to those which arise because, as
discussed above, the relevant probabilities are dominantly sensitive
to certain parameter combinations so the same event rates can be
predicted with for different sets of parameters\footnote{The
distinction between correlations and degeneracies is somehow arbitrary.
In Ref.~\cite{lindner10}, they label the uncertainty of degeneracy or
correlation depending on whether it does or it does not generate a
disjoint oscillation parameter region.}.  

As seen in the figure the limit on $\theta_{13}$ from superbeam
experiments is strongly affected by parameter correlations and
degeneracies, whereas reactor experiments provide a ``clean''
measurement of $\theta_{13}$ whose precision is dominated by
statistics and systematics (see Ref.~\cite{lindner10} for details). 

The results from Ref.~\cite{lindner10} show also that a non zero value
of $\theta_{13}$ close to their present bound would be established in 
any of these experiments. However none of them on their own can give
any information on the CP-phase $\delta_\text{CP}$ and on the mass
hierarchy because of the correlations. Some information on these
parameters can only be obtained if $\theta_{13}$ is close to their
present bound and the results from several of the experiments are
combined.


\subsection{Far Future LBL Experiments}
\label{sec:futlblfar}

In the longer term future one can think of proton driver upgrades of
existing proton sources. For example if the proton sources at either
Fermilab or J-PARC or both were upgraded to 2 or even 4 MW, the mass
ordering could be determined in a larger region of parameter space.
Furthermore a larger detector as well as running both in the neutrino
and the antineutrino mode are considered a requirement to be able to
detect CP violation.

At CERN a proposal to build a neutrino superbeam using the Super
Proton Linac (SPL) which would provide a 2.2 GeV proton beam of 4 MW
has been studied~\cite{nufactrev2}. This would be a low energy
conventional broad band $\nu_\mu$ beam aimed at a large water
Cherenkov detector located in Frejus, 130 km away. In principle at
this short distance matter effects are very small and they do not
affect the precision at which the experiment could observe CP
violation. On the other hand, for the same reason, the experiment
cannot yield any information on the mass ordering.

Another proposal for the search for oscillation in a higher energy 
broad band beam is the Brookhaven proposal~\cite{bnlnuprop}. With 
this beam aimed at a large water Cherenkov detector located at a
baseline of about 2500 km the experiment would be able to see matter
effects at the high energy part of the neutrino spectrum as well CP
violating effects at lower energies~\cite{bnlnuwg}. 

Ultimately the presence of backgrounds that fake $\nu_e$ CC
interactions, together with a small $\nu_e$ component in the initial
beam, make it difficult for experiments using conventional neutrino 
beams to probe very small oscillation amplitudes, below the
$0.01-0.001$ range. This limitation motivates new types of neutrino
facilities that provide $\nu_e$ beams, permitting the search for
$\nu_e \to \nu_\mu$ oscillations, and if the beam energy is above the
$\nu_\tau$ CC interaction threshold, the search for $\nu_e \to
\nu_\tau$ oscillations.  Neutrino Factory and Beta Beam facilities
both provide $\nu_e$ (and $\bar\nu_e$) beams, but with somewhat
different beam properties.

\subsubsection{$\beta$ Beams}

The idea of a Beta Beam facility was first proposed by P.~Zucchelli in
2002~\cite{betabeams}. As the name suggests, it employs beams of
beta-unstable nuclides. By accelerating these ions to high energy and
storing them in a decay ring a very pure beam of electron neutrinos
(or antineutrinos) can be produced. As the kinematics of the beta
decay is well understood, the energy distribution of the neutrinos can
be predicted to a very high accuracy. Furthermore, as the energy of
the beta decay is low compared with that for muon decay, the resulting
neutrino beam has a small divergence.

For low-$Z$ beta-unstable nuclides, typical decay times are measured
in seconds. Thus, there is not high a premium on rapid acceleration
and conventional (or even existing) accelerators could be used for
acceleration in a Beta Beam facility. Two ion species, both having
lifetimes on the order of 1 s, have been identified as optimal
candidates: \Nuc{6}{He} for producing antineutrinos and \Nuc{18}{Ne}
for neutrinos. Also the possibility of generating a monochromatic
neutrino beam using recently discovery isotopes that decay fast
through electron capture~\cite{pepebb} has been studied. Recently the
capabilities of a very long baseline Beta Beam experiment using
radioactive \Nuc{8}{B} and \Nuc{8}{Li} as the source isotopes for the
$\nu_e$ and $\bar\nu_e$ beta-beam has also been
explored~\cite{Agarwalla:2006vf}.

The original Beta Beam proposal~\cite{autin} was to use the CERN SPS
to accelerate the ions in combination with a large detector in the
Frejus tunnel in France which would yield mean neutrino and
antineutrino energies of 0.2~GeV and 0.3~GeV ($L=130$ km) 
respectively. However the expected signal rates are relatively modest.
In addition, it has been pointed out~\cite{jj-beta} that the neutrino
energies are comparable to the target nucleon kinetic energies due to
Fermi motion, and therefore there is no useful spectral information in
the low energy Beta Beam measurements. Hence, the useful information
is restricted to the measured muon neutrino (and antineutrino)
appearance rates. Hence higher energy scenarios are being
considered~\cite{jj-beta,jj-beta2}.  A medium energy beta beam
experiment could be obtained using the Fermilab Tevatron for
acceleration (yielding mean neutrino and antineutrino energies of
1.2~GeV and 1.9~GeV respectively), and a 1~megaton water Cherenkov
detector in the Soudan mine ($L=730$ km).  A high energy beta beam
experiment would require the CERN LHC to accelerate the ions (mean
neutrino and antineutrino energies of 5~GeV and 7.5~GeV respectively)
and a very long baseline, $L=3000$ km.

\begin{figure}\centering
    \includegraphics*[width=3.5in]{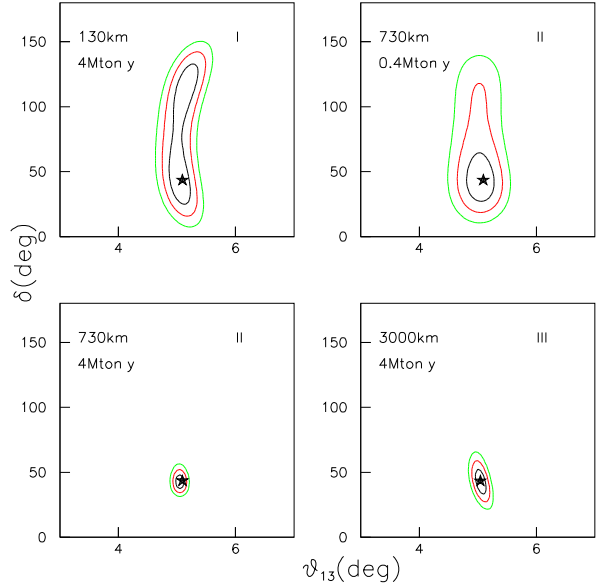}
    \caption{\label{fig:betabeam}%
      (Low-Energy (I), Medium-Energy (II), and High-Energy (III) Beta
      Beam sensitivities. The estimated $\,1\sigma, 2\,\sigma$ and
      $3\,\sigma$ contours are shown for the setups described in the
      text.  See Ref.~\cite{jj-beta}.}
\end{figure}

The sensitivity at these Beta Beam scenarios is illustrated in
Fig.~\ref{fig:betabeam} from Ref.~\cite{jj-beta} 

The figure shows, for the three scenarios, the $1\sigma$, $2\sigma$,
and $3\sigma$ contours in the ($\theta_{13},\, \delta_\text{CP}$)
plane. Note that the expected sensitivity for the medium energy case
with a ``small'' water Cherenkov detector is comparable to the low
energy case with the megaton water Cherenkov detector. However, the
medium energy sensitivity is dramatically improved with the much
bigger detector. The further improvement obtained by going to LHC
energies seems to be marginal. Given the likelihood that the LHC would
not be available as a Beta Beam accelerator for a very long time,
perhaps the most interesting scenario is the medium energy one.

Preliminary results on the capabilities of this intermediate energy
beta beam~\cite{jj-beta2} seems to indicate that the problem of 
discrete ambiguities due to degeneracies and their bias in the
determination of $\theta_{13}$ and $\delta_\text{CP}$ can be solved 
down to somewhat smaller values of $\theta_{13}$ ($\sin^2
2\theta_{13}\gtrsim \mathcal{O}(5 \times10^{-5})$) than with a
Superbeam.

\subsubsection{Neutrino Factories}

For a Neutrino Factory, the production beam is a high intensity proton
beam of moderate energy (beams of $2-50$ GeV have been considered by
various groups) that impinges on a target, typically a high-$Z$
material. The collisions between the proton beam and the target nuclei
produce a secondary pion beam that quickly decays into a longer-lived
muon beam. The remainder of the Neutrino Factory is used to condition
the muon beam, accelerate it rapidly to the desired final energy of a
few tens of GeV, and store it in a decay ring having a long straight
section oriented such that decay neutrinos produced there will hit a
detector located thousands of kilometers from the source.

At a Neutrino Factory in which, for example, positive muons are
stored, the initial beam consists of 50\% $\nu_e$ and 50\%
$\bar\nu_\mu$. The energy spectrum of both type of neutrinos at the
detector site is different for the two flavors and it is given in
terms of the very well known $\mu$ decay distribution. Therefore it
can be predicted with very high precision.

In the absence of oscillations, the $\nu_e$ CC interactions produce
electrons and the $\bar\nu_\mu$ CC interactions produce positive
muons.  Note that the charge of the final state lepton tags the flavor
of the initial neutrino or antineutrino.  In the presence of $\nu_e
\to \nu_\mu$ oscillations, the $\nu_\mu$ CC interactions produce
negative muons (\textit{i.e.}, wrong-sign muons).  This is a very
clean experimental signature since, with a segmented magnetized
iron-scintillator sampling calorimeter for example, it is
straightforward to suppress backgrounds to 1 part in $10^4$ of the
total CC interaction rate, or better.  This means that at a Neutrino
Factory backgrounds to the $\nu_e \to \nu_\mu$ oscillation signal are
extremely small. The full statistical sensitivity can therefore be
exploited down to values of $\sin^2 2\theta_{13}$ approaching
$10^{-4}$ before backgrounds must be subtracted and further advances
in sensitivity scale like $\sqrt{N}$ rather than $N$.  This enables
Neutrino Factories to go beyond the sensitivities achievable by
conventional neutrino Superbeams and $\beta$ beams, by about two
orders of magnitude.

\begin{figure}\centering
    \includegraphics*[width=11cm]{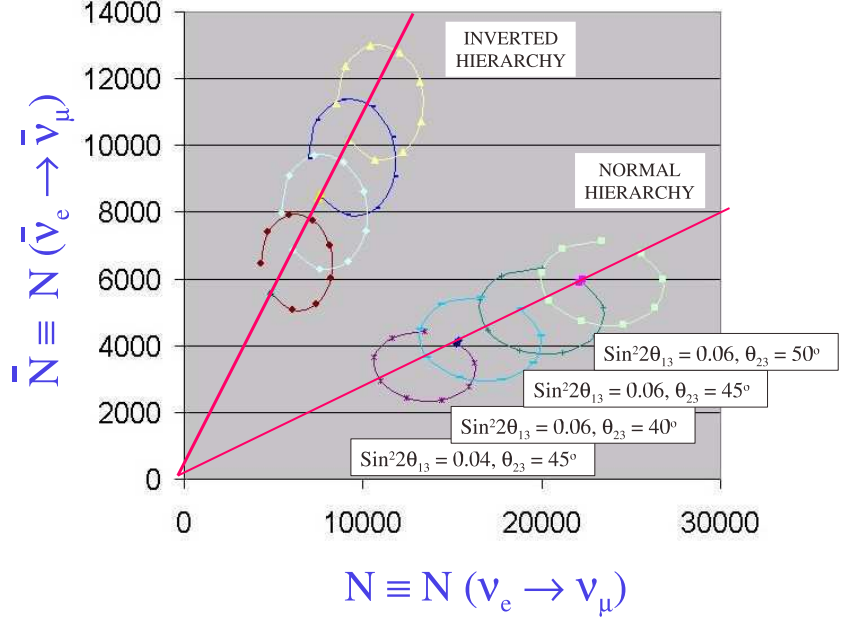}
    \caption{\label{fig:nufactfig}%
      The predicted number of wrong-sign muon events when negative
      muons are stored in the Neutrino Factory, versus the
      corresponding rate when positive muons are stored, shown as a
      function of $\theta_{13}$, $\theta_{23}$, $\delta_\text{CP}$ and
      the assumed mass hierarchy, as labeled. The calculation
      corresponds to a 16~GeV Neutrino Factory with a baseline of
      2000~km, and 10~years of data taking with a 100~kton detector
      and $2 \times 10^{20} \, \mu^+$ and $2 \times 10^{20} \, \mu^-$
      decays in the beam-forming straight section per year. The
      ellipses show how the predicted rates vary as the CP phase
      $\delta_\text{CP}$ varies. See Ref.~\cite{nufbbreport} for
      details.}
\end{figure}

In practice, to measure $\theta_{13}$, determine the mass hierarchy,
and search for CP violation, the analysis of the wrong-sign muon rates
must be performed allowing all of the oscillation parameters to
simultaneously vary within their uncertainties which leads us again to
the possible problem of parameter degeneracies. As an illustration we
show in Fig.~\ref{fig:nufactfig} (from Ref.~\cite{nufbbreport}), as a
function of $\theta_{13}$, $\theta_{23}$, $\delta_\text{CP}$ and the
assumed mass ordering, the predicted number of wrong-sign muon events
when negative muons are stored in the Neutrino Factory, versus the
corresponding rate when positive muons are stored.
Given the large statistics expected the statistical errors would be 
barely visible if plotted. Thus for the parameter region illustrated
by the figure, determining the mass hierarchy (which diagonal line is
the measured point closest to) will be straightforward.  But
determining the exact values for the mixing angles and 
$\delta_\text{CP}$ is more complicated because of the degeneracies. 

To eliminate the false solutions, event samples other than $\nu_e \to
\nu_\mu$ transitions tagged by wrong-sign muons will be important. For
example $\bar\nu_\mu \to \bar\nu_e$ oscillations produce wrong-sign
electrons, $\bar\nu_\mu \to \bar\nu_\tau$ oscillations produce events
tagged by a $\tau^+,$ and $\nu_e \to \nu_\tau$ oscillations produce
events tagged by a $\tau^-$.  Hence, there is a variety of information
that can be used to measure or constrain neutrino oscillations at a
Neutrino Factory, namely the rates and energy distributions of all
these different event types.  If these measurements are made when
there are alternately positive and negative muons decaying in the
storage ring, there are a total of 12~spectra that can be used to
extract information about the oscillations.

The overall conclusion of the different
simulations~\cite{golden,silver,silver2,lindspec,magicbl1,bueno1} (see
also Ref.~\cite{nufactrev,nufactrev2,nufactrev3} for detailed reviews)
is that at a Neutrino Factory ${\sin^2 2\theta_{13}}$ can be measured,
the neutrino mass hierarchy determined, and a search for CP violation
in the lepton sector made for all values of $\sin^2 2\theta_{13}$ down
to \textit{O}$(10^{-4}$), or even a little less.


\subsection{Future Atmospheric Neutrino Experiments}
\label{sec:futatm}

As seen in previous sections, large next generation underground water 
Cherenkov detectors are proposed in US, in Japan and in
Europe~\cite{uno1,uno2,T2K,hyperk}. These large megaton class
detectors are proposed as multi-purpose detectors that probe physics
beyond the sensitivities of the highly successful Super-Kamiokande
detector utilizing a well- tested technology.  The physics goals of
these detectors include: nucleon decay searches, observation of
neutrinos from supernova explosions, observation of supernova relic
neutrinos, and precision measurements of neutrino oscillation
parameters using atmospheric, solar and accelerator produced
neutrinos. 

There are also plans to build a 30-50 kton magnetized tracking iron
calorimeter detector in India within the India-based Neutrino
Observatory (INO) project~\cite{INO} with the primary goal of studying
the oscillations of atmospheric $\nu_\mu$ and $\bar\nu_\mu$. This
detector is planned to have efficient muon charge identification, high
muon energy resolution ($\sim 5\% $) and muon energy threshold of
about 2 GeV.  Although there are no details studies of the
capabilities of such detector, its sensitivity should be comparable to
that evaluated for an earlier proposal for a similar detector, 
MONOLITH~\cite{monolith}).  

It is also worth noticing that presently running experiments whose
main goal is not the study of atmospheric neutrinos will also have
some reasonable samples of atmospheric events. For example SNO despite
its relatively small size, thanks to its depth and flat overburden,
can measure the atmospheric neutrino flux using through-going muons
even above the detector horizon (up to $\cos \eta = 0.4$) without any
contamination from cosmic ray muons.  These data above the horizon can
tell us the unoscillated flux of neutrinos and therefore reduce the
uncertainty of atmospheric neutrino flux models.  In addition, the
charged current interactions of antineutrinos within SNO should
produce additional neutrons compared to neutrinos, and it may be
possible to make a crude measurement of the relative rates of
neutrinos versus antineutrinos.  Also MINOS~\cite{MINOS} is an iron
magnetized calorimeter and thus, has muon charge identification
capabilities for multi-GeV atmospheric muon events. In 2007 they
reported a total of 140 neutrino-induced muon events and an observed
charge ratio in agreement with expectations~\cite{MINOSatm}.  After 5
years of data-taking MINOS is expected to collect about 440
atmospheric $\nu_\mu$ and about 260 atmospheric $\bar\nu_\mu$
multi-GeV contained events.

\begin{figure}\centering
    \includegraphics[height=2.7in]{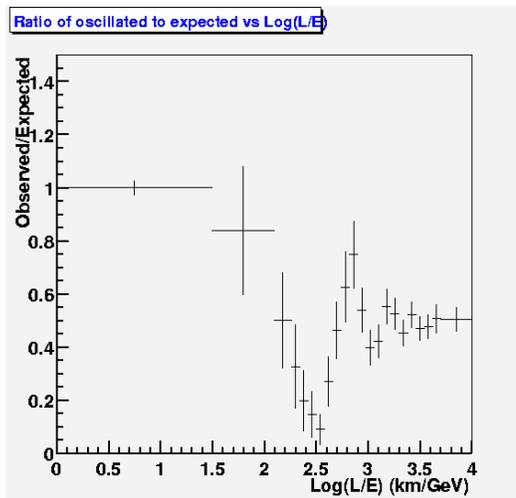}
    \caption{\label{fig:loeuno}%
      Simulated oscillation pattern observable by
      UNO~\cite{uno1,uno2}.}
\end{figure}

The wide dynamic range of neutrino energies and baselines in the
atmospheric sector mean that these future atmospheric neutrino
experiments will provide their own tests of the oscillation model with
the corresponding precision measurement of the dominant oscillation
parameters $\Delta m^2_{32}$ and $\theta_{23}$. As an illustration of
their precision we show in Fig.~\ref{fig:loeuno} the oscillation
pattern which could be observed by the UNO detector using atmospheric
neutrino events.

Furthermore, as discussed in Sec.~\ref{sec:3nuprobs} atmospheric
neutrinos are in principle sensitive to $\theta_{13}$ and the neutrino
mass hierarchy due to Earth matter effects in the $e$-like events. In
addition effects from the solar parameters $\theta_{12}$ and $\Delta
m_{21}^2$ on $e$-like events in the sub-GeV energy range provide
sensitivity to the octant of $\theta_{23}$ and principle even on
$\delta_\text{CP}$.  In general, the precision expected for most of
these effects from atmospheric data on its own is smaller than the one
achievable at long baseline experiments, in particular in the
determination of $\theta_{13}$, $\delta_\text{CP}$, $|\Delta
m_{31}^2|$, or $\sin^22\theta_{23}$. One reason is that atmospheric
neutrino experiments are limited by systematical uncertainties, such
as the theoretical uncertainties on the atmospheric neutrino fluxes
and the interaction cross sections.  A recent analysis of the
capabilities of a future large water Cherenkov detector to determine
these subdominant effects can be found in Ref.~\cite{kajita}. 

\begin{figure}\centering
    \includegraphics[width=0.95\textwidth]{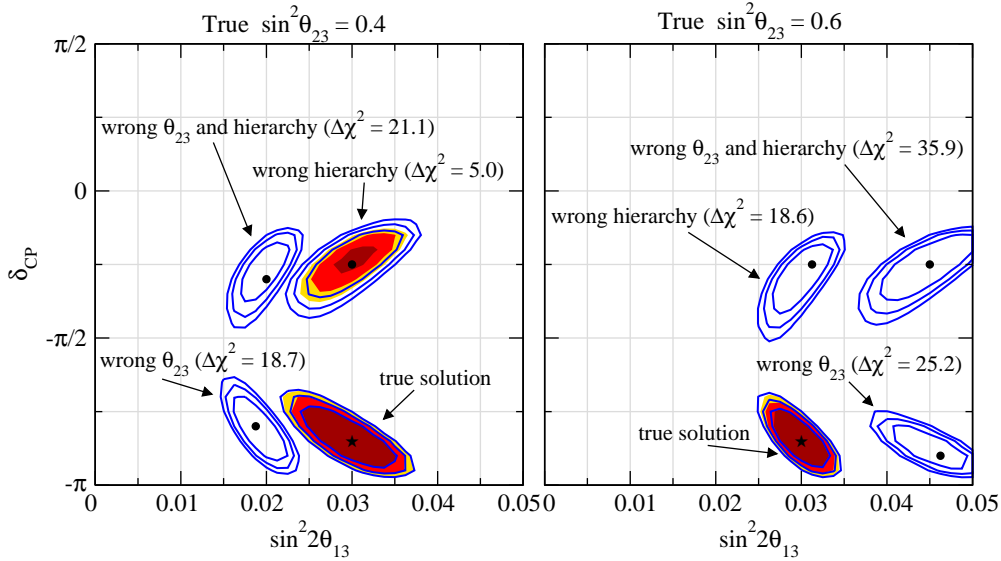}
    \caption{\label{fig:lbl+atm}%
      Allowed regions in the $(\sin^22\theta_{13},\,
      \delta_\text{CP})$ plane at $2\sigma$, 99\%, and $3\sigma$ CL
      (2~d.o.f.) of the true and all degenerate solutions for
      $\sin^22\theta_{13}^\text{true} = 0.03$, $\delta^\text{true} =
      -0.85\pi$, and $\sin^2\theta_{23}^\text{true} = 0.4$ (left) and
      $\sin^2\theta_{23}^\text{true} = 0.6$ (right). The solid curves
      correspond to LBL data only, and the shaded regions correspond
      to LBL+ATM data. The true best fit point is marked with a star,
      the best fit points of the degenerate solutions are marked with
      dots, and the corresponding $\Delta\chi^2$-values of LBL+ATM
      data are given in the figure. The true mass ordering is the
      normal hierarchy.}
\end{figure}

However, due to these effects atmospheric neutrino data can have an
important role in breaking the parameter degeneracies (see
Sec.~\ref{sec:futlblmot}) which appear in the analysis of long
baseline experiments~\cite{atm+lbl,Campagne:2006yx}, in particular the
degeneracies associated to the octant of $\theta_{23}$ and the
sign($\Delta m^2_{31}$).  As an illustration of their potential we
show in Fig.~\ref{fig:lbl+atm} the allowed regions from the analysis
of the phase~II of the T2K experiment assuming a 4~MW superbeam
produced at the J-PARC accelerator, and the 1~Mt
Hyper-Kamiokande~\cite{hyperk} detector serving as the far detector for
the LBL experiment as well as providing the high statistics
atmospheric neutrino data.  The figure shows the allowed regions in
the $(\sin^22\theta_{13},\, \delta_\text{CP})$ plane for an
example-point with the true values $\sin^22\theta_{13} = 0.03$,
$\delta_\text{CP} = -0.85\pi$, and non-maximal values of
$\theta_{23}^\text{true}$. Apart from the true solution, three
degenerate regions are present, corresponding to the wrong octant of
$\theta_{23}$, the wrong sign of $\Delta m_{31}^2$, and the wrong
octant as well as the wrong hierarchy.

This four-fold degeneracy can be lifted to large extent if LBL data is
combined with data from atmospheric neutrinos. As seen in the figure
the degenerate solutions corresponding to the wrong octant of
$\theta_{23}$ are highly disfavored by the inclusion of ATM data.
Furthermore, also the solution with the wrong mass ordering gets
disfavored in the combined analysis, although in this case the ability
to resolve this degeneracy is more subtle. This is due to the fact
that in a water-Cherenkov detector the distinction between neutrino
and antineutrino events is not possible (on an event-by-event basis).
This makes more difficult the determination of the mass ordering which
is only feasible due to the difference of fluxes and cross sections
for neutrinos and antineutrinos.  The ability to resolve the mass
ordering is expected to be much better if the atmospheric neutrino
detector is a magnetized calorimeter which has the capability of
discriminating the charge of the neutrino-induced
muon~\cite{PalPet,monolith}.

%% file: sec.numass.tex
\section{Direct Determination of $m_\nu$}
\label{sec:numass}

Oscillation experiments have provided us with important information on
the differences between the neutrino masses-squared, $\Delta
m^2_{ij}$, and on the leptonic mixing angles, $U_{ij}$. But they are
insensitive to the absolute mass scale for the neutrinos, $m_i$. 

Of course, the results of an oscillation experiment do provide a lower
bound on the heavier mass in $\Delta m^2_{ij}$, $|m_i|\geq\sqrt{\Delta
m^2_{ij}}$ for $\Delta m^2_{ij}>0$. But there is no upper bound on
this mass. In particular, the corresponding neutrinos could be
approximately degenerate at a mass scale that is much higher than
$\sqrt{\Delta m^2_{ij}}$.  Moreover, there is neither upper nor lower
bound on the lighter mass $m_j$. In this section we briefly summarize
the most sensitive probes of the absolute mass scale for the
neutrinos.


\subsection{Kinematic Constraints from Weak Decays}

It was Fermi who first proposed a kinematic search for the neutrino
mass from the hard part of the beta spectra in \Nuc{3}{H} beta decay
$\Nuc{3}{H} \to \Nuc{3}{He} + e^- + \bar\nu_e$. In the absence of
leptonic mixing this search provides a measurement of the electron
neutrino mass.

\Nuc{3}{H} beta decay is a superallowed transition, which means that 
the nuclear matrix elements do not generate any energy dependence, so
that the electron spectrum is given by the phase space alone
\begin{equation}
    \label{eq:betadec}
    \frac{dN}{dE} = C \, p \, E \, (Q-T)
    \sqrt{(Q-T)^{2} - m_{\nu_e}^{2}} F(E) 
    \equiv R(E) \sqrt{(E_0 - E)^2 - m_{\nu_e}^2} \,.
\end{equation}
where $E = T + m_e$ is the total electron energy, $p$ its momentum, 
$Q\equiv E_0-m_e$ is the maximum kinetic energy of the electron and
$F(E)$ is the Fermi function which incorporates final state Coulomb
interactions. In the second equality we have included in a $R(E)$ all
the $m_\nu$-independent factors.

Plotted in terms of the Curie function $K(T) \equiv
\sqrt{\frac{dN}{dE} \, \frac{1}{pEF(E)}}$ a non-vanishing neutrino
mass $m_\nu$ provokes a distortion from the straight-line T-dependence
at the end point: for $m_\nu = 0 \Rightarrow T_\text{max} = Q$ whereas
for $m_{\nu_e} \neq 0 \Rightarrow T_\text{max} = Q - m_{\nu_e}$.
\Nuc{3}{H} beta decay has a a very small energy release $Q = 18.6$~KeV
which makes it particularly sensitive to this kinematic effect.

At present the most precise determination from the
Mainz~\cite{tritmainz} and Troitsk~\cite{trittroitsk} experiments 
give no indication in favor of $m_{\nu_e}\neq 0$ and one sets an
upper limit 
\begin{equation}
    \label{eq:nuelim}
    m_{\nu_e}<2.2~\text{eV}
\end{equation}
at 95\% confidence level (CL). For the other flavors the present
limits are~\cite{pdg}
\begin{align}
    m_{\nu_\mu} &< 190~\text{keV (90\% CL)}
    & \text{from}\qquad & \pi^- \to \mu^- + \bar\nu_\mu \,,
    \\
    m_{\nu_\tau} &< 18.2~\text{MeV (95\% CL)}
    & \text{from}\qquad & \tau^- \to n\pi + \nu_\tau \,.
\end{align}

In the presence of mixing these limits have to be modified and in
general they involve more than one flavor parameter. For neutrinos
with small mass differences the distortion of the beta spectrum is
given by the weighted sum of the individual spectra~\cite{betamix}: 
\begin{equation}
    \label{eq:spectrum2}
    \frac{dN}{dE}
 = R(E) \sum_i|U_{ei}|^2
    \sqrt{(E_0 - E)^2 - m_i^2} \, \Theta(E_0 - E - m_i) \,.
\end{equation}
The step function, $\Theta(E_0 - E - m_i)$, reflects the fact that a
given neutrino can only be produced if the available energy is larger
than its mass. According to Eq.~\eqref{eq:spectrum2}, there are two
important effects, sensitive to the neutrino masses and mixings, on
the electron energy spectrum: (i) Kinks at the electron energies
$E_e^{(i)} = E \sim E_0 - m_i$ with sizes that are determined by
$|U_{ei}|^2$; (ii) A shift of the end point to $E_\text{ep} = E_0 -
m_1$, where $m_1$ is the lightest neutrino mass. The situation is
slightly more involved when the finite energy resolution of the
experiment is considered~\cite{vissanitrit,smirtrit}. 

In general for most realistic situations the distortion of the
spectrum can be effectively described by a single parameter, $m_\beta$
if for all neutrino states $E_0-E=Q_0-T\gg m_i$. In this case one can
expand Eq.~\eqref{eq:spectrum2} as:
\begin{equation} \begin{aligned}
    \frac{dN}{dE}
    &\simeq R(E) \sum_i |U_{ei}|^2 (E_0 - E) \left(
    1 - \frac{m_i^2}{2 (E_0 - E)} \right)
    \\[2mm]
    &= R(E) \, \sum_i |U_{ei}|^2 \,(E_0-E) \left(
    1 - \frac{1}{2 (E_0 - E)}
    \frac{\sum_i |U_{ei}|^2 m^2_i}{\sum_i |U_{ei}|^2} \right)
    \\[2mm]
    &\simeq R(E) \sum_i |U_{ei}|^2 
    \sqrt{(E_0 - E)^2 - m_\beta^2} \,,
\end{aligned} \end{equation}
with
\begin{equation}
    \label{eq:tritium}
    m^2_\beta = \frac{\sum_i m^2_i |U_{ei}|^2}{\sum_i |U_{ei}|^2}
    = \sum_i m^2_i |U_{ei}|^2 \,,
\end{equation}
where the second equality holds if unitarity is assumed.  So the
distortion of the end point of the spectrum is described by a single
parameter which is bounded to be
\begin{equation}
    m_\beta = \sqrt{\sum_i m^2_i |U_{ei}|^2} < 2.2~\text{eV} \,.
\end{equation}
A new experimental project, KATRIN~\cite{katrin}, is under
construction with an estimated sensitivity limit: $m_\beta \sim 0.3$
eV.


\subsection{Neutrinoless Double Beta Decay}

\begin{figure}\centering
    \includegraphics[width=60mm]{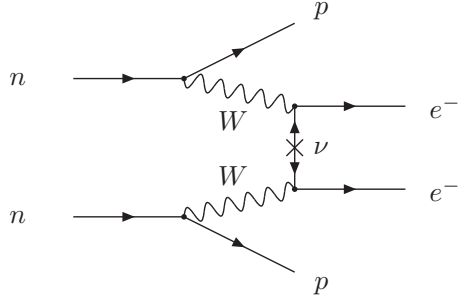}
    \caption{\label{fig:diagram_0nbb}%
      Feynman diagram for neutrinoless double-beta decay.}
\end{figure}

Direct information on neutrino masses can also be obtained from 
neutrinoless double beta decay ($0\nu\beta\beta$) searches: 
\begin{equation}
    \label{eq:bb}
    (A,Z) \to (A,Z+2) + e^- + e^-.
\end{equation}
Schematically in the presence of neutrino masses and mixing the
process in Eq.~\eqref{eq:bb} can be induced by the diagram shown in
Fig.~\ref{fig:diagram_0nbb}. The amplitude of this process is
proportional to the product of the two leptonic currents
\begin{equation}
    {M}_{\alpha\beta}
    \propto \left[ \bar{e} \gamma_\alpha (1-\gamma_5) \nu_e \right]
    \, \left[ \bar{e} \gamma_\beta (1-\gamma_5) \nu_e \right] \,.
\end{equation}
which can only lead to a neutrino propagator from the contraction 
$\langle 0 \,|\, \nu_e(x) \nu_e(y)^T \,|\, 0 \rangle$.  If the
neutrino is a Dirac particle $\nu_e$ field annihilates a neutrino
states and creates an antineutrino state which are different.
Therefore the contraction $\langle 0 \,|\, \nu_e(x) \nu_e(y)^T \,|\, 0
\rangle = 0$ and $M_{\alpha\beta} = 0$. On the contrary, if $\nu_e$ is
a Majorana particle, neutrino and antineutrino are the same state and
$\langle 0 \,|\, \nu_e(x) \nu_e(y)^T \,|\, 0 \rangle \ne 0$.

Thus in order to induce the $0\nu\beta\beta$ decay, $\nu_e$ must be a
Majorana particle. This is also obvious as the process~\eqref{eq:bb}
violates L by two units.  The opposite also holds, if $0\nu\beta\beta$
decay is observed, neutrinos must be massive Majorana
particles~\cite{blackbox}.

However, Majorana neutrino masses are not the only mechanism which can
induce neutrinoless double beta decay.  In general, in models beyond
the standard model there may be other sources of total lepton number
violation which can induce $0\nu\beta\beta$ decay.  Consequently the
observation or limitation of the neutrinoless double beta decay
reaction rate can only be related to a bound on the neutrino mass and
mixing under some assumption about the source of total lepton number
violation in the model. 

For the case in which the only effective lepton number violation at
low energies is induced by the Majorana mass term for the neutrinos, 
the rate of $0\nu\beta\beta$ decay is proportional to the
\emph{effective Majorana mass of $\nu_e$},
\begin{equation}
    m_{ee} = \left| \sum_i m_i U_{ei}^2 \right|
\end{equation}
which, in addition to the masses and mixing parameters that affect the
tritium beta decay spectrum, depends also on the leptonic CP violating
phases.
 
Experimentally, what it is measured is the half-life of the decay.  In
$0\nu\beta\beta$ decay, the experimental signal is two electrons in
the final state, whose energies add up to the Q-value of the nuclear
transition while for the double beta decay with neutrinos 
($2\nu\beta\beta$) (which constitute an intrinsic background) the
energy spectrum of both electrons will be continuous as part of the Q
is carried by the outgoing neutrinos. Also the decay rates for
$0\nu\beta\beta$ and $2\nu\beta\beta$ have very different dependence
on the available Q being the dependence much weaker for
$0\nu\beta\beta$.  For this reason, the sensitivity is better for
isotopes with a high Q-value.

In the case that the only source of lepton number violation at low
energies is induced by the Majorana neutrino mass, the decay half-life
is given by:
\begin{equation}
    \label{eq:0bbrate}
    (T^{0\nu}_{1/2})^{-1}
    = G^{0\nu} \left| M^{0\nu} \right|^2 \left(
    \frac{m_{ee}}{m_e} \right)^2
\end{equation}
where $ G^{0\nu}$ is the phase space integral and $| M^{0\nu} \mid$ is
the nuclear matrix element of the transition. 

The strongest bound from $0\nu\beta\beta$ decay was imposed by the 
Heidelberg-Moscow group~\cite{hmlimit} which used 11 kg of enriched
Ge. After 53.9 kg~yr of data taking they found no signal which
allowed them to set a bound on the half-life of $T^{0\nu}_{1/2} > 1.9
\times 10^{25}$~yr (90\% CL).  This implies (for a given prediction
of the nuclear matrix element):
\begin{equation}
    m_{ee} < 0.26~(0.34)~\text{eV}
    \quad \text{at 68\% (90\%) CL.}
\end{equation}
Taking into account the possible uncertainties in the prediction of 
the nuclear matrix elements, the bound may be weaken by a factor of
about 3~\cite{bbteoreview}.

Despite the result of the Heidelberg-Moscow experiment is that no
positive signal was observed, a subgroup of the collaboration found 
a small peak at some value of $Q$~\cite{klapeak1,klapeak2} which, if
real, would imply a non vanishing range for the effective Majorana
mass between $0.2$--$0.6$ eV (the range might be widened by nuclear
matrix uncertainties).  However, the statistical analysis used, as
well as the assumed background subtraction in order to establish
this evidence at the claimed CL has been subject of severe
criticisms~\cite{ferstru,aalseth,harney,zde02} which renders the
claimed signal as controversial.  

Currently only two large scale experiments are running. CUORICINO at
the Gran Sasso Underground Laboratory in Italy which uses uses
bolometers running at very low temperature and searches for
\Nuc{120}{Te} decay with a Q-value of 2530 keV.  The obtained
half-life limit~\cite{CUORICINO} $T^{0\nu}_{1/2}(\Nuc{120}{Te}) > 2.2
\times 10^{24}$~yr (90\% CL) implies and upper bound on the effective
Majorana neutrino mass $m_{ee} < 0.2$--$1.1$ eV. The second
experiment, NEMO-3~\cite{NEMO3} in the Frejus Underground Laboratory,
is built in form of time projection chambers where the double beta
emitter is either the filling gas of the chamber or is included in
thin foils.  It has obtained a half-life bound for \Nuc{100}{Mo}
$T^{0\nu}_{1/2}(\Nuc{100}{Mo}) > 5.6 \times 10^{23}$~yr (90\% CL)
which result in an upper effective Majorana mass bound $m_{ee} <
0.6$--$2$ eV. 

A series of new experiments is planned with sensitivity of up to
$m_{ee} \sim 0.01$ eV. For a review of the proposed experimental 
techniques see Ref.~\cite{bbexpreview,zuber}. 


\subsection{Cosmological Bounds}

Neutrinos, like any other particles, contribute to the total energy
density of the Universe. Furthermore light neutrinos are relativistic
through most of the evolution of the Universe. As a consequence they
can play a relevant role in large scale structure formation and leave
clear signatures in several cosmological observables.  For an
excellent recent review on the cosmological effects of neutrino masses
we refer the reader to the report of J.~Legourges ad
S.~Pastor~\cite{pastor}.

The main effect of neutrinos in cosmology is to suppress the growth of
fluctuations on scales below the horizon when they become non
relativistic. As a consequence a massive neutrino of a fraction of eV 
would produce a significant suppression in the clustering on small
cosmological scales. Because of this effect it is possible to infer 
constraints, although indirect, on the neutrino masses by comparing 
the most recent cosmological data with the theoretical predictions. 

The relevant quantity in these studies is the total neutrino energy 
density in our Universe, $\Omega_\nu h^2$ (where $h$ is the Hubble 
constant normalized to $H_0 = 100 ~\text{km} ~\text{s}^{-1}
~\text{Mpc}^{-1}$). At present $\Omega_\nu h^2$ is related to the
total mass in the form of neutrinos 
\begin{equation} 
    \Omega_{\nu}h^2 = \sum_i m_i / (94 \text{eV}) \,.
\end{equation}
Therefore cosmological data mostly gives information on the sum of the
neutrino masses $M_\nu = \sum_i m_{\nu_i}$ and has very little to say
on their mixing structure.

The recent precise astrophysical and cosmological observations have
started to provide us with indirect upper limits on absolute neutrino
masses which are competitive with those from laboratory experiments.
The most relevant data come from the Large Scale Structures (LSS) as
obtained from large redshift surveys of galaxies by the 2 degree Field
survey~\cite{2dF} and the Sloan Digital Sky Survey~\cite{SDSS} and
from Cosmic Microwave Background (CMB) anisotropies which at present
are most precisely determined by the WMAP experiment~\cite{WMAP}. 
Additional information can also be extracted from the so-called
Lyman-$\alpha$ forest~\cite{Lyman1,Lyman2}. This corresponds to the
Lyman-$\alpha$ absorption of photons traveling from distant quasars
($z\sim 2-3$) by the neutral hydrogen in the intergalactic medium. 

\begin{figure}\centering
    \includegraphics[width=4.in]{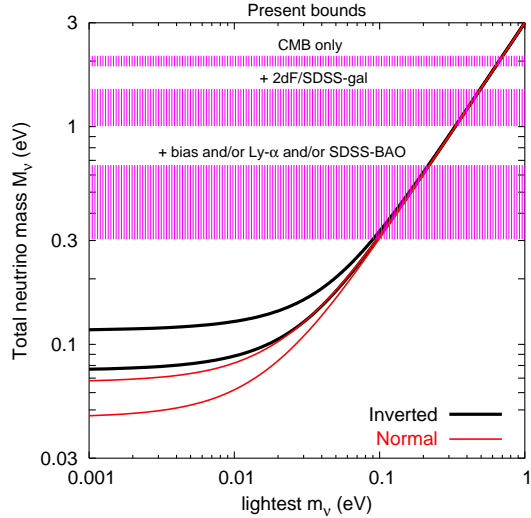}
    \caption{\label{fig:current}%
      Current upper bounds (95\% CL) from cosmological data on the sum
      of neutrino masses, compared to the values in agreement at a
      $3\sigma$ level with neutrino oscillation data. From
      Ref.~\cite{pastor}}
\end{figure}

As discussed in detail in Ref.~\cite{pastor} a single cosmological
bound on neutrino masses does not exist. The bound depends on the data
included in analysis as well as on details on the biases assumed and
on the statistical treatment.  For illustration we show in 
Fig.~\ref{fig:current} the graphical summary given in
Ref.~\cite{pastor}.  In this figure the cosmological bounds obtained
by different groups using a given set of data correspond to the
horizontal bands.  The thickness of these bands roughly describe the
spread of values obtained from similar cosmological data: $2-3$ eV for
CMB only, $0.9-1.7$ eV for CMB and 2dF/SDSS-gal or $0.3-0.9$ eV with
the inclusion of a measurement of the bias and/or Lyman-$\alpha$
forest data and/or the SDSS measurement of the baryon oscillation
peak. 
One can see from Fig.~\ref{fig:current} that current cosmological data
is only sensitive degenerate neutrino spectra with three neutrino 
states of the same mass, $M_\nu/3$. Using only the combined results of
CMB and galaxy clustering data from 2dF and/or SDSS a conservative
upper bound of $\sim$ 1 eV can be imposed.  The addition of more data
leads to an improvement of the bounds, which reach the lowest values
when data from Lyman-$\alpha$ and/or the SDSS measurement of the
baryon oscillation peak are included or the bias is fixed.  Also,
there are analysis in the literature for which the cosmological bounds
are stronger than the 0.3 eV lowest value of the ranges quoted in the 
figure, such as in Ref.~\cite{othercosmo}. They make use of the data
of LSS derived from the Lyman-$\alpha$ forest data and there is no
consensus in the literature whether these can be used in the presence
of massive neutrinos.  

%% file: sec.lsnd.tex
\section{Extended Models for LSND}
\label{sec:lsnd}

For years the most troublesome piece of experimental evidence in
neutrino physics has been the result of the LSND experiment, which
observed a small appearance of electron anti-neutrinos in a muon
anti-neutrino beam at a value of $L/E$. The main reason for this is
that the mass-squared differences required to explain the
solar+KamLAND, atmospheric+LBL and LSND experimental results in terms
of neutrino oscillations differ from one another by various orders of
magnitude. Consequently, there is no consistent way to explain all
these three signals invoking only oscillations among the three known
neutrinos. The argument for this statement is very simple. With three
neutrinos, there are only two independent mass-squared differences,
since by definition:
\begin{equation}
    \Delta m^2_{31} = \Delta m^2_{21} + \Delta m^2_{32} \,.
\end{equation}
This relation cannot be satisfied by three $\Delta m^2_{ij}$ that are
of different orders of magnitude. 
Therefore, in order to explain the LSND anomaly one had to invoke an
extension of the three-neutrino mixing scenario, introducing either a
mechanism to generate a third mass-square difference or a new form of
flavor transition beyond oscillations.

After publication of the MiniBooNE result~\cite{miniboonelast}, which
excluded the LSND signal at 98\% CL in the context of two-neutrino
$\nu_\mu \to \nu_e$ oscillations, some of these models have lost their
main motivation, whereas other can be seen as ways to accommodate both
results. In either case they all represent viable extensions of the
minimal three-neutrino scenario, and in this section we summarize
their present phenomenological status.


\subsection{Four-Neutrino Mixing}
\label{sec:fourmix}

One of the simplest extensions that one can think of to generate a
third $\Delta m^2$ is to add a fourth neutrino to the SM. As discussed
in Sec.~\ref{sec:standard}, the measurement of the decay width of the
$Z^0$ boson into neutrinos makes the existence of three, and only
three, light (that is, $m_\nu \lesssim m_Z/2$) active neutrinos an
experimental fact.  Therefore, the fourth neutrino must not couple to
the standard electroweak current, that is, it must be sterile.

One of the most important issues in the context of four-neutrino
scenarios is the four-neutrino mass spectrum.  In this section we
review the status of four neutrino models in which the third mass
difference is much larger than the ones required to explain the
solar+KamLAND and the atmospheric+LBL results.  The original
motivation for these models was that the third mass difference was the
one required to explain the LSND result. However, these scenarios
cannot explain \emph{both} a positive LSND result and the negative
result of the search at MiniBooNE, so they are no longer a viable
solution of the LSND puzzle.  On the other hand, the existence of
sterile neutrinos is an interesting subject by itself, irrespective of
LSND, and four-neutrino models are the simplest extension of the
Standard Model to include them. Therefore, in the following we will
study the constraints on these models from the existing data.

\begin{figure}\centering
    \includegraphics[width=0.9\textwidth]{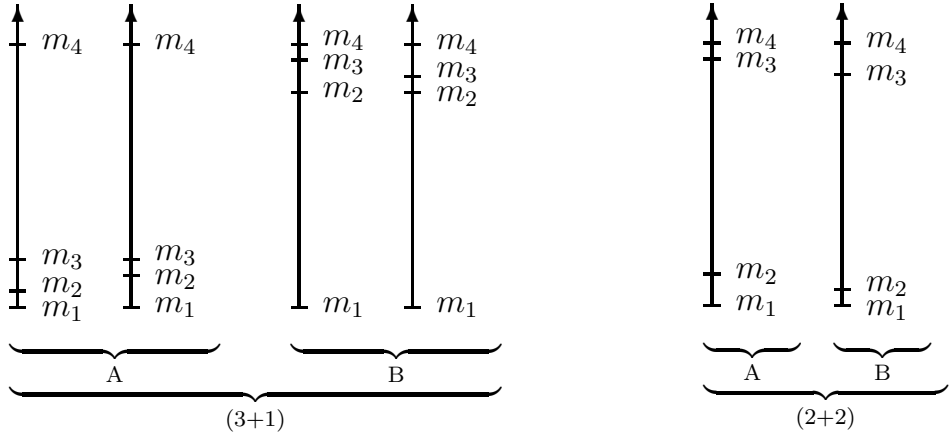}
    \caption{\label{fig:4mass}%
      The six types of 4-neutrino mass spectra. The different
      distances between the masses on the vertical axes represent the
      different scales of mass-squared differences required to explain
      solar, atmospheric and LSND data with neutrino oscillations.}
\end{figure}

There are six possible four-neutrino schemes, shown in
Fig.~\ref{fig:4mass}, that can accommodate the results from solar and
atmospheric neutrino experiments and contain a third much larger
$\Delta m^2$.  They can be divided in two classes: (3+1) and (2+2). In
the (3+1) schemes, there is a group of three close-by neutrino masses
that is separated from the fourth one by the larger gap.  In (2+2)
schemes, there are two pairs of close masses separated by the large
gap. The main difference between these two classes is that in (2+2)
models the extra sterile state cannot be simultaneously decoupled from
\emph{both} solar and atmospheric oscillations, whereas in (3+1)
models the mixing between the sterile neutrino and the three active
ones can be reduced at will. In other words, in (3+1) schemes it is
possible to recover the usual three-neutrino scenario as a limiting
case, whereas (2+2) schemes have unique phenomenological implications.

For what concerns the mixing parameters, we emphasize that the mixing
matrix describing CC interactions in these schemes is a $3\times 4$
matrix.  The reason is that there are three charged lepton mass
eigenstates ($e,\, \mu,\, \tau$) and four neutrino mass eigenstates
($\nu_1,\, \nu_2,\, \nu_3,\, \nu_4$).  As discussed in
Sec.~\ref{sec:lepmix}, if we choose an interaction basis where the
charged leptons are the mass eigenstates, then the CC mixing matrix
$U$ is a sub-matrix of the $4\times 4$ unitary matrix $V^\nu$ that
rotates the neutrinos from the interaction basis to the mass basis,
where the line corresponding to $\nu_s$ is removed. 

\subsubsection{Status of (3+1) schemes}
\label{sec:3+1}

Well before the MiniBooNE experiment, it has been largely discussed in
the literature that the (3+1)-spectra are strongly disfavored by the
data from other SBL laboratory experiments~\cite{3+1bounds1,
3+1bounds2, 3+1bounds3, 3+1bounds4, 3+1bounds5, 3+1bounds6,
3+1bounds7, 3+1michele1, 3+1michele2, 3+1michele3, 3+1michele4} as
explanations of the LSND result. 
This statement is based on a joint analysis of SBL experiments, which
once combined with solar and atmospheric neutrino data severely
constrain the simultaneous mixing of the sterile state with both
$\nu_e$ and $\nu_\mu$.  We summarize here these arguments and the
present phenomenological status.

The probability $P_{\nu_\mu\to\nu_e}$ that is driven by the large $\Delta
m^2_{41}$ (and which is relevant for LSND as well as KARMEN and NOMAD)
is given by 
\begin{equation}
    \label{eq:Pmue}
    P_{\nu_\mu \to \nu_e} = P_{\bar\nu_\mu \to \bar\nu_e} =
    4 \, |U_{e4} U_{\mu 4}|^2 \, \sin^2 \frac{\Delta m^2_{41} L}{4E} \,,
\end{equation}
where $L$ is the distance between source and detector. Here solar and
atmospheric splittings have been neglected since they are too small to
give any observable effect at the relevant $L/E$'s. In this
approximation $\Delta m^2_{41} = \Delta m^2_{42} = \Delta m^2_{43}$
for schemes (3+1)A, and $\Delta m^2_{41} = \Delta m^2_{31} = \Delta
m^2_{21}$ for schemes (3+1)B. 

The LSND experiment gave an allowed region in the $(\Delta m^2_{41},\,
|U_{e4} U_{\mu4}|^2)$ plane which can be directly obtained from the
two-neutrino oscillation region shown in Fig.~\ref{fig:lsnd} with the
identifications $\Delta m^2 \to \Delta m^2_{41}$ and $\sin^22\theta
\to 4 |U_{e4} U_{\mu4}|^2$. In the same way, the KARMEN and MiniBooNE
experiments gave an excluded region in the same plane which can be
immediately derived from Fig.~\ref{fig:lsnd}. As can be seen,
practically all the LSND region is now excluded.

\begin{figure}\centering
    \includegraphics[width=0.95\textwidth]{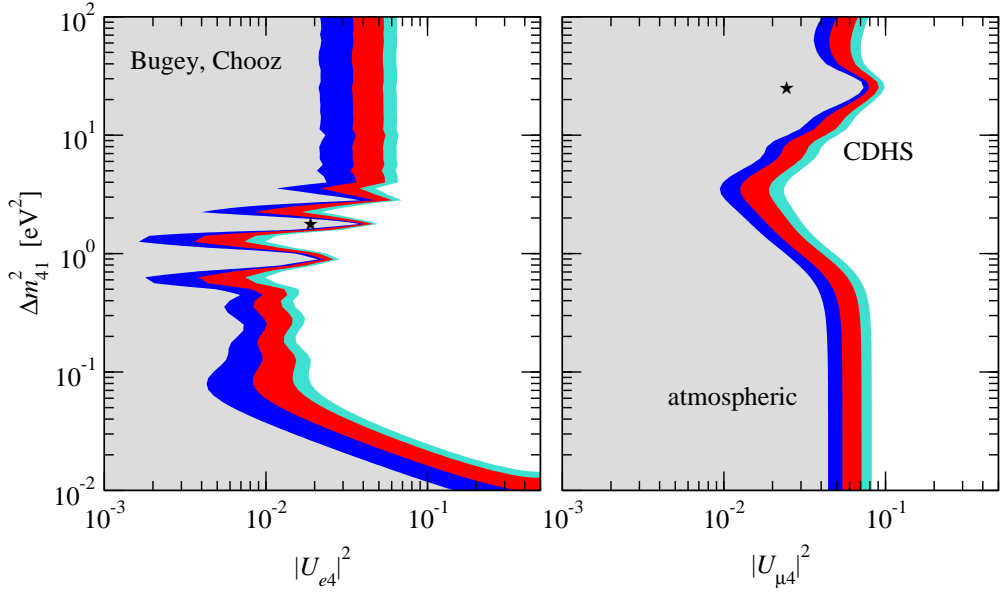}
    \caption{\label{fig:disapp}%
      Bounds on $|U_{e4}|^2$ (left panel) and on $|U_{\mu 4}|^2$
      (right panel) as a function of $\Delta m^2_{41}$. Different
      contours correspond to the two-dimensional allowed regions at
      90\%, 95\%, 99\% and $3\sigma$ CL. The best fit point is marked
      with a star.}
\end{figure}

Further constraints on $|U_{e4} U_{\mu4}|^2$ can be obtained by
combining the bounds on $|U_{e4}|$ and $|U_{\mu4}|$ from reactor and
accelerator experiments in combination with the information from solar
and atmospheric neutrinos. The strongest constraints in the relevant
$\Delta m^2_{41}$ region are given by the Bugey and CDHS experiments.
Using their limits on the survival probabilities one finds
\begin{align}
    \label{eq:bugey31}
    \begin{split}
	4 |U_{e4}|^2 \left( 1 - |U_{e4}|^2 \right) 
	&< D_e^\text{Bugey}(\Delta m^2_{41}) \,,
	\\
	D_e^\text{Bugey} &< 0.001-0.1
	\quad\text{for}\quad
	0.1 \lesssim \Delta m^2_{41}/\eVq \lesssim 10 \,,
    \end{split}
    \\
    \label{eq:cdhs31}
    \begin{split}
	4 |U_{\mu4}|^2 \left( 1 - |U_{\mu4}|^2 \right)
	&< D_\mu^\text{CDHS}(\Delta m^2_{41}) \,,
	\\
	D_\mu^\text{CDHS} &< 0.05-0.1
	\quad\text{for}\quad
	\Delta m^2_{41} \gtrsim 0.5~\eVq \,,
    \end{split}
\end{align}
while for lower $\Delta m^2_{41}$ the CDHS bound weakens considerably.
These bounds are plotted in Fig.~\ref{fig:disapp} as a function of
$\Delta m^2_{41}$.
In principle, the inequalities in Eq.~\eqref{eq:bugey31} and
Eq.~\eqref{eq:cdhs31} can be satisfied with either small mixing
parameters $|U_{\alpha 4}|^2 \lesssim D_\alpha / 4$, or close to
maximal mixing, $|U_{\alpha 4}|^2 \gtrsim 1-D_\alpha / 4$. This is the
generalization to these schemes of the symmetry $\theta
\leftrightarrow \frac{\pi}{2} - \theta$ of two-neutrino vacuum
oscillations (see Sec.~\ref{sec:oscvac}). Solar and atmospheric data
are invoked in order to resolve this ambiguity. Since in any of the
allowed regions for solar oscillations $P^\odot_{ee} \lesssim 0.5$,
only the small values of $|U_{e4}|$ are possible. For atmospheric
neutrinos one can use the fact that oscillations with the large
$\Delta m^2_{41}$ would wash-out the up-down asymmetry and this
wash-out grows with the projection of the $\nu_\mu$ over states
separated by the large $\Delta m^2_{41}$, which is controlled by the
mixing parameter $|U_{\mu4}|$. Consequently only the small values of
$|U_{\mu4}|$ are allowed. Thus naively one obtains the bound 
\begin{equation}
    4\, |U_{e4} U_{\mu4}|^2 < 0.25
    \, D_e^\text{Bugey}(\Delta m^2_{41})
    \, D_\mu^\text{CDHS}(\Delta m^2_{41}).
\end{equation}

\begin{figure}\centering
    \includegraphics[width=100mm]{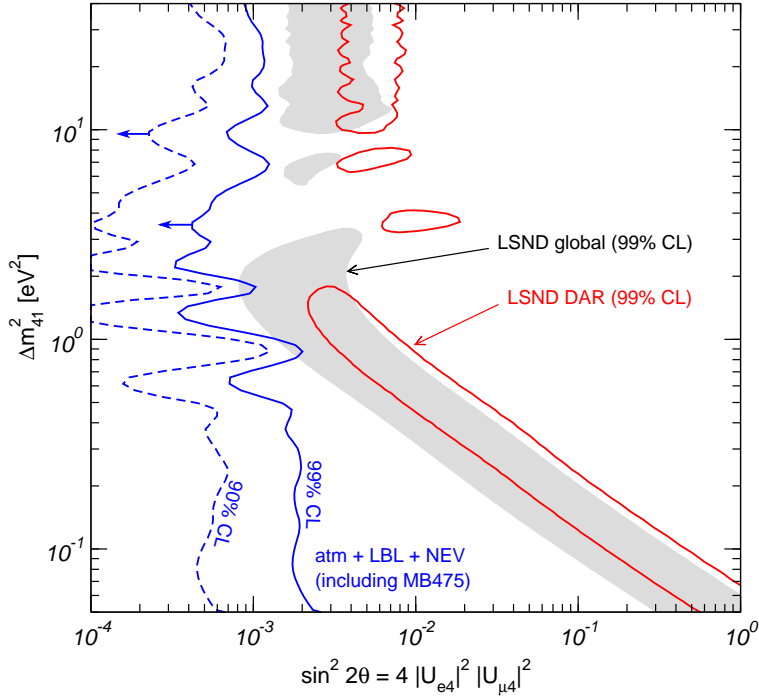}
    \caption{\label{fig:3plus1}%
      Upper bounds on $\sin^22\theta = 4|U_{e4} U_{\mu4}|^2$ from the
      analysis of atmospheric+LBL+NEV data, in the context of
      (3+1)-schemes. Also shown are the regions allowed at 99\% CL
      from the global LSND data and the decay-at-rest (DAR) LSND data.
      Adapted from Ref.~\cite{Maltoni:2007zf}.}
\end{figure}

A detailed and statistically meaningful evaluation of the final
combined limit, including the results of atmospheric+LBL data together
with all the short-baseline experiments observing \emph{no evidence}
(NEV), has been presented in Ref.~\cite{Maltoni:2007zf} and is
summarized in Fig.~\ref{fig:3plus1}. 
The inclusion of the recent results from MiniBooNE at $E_\nu \ge
475$~MeV (MB475) further constrain the possible value of the
sterile-active admixture $4|U_{e4} U_{\mu4}|^2$, hence enhancing the
rejection of these schemes.
A statistical analysis using the \emph{parameter goodness of fit} (PG)
proposed in~\cite{Maltoni:2003cu} gives $\chi^2_\text{PG} = 24.7$ for
2 d.o.f., corresponding to a $4.6\sigma$ rejection ($\text{PG} =
4\times 10^{-6}$) of the (3+1) hypothesis.
These results show that (3+1) schemes are now ruled out as a possible
explanation of the LSND result~\cite{Maltoni:2007zf}.
In addition, it should be noted that the low-energy excess observed by
MiniBooNE at $E_\nu \le 475$~MeV cannot be explained in terms of
oscillations with only one large mass-squared difference, thus adding
another problem to these models in case this excess is confirmed to be
a real signal.

\subsubsection{(2+2) schemes: active-sterile admixtures}

The main feature of (2+2)-spectra is that either solar or atmospheric 
oscillations must involve the sterile neutrino. Such oscillations are,
however, disfavored for both the solar and atmospheric neutrinos.  One
expects then that the (2+2) schemes are disfavored. However, as first
discussed by Ref.~\cite{carlo4,ourfour1}, within (2+2) schemes,
oscillations into pure active or pure sterile states are only limiting
cases of the most general possibility of oscillations into an
admixture of active and sterile neutrinos. One can wonder then whether
some admixture of active-sterile oscillations gives an acceptable
description of both solar and atmospheric data.

For the phenomenology of neutrino oscillations, the (2+2)A and (2+2)B
schemes are equivalent up to the relabeling of the mass eigenstates
(or, equivalently, of the mixing angles). Thus in what follows we
consider the scheme B, where the mass spectrum presents the following
hierarchy:
\begin{equation}
    \Delta m^2_\odot = \Delta m^2_{21}
    \ll \Delta m^2_\text{atm} = \Delta m^2_{43}
    \ll \Delta m^2_{41} \simeq
    \Delta m^2_{42} \simeq \Delta m^2_{31} \simeq \Delta m^2_{32} \,.
\end{equation}

Choosing a convention that is convenient for the study of solar and
atmospheric neutrinos, the matrix $V^\nu_{\alpha i}$
($\alpha=e,s,\mu,\tau$) can be written as follows:
\begin{equation}
    V^\nu = R_{24} \, \tilde{R}_{23} \, R_{34} \,
    \tilde{R}_{14} \, \tilde{R}_{13} \, R_{12} \,,
\end{equation}
where $\tilde{R}_{ij}$ represents a complex rotation of an angle
$\theta_{ij}$ and a phase $\delta_{ij}$ in the $ij$ plane, while
$R_{ij}$ is an ordinary rotation by an angle $\theta_{ij}$. Since the
parametrization of the leptonic mixing matrix $U$ and of the
corresponding lines in $V^\nu$ are the same, and in particular involve
six mixing angles and three phases, we will concentrate below on the
allowed values of the $4\times4$ neutrino mixing matrix $V^\nu$.

This general form can be further simplified by taking into account the
negative results from the reactor experiments (in particular the Bugey
experiment) which for large $\Delta m^2_{41}$ imply that
\begin{equation}
    |V^\nu_{e3}|^2 + |V^\nu_{e4}|^2 
    = c_{14}^2 s^2_{13} + s_{14}^2 \lesssim 10^{-2}.
\end{equation}
So for our purposes, the two angles $\theta_{13}$ and $\theta_{14}$
can then be safely neglected and the two Dirac phases $\delta_{13}$
and $\delta_{14}$ also disappear from the equations. On the other
hand, the third phase $\delta_{23}$ remains, and the $V^\nu$ matrix
takes the effective form:
\begin{equation}
    \label{eq:Umatrix}
    V^\nu =
    \begin{pmatrix}
	c_{12} & s_{12} & 0 & 0
	\\
	-s_{12} c_{23} c_{24}
	& c_{12} c_{23} c_{24}
	& \tilde{s}_{23} c_{24} c_{34} - s_{24} s_{34}
	& \tilde{s}_{23} c_{24} s_{34} + s_{24} c_{34}
	\\
	s_{12} \tilde{s}_{23}^\star
	& -c_{12} \tilde{s}_{23}^\star
	&  c_{23} c_{34}
	&  c_{23} s_{34}
	\\
	s_{12} c_{23} s_{24}
	& -c_{12} c_{23} s_{24}
	& -\tilde{s}_{23} s_{24} c_{34} - c_{24} s_{34}
	& -\tilde{s}_{23} s_{24} s_{34} + c_{24} c_{34}
    \end{pmatrix}
\end{equation}
where $\tilde{s}_{23} = s_{23} e^{i\delta_{23}}$. The full parameter
space relevant to solar and atmospheric neutrino oscillation can be
covered with the four angles $\theta_{ij}$ in the first quadrant,
$\theta_{ij} \in [0,\, \pi/2]$, while the Dirac phase $\delta_{23}$
spans the full range $\delta_{23} \in [0,\, 2\pi]$.

In this scheme solar neutrino oscillations are generated by the
mass-squared difference between $\nu_2$ and $\nu_1$ while atmospheric
neutrino oscillations are generated by the mass-squared difference
between $\nu_3$ and $\nu_4$.  It is clear from Eq.~\eqref{eq:Umatrix}
that the survival of solar $\nu_e$'s depends mainly on $\theta_{12}$
while atmospheric $\nu_e$'s are not affected by the four-neutrino
oscillations in the approximation $\theta_{13}=\theta_{14}=0$ and
neglecting the effect of $\Delta m^2_{21}$ in the range of atmospheric
neutrino energies. The survival probability of atmospheric $\nu_\mu$'s
depends mainly on the $\theta_{34}$.

Thus solar neutrino oscillations occur with a mixing angle
$\theta_{12}$ between the states
\begin{equation}
    \label{eq:nusol}
    \nu_e\to \nu_\alpha \quad \text{with} \quad
    \nu_\alpha = c_{23} c_{24}\, \nu_s 
    + \sqrt{1 - c_{23}^2 c_{24}^2}\, \nu_a \,,
\end{equation}
where $\nu_a$ is a linear combination of $\nu_\mu$ and $\nu_\tau$,
\begin{equation}
    \label{eq:defnua}
    \nu_a = \frac{1}{\sqrt{1-c_{23}^2 c_{24}^2}}
    (\tilde{s}_{23}^\star \nu_\mu + c_{23} s_{24} \nu_\tau) \,.
\end{equation}
We remind the reader that $\nu_\mu$ and $\nu_\tau$ cannot be 
distinguished in solar neutrino experiments, because their matter
potential and their interaction in the detectors are equal, due to
only NC weak interactions. Thus solar neutrino oscillations cannot
depend on the mixing angle $\theta_{34}$ and depend on $\theta_{23}$
and $\theta_{24}$ through the combination $c_{23}^2 c_{24}^2$.

Atmospheric neutrino oscillations, \textit{i.e.} oscillations with the
mass difference $\Delta m^2_{43}$ and mixing angle $\theta_{34}$,
occur between the states
\begin{equation}
    \label{eq:nuatm}
    \nu_\beta \to \nu_\gamma
    \quad\text{with}\quad\left\lbrace
    \begin{aligned}
	\nu_\beta &= \tilde{s}_{23} c_{24} \nu_s 
	+ c_{23} \nu_\mu - \tilde{s}_{23} s_{24} \nu_\tau \,,
	\\
	\nu_\gamma &= s_{24} \nu_s + c_{24} \nu_\tau \,.
    \end{aligned}\right.
\end{equation}
We learn that the mixing angles $\theta_{23}$ and $\theta_{24}$
determine two projections. First, the projection of the sterile
neutrino onto the states in which the solar $\nu_e$ oscillates is
given by 
\begin{equation}
    \eta_s \equiv c_{23}^2 c_{24}^2
    = 1 - |V^\nu_{a1}|^2 - |V^\nu_{a2}|^2 =
    |V^\nu_{s1}|^2 + |V^\nu_{s2}|^2 \,.
\end{equation}
Second, the projection of the $\nu_\mu$ over the solar neutrino 
oscillating states is given by
\begin{equation} 
    \label{eq:muatm}
    d_\mu \equiv s^2_{23} = |V^\nu_{\mu 1}|^2 + |V^\nu_{\mu 2}|^2 
    = 1 - |V^\nu_{\mu 3}|^2 - |V^\nu_{\mu 4}|^2
\end{equation}
One expects $s_{23}$ to be small in order to explain the atmospheric
neutrino deficit. We will see that this is indeed the case.
Furthermore, the negative results from the CDHS and CCFR searches for 
$\nu_\mu$-disappearance also constrain such a projection to be smaller
than 0.2 at 90\% CL for $\Delta m^2_{41} \gtrsim 0.4~\eVq$.

We distinguish the following limiting cases:
\begin{enumerate}
  \item[(i)] If $c_{23} = 1$ then $V^\nu_{\mu 1} = V^\nu_{\mu 2}=0$.
    The atmospheric $\nu_\mu=\nu_\beta$ state oscillates into a state
    $\nu_\gamma=c_{24}\nu_\tau +s_{24}\nu_s$. We will denote this case
    as \emph{restricted}.  In particular:
    \begin{itemize}
      \item If $c_{23} = c_{24} = 1$, $V^\nu_{a1} = V^\nu_{a2} = 0$
	($V^\nu_{s3} = V^\nu_{s4}=0$) and we have the limit of pure
	two-generation solar $\nu_e \to \nu_s$ transitions and
	atmospheric $\nu_\mu \to \nu_\tau$ transitions.
	
      \item If $c_{24} = 0$ then $V^\nu_{s1} = V^\nu_{s2}=0$ and
	$V^\nu_{\tau 3} = V^\nu_{\tau 4} = 0$, corresponding to the
	limit of pure two-generation solar $\nu_e\to\nu_\tau$
	transitions and atmospheric $\nu_\mu\to\nu_s$ transitions.
    \end{itemize}
    
  \item[(ii)] If $c_{23} = 0$ then $V^\nu_{s1} = V^\nu_{s2} = 0$
    corresponding to the limit of pure two-generation solar $\nu_e \to
    \nu_a$ with $a = \mu$ and there are no atmospheric neutrino
    oscillations as the projection of $\nu_\mu$ over the relevant
    states cancels out ($V^\nu_{\mu 3} = V^\nu_{\mu 4} = 0$).
\end{enumerate}

Notice that in the restricted case $\theta_{23} = 0$ the Dirac phase
$\delta_{23}$ also vanishes from the equations, and we have
effectively two-neutrino oscillations for both the solar ($\nu_e \to
\nu_\alpha$) and atmospheric ($\nu_\mu \to \nu_\gamma$) cases. 

To summarize, solar neutrino oscillations depend on the new mixing
angles only through the product $c_{23} c_{24}$ and therefore the
analysis of the solar neutrino data in four-neutrino mixing schemes is
equivalent to the two-neutrino analysis but taking into account that
the parameter space is now three-dimensional $(\Delta m^2_{21},\,
\tan^2\theta_{12},\, c^2_{23} c^2_{24})$.  Atmospheric neutrino
oscillations are affected independently by the angles $\theta_{23}$
and $\theta_{24}$ and by the phase $\delta_{23}$, and the analysis of
the atmospheric neutrino data in the four-neutrino mixing schemes is
equivalent to the two-neutrino analysis, but taking into account that
the parameter space is now five-dimensional $(\Delta m^2_{43},\,
\theta_{34},\, c^2_{23},\, c^2_{24},\, \delta_{23})$. Furthermore the
allowed ranges of active-sterile oscillations depends on the assumed
\Nuc{8}{B} neutrino flux~\cite{b8barger,ourbobe,ourpostnu04}. Allowing
for \Nuc{8}{B} neutrino flux larger than the SSM expectation result
into a less stringent limit on the active-sterile admixture.

\begin{figure}\centering
    \includegraphics[width=0.98\textwidth]{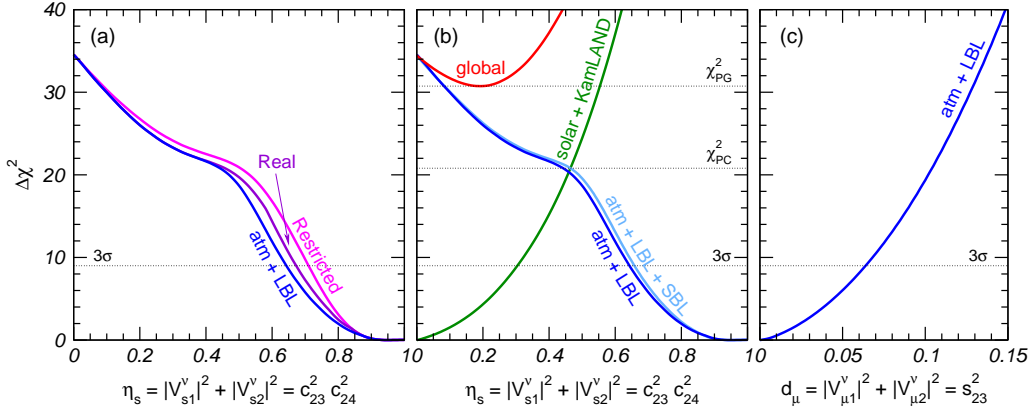}
    \caption{\label{fig:chi24}%
      $\Delta\chi^2$ as a function of the active-sterile admixture
      $\eta_s = |V^\nu_{s1}|^2 + |V^\nu_{s2}|^2$ and the parameter
      $d_\mu = |V^\nu_{\mu 1}|^2 + |V^\nu_{\mu 2}|^2$ from the
      analysis of solar+KamLAND data (green line) and atmospheric+LBL
      data (blue lines) in (2+2)-schemes. In the left panels we show
      the atmospheric+LBL $\chi^2$ for the \emph{restricted}
      ($\theta_{23} = 0$), \emph{real} ($\delta_{23} = \{0,\, \pi\}$)
      and general cases. The red line in the central panel show the
      increase in the $\chi^2$ once the solar+KamLAND and
      atmospheric+LBL data are combined together.}
\end{figure}

As an illustration we show in the central panel of
Fig.~\ref{fig:chi24} (green line) the shift in $\chi^2$ for the
analysis of solar (+~KamLAND) neutrino data as a function of the
active-sterile admixture $\eta_s = |V^\nu_{s1}|^2 + |V^\nu_{s2}|^2 =
c_{23}^2 c_{24}^2$. From the figure we conclude that the solar data
favor pure $\nu_e \to \nu_a$ oscillations but sizable active-sterile
admixtures are still allowed. In this curve the \Nuc{8}{B} flux
allowed to take larger values than in the SSM which, as discussed
above so the active-sterile bound is as model independent as possible.

Similar analysis can be performed for the atmospheric neutrino
data~\cite{2+2ours} to obtain the allowed regions for the oscillation 
parameters $\Delta m^2_{43}$ and $\tan^2\theta_{34}$ from the global
analysis for different values of $\theta_{23}$ and $\theta_{24}$ (or,
equivalently, of the projections $d_\mu = |V^\nu_{\mu 1}|^2 +
|V^\nu_{\mu 2}|^2$ and $\eta_s = |V^\nu_{s1}|^2 + |V^\nu_{s2}|^2$).
The global minimum corresponds to almost pure atmospheric $\nu_\mu -
\nu_\tau$ oscillations and the allowed regions become considerably
smaller for increasing values of the mixing angle $\theta_{23}$, which
determines the size of the projection of $\nu_\mu$ over the neutrino
states oscillating with $\Delta m^2_{43}$, and for increasing values
of the mixing angle $\theta_{24}$, which determines the active-sterile
admixture in which the \emph{almost-$\nu_\mu$} oscillates.  Therefore
the atmospheric neutrino data give an upper bound on both mixings
which further implies a lower bound on the combination $\eta_s =
|V^\nu_{s1}|^2 + |V^\nu_{s2}|^2 = c_{23}^2 c_{24}^2$. The same
combination is limited from above by the solar neutrino data. 

In the left panel of Fig.~\ref{fig:chi24} we show the shift in
$\chi^2$ for the analysis of the atmospheric (+~LBL) data as a
function of the active-sterile admixture $\eta_s = |V^\nu_{s1}|^2 +
|V^\nu_{s2}|^2 = c_{23}^2 c_{24}^2$, in the general case (in which the
analysis is optimized with respect to both the parameter $d_\mu =
|V^\nu_{\mu 1}|^2 + |V^\nu_{\mu 2}|^2 = s^2_{23}$ and the Dirac phase
$\delta_{23}$) as well as the \emph{real} case (when only the values
$\delta_{23} = \{ 0,\, \pi \}$ are considered, as in our previous
analyses) and the \emph{restricted} case ($\theta_{23} = 0$).
Similarly, in the right panel of Fig.~\ref{fig:chi24} we show the
shift in $\chi^2$ as a function of the parameter $d_\mu = |V^\nu_{\mu
1}|^2 + |V^\nu_{\mu 2}|^2 = s^2_{23}$.

In summary, the analysis of the solar data favors the scenario in 
which the solar oscillations in the $1-2$ plane are $\nu_e-\nu_a$
oscillations, and gives an upper bound on the projection of $\nu_s$ on
this plane. On the other hand, the analysis of the atmospheric data
prefers the oscillations of the $3-4$ states to occur between a
close-to-pure $\nu_\mu$ and an active ($\nu_\tau$) neutrino and gives
an upper bound on the projection of the $\nu_s$ over the $3-4$ states
or, equivalently, a lower bound on its projection over the $1-2$
states. From Fig.~\ref{fig:chi24} we see that the exclusion curves
from the analyses solar+KamLAND data and atmospheric+LBL data (with or
without the inclusion of SBL experiments) are in strong disagreement,
and only overlap at $\chi^2_\text{PC} = 20.8$ (\textit{i.e.}, at the
$4.6\sigma$ level). More quantitatively we find $\chi^2_\text{PG} =
30.7$, which corresponds to a $5.5\sigma$ rejection ($\text{PG} =
3\times 10^{-8}$) of the (2+2) hypothesis.
These conclusions are unaffected by the recent MiniBooNE result.


\subsection{Five-Neutrino and Six-Neutrino Mixing}
\label{sec:fivemix}

Five-neutrino schemes of the (3+2) type are a straight-forward
extension of (3+1) schemes. In addition to the cluster of the three
neutrino mass states accounting for ``solar'' and ``atmospheric'' mass
splittings now two states at the eV scale are added, with a small
admixture of $\nu_e$ and $\nu_\mu$ to account for the LSND signal.
In the Appendix of Ref.~\cite{3+1bounds7} it was suggested that such
models could somewhat relax the tension existing between
short-baseline experiments and the LSND data. In Ref.~\cite{3+2} a
complete analysis was performed, finding that indeed the disagreement
between LSND and null-result experiments is reduced. This possibility
was recently reconsidered in Ref.~\cite{Maltoni:2007zf} at the light
of the new MiniBooNE data, and found to be strongly disfavored due to
the same tension between SBL and LSND data already discussed for (3+1)
models. Furthermore, according to Ref.~\cite{Maltoni:2007zf} the
addition of a third sterile neutrino does not significantly improve
the quality of the global fit. In this section we will review these
arguments in some detail and summarize the present status of (3+2) and
(3+3) models.

As already mentioned, MiniBooNE data (MB) are consistent with zero (no
excess) above 475~MeV, whereas below this energy a $3.6\sigma$ excess
of $96 \pm 17 \pm 20$ events is observed. Whether this excess comes
indeed from $\nu_\mu\to\nu_e$ transitions or has some other origin is
under investigation~\cite{miniboonelast}. Lacking any explanation in
terms of backgrounds or systematical uncertainties, in the following
we will present the results obtained using both the full energy range 
from 300~MeV to 3~GeV (``MB300'') and for the restricted range from
475~MeV to 3~GeV (``MB475'').

\subsubsection{(3+2) schemes}
\label{sec:app-32}

In these schemes the appearance data (LSND, KARMEN, NOMAD, and MB) can
be described using the SBL approximation $\Delta m^2_{21} \approx
\Delta m^2_{31} \approx 0$ of the relevant appearance probability
which is given by
\begin{multline}
    \label{eq:5nu-prob}
    P_{\nu_\mu\to\nu_e} =
    4 \, |U_{e4}|^2 |U_{\mu 4}|^2 \, \sin^2 \phi_{41} +
    4 \, |U_{e5}|^2 |U_{\mu 5}|^2 \, \sin^2 \phi_{51}
    \\
    + 8 \,|U_{e4}U_{\mu 4}U_{e5}U_{\mu 5}| \,
    \sin\phi_{41}\sin\phi_{51}\cos(\phi_{54} - \delta) \,,
\end{multline}
with the definitions
\begin{equation}
    \label{eq:5nu-def}
    \phi_{ij} \equiv \frac{\Delta m^2_{ij}L}{4E} \,,
    \qquad \delta \equiv
    \arg\left(U_{e4}^* U_{\mu 4} U_{e5} U_{\mu 5}^* \right) \,. 
\end{equation}
Eq.~\eqref{eq:5nu-prob} holds for neutrinos (NOMAD and MB); for
anti-neutrinos (LSND and KARMEN) one has to replace $\delta \to
-\delta$. Note that Eq.~\eqref{eq:5nu-prob} is invariant under the
transformation $4\leftrightarrow 5$ and $\delta \leftrightarrow
-\delta$, and depends only on the combinations $|U_{e4}U_{\mu 4}|$ and
$|U_{e5}U_{\mu 5}|$.

\begin{figure}\centering 
    \includegraphics[width=0.98\textwidth]{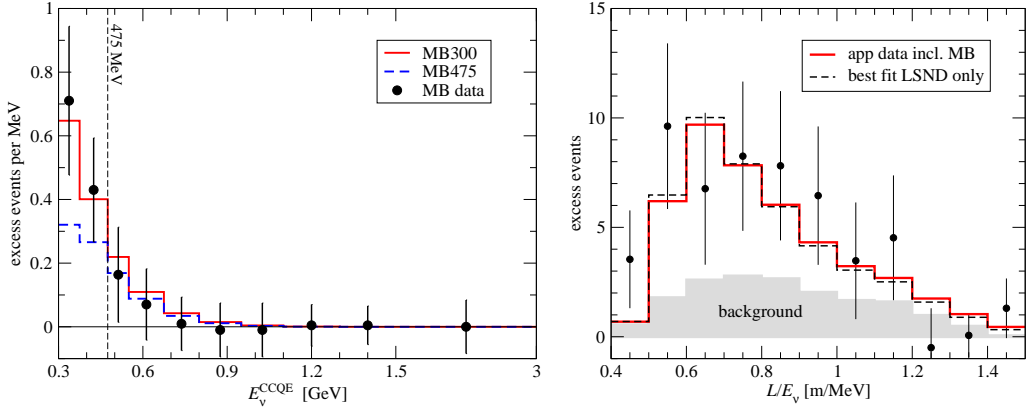}
    \caption{\label{fig:spectrum}%
      Spectral data for the MiniBooNE (left) and LSND (right)
      experiments. The histograms show the prediction at the best fit
      points in (3+2) mass schemes for SBL appearance data LSND,
      KARMEN, NOMAD, MB. Adapted from Ref.~\cite{Maltoni:2007zf}.}
\end{figure}

An important observation is that non-trivial values of the complex 
phase $\delta$ lead to CP violation, and hence in (3+2) schemes much
more flexibility is available to accommodate the results of LSND
(anti-neutrinos) and MB (neutrinos). As a consequence MB is perfectly
compatible with LSND in the (3+2) framework. In
Fig.~\ref{fig:spectrum} we show the prediction for MB and LSND at the
best fit points of the combined MB, LSND, KARMEN, NOMAD analysis. As
can be seen from this figure, MB data can be fitted very well while
simultaneously explaining the LSND evidence. Furthermore, in this case
also the low energy MB data can be explained, and therefore, in
contrast to (3+1) schemes, (3+2) oscillations offer an appealing
possibility to account for this excess. The parameter values at the
best fit points are given in Table~\ref{tab:bfp}; for both MB475 and
MB300 a goodness-of-fit of 85\% is obtained, showing that MB is in
very good agreement with global SBL appearance data including LSND.

\begin{table}\centering
    \begin{tabular}{l@{\quad}cc@{\quad}cc@{\quad}c@{\quad}c}
	\hline\hline
	data set
	& $|U_{e4} U_{\mu 4}|$ & $\Delta m^2_{41}$ 
	& $|U_{e5} U_{\mu 5}|$ & $\Delta m^2_{51}$
	& $\delta$ & $\chi^2_\text{min} / \text{dof}$
	\\
	\hline
	appearance (MB475)
	& 0.044 & 0.66
	& 0.022 & 1.44
	& $1.12\pi$ & $16.9 / 24$
	\\
	appearance (MB300)
	& 0.31 & 0.66
	& 0.27 & 0.76
	& $1.01\pi$ & $18.5 / 26$
	\\
	\hline\hline
    \end{tabular}
    \caption{\label{tab:bfp}%
      Parameter values and $\chi^2_\text{min} / \text{dof}$ of the
      best fit points for SBL appearance data from LSND, KARMEN, NOMAD
      and MB in (3+2) schemes. Mass-squared differences are given in
      \eVq. Results are shown without (MB475) and including (MB300)
      the low energy data from MB. Taken from
      Ref.~\cite{Maltoni:2007zf}.}
\end{table}

\begin{figure}\centering
    \includegraphics[width=0.98\textwidth]{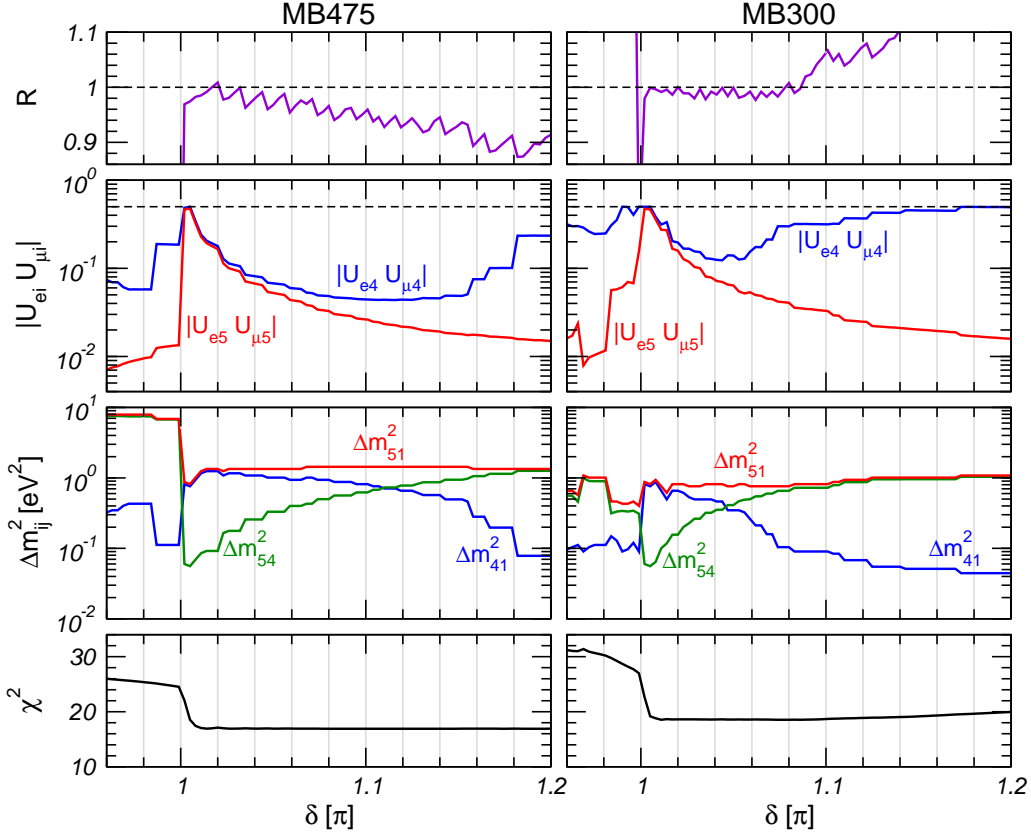}
    \caption{\label{fig:CP}%
      Fit to appearance data LSND, KARMEN, NOMAD, and MB475 (left) or
      MB300 (right) in (3+2) schemes. We show the $\chi^2$, the mass
      splittings $\Delta m^2_{i1}$, the mixing angles $|U_{ei} U_{\mu
      i}|$ and the ratio $R = \big( |U_{e4} U_{\mu 4}| \Delta m^2_{41}
      \big) \,\big/\, \big( |U_{e5} U_{\mu 5}| \Delta m^2_{51} \big)$
      as a function of the complex phase $\delta$ defined in
      Eq.~\eqref{eq:5nu-def}. Adapted from
      Ref.~\cite{Maltoni:2007zf}.}
\end{figure}

The possibility of explaining the MiniBooNE low-energy events and
simultaneously fitting all appearance experiments in these schemes 
(unlike in 3+1) can be understood by comparing the energy dependence
of the corresponding transition probability, $P_{\nu_\mu\to\nu_e}$.
For the four-neutrino case, Eq.~\eqref{eq:Pmue} has a high-energy tail
which drops as $1/E^2$, and there is no way to suppress it except by
reducing the mixing angle $|U_{e4} \, U_{\mu 4}|$. However, doing so
unavoidably suppresses $P_{\nu_\mu\to\nu_e}$ also in the low-energy
region, and this is a problem for the MB300 data set. This is why
(3+1) models cannot account for the low-energy excess observed by
MiniBooNE.
Conversely, in five-neutrino models the transition probability is
given by Eq.~\eqref{eq:5nu-prob}, and the high-energy limit has a more
complex behavior:
\begin{multline}
    \label{eq:5nu-he1}
    P_{\nu_\mu\to\nu_e} \propto \frac{1}{E^2}
    \Big[ \left( |U_{e4} \, U_{\mu 4}| \Delta m^2_{41} \right)^2
    + \left( |U_{e5} \, U_{\mu 5}| \Delta m^2_{51} \right)^2
    \\
    + \left( |U_{e4} \, U_{\mu 4}| \Delta m^2_{41} \right)
    \left( |U_{e5} \, U_{\mu 5}| \Delta m^2_{51} \right)
    \cos\delta \Big] \,,
\end{multline}
which for positive $\Delta m^2_{41}$ and $\Delta m^2_{51}$ identically
vanishes if and only if:
\begin{equation}
    \label{eq:5nu-he2}
    |U_{e4} \, U_{\mu 4}| \Delta m^2_{41} =
    |U_{e5} \, U_{\mu 5}| \Delta m^2_{51}
    \qquad\text{and}\qquad
    \delta = \pi \,.
\end{equation}
Therefore, within five-neutrino models it is possible to suppress
$P_{\nu_\mu\to\nu_e}$ at high-energy while keeping it sizable in the
low-energy region, where the expansion which led to
Eq.~\eqref{eq:5nu-he1} does not hold. This explains how (3+2) schemes
accommodate the low energy events. From Table~\ref{tab:bfp} and from
the upper panels of Fig.~\ref{fig:CP} it is immediate to see that
Eq.~\eqref{eq:5nu-he2} is realized with excellent accuracy around the
appearance best-fit point for MB300 data. Interestingly, also the
appearance best-fit for MB475 data closely respects this relations.

Let us now turn to the issue of reconciling MiniBooNE with LSND. First
of all, note that the condition~\eqref{eq:5nu-he2} suppresses 
$P_{\nu_\mu \to \nu_e}$ at high energies for both neutrinos and
antineutrinos. At lower energies a non-trivial way to keep this
probability small can be obtained by requiring $\cos(\phi_{54} -
\delta) = -1$. Then one has
\begin{equation}
    \label{eq:P1}
    P_{\nu_\mu\to\nu_e} = 4 \left( |U_{e4} U_{\mu 4}| \sin\phi_{41} 
    - |U_{e5} U_{\mu 5}| \sin\phi_{51} \right)^2 \,,
    \quad \cos(\phi_{54} - \delta) = -1 \,.
\end{equation}
Hence, $P_{\nu_\mu\to\nu_e}$ is small for $|U_{e4} U_{\mu 4}| \approx
|U_{e5} U_{\mu 5}|$ and $\phi_{54} \ll 1$. This is precisely the
behavior shown in Fig.~\ref{fig:CP}: when $\delta$ approaches $\pi$
from above, $\Delta m^2_{54}$ becomes small and the $|U_{ei} U_{\mu
i}|$ approach each other. Writing $\delta = \pi + \epsilon$ one has
$\cos(\phi_{54} - \delta) \approx -1 + \mathcal{O}(\phi_{54}^2,\,
\epsilon^2)$, Eq.~\eqref{eq:P1} is valid, and the oscillation
probability is suppressed in MB. 
Now the question arises whether large enough values for
$P_{\bar\nu_\mu\to\bar\nu_e}$ can be achieved in order to explain
LSND. The difference of anti-neutrino and neutrino probabilities is
given by
\begin{equation} \begin{split}
    \label{eq:DP}
    P_{\bar\nu_\mu\to\bar\nu_e} - P_{\nu_\mu\to\nu_e}
    &= 16 \, |U_{e4} \, U_{\mu 4} \, U_{e5} \, U_{\mu 5}| \,
    \sin\phi_{41} \, \sin\phi_{51} \, 
    \sin\phi_{54} \, \sin\epsilon
    \\[1mm]
    &\approx 16 \, |U_{e4} \, U_{\mu 4} \, U_{e5} \, U_{\mu 5}| \,
    \sin^2(\phi_{51}) \, \phi_{54} \epsilon \,,
\end{split} \end{equation}
where in the last step $\phi_{54},\epsilon \ll 1$ has been used. Since
$\phi_{54}$ and $\epsilon$ are small, the other factors have to be as
large as possible in order to get a sufficient probability for LSND.
Indeed, for $\Delta m^2_{51} \approx 1$~eV$^2$ one has
$\sin^2\phi_{51} \approx 1$, and also $|U_{ei} \, U_{\mu i}|$ grow for
$\epsilon \to 0$ (see Fig.~\ref{fig:CP}). Once the maximal values
allowed by unitarity, $|U_{e4} \, U_{\mu 4}| = |U_{e5} \, U_{\mu 5}| =
1/2$, are reached the LSND probability is given roughly by
$P_{\bar\nu_\mu\to\bar\nu_e} \sim 4 \epsilon^2$, where we used
$P_{\nu_\mu\to\nu_e} \approx 0$ (in order to explain MB) and
$\phi_{54} \approx \epsilon$ (in order to have $\cos(\phi_{54} -
\delta) \approx -1$). Using the experimental value $P_\mathrm{LSND} =
0.0026$ one finds that a fit should be possible for $\epsilon \gtrsim
0.025 \approx 0.008\pi$, in agreement with our results.

The similar structure of the left and right panels of
Fig.~\ref{fig:CP} suggests that this mechanism works equally well for
MB475 and MB300, and fitting the low energy excess in MB does not
affect these considerations. 
Obviously, this explanation is not valid for $\delta < \pi$, since the
CP asymmetry Eq.~\eqref{eq:DP} has the wrong sign to reconcile LSND
and MB. As visible in Fig.~\ref{fig:CP}, the fit jumps into a quite
different solution, which anyway gives a poor $\chi^2$.

Concerning the constraints from disappearance experiments, in the
(3+2) schemes the relevant survival probability in the SBL
approximation is given 
\begin{multline}
    P_{\nu_\alpha\to\nu_\alpha} = 1
    - 4\left(1 - \sum_{i=4,5} |U_{\alpha i}|^2 \right)
    \sum_{i=4,5} |U_{\alpha i}|^2 \, \sin^2\phi_{i1}
    \\
    - 4\, |U_{\alpha 4}|^2|U_{\alpha 5}|^2 \, \sin^2\phi_{54}
\end{multline}
where $\phi_{ij}$ is given in Eq.~\eqref{eq:5nu-def}. Similar as in
the (3+1) case, also for (3+2) schemes atmospheric neutrino data
provide an important constraint on $\nu_\mu$ oscillations with sterile
neutrinos. It turns out that in practice the same constraint
$\chi^2_\mathrm{ATM}(d_\mu)$ as in the four-neutrino case applies (see
the right panel of Fig.~\ref{fig:chi24}), where now the definition
$d_\mu = |U_{\mu 4}|^2 + |U_{\mu 5}|^2$ has to be used.

\begin{figure}\centering 
    \includegraphics[width=0.98\textwidth]{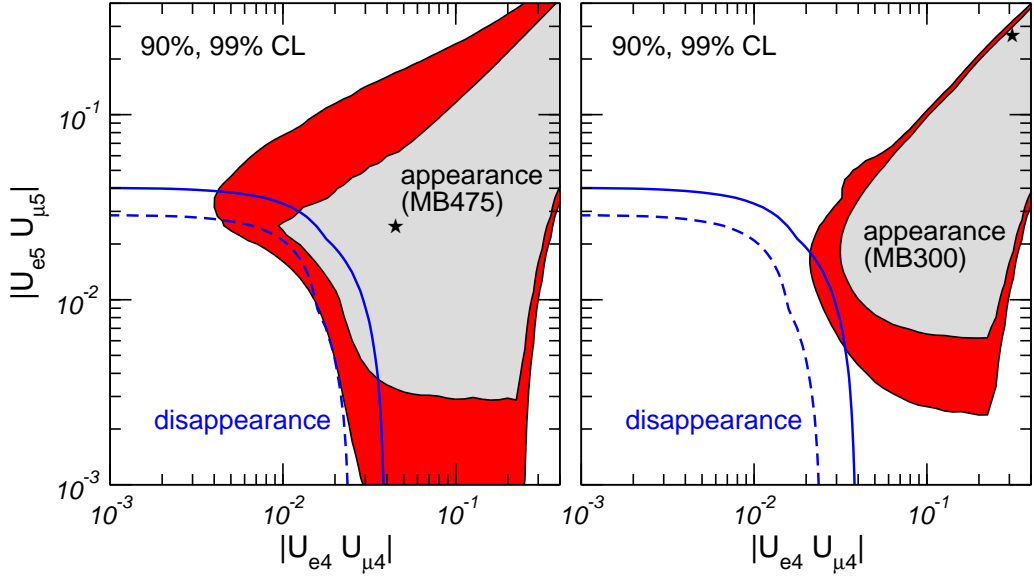}
    \caption{\label{fig:app-vs-dis}%
      Allowed regions at 90\% and 99\%~CL in (3+2) schemes for
      appearance data (shaded regions) and disappearance data (dashed
      and solid curves) projected onto the plane of $|U_{e4} U_{\mu
      4}|$ and $|U_{e5} U_{\mu 5}|$. In the left panel the two lowest
      energy data points in MB have been omitted (MB475), whereas in
      the right panel the full MB energy range has been used in the
      fit (MB300). Taken from Ref.~\cite{Maltoni:2007zf}.}
\end{figure}

As discussed above, and clearly shown in Table~\ref{tab:bfp}, rather
large values of $|U_{e4} \, U_{\mu 4}|$ and $|U_{e5} \, U_{\mu 5}|$
are needed in order to reconcile the negative result of MiniBooNE with
the positive signal of LSND. This is particularly true when the full
MB300 data set is used. However, even in (3+2) schemes short-baseline
experiments pose stringent bounds on the mixing angles $|U_{ei}|$ and
$|U_{\mu i}|$, in close analogy with (3+1) models described in
Sec.~\ref{sec:3+1}. Therefore, one expects that reconciling appearance
and disappearance data will be a problem also within (3+2) models.
This tension is illustrated in Fig.~\ref{fig:app-vs-dis}, where the
projections of the allowed regions in the plane of the appearance
amplitudes $|U_{e4} U_{\mu 4}|$ and $|U_{e5} U_{\mu 5}|$ are shown. 
Indeed the opposite trend of the two data sets is clearly visible,
especially when the low energy excess in MB is included (right panel).

In order to quantify the disagreement between appearance and
disappearance data, one can aply the PG test described
in~\cite{Maltoni:2003cu} using the $\chi^2_\text{PG}$ values, for
global data without MB, with MB475, and with MB300:
\begin{equation} \begin{aligned} \label{eq:5nuPG}
    \chi^2_\text{PG} &= 17.5\,, \quad
    &\text{PG} &= 1.5\times 10^{-3} \qquad \text{(no MB)}
    \\
    \mbox{APP vs DIS:} \qquad
    \chi^2_\text{PG} &= 17.2 \,, \quad 
    & \text{PG} &= 1.8\times 10^{-3} \qquad \text{(MB475)}
    \\
    \chi^2_\text{PG} &= 25.1 \,, \quad
    & \text{PG} &= 4.8\times 10^{-5} \qquad \text{(MB300)}
\end{aligned} \end{equation}
From these numbers one can conclude that also in (3+2) schemes the
tension between appearance and disappearance experiments is quite
severe. If MB475 is used the result is very similar to the situation
without MB data implying inconsistency at about $3.1\sigma$, whereas
in case of the full MB300 data the tension becomes significantly worse
(about $4\sigma$), since appearance data are more constraining because
of the need to accommodate LSND as well as the MB excess at low
energies.

\subsubsection{(3+3) six-neutrino mass schemes}

Since there are three active neutrinos it seems natural to consider
also the case of three sterile neutrinos. If all three additional
neutrino states have masses in the eV range and mixings as relevant
for the SBL experiments under consideration, such a model will
certainly have severe difficulties to accommodate standard
cosmology~\cite{Hannestad:2006mi, Dodelson:2005tp}, and one has to
refer to some non-standard cosmological scenario~\cite{Foot:1995bm,
Chu:2006ua, Smith:2006uw, Babu:1991at, Bento:2001xi, Gelmini:2004ah}.

Besides this fact, the results of the search performed in 
Ref.~\cite{Maltoni:2007zf} show that there is only a marginal
improvement of the fit (by 1.7 units in $\chi^2$) for MB475 (3.5 for
MB300) with respect to (3+2), to be compared with 4 additional
parameters in the model. Hence, the conclusion is that that there are
no qualitatively new effects in the (3+3) scheme. The conflict between
appearance and disappearance data remains a problem, and the
additional freedom introduced by four new parameters does not relax
significantly this tension.


\subsection{Violation of CPT}
\label{sec:cptviol}

In Ref.~\cite{cpt1} it was observed that the LSND signal could be
accommodated with the solar and atmospheric neutrino anomalies without
enlarging the neutrino sector if CPT was violated. Once such a drastic
modification of standard physics is accepted, oscillations with four
independent $\Delta m^2$ are possible, two in the neutrino and two in
the anti-neutrino sector. The basic realization behind these proposals
is that the oscillation interpretation of the solar results involves
oscillations of electron neutrinos with $\Delta m^2_\odot \lesssim
10^{-4}~\eVq$ while the LSND signal for short baseline oscillations
with $\Delta m^2_\text{LSND} \gtrsim 10^{-1}~\eVq$ stems dominantly
from anti-neutrinos ($\bar{\nu}_\mu \to \bar{\nu}_e$). If CPT was
violated and neutrino and anti-neutrino mass spectra and mixings were
different~\cite{cpt1,cpt2,cpt4,cpt5,cpt3} both results could be made
compatible in addition to the interpretation of the atmospheric
neutrino data in terms of oscillations of both $\nu_\mu$ and
$\bar{\nu}_\mu$ with $\Delta m^2_\text{atm} \sim 10^{-3}~\eVq$.

Notice that, in principle, these scenarios could accommodate both the
evidence of LSND and the negative results of MiniBooNE obtained
running in the neutrino mode, as the neutrino spectrum may not
oscillate with the large $\Delta m^2$ while the antineutrino does.  
For this scenarios the data from MiniBooNE running in the antineutrino
mode will be most relevant.

In the original spectrum proposed, neutrinos had mass splittings
$\Delta m^2_\odot = \Delta m^2_{21} \ll \Delta m^2_{31} = \Delta
m^2_\text{atm}$ to explain the solar and atmospheric observations,
while for anti-neutrinos $\Delta m^2_\text{atm} = \Delta\bar{m}^2_{21}
\ll \Delta\bar{m}^2_{31} = \Delta m^2_\text{LSND}$. Within this
spectrum the mixing angles could be adjusted to obey the relevant
constraints from laboratory experiments, mainly due to the
non-observation of reactor $\bar{\nu}_e$ at short distances and a
reasonable description of the data could be
achieved~\cite{cpt3,solveig,strumia2}.  In general, stronger
constraints on the possibility of CPT violation arise, once a specific
source of CPT violation which involves other sectors of the theory is
invoked~\cite{cptirina,cptirina2}.

On pure phenomenological grounds, the first test of this scenario came
from the KamLAND experiment since the suggested CPT-violating neutrino
spectrum allowed to reconcile the solar, atmospheric and LSND
anomalies, but, once the constraints from reactor experiments were
imposed, no effect in KamLAND was predicted. The observation of a
deficit in KamLAND at $3.5\sigma$ CL clearly disfavored these
scenarios.  Furthermore, KamLAND results demonstrate that
$\bar{\nu}_e$ oscillate with parameters consistent with the LMA
$\nu_e$ oscillation solution of the solar anomaly. This fact by itself
can be used to set constraints on the possibility of CPT
violation~\cite{cptjb,strumia2}.  

The present situation is that the results of solar experiments in
$\nu$ oscillations, together with the results from KamLAND and the
bounds from other $\bar\nu$ reactor experiments show that both
neutrinos and anti-neutrinos oscillate with $\Delta m^2_\odot,\,
\Delta m^2_\text{reac} \leq 10^{-3}~\eVq$. Adding this to the evidence
of oscillations of both atmospheric neutrinos and anti-neutrinos with
$\Delta m^2_\text{atm} \sim 10^{-3}~\eVq$, leaves no room for
oscillations with $\Delta m^2_\text{LSND}\sim 1~\eVq$. The obvious
conclusion then is that CPT violation could no-longer explain LSND and
perfectly fit all other data~\cite{strumia2}.

\begin{figure}\centering
    \includegraphics[width=3.5in]{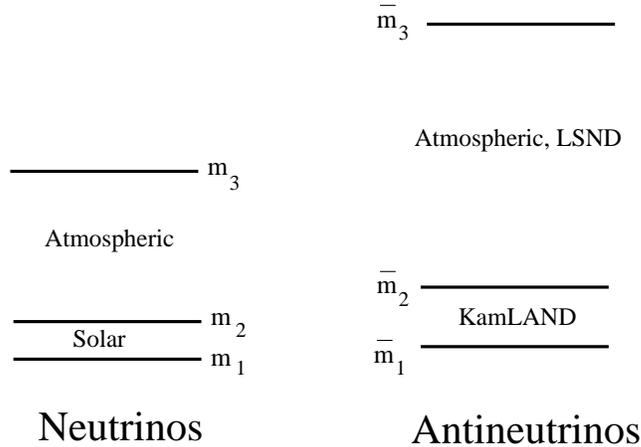}
    \caption{\label{fig:cpt-schemes}%
      Post KamLAND CPT violating neutrino mass spectrum proposed in
      Ref.~\cite{cpt5}.}
\end{figure}

This conclusion relies strongly on the fact that atmospheric
oscillations have been observed for both neutrinos and anti-neutrinos
with the same $\Delta m^2_\text{atm}$. However atmospheric neutrino
experiments do not distinguish neutrinos from anti-neutrinos, and
neutrinos contribute more than anti-neutrinos to the event rates by a
factor $\sim$ 4--2 (the factor decreases for higher energies). Based
on this fact, in Refs.~\cite{cpt5,strumia2} an alternative
CPT-violating spectrum was proposed as shown in
Fig.~\ref{fig:cpt-schemes}. In this scheme only atmospheric neutrinos
oscillate with $\Delta m^2_\text{atm}$ and give most of the
contribution to the observed zenith angular dependence of the deficit
of $\mu$-like events.  Atmospheric $\bar{\nu}_\mu$ dominantly
oscillate with $\Delta m^2_\text{LSND}$ which leads to an almost
constant (energy and angular independent) suppression of the
corresponding events.  For low $\bar{\nu}_\mu$ energies oscillations
with $\Delta m^2_\text{reac}$ can also be a source of zenith-angular
dependence. The claim in Ref.~\cite{cpt5} was that altogether this
suffices to give a good description of the atmospheric data such that
the scheme in Fig.~\ref{fig:cpt-schemes} could still be a viable
solution to all the neutrino puzzles. This conclusion was contradicted
in Ref.~\cite{strumia2} by an analysis of atmospheric and K2K data.
However, according to the authors in Ref.~\cite{cpt5} an important
point to their conclusion was the consideration of the full $3\nu$ and
$3\bar{\nu}$ oscillations, while the analysis in Ref.~\cite{strumia2}
was made on the basis of a $2\nu + 2\bar{\nu}$ approximation.

In Ref.~\cite{ourcpt} the status of the CPT violating scenario in
Fig.~\ref{fig:cpt-schemes} as explanation to the existing neutrino
anomalies was revisited and we summarize and update here their main 
conclusions.

One can label the states as in Fig.~\ref{fig:cpt-schemes} with $\Delta
m^2_{ij} = m^2_i - m^2_j$ and $\Delta\bar{m}^2_{ij} = \bar{m}^2_i -
\bar{m}^2_j$ and denote by $U$ and $\bar{U}$ the corresponding
neutrino and anti-neutrino mixing matrix (neglecting CP phases).

In the anti-neutrino sector the most constraining results of
$\bar\nu_e$ mixing comes from KamLAND, Bugey and CHOOZ. They provide
information on the $\bar{\nu}_e$ survival probability:
\begin{equation} \begin{split}
    \label{eq:preac}
    P_{ee}^\text{reac}
    &= 1-\bar{c}_{13}^4 \, \sin^22\bar{\theta}_{12}
    \, \sin^2\left(\frac{\Delta\bar{m}^2_{21} L}{4 E} \right) 
    \\
    & \hspace{5mm}
    \sin^22{\bar{\theta}_{13}} \left[ \bar{c}_{12}^2
    \, \sin^2\left( \frac{\Delta\bar{m}^2_{31} L}{4 E} \right)
    + \bar{s}_{12}^2
    \, \sin^2\left( \frac{\Delta\bar{m}^2_{32} L}{4 E} \right) \right]
    \\[3mm]
    & \simeq
    \begin{cases}
	1-\sin^22\bar{\theta}_{13}
	\sin^2\left( \dfrac{\Delta\bar{m}^2_{31}L}{4E} \right)
	& \text{for}\quad \Delta\bar{m}^2_{21} L/E \ll 1
	\\[4mm]
	\bar{s}_{13}^4 + \bar{c}_{13}^4\left[
	1-\sin^22\bar{\theta}_{12} \,
	\sin^2\left(\dfrac{\Delta\bar{m}^2_{21} L}{4E} \right)\right]
	& \text{for}\quad \Delta\bar{m}^2_{31} L/E \gg 1
    \end{cases}
\end{split} \end{equation}

In the scheme under consideration the probability associated with the
$\bar{\nu}_\mu \to \bar{\nu}_e$ signal in LSND is given by
\begin{equation}
    \label{eq:plsnd}
    P_\text{LSND} \equiv \sin^22\theta_\text{LSND}
    \sin^2 \left(\frac{\Delta m^2_\text{LSND} L}{4 E} \right)
    = \bar{s}_{23}^2 \sin^2 2\bar{\theta}_{13}
    \,\sin^2 \left(\frac{\Delta\bar{m}^2_{31} L}{4 E} \right) \,,
\end{equation}
where terms proportional to $\Delta\bar{m}^2_{21}$ which are
irrelevant for LSND distances and energies have been neglected.  In
Eq.~\eqref{eq:plsnd}: 
\begin{equation}
    \label{eq:defparam}
    \Delta m^2_\text{LSND} = \Delta\bar{m}^2_{31} \,,
    \qquad \sin^22\theta_\text{LSND} =
    \bar{s}_{23}^2 \, \sin^2 2\bar{\theta}_{13} \,.
\end{equation}
For the neutrino sector relevant information arises from solar
neutrino experiments and the K2K and MINOS LBL experiments. The
relevant probabilities are given in Eq.~\eqref{eq:p3} for solar
neutrinos and Eq.~\eqref{eq:P3atmmm} for K2K and MINOS.

Finally, the analysis of atmospheric neutrino data involves
oscillations of both neutrinos and anti-neutrinos, and, in the
framework of $3\nu+3\bar{\nu}$ mixing, matter effects become relevant
and its effect has to be quantified by numerically solving the
evolution equations for neutrinos and antineutrinos.

The basic approach to test the status of the scheme in
Fig.~\ref{fig:cpt-schemes} as a possible explanation of the LSND
anomaly together with all other neutrino and anti-neutrino oscillation
data is as follows. First, one performs a global analysis of all the
relevant data, but leaving out LSND data. The goal of this analysis is
to obtain the allowed ranges of parameters $\Delta m^2_\text{LSND}$
and $\sin^2 2\theta_\text{LSND}$ as defined in Eq.~\eqref{eq:defparam}
from this all-but-LSND data set. Then one compares these allowed
regions to the corresponding allowed parameter region from LSND, and
quantify at which CL both regions become compatible.

In this approach one starts by defining the most general $\chi^2$ for
the all-but-LSND data set:
\begin{multline}
    \label{eq:chi1}
    \chi^2_\text{all-but-LSND}(\Delta m^2_{21}, \Delta m^2_{31},
    \theta_{12}, \theta_{13}, \theta_{23} \,|\, \Delta\bar{m}^2_{21},
    \Delta\bar{m}^2_{31}, \bar{\theta}_{12}, \bar{\theta}_{13},
    \bar{\theta}_{23}) =
    \\
    \chi^2_\text{solar}(\Delta m^2_{21}, \theta_{12}, \theta_{13})
    + \chi^2_\text{K2K+MINOS}(\Delta m^2_{31}, \theta_{23}, \theta_{13})
    \\
    + \chi^2_\text{Bugey+CHOOZ+KLAND}(\Delta\bar{m}^2_{21},
    \Delta\bar{m}^2_{31}, \bar{\theta}_{12}, \bar{\theta}_{13})
    \\
    + \chi^2_\text{atmos}(\Delta m^2_{21}, \Delta m^2_{31}, \theta_{12},
    \theta_{23}, \theta_{13} \,|\, \Delta\bar{m}^2_{21},
    \Delta\bar{m}^2_{31}, \bar{\theta}_{12}, \bar{\theta}_{23},
    \bar{\theta}_{13}) \,.
\end{multline}
In order to test the status of the CPT interpretation of the LSND
signal using data independent of the ``tension'' between LSND and
KARMEN results~\cite{church} the constraints from the non-observation
of $\bar{\nu}_\mu \to \bar{\nu}_e$ transitions at KARMEN have not been
included.  

Using all the data described above except from the LSND experiment we
find the following all-but-LSND best fit point:
\begin{equation} \begin{aligned}
    \label{eq:bestfit}
    \;\Delta m^2_{21}\; & = 6.8\times 10^{-5}~\eVq \,, \qquad
    & \;\Delta\bar{m}^2_{21}\; &= 7.6\times 10^{-5}~\eVq \,,
    \\
    |\Delta m^2_{31}| &= 2.4\times 10^{-3}~\eVq \,, \qquad
    & |\Delta\bar{m}^2_{31}| &= 2.2 \times 10^{-3}~\eVq \,,
    \\
    \sin^2\theta_{12} &= 0.30 \,, \qquad
    & \sin^2\bar{\theta}_{12} &= 0.36 \,,
    \\
    \sin^2\theta_{13} &= 0 \,, \qquad
    & \sin^2\bar{\theta}_{13} &= 0 \,,
    \\
    \sin^2\theta_{23} &= 0.5 \,, \qquad
    & \sin^2\bar{\theta}_{23} &= 0.5 \,. 
\end{aligned} \end{equation}
This can be directly compared to the corresponding analysis in the CPT
conserving scenario:\footnote{Note that for consistency with the
CPT-violating analysis here we have neglected the effects of the
finite solar mass splitting on the atmospheric data. This explains why
in Eq.~\eqref{eq:best_cptcons} we find maximal $\theta_{23}$ mixing,
in contrast with Eq.~\eqref{eq:3nuranges} and the discussion in
Sec.~\ref{sec:3nu}.}
\begin{equation} \begin{aligned}
    \label{eq:best_cptcons}
    \;\Delta m^2_{21}\; = \;\Delta\bar{m}^2_{21}\; &= 7.7\times 10^{-5}~\eVq \,,
    \\
    |\Delta m^2_{31}| = |\Delta\bar{m}^2_{31}| &= 2.4\times 10^{-3}~\eVq \,,
    \\
    \sin^2\theta_{12} = \sin^2\bar{\theta}_{12} &= 0.32 \,,
    \\
    \sin^2\theta_{13} = \sin^2\bar{\theta}_{13} &= 0 \,,
    \\
    \sin^2\theta_{23} = \sin^2\bar{\theta}_{23} &= 0.5 \,.
\end{aligned} \end{equation}
Thus allowing for different mass and mixing parameters for neutrinos
and anti-neutrinos, all-but-LSND data choose a best fit point very
close to CPT conservation and maximal $23$ mixing.

\begin{figure}\centering
    \includegraphics[width=5.3in]{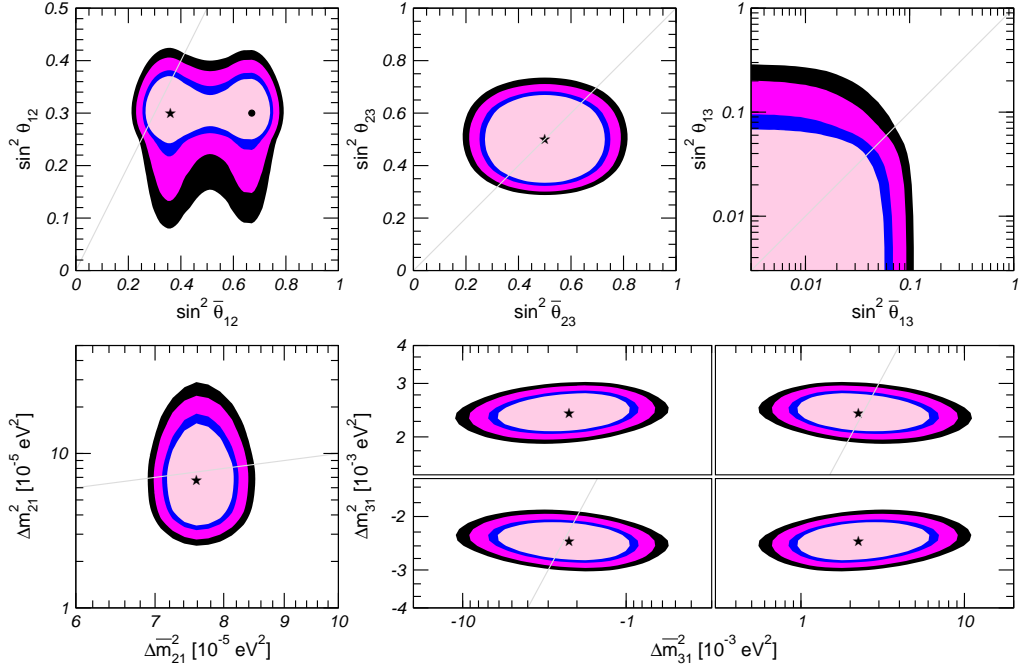}
    \caption{\label{fig:cpt-limits}%
      Allowed regions for neutrino and anti-neutrino mass splittings
      and mixing angles in the CPT violating scenario.  Different
      contours correspond to the two-dimensional allowed regions at
      90\%, 95\%, 99\% and $3\sigma$ CL. The best fit point is marked
      with a star.}
\end{figure}

Next we illustrate the amount of CPT violation which is still viable.
In order to do so we plot in Fig.~\ref{fig:cpt-limits} the allowed
regions for the neutrino and anti-neutrino mass splittings and mixing
angles. From the figure we find that present data constraints all
mixing angles and absolute values of mass-squared differences to be
close to the CPT conserving case. However, as seen in the four lower
left panels, nothing is known about the relative ordering of neutrinos
versus antineutrino states.  CPT scenarios in which neutrinos states
are normally ordered while antineutrino states are inversely ordered
(or vice-versa) cannot be discriminated from the CPT conserving case.

\begin{figure}\centering
    \includegraphics[width=3.5in]{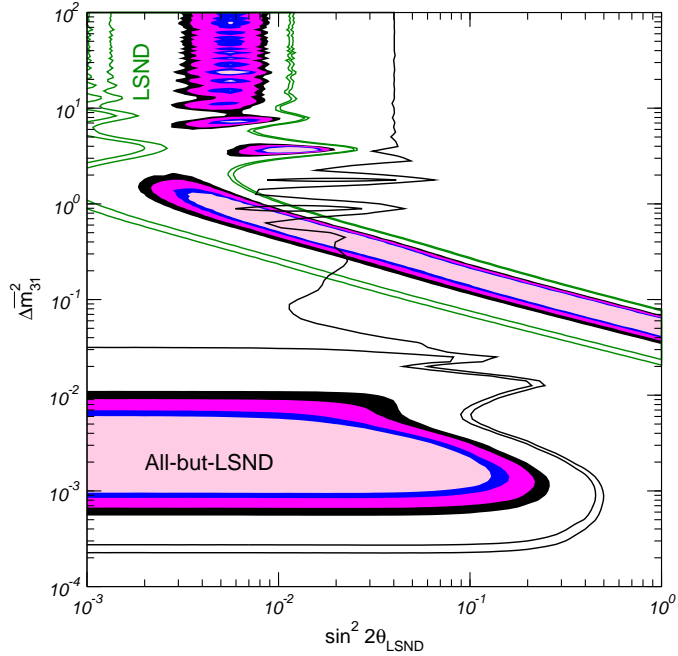}
    \caption{\label{fig:cpt-status}%
      90\%, 95\%, 99\%, and $3\sigma$ CL allowed regions (filled) in
      the ($\Delta\bar{m}^2_{31} = \Delta m^2_\text{LSND}$, $\sin^2
      2\theta_\text{LSND}$) plane required to explain the LSND signal
      together with the corresponding allowed regions from our global
      analysis of all-but-LSND data. The contour lines correspond to
      $\Delta\chi^2 = 26$ and $30$ ($4.7\sigma$ and $5.1\sigma$,
      respectively).}
\end{figure}

Concerning LSND, the results show that values of $\Delta\bar{m}^2_{31}
= \Delta\bar{m}^2_\text{LSND}$ large enough to fit the LSND result do
not appear as part of the $3\sigma$ CL allowed region of the
all-but-LSND analysis which is bounded to $\Delta\bar{m}^2_{31} <
0.01~\eVq$.  The upper bound on $\Delta\bar{m}^2_{31}$ is determined
by atmospheric neutrino data.  It is clear from these results that the
CPT violation scenario cannot give a good description of the LSND data
and simultaneously fit all-but-LSND results. The quantification of
this statement is displayed in Fig.~\ref{fig:cpt-status} where we show
the allowed regions in the ($\Delta\bar{m}^2_{31} = \Delta
m^2_\text{LSND}$, $\sin^2 2\theta_\text{LSND}$) plane required to
explain the LSND signal together with the corresponding allowed
regions from the global analysis of all-but-LSND data.

Fig.~\ref{fig:cpt-status} illustrates that below $3\sigma$ CL there is
no overlap between the allowed region of the LSND analysis and the
all-but-LSND one, and that for this last one the region is restricted
to $\Delta\bar{m}^2_{31} = \Delta m^2_\text{LSND}<0.01~\eVq$.  Only
for $\Delta\chi^2 \gtrsim 25$ values of $\Delta\bar{m}^2_{31}\sim
\mathcal{O}(\eVq)$ become allowed for the all-but-LSND region --~as
determined mainly by the constraints from Bugey~-- and at the
$4.6\sigma$ level ($\chi^2_\text{PC} = 25.2$) an agreement with LSND
becomes possible.

The information most relevant to this conclusion comes from the
atmospheric neutrino events. Within the constraints imposed by solar
and LBL neutrino data, and reactor anti-neutrino experiments,
atmospheric data is precise enough to be sensitive to anti-neutrino
oscillation parameters and cannot be described with oscillations with
the wavelengths required in the CPT violating scenario.


\subsection{Other Extensions}
\label{sec:lsndothers}

Further extensions of the $3\nu$ mixing scenario have been proposed in
the literature with the aim of accommodating the LSND
observation~\cite{4cpt, 4mavas, nudec1, 4nudec, 4nubrane, 3nudec1,
3nudec2}. We briefly summarize the basic features of some of these
proposals.

In Ref.~\cite{4cpt} it is examined the possibility that a 
four-neutrino scenario with $CPT$ violation can explain the LSND
effect and remain consistent with all other data. The authors find
that models with a (3+1) mass structure in the neutrino sector are
viable with the third mass difference between $1-7~\eVq$ while a (2+2)
structure is permitted only in the antineutrino sector and the third
mass difference can extend to all the ranges allowed by LSND, KARMEN
and Bugey. The non-observation of a signature at MiniBooNE in the
neutrino channel cannot rule out any of these models. The observation
or non-observation of a signal in MiniBooNE in the antineutrino mode
could discriminate between these two scenarios.

The proposal that mass-varying neutrinos could provide an explanation
for the LSND signal for $\bar\nu_\mu \to \bar\nu_e$ oscillations was
studied in Ref.~\cite{4mavas}. It is based on the observation of
Ref.~\cite{dark2} that all positive oscillation signals mostly occur
in matter and therefore if there is a contribution to the neutrino
mass depending on the matter density, the effective mass differences
in the different experiments can be different, a fact that can help in
accommodating the LSND results.  They find that three active
mass-varying neutrinos are insufficient to describe all existing
neutrino data including LSND. But mass-varying neutrinos with a (3+1)
mass structure in the neutrino sector can work with a small
active-sterile mixing which is generated solely by MaVaN effects.
These scenarios predict a null MiniBooNE (and are therefore still
viable), a null result at Double-CHOOZ, but positive signals for
underground reactor experiments and for $\nu_\mu \to \nu_e$
oscillations in long-baseline experiments.

Explanations of the LSND evidence for electron antineutrino appearance
based on neutrino decay have been also studied in the
literature~\cite{nudec1,4nudec}. In the most updated of this
proposals~\cite{4nudec} a fourth heavier sterile neutrino is
introduced with a (3+1) mass structure. Such neutrino is produced in
pion and muon decays because of a small mixing with muon neutrinos,
and then decays into a scalar particle and a light neutrino,
predominantly of the electron type giving the observed signature at
LSND. The basic point is that the different $L/E$ dependence of the
flavor conversion probabilities for this scenario compared to the
oscillation case allows for a larger range of compatibility between
the LSND observation and the null result at KARMEN and other SBL
experiments.  In the minimal version, this decay model predicted a
signal in the MiniBooNE experiment corresponding to a transition
probability of the same order as seen in LSND and it is therefore
disfavored by the negative result of MiniBooNE. However, in
Ref.~\cite{4nudec} it was noted that the inclusion of a second sterile
neutrino would open up the possibility of CP-violating effects, in a
similar way to what discussed in Sec.~\ref{sec:fivemix}, hence
potentially allowing to reconcile LSND and MiniBooNE.

The possibility of a new resonance in active-sterile neutrino
oscillations arising in theories with large extra dimensions is
studied in Ref.~\cite{4nubrane}. In these scenarios the fluctuations
in the brane effectively increase the path-length of active neutrinos
relative to the path-length of sterile neutrinos through the
extra-dimensional bulk.  This imparts an energy dependence to the
oscillation amplitude which can lead to the resonant enhancement of
active-sterile neutrino mixing. For energies below the resonance, the
flavor conversion probabilities take the standard oscillation form
while above the resonance, active-sterile oscillations are suppressed.
The authors find that if the resonant energy lies in the range 30~MeV
to 400~MeV, suitably chosen between the BUGEY and CDHS energies, then
all neutrino oscillation data can be accommodated in a consistent 
(3+1) neutrino framework. Such an energy range corresponds to brane
fluctuations with a height to width ratio of $\sim 10^{-8}$. The
resonant energy might be identifiable in either the LSND spectral data
and the muon neutrino disappearance from a stopped-pion source, or, as
a matter of fact, in the MiniBooNE data itself. 

In Ref.~\cite{3vli} it was postulated the existence of
Lorentz-violating, CPT-conserving interactions which could allow
three-neutrino solutions to the LSND anomaly that are also consistent
with all other neutrino data.  They found that a highly non-trivial
energy dependence of the Lorentz-violating interactions is required to
accommodate all the data.  The non observation of a signal at
MiniBooNE imposes even stronger constrains on the energy dependence of
these interactions. 

%% file: sec.conclu.tex
\section{Summary and Conclusions}
\label{sec:conclu}

In this review we have presented the progress on some fronts of the
phenomenology of massive neutrinos. The present experimental situation
concerning the searches for neutrino flavor mixing with 
respectively, solar, atmospheric, reactor and accelerator neutrinos
was presented in Sec.~\ref{sec:expe}. From this data we have learned
that:
\begin{itemize}
  \item Solar $\nu_e's$ convert to $\nu_{\mu}$ or $\nu_\tau$ with 
    confidence level (CL) of more than $7\sigma$.
    
  \item KamLAND find that reactor $\bar{\nu}_e$ disappear over
    distances of about 180 km and they observe a distortion of their 
    energy spectrum. Altogether their evidence has more than $3\sigma$
    CL.
    
  \item The evidence of atmospheric (ATM) $\nu_\mu$  disappearing is
    now at $> 15 \sigma$, most likely converting to $\nu_\tau$.
    
  \item K2K observe the disappearance of accelerator $\nu_\mu$'s at
    distance of 250 km and find a distortion of their energy spectrum 
    with a CL of 2.5--4$\sigma$.
    
  \item MINOS observes the disappearance of accelerator $\nu_\mu$'s at
    distance of 735 km and find a distortion of their energy spectrum 
    with a CL of $\sim 5\sigma$.
    
  \item LSND found evidence for $\bar{\nu}_\mu \rightarrow
   \bar{\nu}_e$. This evidence has not been confirmed by the recent
   MiniBooNE search for $\nu_\mu\rightarrow \nu_e$ at 98\% CL.

\end{itemize}
These results imply that neutrinos are massive and the Standard Model
has to be extended at least to include neutrino masses. 

In Sec.~\ref{sec:theory} we have presented the low energy formalism
for adding neutrino masses to the SM which would allow for the
observed leptonic mixing and we have described the phenomenology
associated with neutrino oscillations in vacuum and in matter needed
for the interpretation of the data. 

The minimum joint description of solar, atmospheric, long baseline and
reactor data requires the mixing of the three known neutrinos. In
Sec.~\ref{sec:3nu} we have presented an update of the three-neutrino
oscillation interpretation of the existing bulk of neutrino data (with
the exception of the LSND result). The derived ranges for the two mass
differences at $1\sigma$ ($3\sigma$) are:
\begin{equation} \begin{aligned}
    \Delta m^2_{21}
    &= 7.67 \,_{-0.21}^{+0.22} \,\left(_{-0.61}^{+0.67}\right)
    \times 10^{-5}~\eVq \,,
    \\
    \Delta m^2_{31} &=
    \begin{cases}
	-2.37 \pm 0.15 \,\left(_{-0.46}^{+0.43}\right)
	\times 10^{-3}~\eVq & \text{(inverted hierarchy)} \,,
	\\[1mm]
	+2.46 \pm 0.15 \,\left(_{-0.42}^{+0.47}\right)
	\times 10^{-3}~\eVq & \text{(normal hierarchy)}
    \end{cases}
\end{aligned} \end{equation}
while the leptonic mixing matrix at the $3\sigma$ level is:
\begin{equation}
    |U|_{3\sigma} =
    \begin{pmatrix} 
	0.77 \to 0.86 & ~\quad~ 0.50 \to 0.63 & ~\quad~ 0.00 \to 0.22 \\
	0.22 \to 0.56 & ~\quad~ 0.44 \to 0.73 & ~\quad~ 0.57 \to 0.80 \\
	0.21 \to 0.55 & ~\quad~ 0.40 \to 0.71 & ~\quad~ 0.59 \to 0.82
    \end{pmatrix} \,.
\end{equation}
This minimal picture of three massive neutrinos, although can give a
satisfactory description of the data is still incomplete. We have no
direct evidence of the mixing angle $\theta_{13}$, we ignore if CP is
a symmetry of the leptonic sector, we do not know the ordering of the
mass states, and most importantly we do not know if neutrinos are
Majorana or Dirac particles.  

Also, although oscillations have allowed us to establish that
neutrinos have mass they only provide a lower bound on the heaviest
neutrino mass, but give no information on the mass of the lightest
state which sets the absolute neutrino mass scale. As described in
Sec.~\ref{sec:numass} at present, the most precise and model
independent bound on scale of neutrino masses still arises from the
non observation of any distortion at the end-point of the energy
spectrum of the electrons emitted in tritium beta decay which imply
that $\sqrt{\sum_i m^2_i |U_{ei}|^2} <2.2~\text{eV}$ at 95\%
confidence level. 

We have argued and show how, despite all these limitations, the
attained precision in the observed signals is already good enough to
allow us the use of the existing data to probe physics beyond neutrino
masses and mixing. In Sec.~\ref{sec:fluxes} we have shown how the
independent determination of the relevant neutrino oscillation
parameters from non-solar and non-atmospheric neutrino experiments
makes it possible to extract the solar and atmospheric neutrino fluxes
directly from the corresponding neutrino data with an accuracy
comparable, to that of the theoretical predictions from the solar
model and from atmospheric flux calculations.

Using the good description of neutrino data in terms of neutrino
oscillations, it is also possible to constraint other exotic forms of
new physics as shown in Secs.~\ref{sec:npatm} and~\ref{sec:npsolar}.
We have presented updated bounds on the violation of some fundamental
symmetries like Lorentz Invariance, the Equivalence Principle and CPT.
These forms of new physics, if non-universal, can also induce neutrino
flavor oscillations whose main differentiating characteristic is a
different energy dependence of the oscillation wavelength.  We have
also discussed the present constraints on the presence of non-standard
neutrino interactions affecting the solar and atmospheric neutrino
propagation in matter as well as on effects associated with mass
varying neutrinos models. Under certain assumptions these results
represent the most stringent bounds on these forms of new physics.

Another hint for neutrino masses came from the LSND experiment.  The
simplest interpretation of the LSND data is that there are
$\nu_e\to\nu_\mu$ oscillations with $\Delta m^2_\text{LSND} =
\mathcal{O}(1~\eVq)$ and $\sin^22\theta_\text{LSND} =
\mathcal{O}(0.003)$.

The fact that $\Delta m^2_\text{LSND} \gg \Delta m^2_{31},\, \Delta
m^2_{21}$ means that this result cannot be included within the
framework of oscillations among the three active neutrinos alone. 
Therefore a further extension of the SM model is needed.  We present
in Sec.~\ref{sec:lsnd} the updated status of some extensions including
a fourth sterile neutrino or three neutrinos with violation of CPT
symmetry. The conclusion is that these extensions are either ruled out
or strongly disfavor as possible explanations of the LSND signal.

At the time of the finishing of this review the results of the
MiniBooNE experiment, which was especially designed to make a
conclusive statement about the LSND's neutrino oscillation evidence,
were made public.  In their search for $\nu_\mu \to \nu_e$ they found
no evidence of the expected signal corresponding to the LSND evidence,
thus excluding the LSND claim at 98\% CL in the 2-$\nu$ oscillation
framework.  This result puts further constrains on the possible
mixings of an additional sterile neutrino, while in principle could be
accommodated together with the LSND evidence if CPT was violated.
However CPT violation is independently excluded well beyond $3\sigma$
as explanation to the LSND data because (i) KamLAND finds that reactor
$\bar{\nu}_e$ oscillate with wavelength and amplitude in good
agreement with the expectations from the LMA solution of the solar
$\nu_e$, and (ii) both ATM neutrinos and antineutrinos have to
oscillate with similar wavelengths and amplitudes to explain the ATM
data.

In summary, neutrino physics has provided us with the first evidence
of physics beyond the standard model, it has allowed us to test many
theoretical ideas and given experimental guidance in the construction
of new ones. But many questions remain open and the only way to answer
them is by new experiments, some of which are, fortunately, on the
way.

%% file: sec.appatm.tex
\section{Atmospheric Neutrino Analysis}
\label{sec:appatm}

In this appendix we describe and update the details entering into the
simulation of the atmospheric neutrino event rates of Super-Kamiokande 
used in the analysis presented in this review as well as the
corresponding statistical treatment. 


\subsection{Event Rates}

Underground experiments can record atmospheric neutrinos by direct
observation of their charged current interaction inside the detector.
These so-called contained events can be classified into \emph{fully
contained} events, when the charged lepton (either electron or muon)
produced by the neutrino interaction does not escape the detector, and
\emph{partially contained muons}, when a muon is produced inside but
then leaves the detector. In the simulation of Super--Kamiokande used
in this review the fully contained events are further divided into
three data samples, based on the energy of the charged lepton:
\begin{itemize}
  \item $\text{sub-GeV}_\text{low}$, with lepton momentum below
    400~MeV;
  \item $\text{sub-GeV}_\text{high}$, with higher lepton momentum but
    visible energy below 1.33~GeV;
  \item $\text{multi-GeV}$, with visible energy above this cutoff.
\end{itemize}
In addition, both fully and partially contained events are also
divided into 10 zenith bins, based on the reconstructed direction of
the charged lepton. Thus in this simulation there id a total of 70
data bins for contained events.

The expected number of events in each bin can be obtained as:
\begin{multline} \label{eq:contained}
    N_\text{bin}(\vec\omega) = n_t T \sum_{\alpha,\beta,\pm} 
    \int_0^\infty dh \int_{-1}^{+1} dc_\nu
    \int_{E_\text{min}}^\infty dE_\nu \int_{E_\text{min}}^{E_\nu} dE_l
    \int_{-1}^{+1} dc_a \int_0^{2\pi} d\varphi_a
    \\
    \frac{d^3 \Phi_\alpha^\pm}{dE_\nu \, dc_\nu \, dh}(E_\nu, c_\nu, h)
    \, P_{\alpha\to\beta}^\pm(E_\nu, c_\nu, h \,|\, \vec\omega)
    \, \left[ \frac{d^2\sigma_\beta^\pm}{dE_l \, dc_a}
    \, \pi_\text{ring} \right](E_\nu, E_l, c_a)
    \\
    \times \varepsilon_\beta^\text{bin}(E_l, c_l(c_\nu, c_a, \varphi_a))
    \,,
\end{multline}
where $P_{\alpha\to\beta}^+$ ($P_{\alpha\to\beta}^-$) is the
$\nu_\alpha \to \nu_\beta$ ($\bar{\nu}_\alpha \to \bar{\nu}_\beta$)
conversion probability for given values of the neutrino energy
$E_\nu$, the cosine $c_\nu$ of the angle between the incoming neutrino
and the vertical direction, and the production altitude $h$. In the
Standard Model one has $P_{\alpha\to\beta}^\pm = \delta_{\alpha\beta}$
for all $\alpha$ and $\beta$, whereas when neutrino oscillations are
considered the expression for $P_{\alpha\to\beta}^\pm$ depends also on
the conversion model and on its parameters $\vec\omega$.
In Eq.~\eqref{eq:contained} $n_t$ is the number of targets, $T$ is the
experiment running time, $\Phi_\alpha^+$ ($\Phi_\alpha^-$) is the flux
of atmospheric neutrinos (antineutrinos) of type $\alpha$ and
$\sigma_\beta^+$ ($\sigma_\beta^-$) is the charged-current neutrino-
(antineutrino-) nucleon interaction cross section.
The variable $E_l$ is the energy of the final lepton of type $\beta$,
while $c_a$ and $\varphi_a$ parametrize the opening angle between the
incoming neutrino and the final lepton directions as determined by the
kinematics of the neutrino interaction.
The factor $\pi_\text{ring}$ is introduced only for fully-contained
events, and accounts for the probability that the event is tagged as
single-ring.
Finally, $\varepsilon_\beta^\text{bin}$ gives the probability that a
charged lepton of type $\beta$, energy $E_l$ and direction $c_l$
contributes to the given bin.

In general, Sub-GeV events arise from neutrinos of several hundreds of
MeV, while multi-GeV and partially-contained events are originated by
neutrinos with energies of several GeV. Higher energy neutrinos and
anti-neutrinos of type $\mu$ can be detected indirectly by observing
the muons produced by charged-current interactions in the vicinity of
the detector: the so called \emph{upgoing muons}. If the muon stops
inside the detector, it will be called a ``stopping'' muon, while if
the muon track crosses the full detector the event is classified as a
``through--going'' muon. On average, stopping muons arise from
neutrinos with energies around ten GeV, while through--going muons are
originated by neutrinos with energies around hundred GeV.

Stopping and through--going muon samples are also divided into 10
zenith bins; however, only leptons coming from below the horizon are
considered, since downgoing events are dominated by the much higher
background of primary muons which penetrate the mountain above the
detector.  Again, the expected number of events in each bin is given
by:
\begin{multline} \label{eq:upgoing}
    N_\text{bin}(\vec\omega) = \rho_\text{rock} T \sum_{\alpha,\pm} 
    \int_0^\infty dh \int_{-1}^{+1} dc_\nu
    \int_{E_\text{min}}^\infty dE_\nu 
    \int_{E_\text{min}}^{E_\nu} dE^0_\mu \int_{E_\text{min}}^{E^0_\mu} dE^\text{fin}_\mu
    \int_{-1}^{+1} dc_a \int_0^{2\pi} d\varphi_a
    \\
    \frac{d^3 \Phi_\alpha^\pm}{dE_\nu \, dc_\nu \, dh}(E_\nu, c_\nu, h)
    \, P_{\alpha\to\mu}^\pm(E_\nu, c_\nu, h \,|\, \vec\omega)
    \, \frac{d^2\sigma_\mu^\pm}{dE^0_\mu \, dc_a}(E_\nu, dE^0_\mu, c_a)
    \\
    \times R_\text{rock}(E^0_\mu,E^\text{fin}_\mu)
    \, \mathcal{A}_\text{eff}^\text{bin}(E^\text{fin}_\mu,
    c_l(c_\nu, c_a, \varphi_a)) \,,
\end{multline}
where $\rho_\text{rock}$ is the density of targets in standard rock,
$R_\text{rock}$ is the effective muon range~\cite{ls} for a muon which
is produced with energy $E^0_\mu$ and reaches the detector with energy
$E^\text{fin}_\mu$, and $\mathcal{A}_\text{eff}^\text{bin}$ is the
effective area. The other variables and physical quantities are the
same as for contained events.

\subsubsection{Atmospheric neutrino fluxes}

The simulations used in this review use the latest three--dimensional
calculation of the atmospheric neutrino flux performed by Honda and
presented in Ref.~\cite{honda}. The triple-differential neutrino
fluxes which appear in Eqs.~\eqref{eq:contained}
and~\eqref{eq:upgoing} can be written as:
\begin{equation}
    \frac{d^3 \Phi_\alpha^\pm}{dE_\nu \, dc_\nu \, dh}(E_\nu, c_\nu, h)
    \equiv
    \kappa_\alpha^\pm(E_\nu, c_\nu, h)
    \int_0^{2\pi} 
    \frac{d^3 \Phi_\alpha^\pm}{dE_\nu \, dc_\nu \, d\varphi_\nu}
    (E_\nu, c_\nu, \varphi_\nu) \, d\varphi_\nu \,,
\end{equation}
where the first term is the altitude distribution, normalized to 1,
and the second term is the integral over the azimuth angle of the
neutrino flux. Both quantities are described in detail in
Ref.~\cite{honda}, and are available as data tables in the authors'
web page.

Note that the neutrino fluxes, in particular in the sub-GeV range,
depend considerably on the solar activity.  In order to take this fact
into account in the simulation averaged neutrino flux are used and
they are defined as follows,
\begin{equation}
    {\Phi}_\alpha^\pm
    \equiv c_\text{max} \Phi_{\alpha,\pm}^\text{max} 
    + c_\text{min} \Phi_{\alpha,\pm}^\text{min} \;, 
\end{equation}
where $\Phi_{\alpha,\pm}^{max}$ and $\Phi_{\alpha,\pm}^{min}$ are the
atmospheric neutrino fluxes when the Sun is most active (solar
maximum) and quiet (solar minimum), respectively.  The coefficients
$c_\text{min}$ and $c_\text{max} \equiv (1 - c_\text{min})$ are
determined according to the running period of each experiment,
assuming that the flux changes linearly with time between solar
maximum and minimum.  For Super-Kamiokande phase I we use
$c_\text{min} = 47\%$ and $c_\text{max} = 53\%$.

\subsubsection{Cross sections and single-ring acceptances}

One important ingredient in the calculation of the expected rates is
the charged-current neutrino-nucleon interaction cross section,
$\sigma_\text{CC}$.  In order to determine accurately the expected
event rates for the various data samples, the contributions to the
cross section from the exclusive channels of lower multiplicity,
quasi-elastic (QE) scattering and single pion (1$\pi$) production are
considered separately, and additional channels are included as part of
the deep inelastic (DIS) cross
section~\cite{LlewellynSmith:1971zm,Lipari:1994pz}:
\begin{equation}
    \sigma_\text{CC} = \sigma_\text{QE}
    + \sigma_{1\pi} + \sigma_\text{DIS} \,.
\end{equation}
For fully-contained events, each exclusive channel is further
multiplied by the corresponding probability that the event will be
tagged as single-ring:
\begin{equation} 
    \sigma_\text{CC} \, \pi_\text{ring}
    = \sigma_\text{QE} \, \pi_\text{QE}^\text{ring}
    + \sigma_{1\pi} \, \pi_{1\pi}^\text{ring}
    + \sigma_\text{DIS} \, \pi_\text{DIS}^\text{ring} \,.
\end{equation}

Details of the values used in the calculations for the QE and DIS 
cross sections can be found in Ref.~\cite{review}.
For single pion production the present simulation uses the model of
Fogli and Nardulli~\cite{Fogli:1979cz} which includes hadronic masses
below $W=1.4$~GeV. In order to correctly account for the finite
scattering angle between the incoming neutrino and the final lepton,
it is assumed that the whole process occur via a $\Delta$-resonance,
and then one can write:
\begin{equation}
    \frac{d^2 \sigma_{1\pi}}{dE_l \, dc_a} =
    \left( \frac{d\sigma_{1\pi}}{dE_\nu} \right) \, 
    \frac{N(E_\nu, E_l)}{(W_c - M_\Delta)^2 + \Gamma_\Delta^2 / 4} \,,
\end{equation}
where $(d\sigma_{1\pi} / dE_\nu)$ is taken from
Refs.~\cite{Fogli:1979cz,Nakahata:1986zp}, $W_c = \sqrt{m_N^2 + 2 m_N
(E_\nu - E_l) + q^2}$ is the invariant mass of the final hadronic
system, $q^2 = m_l^2 - 2 E_\nu (E_l - p_l c_a)$ is the momentum
transfer, and $N(E_\nu, E_l)$ is a normalization factor which ensures
that the integral of the Breit-Wigner factor over the physical region
is always one.

Concerning the single-ring acceptances:
\begin{itemize}
  \item $\pi_\text{QE}^\text{ring}$ is shown in Fig.~6.7 of
    Ref.~\cite{Kameda} and in Fig.~5.11 of Ref~\cite{Ishitsuka}. We
    approximate these plots by $0.96 / (1 + 0.03 p_l^{1.38})$ for
    $e$-like events and by the constant value $0.97$ for $\mu$-like
    events.
    
  \item $\pi_{1\pi}^\text{ring}$ is constructed as the probability that
    the energy of the final pion is larger than $m_\pi
    \gamma_\text{cut}$. This choice is motivated by the fact that a
    very energetic $\pi^\pm$ is more likely to produce a visible track
    (hence leading to a multi-ring event) than a low energy one. After
    comparison with the SK acceptances, kindly provided to us by the
    Super-Kamiokande collaboration, we choose the empirical value
    $\gamma_\text{cut} = 2.1$.
    
  \item $\pi_\text{DIS}^\text{ring}$ is constructed in a similar way,
    as the probability that the energy of \emph{all} the pions in the
    final state is larger than $n_\pi m_\pi \gamma_\text{cut}$. The
    mean multiplicity of pions, $n_\pi$, is estimated from the result
    of Fermilab 15-foot hydrogen bubble chamber
    experiment~\cite{Barish}. Again we choose $\gamma_\text{cut} =
    2.1$.
\end{itemize}
Although we are aware that this construction is nothing more than a
toy model, it allows us to express the single-ring cut in terms of the
\emph{relevant physical quantities}, \textit{i.e.} the kinematic
variables of the hadronic system. It is impressive that with the
\textit{ad-hoc} choice of a single parameter, $\gamma_\text{cut}$, we
can reproduce with good accuracy the SK acceptances (which are
presented as functions of the neutrino energy), the momentum
distributions of single-ring events shown in Fig.~6.3 of
Ref.~\cite{Ishitsuka}, and the fractions of neutrino interaction modes
given in Table 6.2 of the same reference.

\subsubsection{Detector efficiencies, effective muon range and effective
  area}

The final necessary ingredient for the calculation of the expected rates
for contained events is the detector efficiencies,
$\varepsilon_\beta^\text{bin}$, for each data sample and zenith bin.
The single-ring cut has already been discussed in the previous
section. For the remaining cuts, we write:
\begin{equation}
    \varepsilon_\beta^\text{bin}(E_l, c_l)
    = \varepsilon_\beta^\text{thres}(E_l)
    \; \varepsilon_\beta^\text{cont}(E_l, c_l)
    \; \varepsilon_\text{bin}^\text{zen}(c_l)
\end{equation}
where:
\begin{itemize}
  \item $\varepsilon_\beta^\text{thres}(E_l)$ accounts for the cuts on
    lepton momentum and visible energy: $100~\text{MeV} < p_l <
    400~\text{MeV}$ for $\text{sub-GeV}_\text{low}$ electrons,
    $200~\text{MeV} < p_l < 400~\text{MeV}$ for
    $\text{sub-GeV}_\text{low}$ muons, $400~\text{MeV} < p_l <
    1.2~\text{GeV}$ for $\text{sub-GeV}_\text{high}$ events, and $p_l
    > 1.2~\text{GeV}$ for $\text{multi-GeV}$ events. No cut of this
    type is needed for partially-contained events. The cuts at
    $1.2~\text{GeV}$ on lepton momentum nicely mimic the cut at
    $1.33~\text{GeV}$ on visible energy introduced by the
    Super-Kamiokande collaboration to separate sub-GeV and multi-GeV
    events. Following Ref.~\cite{skatmlast} we take into account the
    finite momentum resolution: $0.6\% + 2.6\% \big/
    \sqrt{p_l~[\text{GeV}/c]}$ for single-ring electrons and $1.7\% +
    0.7\% \big/ \sqrt{p_l~[\text{GeV}/c]}$ for single-ring muons.
    
  \item $\varepsilon_\beta^\text{cont}(E_l, c_l)$ describes the
    probability that a muon of given energy and direction produces a
    fully-contained or a partially-contained event. For $e$-like
    events this function is identically one.
    
  \item $\varepsilon_\text{bin}^\text{zen}(c_l)$ gives the probability
    that the final lepton contributes to the zenith bin under
    consideration. The division into zenith bins is performed
    according to the cosine $c_l$ of the angle between the charged
    lepton trajectory and the vertical direction. This angle is
    related to the neutrino direction $c_\nu$ and the scattering angle
    $\{ c_a,\, \varphi_a \}$ by the relation $c_l = c_\nu c_a - s_\nu
    s_a \cos\varphi_a$. Also in this case we take into account the
    finite angular resolution of the detector: $3.0^\circ$ for
    single-ring electrons, $1.8^\circ$ for single-ring FC muons,
    $2.8^\circ$ for partially-contained muons, and $1.0^\circ$ for
    upgoing events.
\end{itemize}

Concerning the calculation of upgoing events, the effective muon range
$R_\text{rock}$ which appears in Eq.~\eqref{eq:upgoing} is defined in
analogy with Eq.~(23) of Ref.~\cite{ls}:
\begin{equation}
    R_\text{rock}(E^0_\mu,E^\text{fin}_\mu) = \int_0^\infty
    F_\text{rock}(E^0_\mu,E^\text{fin}_\mu,L) \, dL
\end{equation}
where $F_\text{rock}(E^0_\mu,E^\text{fin}_\mu,L)$ gives the energy
distribution for a muon initially produced with energy $E^0_\mu$ after
traveling a distance $L$ in rock. The effective area
$\mathcal{A}_\text{eff}^\text{bin}$ is defined as:
\begin{align}
    \label{eq:area_thru}
    \mathcal{A}_\text{eff}^\text{thru}(E_\mu,c_l)
    &= \int_{L_\text{min}}^\infty
    S_\text{water}(E_\mu,L)
    \, \frac{dA_\text{SK}}{dL}(L,c_l) \, dL \,,
    \\
    \begin{split} \label{eq:area_stop}
	\mathcal{A}_\text{eff}^\text{stop}(E_\mu,c_l)
	&= \int_{L_\text{min}}^\infty
	\big[ S_\text{water}(E_\mu,L_\text{min})
	- S_\text{water}(E_\mu,L) \big]
	\, \frac{dA_\text{SK}}{dL}(L,c_l) \, dL
	\\
	~ &= S_\text{water}(E_\mu,L_\text{min})
	\, A_\text{SK}(L_\text{min},c_l)
	- \mathcal{A}_\text{eff}^\text{thru}(E_\mu,c_l) \,,
    \end{split}
\end{align}
where $A_\text{SK}(L,c_l)$ is the projected area of the detector that
corresponds to trajectories with internal path length longer than $L$,
as given in Eq.~(13) of Ref.~\cite{Lipari:1998rf}, and
\begin{equation}
    S_\text{water}(E_\mu,L) = \int_0^\infty
    F_\text{water}(E_\mu,E',L) \, dE' \leq 1
\end{equation}
is the probability that a muon with initial energy $E_\mu$ travels for
at least a distance $L$ in water before losing all its energy. For
Super-Kamiokande the minimum track length required to trigger an event
is $L_\text{min} = 7$~m. These expressions generalize those introduced
in Ref.~\cite{ourold1,ourold2}, where the statistical fluctuations
during muon propagation in the Earth were neglected. The old formulas
can be recovered from Eqs.~\eqref{eq:area_thru}
and~\eqref{eq:area_stop} by setting $S_\text{water}(E_\mu,L) =
\theta[L_\text{path}(E_\mu) - L]$. More details on our calculations of
muon propagation in matter can be found in Ref.~\cite{ouricecube}.

\subsubsection{Event Distributions}

\begin{figure}
    \includegraphics[width=0.95\textwidth]{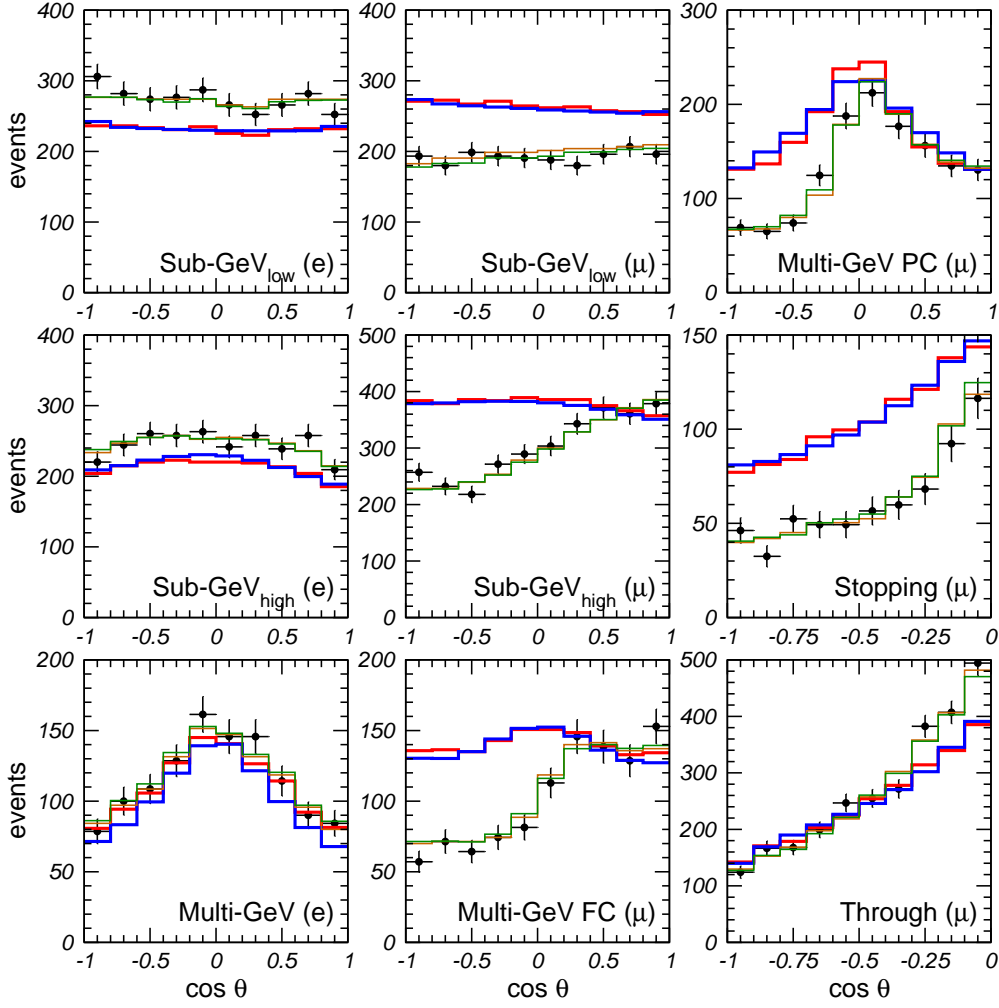}
    \caption{\label{fig:two-compare}%
      Comparison of our calculations with Super-Kamiokande ones. The
      blue and green lines show our predictions for no-oscillations
      and for $\Delta m^2 = 2.5\times 10^{-3}~\text{eV}^2$ and $\theta
      = 45^\circ$, respectively. The red and orange lines are the
      corresponding calculations of the Super-Kamiokande
      collaborations. Black crosses show the experimental results with
      their statistical errors.}
\end{figure}

In order to verify the quality of the simulation, in
Fig.~\ref{fig:two-compare} we compare the simulated predictions of the
event number for the different SK data samples with those of the
Super-Kamiokande collaboration, taken from Ref.~\cite{skII-litch}. The
blue line gives the expected number of events for no oscillations, and
should be compared with the red line. As can be seen, the simulation
used in this review agrees quite well with those of the SK
collaboration for all the data samples, with the exception of some
small deformations in multi-GeV $e$-like events and in
partially-contained events. In order to compensate for these
discrepancies, we rescale our Monte-Carlo for no-oscillations so to
match exactly the Super-Kamiokande prediction. After this correction,
we turn to the oscillation case and we plot the expected number of
events for $\Delta m^2 = 2.5\times 10^{-3}~\text{eV}^2 $ and $\theta =
45^\circ$ (green line). This line is in excellent agreement with the
Super-Kamiokande expectations (orange line), which we take from
Ref.~\cite{skII-litch}.


\subsection{Details of the $\chi^2$ calculation}

The basic idea of the pull method consists in parametrizing the
systematic errors and the theoretical uncertainties in terms of a set
of variables $\{\xi_i\}$, called \emph{pulls}, which are then treated
on the same footing as the other parameters of the model. The $\chi^2$
function can be decomposed into the sum of two parts:
\begin{equation} \label{eq:chisq-full}
    \chi^2(\vec\omega, \vec\xi) = 
    \chi^2_\text{data}(\vec\omega, \vec\xi) +
    \chi^2_\text{pulls}(\vec\xi),
\end{equation}
where $\vec\omega$ denotes the parameters of the model,
$\chi^2_\text{data}$ is the usual term describing the deviation of the
experimental results from their theoretical predictions, and the extra
term $\chi^2_\text{pulls}$ provides proper penalties to account for
deviations of the systematics and the theoretical inputs from their
nominal value.

For the Super-Kamiokande experiment $\chi^2_\text{pulls}(\vec\xi)$ can
be properly written as a positive quadratic function of $\xi_i$.  It
is convenient to define the pulls in such a way that for each source
of systematics or theoretical input $i$ the value $\xi_i=0$
corresponds to the ``expected value'' reported by the collaboration or
predicted by the theory, and $\xi_i=\pm 1$ corresponds to a $1\sigma$
deviation. Since pulls describing different systematics and
theoretical inputs are assumed to be uncorrelated, the expression of
$\chi^2_\text{pulls}$ is very simple:
\begin{equation}
    \chi^2_\text{pulls}(\vec\xi) = 
    \sum_{i,\text{theory}} \xi_i^2 + \sum_{i,\text{syst}} \xi_i^2.
\end{equation}

The form of $\chi^2_\text{data}$ depends on the expected distribution
of the experimental results.  The outcome of the Super-Kamiokande
experiment is the number of events observed in each energy and
zenith-angle bin, which follows a Poisson distribution. However, since
the number of events in each bin is large, $\chi^2_\text{data}$ can be
well approximated by a quadratic function of the differences between
observed and expected rates of events:
\begin{equation}
    \chi^2_\text{data}(\vec\omega, \vec\xi) = \sum_n
    \left(
	\frac{R_n^\text{th}(\vec\omega, \vec\xi)
	  - R_n^\text{ex}}{\sigma_n^\text{stat}}
    \right)^2
\end{equation}
where $R_n^\text{th}$ ($R_n^\text{ex}$) is the ratio between the
expected (observed) number of events and the theoretical Monte Carlo
for the case of no oscillations.  Note that the dependence of
$\chi^2_\text{data}$ on both the parameters $\vec\omega$ and the pulls
$\vec\xi$ is entirely through $R_n^\text{th}(\vec\omega, \vec\xi)$. In
the pull approach, $\vec\omega$ and $\vec\xi$ play a very similar
role, and in principle should be treated in the same way.  However,
for the Super-Kamiokande experiment the bounds on $\vec\xi$ implied by
$\chi^2_\text{pulls}$ are in general significantly stronger than those
implied by $\chi^2_\text{data}$, and is therefore a good approximation
to retain the dependence $\chi^2_\text{data}$ on $\vec\xi$ only to the
lowest orders. This is done by expanding $R_n^\text{th} (\vec\omega,
\vec\xi)$ in powers of $\xi_i$ up to the first order:
\begin{equation} \label{eq:approx}
  R_n^\text{th}(\vec\omega, \vec\xi) \approx
  R_n^\text{th}(\vec\omega)
  \left[ 1 + \sum_i \pi_n^i(\vec\omega)\, \xi_i \right],
  \quad \text{where} \quad 
  \left\{ \begin{aligned}
    R_n^\text{th}(\vec\omega) &\equiv R_n^\text{th}(\vec\omega, 0), \\[1mm]
    R_n^\text{th}(\vec\omega) \, \pi_n^i(\vec\omega) & \equiv
    \left. \frac{\partial R_n^\text{th}(\vec\omega,
	  \vec\xi)}{\partial \xi_i} \right|_{\vec\xi=0}.
    \end{aligned}\right.
\end{equation}
It is easy to prove~\cite{Fogli:2002pt} that under the
approximation~\eqref{eq:approx} the pull definition given
in~\eqref{eq:chisq-full} is mathematically equivalent to the usual
covariance definition of the $\chi^2$.

In the present work, we have neglected the dependence of $\pi_n^i$ on
the neutrino parameters $\vec\omega$. With this approximation, we can
write:
\begin{equation} \label{eq:chisq-work}
    \chi^2(\vec\omega) = \min_{\vec\xi} 
    \left[ \sum_n \left( 
	\frac{R_n^\text{th}(\vec\omega)
	\left[ 1 + \sum_i \pi_n^i\, \xi_i \right]
	- R_n^\text{ex}}
	{\sigma_n^\text{stat}} \right)^2 
	+ \sum_{i,\text{theory}} \xi_i^2
	+ \sum_{i,\text{syst}} \xi_i^2
    \right]
\end{equation}
where we have introduced the function $\chi^2(\vec\omega) =
\min_{\{\xi_i\}} \chi^2(\vec\omega,\vec\xi)$. It is clear from
Eq.~\eqref{eq:chisq-work} that in the present approach the systematic
and theoretical uncertainties are completely characterized by the set
of quantities $\{\pi_n^i\}$, which describe the strength of the
``coupling'' between the pull $\xi_i$ and the observable
$R^\text{th}_n$. In the rest of this section we will discuss in detail
how we have parametrized and taken into account the various sources of
uncertainty.

\subsubsection{Theoretical uncertainties}

\begin{table}
    \catcode`?=\active \def?{\hphantom{0}}
    \newcommand{\pul}[2]{\pi^\text{#1}_\text{#2}}
    \newcolumntype{C}{>{~$}c<{$~}}
    \setlength{\extrarowheight}{-1pt}
    \begin{tabular}{|>{~}lr<{~}|CCCC|CCC|}
	\hline
	\multicolumn{2}{|c|}{Sample}
	& \pul{flux}{tilt} & \pul{flux}{zenith} & \pul{flux}{anti} & \pul{flux}{ratio}
	& \pul{cross}{QE} & \pul{cross}{$1\pi$} & \pul{cross}{DIS}
	\\
	\hline
	& ?1 & -1.68 & -0.16 & +0.63 & -1.00 & +0.81 & +0.10 & +0.09 \\ 
	& ?2 & -1.68 & -0.14 & +0.63 & -1.00 & +0.81 & +0.10 & +0.09 \\ 
	& ?3 & -1.68 & -0.11 & +0.62 & -1.00 & +0.81 & +0.11 & +0.09 \\ 
	& ?4 & -1.68 & -0.09 & +0.62 & -1.00 & +0.81 & +0.11 & +0.09 \\ 
	\Turn{$\text{sub-GeV}_\text{low}$ $e$}
	& ?5 & -1.68 & -0.06 & +0.63 & -1.00 & +0.80 & +0.11 & +0.09 \\ 
	& ?6 & -1.68 & -0.04 & +0.63 & -1.00 & +0.81 & +0.11 & +0.09 \\ 
	& ?7 & -1.68 & -0.01 & +0.63 & -1.00 & +0.81 & +0.10 & +0.09 \\ 
	& ?8 & -1.68 & +0.02 & +0.64 & -1.00 & +0.81 & +0.10 & +0.09 \\ 
	& ?9 & -1.69 & +0.04 & +0.64 & -1.00 & +0.82 & +0.09 & +0.09 \\ 
	& 10 & -1.69 & +0.07 & +0.65 & -1.00 & +0.83 & +0.09 & +0.09 \\ 
	\hline
	& ?1 & -0.75 & -0.64 & +0.52 & -1.00 & +0.61 & +0.28 & +0.11 \\ 
	& ?2 & -0.74 & -0.49 & +0.51 & -1.00 & +0.60 & +0.28 & +0.12 \\ 
	& ?3 & -0.74 & -0.35 & +0.51 & -1.00 & +0.59 & +0.29 & +0.12 \\ 
	& ?4 & -0.73 & -0.21 & +0.50 & -1.00 & +0.58 & +0.29 & +0.13 \\ 
	\Turn{$\text{sub-GeV}_\text{high}$ $e$}
	& ?5 & -0.73 & -0.08 & +0.50 & -1.00 & +0.57 & +0.30 & +0.13 \\ 
	& ?6 & -0.72 & +0.05 & +0.50 & -1.00 & +0.57 & +0.30 & +0.13 \\ 
	& ?7 & -0.72 & +0.19 & +0.50 & -1.00 & +0.58 & +0.29 & +0.13 \\ 
	& ?8 & -0.73 & +0.32 & +0.50 & -1.00 & +0.59 & +0.29 & +0.13 \\ 
	& ?9 & -0.73 & +0.47 & +0.51 & -1.00 & +0.60 & +0.28 & +0.12 \\ 
	& 10 & -0.74 & +0.63 & +0.52 & -1.00 & +0.61 & +0.28 & +0.12 \\ 
	\hline
	& ?1 & +0.29 & -0.84 & +0.27 & -1.00 & +0.49 & +0.23 & +0.28 \\ 
	& ?2 & +0.32 & -0.64 & +0.27 & -1.00 & +0.48 & +0.22 & +0.29 \\ 
	& ?3 & +0.35 & -0.44 & +0.26 & -1.00 & +0.47 & +0.22 & +0.31 \\ 
	& ?4 & +0.41 & -0.25 & +0.26 & -1.00 & +0.45 & +0.21 & +0.33 \\ 
	\Turn{multi-GeV $e$}
	& ?5 & +0.48 & -0.08 & +0.25 & -1.00 & +0.44 & +0.21 & +0.35 \\ 
	& ?6 & +0.48 & +0.08 & +0.25 & -1.00 & +0.43 & +0.21 & +0.36 \\ 
	& ?7 & +0.41 & +0.25 & +0.26 & -1.00 & +0.45 & +0.21 & +0.33 \\ 
	& ?8 & +0.36 & +0.44 & +0.26 & -1.00 & +0.47 & +0.22 & +0.31 \\ 
	& ?9 & +0.33 & +0.63 & +0.27 & -1.00 & +0.48 & +0.22 & +0.30 \\ 
	& 10 & +0.30 & +0.84 & +0.27 & -1.00 & +0.49 & +0.23 & +0.29 \\ 
	\hline
    \end{tabular}
    \caption{\label{tab:theo-fce}%
      Coupling factors of the flux and cross-section pulls with
      fully-contained $e$-like events.}
\end{table}

\begin{table}
    \catcode`?=\active \def?{\hphantom{0}}
    \newcommand{\pul}[2]{\pi^\text{#1}_\text{#2}}
    \newcolumntype{C}{>{~$}c<{$~}}
    \setlength{\extrarowheight}{-1pt}
    \begin{tabular}{|>{~}lr<{~}|CCCC|CCC|}
	\hline
	\multicolumn{2}{|c|}{Sample}
	& \pul{flux}{tilt} & \pul{flux}{zenith} & \pul{flux}{anti} & \pul{flux}{ratio}
	& \pul{cross}{QE} & \pul{cross}{$1\pi$} & \pul{cross}{DIS}
	\\
	\hline
	& ?1 & -1.38 & -0.19 & +0.64 & +1.00 & +0.78 & +0.14 & +0.08 \\ 
	& ?2 & -1.38 & -0.12 & +0.64 & +1.00 & +0.77 & +0.14 & +0.08 \\ 
	& ?3 & -1.38 & -0.05 & +0.63 & +1.00 & +0.77 & +0.14 & +0.09 \\ 
	& ?4 & -1.38 & +0.01 & +0.63 & +1.00 & +0.77 & +0.15 & +0.09 \\ 
	\Turn{$\text{sub-GeV}_\text{low}$ $\mu$}
	& ?5 & -1.38 & +0.08 & +0.62 & +1.00 & +0.76 & +0.15 & +0.09 \\ 
	& ?6 & -1.38 & +0.14 & +0.62 & +1.00 & +0.75 & +0.16 & +0.09 \\ 
	& ?7 & -1.37 & +0.20 & +0.61 & +1.00 & +0.74 & +0.16 & +0.10 \\ 
	& ?8 & -1.37 & +0.26 & +0.60 & +1.00 & +0.73 & +0.17 & +0.10 \\ 
	& ?9 & -1.37 & +0.31 & +0.60 & +1.00 & +0.72 & +0.18 & +0.10 \\ 
	& 10 & -1.37 & +0.37 & +0.59 & +1.00 & +0.71 & +0.19 & +0.10 \\ 
	\hline
	& ?1 & -0.71 & -0.66 & +0.46 & +1.00 & +0.60 & +0.27 & +0.12 \\ 
	& ?2 & -0.70 & -0.49 & +0.46 & +1.00 & +0.60 & +0.28 & +0.13 \\ 
	& ?3 & -0.70 & -0.32 & +0.46 & +1.00 & +0.60 & +0.27 & +0.13 \\ 
	& ?4 & -0.69 & -0.14 & +0.46 & +1.00 & +0.60 & +0.27 & +0.14 \\ 
	\Turn{$\text{sub-GeV}_\text{high}$ $\mu$}
	& ?5 & -0.68 & +0.02 & +0.46 & +1.00 & +0.59 & +0.26 & +0.14 \\ 
	& ?6 & -0.68 & +0.18 & +0.44 & +1.00 & +0.58 & +0.27 & +0.14 \\ 
	& ?7 & -0.68 & +0.32 & +0.44 & +1.00 & +0.57 & +0.29 & +0.14 \\ 
	& ?8 & -0.68 & +0.45 & +0.43 & +1.00 & +0.56 & +0.29 & +0.14 \\ 
	& ?9 & -0.69 & +0.57 & +0.44 & +1.00 & +0.57 & +0.29 & +0.14 \\ 
	& 10 & -0.69 & +0.70 & +0.45 & +1.00 & +0.58 & +0.28 & +0.14 \\ 
	\hline
	& ?1 & +0.23 & -0.85 & +0.24 & +1.00 & +0.51 & +0.22 & +0.27 \\ 
	& ?2 & +0.22 & -0.65 & +0.23 & +1.00 & +0.50 & +0.22 & +0.27 \\ 
	& ?3 & +0.22 & -0.46 & +0.22 & +1.00 & +0.50 & +0.22 & +0.27 \\ 
	& ?4 & +0.22 & -0.25 & +0.22 & +1.00 & +0.49 & +0.23 & +0.28 \\ 
	\Turn{multi-GeV $\mu$}
	& ?5 & +0.25 & -0.03 & +0.22 & +1.00 & +0.49 & +0.21 & +0.30 \\ 
	& ?6 & +0.26 & +0.15 & +0.19 & +1.00 & +0.48 & +0.22 & +0.30 \\ 
	& ?7 & +0.24 & +0.31 & +0.20 & +1.00 & +0.48 & +0.23 & +0.29 \\ 
	& ?8 & +0.23 & +0.47 & +0.22 & +1.00 & +0.49 & +0.22 & +0.28 \\ 
	& ?9 & +0.23 & +0.65 & +0.23 & +1.00 & +0.50 & +0.22 & +0.28 \\ 
	& 10 & +0.24 & +0.85 & +0.24 & +1.00 & +0.50 & +0.22 & +0.28 \\ 
	\hline
    \end{tabular}
    \caption{\label{tab:theo-fcm}%
      Coupling factors of the flux and cross-section pulls with
      fully-contained $\mu$-like events.}
\end{table}

\begin{table}
    \catcode`?=\active \def?{\hphantom{0}}
    \newcommand{\pul}[2]{\pi^\text{#1}_\text{#2}}
    \newcolumntype{C}{>{~$}c<{$~}}
    \setlength{\extrarowheight}{-1pt}
    \begin{tabular}{|>{~}lr<{~}|CCCC|CCC|}
	\hline
	\multicolumn{2}{|c|}{Sample}
	& \pul{flux}{tilt} & \pul{flux}{zenith} & \pul{flux}{anti} & \pul{flux}{ratio}
	& \pul{cross}{QE} & \pul{cross}{$1\pi$} & \pul{cross}{DIS}
	\\
	\hline
	& ?1 & +1.42 & -0.86 & +0.29 & +1.00 & +0.22 & +0.16 & +0.63 \\ 
	& ?2 & +1.46 & -0.67 & +0.28 & +1.00 & +0.21 & +0.16 & +0.63 \\ 
	& ?3 & +1.56 & -0.47 & +0.28 & +1.00 & +0.21 & +0.15 & +0.64 \\ 
	& ?4 & +1.73 & -0.25 & +0.28 & +1.00 & +0.17 & +0.14 & +0.69 \\ 
	\Turn{partially-contained $\mu$}
	& ?5 & +1.78 & -0.05 & +0.25 & +1.00 & +0.16 & +0.12 & +0.72 \\ 
	& ?6 & +1.66 & +0.11 & +0.23 & +1.00 & +0.18 & +0.13 & +0.70 \\ 
	& ?7 & +1.52 & +0.29 & +0.25 & +1.00 & +0.19 & +0.14 & +0.67 \\ 
	& ?8 & +1.44 & +0.47 & +0.27 & +1.00 & +0.21 & +0.15 & +0.64 \\ 
	& ?9 & +1.41 & +0.67 & +0.28 & +1.00 & +0.21 & +0.15 & +0.64 \\ 
	& 10 & +1.40 & +0.86 & +0.30 & +1.00 & +0.21 & +0.15 & +0.63 \\ 
	\hline
	& ?1 & +1.88 & -0.93 & +0.32 & +1.00 & +0.16 & +0.11 & +0.72 \\ 
	& ?2 & +1.87 & -0.83 & +0.32 & +1.00 & +0.17 & +0.12 & +0.72 \\ 
	& ?3 & +1.88 & -0.73 & +0.31 & +1.00 & +0.17 & +0.12 & +0.72 \\ 
	& ?4 & +1.90 & -0.63 & +0.31 & +1.00 & +0.17 & +0.11 & +0.72 \\ 
	\Turn{stopping $\mu$}
	& ?5 & +1.96 & -0.53 & +0.31 & +1.00 & +0.17 & +0.11 & +0.72 \\ 
	& ?6 & +2.03 & -0.43 & +0.31 & +1.00 & +0.16 & +0.11 & +0.73 \\ 
	& ?7 & +2.12 & -0.32 & +0.32 & +1.00 & +0.14 & +0.11 & +0.75 \\ 
	& ?8 & +2.17 & -0.21 & +0.31 & +1.00 & +0.12 & +0.10 & +0.78 \\ 
	& ?9 & +2.15 & -0.11 & +0.29 & +1.00 & +0.12 & +0.09 & +0.79 \\ 
	& 10 & +2.11 & -0.02 & +0.26 & +1.00 & +0.13 & +0.09 & +0.79 \\ 
	\hline
	& ?1 & +4.53 & -0.95 & +0.32 & +1.00 & +0.03 & +0.01 & +0.96 \\ 
	& ?2 & +4.47 & -0.85 & +0.32 & +1.00 & +0.03 & +0.01 & +0.96 \\ 
	& ?3 & +4.46 & -0.75 & +0.32 & +1.00 & +0.03 & +0.01 & +0.96 \\ 
	& ?4 & +4.46 & -0.65 & +0.32 & +1.00 & +0.03 & +0.01 & +0.96 \\ 
	\Turn{through-going $\mu$}
	& ?5 & +4.47 & -0.55 & +0.31 & +1.00 & +0.03 & +0.01 & +0.97 \\ 
	& ?6 & +4.46 & -0.45 & +0.31 & +1.00 & +0.02 & +0.01 & +0.97 \\ 
	& ?7 & +4.44 & -0.34 & +0.30 & +1.00 & +0.02 & +0.01 & +0.97 \\ 
	& ?8 & +4.42 & -0.24 & +0.29 & +1.00 & +0.02 & +0.01 & +0.97 \\ 
	& ?9 & +4.41 & -0.14 & +0.28 & +1.00 & +0.03 & +0.01 & +0.96 \\ 
	& 10 & +4.44 & -0.05 & +0.26 & +1.00 & +0.03 & +0.01 & +0.96 \\ 
	\hline
    \end{tabular}
    \caption{\label{tab:theo-upm}%
      Coupling factors of the flux and cross-section pulls with
      partially-contained and upgoing $\mu$-like events.}
\end{table}

Theoretical uncertainties arise from our limited knowledge of the
atmospheric neutrino fluxes and cross-sections. The corresponding
coefficients $\pi_n^i$ have been calculated assuming two-neutrino
oscillations with $\Delta m^2 = 2.5\times 10^{-3}~\eVq$ and $\theta =
45^\circ$, and are listed in Tables~\ref{tab:theo-fce},
\ref{tab:theo-fcm} and~\ref{tab:theo-upm}.

We have parametrized flux uncertainties in terms of six pulls:
$\xi^\text{flux}_\text{norm}$, $\xi^\text{flux}_\text{tilt}$,
$\xi^\text{flux}_\text{zenith}$, $\xi^\text{flux}_\text{anti}$,
$\xi^\text{flux}_\text{ratio}$ and $\xi^\text{flux}_\text{multi}$.
\begin{itemize}
  \item $\xi^\text{flux}_\text{norm}$ (20\%) is a total normalization
    error, with the same coupling $\pi^\text{flux}_\text{norm} = 1$ to
    all the data samples;

  \item $\xi^\text{flux}_\text{tilt}$ (5\%) is a``tilt'' factor which
    parametrizes possible deformations of the flux energy spectrum:
    \begin{equation}
	\Phi_\delta(E) = \Phi_0(E) \left( \frac{E}{E_0} \right)^\delta
	\approx \Phi_0(E) \left[ 1 + \delta \ln \frac{E}{E_0} \right] \,.
    \end{equation}
    The uncertainty on the factor $\delta$ is 5\%, and in analogy with
    Ref.~\cite{Kameda} we have chosen $E_0 = 2$~GeV;
    
  \item $\xi^\text{flux}_\text{zenith}$ (5\%) describes the uncertainty
    associated to the up-down asymmetry;
    
  \item $\xi^\text{flux}_\text{anti}$ (5\%) describes the uncertainty
    associated to the $\nu / \bar{\nu}$ asymmetry;

  \item $\xi^\text{flux}_\text{ratio}$ (2.5\%) parametrizes the
    uncertainty on the $\nu_\mu / \nu_e$ ratio.
    
  \item $\xi^\text{flux}_\text{multi}$ (5\%) is an extra normalization
    factor which affects only fully-contained multi-GeV events (both
    $e$-like and $\mu$-like). Its corresponding coupling
    $\pi^\text{flux}_\text{multi}$ is $1$ for FC multi-GeV events and
    $0$ for all the other samples.
\end{itemize}

Concerning cross-section uncertainties, we properly take into account
the contributions to the total number of observed events coming from
three different types of charged-current interactions: quasi-elastic
neutrino scattering (QE), 1-pion production ($1\pi$), and
deep-inelastic scattering (DIS). We neglect for simplicity coherent
scattering on oxygen and neutral-current interactions, which
contribute only marginally to the considered data samples.
For each different type of neutrino interactions we introduce two
pulls:
\begin{itemize}
  \item $\xi^\text{QE}_\text{norm}$, $\xi^{1\pi}_\text{norm}$,
    $\xi^\text{DIS}_\text{norm}$ (15\%) describe the total
    normalization errors;
    
  \item $\xi^\text{QE}_\text{ratio}$, $\xi^{1\pi}_\text{ratio}$,
    $\xi^\text{DIS}_\text{ratio}$ ($-0.5\%$ for electrons and $+0.5\%$
    for muons) parametrize the uncertainty of the $\sigma^i_{\nu_\mu}
    / \sigma^i_{\nu_e}$ ratios.
\end{itemize}

The fraction of events for each data sample originating from the
different types of neutrino interactions is based on our own
calculations, however we have verified that it is consistent with the
numbers listed in Table 8.2 of Ref.~\cite{Kameda} (after neglecting
event misidentification and neutral-current contributions and
rescaling the factors so that the sum is 100\%).

\subsubsection{Systematic uncertainties}

\begin{table}
    \catcode`?=\active \def?{\hphantom{0}}
    \newcommand{\pul}[1]{[\pi\xi]^\text{sys}_\text{#1}}
    \newcommand{\zero}{\text{---}}
    \newcolumntype{L}{>{~}p{24mm}<{~}}
    \newcolumntype{C}{>{~\hfill$}p{18mm}<{$\hfill~}} 
    \setlength{\extrarowheight}{-1pt}
    \begin{tabular}{|L|CCCCC|}
	\hline
	\multicolumn{1}{|c|}{Sample}
	& \pul{hadron} & \pul{$\mu/e$} & \pul{ring} & \pul{f-vol} & \pul{E-cal} \\
	\hline
	sub-GeV $e$     & -0.25\% & -1.1\% & -0.75\% & -0.3\% & -0.4\% \\
	mid-GeV $e$     & -0.25\% & -1.1\% & -0.75\% & -0.3\% & -0.4\% \\
	multi-GeV $e$   & -0.50\% & -1.6\% & -2.75\% & -0.5\% & +0.3\% \\
	sub-GeV $\mu$   & +0.25\% & +1.1\% & +0.75\% & +0.3\% & +0.4\% \\
	mid-GeV $\mu$   & +0.25\% & +1.1\% & +0.75\% & +0.3\% & +0.4\% \\
	multi-GeV $\mu$ & +0.50\% & +1.6\% & +2.75\% & +0.5\% & -0.3\% \\
	part-cont $\mu$ & -0.50\% & +1.6\% & -1.25\% & +0.5\% & -3.3\% \\
	stopping $\mu$  & \zero   & \zero  & -0.30\% & +0.7\% & -0.3\% \\
	through $\mu$   & \zero   & \zero  & -0.30\% & +0.7\% & -0.3\% \\
	\hline
    \end{tabular}
    \\[1mm]
    \begin{tabular}{|L|CCCC|}
	\hline
	\multicolumn{1}{|c|}{Sample}
	& \pul{PC-nrm} & \pul{track} & \pul{up-eff} & \pul{t-sep} \\
	\hline
	part-cont $\mu$ & 5.27\% & \zero & \zero & \zero \\
	stopping $\mu$  & \zero  & 6.4\% & 1.0\% & 1.0\% \\
	through $\mu$   & \zero  & 1.4\% & 1.0\% & \zero \\
	\hline
    \end{tabular}
    \caption{ \label{tab:syst}%
      Coupling factors of the systematics pulls $\pul{hadron}$,
      $\pul{$\mu/e$}$, $\pul{ring}$, $\pul{f-vol}$, $\pul{E-cal}$,
      $\pul{PC-nrm}$, $\pul{FC/PC}$, $\pul{track}$, $\pul{up-eff}$ and
      $\pul{t-sep}$ with the various observables. The coefficients are
      the same for all the bins in a given data sample.}
\end{table}

The systematics uncertainties of the Super-Kamiokande experiment are
derived from Tables 9.2, 9.3, 9.4 and 9.5 of Ref.~\cite{Kameda}. In
Table~\ref{tab:syst} we list the product of the given uncertainty
$\xi_i$ with the corresponding coupling $\pi_n^i$. We include in our
calculations the following sources of systematics:
\begin{itemize}
  \item $\xi^\text{sys}_\text{hadron}$ is the uncertainty associated
    with the simulation of hadronic interactions;
    
  \item $\xi^\text{sys}_{\mu/e}$ describes the errors in the particle
    identification procedure;
    
  \item $\xi^\text{sys}_\text{ring}$ is the uncertainty coming from the
    ring-counting procedure;

  \item $\xi^\text{sys}_\text{f-vol}$ is the uncertainty in the
    fiducial volume determination, introduced by the vertex fitter
    procedure;
    
  \item $\xi^\text{sys}_\text{E-cal}$ is the uncertainty in the energy
    calibration;
    
  \item $\xi^\text{sys}_\text{PC-nrm}$ accounts for all the errors
    which are unique to partially-contained events;
    
  \item $\xi^\text{sys}_\text{track}$ is the uncertainty in the track
    reconstruction of upgoing muons;
    
  \item $\xi^\text{sys}_\text{up-eff}$ is the detection efficiency of
    upgoing muons;

  \item $\xi^\text{sys}_\text{t-sep}$ is the uncertainty in the
    separation of stopping and through-going events.
\end{itemize}